\documentclass[12pt,a4paper]{article}
\usepackage[thinlines]{easytable} 
\usepackage{multirow}
\usepackage{booktabs}
\usepackage{array}
\usepackage{multicol}
\newcounter{fig}

\newcommand{\bea}{\begin{eqnarray}}
\newcommand{\eea}{\end{eqnarray}}
\newcommand{\be}{\begin{equation}}
\newcommand{\ee}{\end{equation}}

\newcommand{\re}[1]{(\ref{#1})}

\newcommand{\CK}{\ensuremath{\mathcal{K}}}

\newcommand{\CL}{\ensuremath{\mathcal{L}}}

\newcommand{\CE}{\ensuremath{\mathcal{E}}}
\newcommand{\CA}{\ensuremath{\mathcal{A}}}

\newcommand{\pa}{\partial}

\usepackage{ulem}
\usepackage[utf8]{inputenc}
\RequirePackage[T2A]{fontenc}
\RequirePackage[english]{babel}
\RequirePackage{indentfirst}
\RequirePackage{amsmath,amssymb,amsfonts}
\RequirePackage{titlesec}
\RequirePackage[titles]{tocloft}
\RequirePackage{graphicx}
\RequirePackage{float}
\RequirePackage{caption}
\RequirePackage{wrapfig}
\RequirePackage[colorlinks=true,allcolors=blue,
unicode=true,hypertexnames=false]{hyperref}
\hypersetup{pdfstartview={XYZ null null 1.25}}

\usepackage[left=2cm,right=2cm,top=2cm,bottom=2cm]{geometry}

\title{Fractional Hopfions in the Faddeev-Skyrme model with a symmetry breaking potential}
\author{
    {\large A.~Samoilenka}$^{\dagger}$
    and {\large Ya. Shnir}$^{\dagger \star}$
    \\ \\
    \\ $^{\dagger}${\small Department of Theoretical Physics and Astrophysics}\\
    {\small Belarusian State University, Minsk 220004, Belarus}
    \\ $^{\star}${\small BLTP, JINR, Dubna, Russia}
}

\begin{document}
\maketitle

\begin{abstract}
We construct new solutions of the Faddeev-Skyrme model with a symmetry breaking potential admitting
$S^1$ vacuum. It includes, as a limiting case, the usual $SO(3)$ symmetry breaking mass term, another
limit corresponds to the potential $m^2 \phi_1^2$, which gives a mass to the corresponding
component of the scalar field.
However we find that the spacial distribution of the energy density of these solutions
has more complicated structure, than in the case of the usual Hopfions, typically it represents
two separate linked tubes with different thicknesses and positions.
In order to classify these configurations
we define a counterpart of the usual position curve, which
represents a collection of loops $\mathcal{C}_1, \mathcal{C}_{-1}$ corresponding to the preimages
of the points $\vec \phi = (\pm 1 \mp \mu, 0,0)$, respectively. Then the Hopf invariant can be defined as
$Q= {\rm link} (\mathcal{C}_1,\mathcal{C}_{-1})$. In this model, in the sectors of degrees $Q=5,6,7$
we found solutions of new type, for which one or both of these tubes represent
trefoil knots. Further, some of these solutions possess different types of curves  $\mathcal{C}_1$ and $\mathcal{C}_{-1}$.
\end{abstract}

\section{Introduction}
One of the most interesting topological solitons, which appear in the scalar
field theories in 3+1 dimensional space-time, are so-called "Hopfions"
\cite{Faddeev-Hopf,Faddeev:1996zj}.
These stable field configurations, whose topology is defined by the first Hopf map  ${S}^3 \mapsto {S}^2$,
provide a first, and the best known example of knot solitons in field theory.

It was suggested later that stable knotted vortex configurations of that type
may exist in a system of multi-component superconductors \cite{Babaev2009,Garaud:2011}, in two-condensate  Ginzburg-Landau model
with oppositely charged components \cite{Babaev:2001zy}, or in the single-component
Bose-Einstein condensate with trapping potential \cite{Kartashov:2014kva}. Recently, there was a significant interest in construction
of the Hopfion-like solutions in liquid crystals \cite{Ackerman:2017lse}.
It was pointed out that the knotted Hopfion configuration may exist in frustrated magnets
\cite{Sutcliffe:2017aro}.
In low energy QCD the solutions which represent knots of the
gluon fields, may describe glueballs \cite{Faddeev:1998eq,Faddeev:2003aw}.  It was pointed out that the
Faddeev-Skyrme model emerges as a low-energy limit
of scalar QED with a certain type potential  \cite{Gorsky:2013doa}.
Although knot solutions not always enjoy the topological stability
\cite{Speight:2008na}, extended solitons which are formed from vortex lines taking the form of knots or links,
are of considerable interest.

The structure of the Faddeev-Skyrme model looks similar to the usual Skyrme model \cite{skyrme1,skyrme2},
in 3+1 dimensions the corresponding Lagrangian includes the usual $\sigma$-model term, the Skyrme term,
which is quartic in derivatives of the field, and an optional potential term, which does not contain the derivatives of the field.
A very interesting feature of the Skyrme model  is that
the asymptotic form of the field of the Skyrmion solutions and therefore, the character of interaction
between them,  strongly depends to the form of the
potential. Thus, a particular choice of the potential, which in the original Skyrme model
was introduced to provide a mass to the pion field, may strongly affect the structure of multisoliton configurations
\cite{Battye:2004rw,Battye:2009ad}. The explicit form of the potential becomes even more important in the
self-dual modifications of the Skyrme model, proposed to construct the soliton solutions which satisfy
the first-order Bogomol'nyi-type equation \cite{Adam:2010fg,Adam:2010ds,Sutcliffe:2010et}.

In the last few years a number of
modified versions of the Skyrme model were proposed, in particular the model with an additional potential term
that is quartic in the pion field was considered in \cite{Gillard:2015eia}.
These modifications are mainly motivated by the fact that there is a
large disagreement between the experimental data
for binding energies of hadrons and predictions of the Skyrme model, considered as an effective low energy theory
of hadrons, the binding energy of nuclei do not exceed 1$\%$
of their masses while the binding energies of Skyrmions are on the level of 10$\%$.
It was noticed that a combination of the
a sextic in derivatives of the pion field term and/or
the higher-order potential may significantly reduce the binding energy of the
solitons \cite{Gudnason:2016tiz,Gudnason:2016cdo}.
On the other hand, various symmetry breaking potentials were considered in the Skyrme model to construct
half-Skyrmions \cite{Gudnason:2015nxa} or various vortex strings \cite{Gudnason:2016yix}. Clearly, all these modifications
preserve the topological properties of the Skyrme model.

Despite the qualitative similarity between the Skyrme model and the Faddeev-Skyrme model, the effect of the potential term
on the structure of the multisoliton configurations not yet been studied in detail.
In the
paper by Foster \cite{Foster:2010zb} the Faddeev-Skyrme model with an $SO(3)$ symmetry breaking potential,
which is similar to the usual pion
mass potential of the Skyrme model, was considered in the limit of the infinite mass. The effect of this term
for the axially symmetric configurations of lowest degrees was investigated in earlier work \cite{Kundu:1982bc}

Note that potential of that type,  was also used to stabilize
the classically isospinning Hopfions \cite{Harland:2013uk,Battye:2013xf}. The double vacuum potential in the Faddeev-Skyrme model
with a potential term with two discrete vacua was considered in \cite{Kobayashi:2013bqa}.

In this paper we investigate the effect of a symmetry breaking potential admitting
$S^1$ vacuum on the soliton solutions of the Faddeev-Skyrme model. We perform full 3d numerical computations to find the corresponding
field configurations in the sectors of degrees up to $Q=7$.

\section{The model}
The Lagrangian density of the Faddeev-Skyrme model in (3+1)-dimensional flat space can be written in terms of the real
triplet of scalar fields $\vec\phi=(\phi_1,\phi_2,\phi_3)$ as \cite{Faddeev-Hopf,Faddeev:1996zj}
\be
\label{lagr}
\CL=\int d^3 x\left[c_2\pa_\mu \vec\phi\cdot\pa^\mu \vec\phi -
\frac{c_4}{2}\left(\pa_\mu \vec\phi\times\pa_\nu \vec\phi\right)^2 - V(\phi)\right]
\ee
where $c_2,c_4$ are some real positive parameters and $g^{\mu\nu}=diag(1,-1,-1,-1)$.

As in the usual non-linear $O(3)$ sigma model,
the scalar field is constrained to the unit sphere $S^2$: ~$|\vec{\phi}|=1$.
The topological restriction on the field $\vec \phi$ is that it approaches its
vacuum value at spacial boundary, i.e. $\vec \phi_\infty = (0,0,1)$.
This allows a one-point compactification of
the domain space $\mathbb{R}^3$ to $S^3$ and
the field of the finite energy solutions of the
model, the Hopfions, is a map $\phi^a:\mathbb{R}^3 \mapsto S^2$ which belongs
to an equivalence class characterized by the homotopy group $\pi_3(S^2)=\mathbb{Z}$.

Mathematically, the definition of the Hopf invariant
can be given by defining an area form $\omega$, which is a generator of the second cohomology group
$H^2(S^3)$ on the target space $S^2$.
Then the Hopf map $\phi: S^3 \mapsto S^2$ has an induced pullback of the
cohomology group $\phi^*:H^2(S^2) \mapsto H^2(S^3)$.
Since the second cohomology group of $S^3$ is trivial $H^2(S^3)=0$,
the pullback $F=\phi^* \omega$ of the area two-form $\omega$ by $\phi$ is exact,
$F=dA$ where $A$ is a one-form. Explicitly, the corresponding  Hopf invariant is defined
as the integral of a Chern-Simons three-form over the space $S^3$
\be
Q=\frac{1}{8\pi^2}\int\limits_{S^3}\!\! F \wedge A \in \mathbb{Z}
\label{Hopf-charge}
\ee
where $F=\frac12 F_{\mu\nu} dx^\mu \wedge dx^\nu$ is a 2-form with components
\be
F_{\mu\nu} = \vec\phi \cdot \partial_\mu \vec \phi \times  \partial_\nu \vec\phi
\ee
Note that the Hopf invariant cannot be written in terms of $\phi$ as a local density.
More simple geometrical definition of this invariant is as
the linking number of two loops on the domain space $S^3$, which are the preimages of two distinct points
on the target space $S^2$.

The static energy functional of the  model \re{lagr} is
\be\label{hopfenergy}
E=\int d^3 x \mathcal{E} = \int d^3 x\left[c_2\pa_i\vec\phi\cdot\pa_i\vec\phi +
\frac{c_4}{2}\left(\pa_i\vec\phi\times\pa_j\vec\phi\right)^2 + V\right]
\ee
The soliton solutions of the Faddeev-Skyrme model,  correspond to the stationary points of this  functional.

Similarity of the Lagrangian \re{lagr} to the planar Skyrme model \cite{BB,Bsk}
suggests that the model can support
extended stringlike solutions, which can be constructed from baby Skyrmions located in the plane transverse to
the direction of the string \cite{Kobayashi:2013xoa}. Intuitively, the topological charge of such a configuration can be given
by the product
of the winding number of the planar Skyrmions and the number of the twists of the string in the extra spatial direction.
Thus, the solutions we expect to find may in some way resemble the elastic rods that can bend, twist and  stretch. Physically, field
configurations of that type can be considered as a vortex, which is bended  and twisted a few times \cite{Harland:2010wc}. Then the
identification of the end points of the vortex yields the loop, which can transform itself into a knot to minimise its energy.

Another peculiar feature of the Faddeev-Skyrme model is that
for a given degree $Q$, there are usually several different stable static soliton solutions of rather
similar energy \cite{Sutcliffe:2007ui}. The number of solutions seems to grow with $Q$, thus the identification of a global minimum of the
energy functional in a given sector becomes rather involved.

In the absence of the potential term the corresponding solutions are well known, they are constructed in the sectors of degrees up
to $Q=29$ or even higher \cite{Sutcliffe:2007ui,Jennings:2014mca}. In the model with the simple
potential term, which breaks global $SO(3)$ symmetry \cite{Kundu:1982bc,Foster:2010zb,Harland:2013uk,Battye:2013xf}, the structure of the
solutions of the massive Faddeev-Skyrme model is not very much different from the massless case. Note also the modification
of the model via inclusion other 4th order on derivative terms does not change this pattern \cite{Gladikowski:1996mb}.

It is known that the energy functional of the  Faddeev-Skyrme model is bounded from
below by the Vakulenko-Kapitanskii inequality \cite{Vakulenko:1979uw}
\be
E\ge c |Q|^{3/4}
\label{VK-bound}
\ee
where $c$ is a positive constant. However, this bound is modified if  a potential term is included in the model
\cite{Harland:2013rxa}.

Note that the relation between the topological charge
of the Hopfions and the energy is not linear as
the Bogomolny-type bound  in the Skyrme model. The physical reason of that is that the topological charge
of Hopfions is produced via knotting of the vortex lines and/or their linking. This difference is important because the linear
dependence between the topological charge and the energy means that the multisoliton
configuration of degree $Q$ can be decomposed into set of
$Q$ interacting solitons of unit charge and the interaction energy is vanishing as the individual solitons become infinitely separated.
This is not the case for the soliton solutions of the usual Faddeev-Skyrme model.

Here we will consider the modification of the usual Faddeev-Skyrme model by an additional
potential term
\be\label{potential}
V=m^2\left(\phi_1\sin\alpha-(1-\phi_3)\cos\alpha\right)^2
\ee
where $m$ is a mass parameter and $\alpha\in [0,\pi]$ is an angular parameter. Depending on the value of this parameter the potential
interpolates between the usual term,
which explicitly breaks global $SO(3)$ symmetry \cite{Kundu:1982bc,Foster:2010zb,Harland:2013uk,Battye:2013xf}, as $\alpha=0$, and
the Heisenberg type potential
\be
V=m^2{\phi_1}^2
\ee
as $\alpha=\pi/2$.

Potential of this form in the usual Skyrme model allows us to construct half-Skyrmion solutions \cite{Gudnason:2015nxa}, as we will
see in the Faddeev-Skyrme model it will also significantly affect the structure of the Hopfions.
Note that the vacuum of the model with the potential \re{potential} is a circle $S^1$ on the target space
$S^2$ and the vacuum boundary
condition remains the same, since  $V\to 0$ as $\vec \phi\to \vec\phi_\infty = (0,0,1)$.
Another remark is that the potential \re{potential} can be equivalently written as
$V=m^2(\phi_1-c)^2$ via an appropriate rotation of the fields in the internal space and by setting $c=\cos\alpha$. Also
the corresponding vacuum boundary condition should be imposed.

Unlike Skyrmions, the location of the Hopfions cannot be identified with a
maximum of the topological charge density
distribution. Instead, it is convenient to look on the position of the maxima of the energy density.
Since it costs a lot of energy to deviate from the vacuum $\vec \phi_\infty$, the corresponding curve usually
follows the positions of the preimage of antipodal point  $\vec\phi_0 = (0,0,-1)$.
This curve is usually referred to as the position curve \cite{Sutcliffe:2007ui}, considering the Hopfions most authors make use of it
to visualize the shape of the configuration.

It is instructive to visualize the linking invariant via plotting not just the position curve but also another
curve in its neighborhood, which allows us to see the Hopf charge as the linking number of these two
curves. Actually, any other preimage curve of $\vec\phi$ will be linked with the preimage of $\vec\phi_0$ $Q$ times.
In the absence of the symmetry breaking potential, or in the model with usual pion-mass type potential \cite{Foster:2010zb},
it is convenient to define this linking curve
as a tube given by an isosurface of the preimage of the vector $\vec \phi= (-1+\mu, 0, 0)$, where typically $\mu=0.1$.
The linking number then is just the number of times the linking curve wraps around the position curve.

The situation is different however in the model \re{lagr} with potential \re{potential}. Then, for the relatively large values of the
mass parameter $m$ the location of the soliton can be identified
as collection of curves of maximal energy, it interpolates between the preimages of antipodal point  $\vec\phi
= (0,0,-1)$ as above for
$\alpha=0$ to the loops, which follow the preimages of two distinct points $\vec\phi = (\pm 1,0,0)$,
as $\alpha=\pi/2$: $\mathcal{C}_1 = \vec \phi^{-1}(1,0,0)$ and $\mathcal{C}_{-1} = \vec \phi^{-1}(-1,0,0)$.
Since these loops are linked $Q$ times, the definition of the linking number now can be related with
the positions of the preimages of these points: $Q= {\rm link} (\mathcal{C}_1,\mathcal{C}_{-1})$.

Note that the first two terms of the energy functional \re{hopfenergy} are invariant with respect to
global $SO(3)$ transformations. Thus, the internal rotations of the triplet $\vec \phi$
allow us to consider the limiting symmetry breaking
potentials $V\sim{\phi_1}^2$, $\sim{\phi_2}^2$ and $\sim{\phi_3}^2$ all on equal footing.

\section{Initial approximation}
The peculiarity of the Faddeev-Skyrme model is that the Hopf index is not a winding number,
the boundary conditions are identical for
configurations in all topological sectors. As usual, the energy minimization scheme needs an appropriate initial configuration
in a given sector. The most effective approach here is related with the
generalization of the rational map approximation, suggested by Sutcliffe \cite{Sutcliffe:2007ui}.
This construction can be nicely described in terms of a degree one map $\mathbb{R}^3 \mapsto S^3 \in \mathbb{C}^2$.
One can consider the complex variables which parameterize the sphere $S^3$ \cite{Sutcliffe:2007ui}
\be
\left(Z_1,Z_0 \right) = \left(\sin f(r) \sin \theta e^{i\varphi};~~ \cos f(r) + i \sin f(r) \cos \theta \right)
\label{rational-Hopf}
\ee
where $f(r)$ is a monotonically decreasing function with the boundary values $f(0)=\pi$ and $f(\infty)=0$.
Clearly,  the coordinates $Z_1,Z_0$
are restricted to the unit sphere $S^3$, $|Z_1|^2 + |Z_2|^2=1$.
Then a torus knot on the domain space can be described as the intersection of a curve $Q(Z_1,Z_0)$ with a unit sphere
$S^3$.

Then the components
of the field $\vec\phi$, which are coordinates on the Riemann sphere $S^2$, are given by the rational map $W: S^3 \in \mathbb{C}^2 \mapsto CP^1$:
\be
W(Z_1,Z_0) = \frac{\phi_1+i\phi_2}{1+\phi_3} = \frac{P(Z_1,Z_0)}{Q(Z_1,Z_0)} \, .
\label{rational-Hopf-2}
\ee
Here the polynomials $P(Z_1,Z_0)$ and $Q(Z_1,Z_0)$  have no common factors and have
no common roots on the two-sphere $S^2$. Thus,
the rational map ansatz \re{rational-Hopf-2} produces a curve in $\mathbb{R}^3$ and
the map $\phi: \mathbb{R}^3 \mapsto S^2$ is equivalent to the map
from a three-sphere to
the complex projective line, $W:S^3 \mapsto CP^1$.

According to the classification of \cite{Sutcliffe:2007ui},
there are three different types of input configurations. The axially symmetric Hopfions are produced by the rational map
\be
W(Z_1,Z_0) =\frac{Z_1^n}{Z_0^m} \, ,
\label{map-unknot}
\ee
in this case the position curve, which is defined as solution of the equation $Z_0=0$,  is just a circle
in the $x$-$y$ plane centered at the origin.
This Hopfion is constructed via embedding of
a two-dimensional charge $n$ planar Skyrmion configuration as a slice of a circle in
three-dimensional space. Since the baby Skyrmions possess an internal phase, the configuration can be twisted by the angle $2\pi m$ as
it travels along the circle. Thus,
the total topological charge of the three dimensional configuration is given by the product of the winding number in the
plane and the number of twists: $Q=mn$. Following \cite{Sutcliffe:2007ui}, we can label the axially symmetric Hopfions of that
type as $Q \mathcal{A}_{mn}$, here the first subscript gives the number of twists and the second
is the winding number of the two-dimensional planar solitons.
In terms of the mathematical knot theory the axially symmetric Hopfions are trivial knots,
so called unknots. They are closed field configurations with two independent winding numbers along two fundamental circles
of the torus.

Note that numerical simulations reveal that in general configurations of higher degree do not possess the axial symmetry, since
increase of number of twists per unit resolution may
break the axial symmetry making the position curve not planar. This happened, for example, for the
charge three Hopfion, the energy minimization transforms the corresponding
axially symmetric initial configuration $\mathcal{A}_{31}$ into
the pretzel-like loop, bending toward the third direction. It was suggested to label configurations of that type as
$\widetilde{\mathcal{A}}_{31}$ to emphasize the deformation \cite{Sutcliffe:2007ui}.

As the value of the Hopf invariant increases, some new possibilities arise. First, we can construct
the Hopfions with  two or more interlinked and disconnected position curves.
These configurations are referred to as links. The Hopf charge of this configuration is not just
a simple sum of the Hopf indexes associated with each individual unknot, it also includes in addition the sum of their secondary
linking numbers due to the inter-linking with the other components.
Configurations of that type are labeled as
$\mathcal{L}_{a,b}^{n,m}$, here the subscripts label the Hopf indexes of the unknots
and the superscript above each subscript counts the secondary linking number, which appears due to
inter-linking with the other components. The total charge of the Hopfion of that type is just the sum of all four indices
\cite{Sutcliffe:2007ui}.

Rational map approximation \re{rational-Hopf-2}
can be used to produce various links in a given topological sector.
Now the denominator of the map
must be reducible to give rise to the linked position curves. For example,
the link of the type $\mathcal{L}_{n,n}^{1,1}$ is generated by the map
\be
W=\frac{Z_1^{n+1}}{Z_1^2-Z_0^2} = \frac{Z_1^n}{2(Z_1-Z_0)}+\frac{Z_1^n}{2(Z_1+Z_0)} \, .
\label{map-link}
\ee
The Hopf index of this configuration is $Q=2n+2$, it corresponds to the two $\mathcal{A}_{n1}$ unknots linked
once.

In order to construct configurations of another type, the knots,
Sutcliffe suggested to consider the rational map of the following form \cite{Sutcliffe:2007ui}
\be
W(Z_1,Z_0) =\frac{Z_1^\alpha Z_0^\beta}{Z_1^a+Z_0^b}
\label{rational-Hopf-3}
\ee
where $\alpha$ is a positive integer and $\beta$ is a non-negative integer. This Hopfion is denoted as $\mathcal{K}_{a,b}$,
in order to produce
a torus knot $a$ and $b$ must be co-prime\footnote{If they are not, the rational map \re{rational-Hopf-3}
is degenerated producing a link.}
and $a>b$. Thus, this configuration possesses $a(b-1)$ crossings, for example the trefoil knot
$\mathcal{K}_{3,2}$ is also the $(3,2)$ torus knot. The Hopf index is given by the crossing number and the number of
times the linking curve wraps around the position curve,
\be
Q= \alpha b + \beta a
\ee
The rational map \re{rational-Hopf-3} allows to produce knotted configurations for any value of $Q$, for example simplest $Q=1$
unknot corresponds to the setting $a=2,b=1,\alpha=1$ and $\beta=0$.

\section{Numerical results}
To find the stationary points of the energy functional \re{hopfenergy} with the potential
\re{potential} we make use of the numerical minimization technique described in \cite{Samoilenka:2015bsf,Samoilenka:2016wys}.
The fields are discretized on the grid with $150^3$, or $200^3$ point with  spatial grid spacings
$\Delta x=0.1$, $0.06$ or $\Delta x=0.05$. The initial configuration were produced via the rational map
approximation as described above.
As a consistency check, we verify that our algorithm correctly reproduces the known results for the Hopfion
configuration of the usual rescaled
massless Faddeev-Skyrme model at $m=0$, it agrees with previously known value within $0.5 \%$ accuracy.
For each solution  we evaluated the value of the Hopf invariant and checked that the
corresponding virial relation $  \CE_2+3 V = \CE_4 $
between the potential, quadratic, and quartic in derivatives terms in \re{hopfenergy},
holds. The estimated errors are of order of $10^{-2}$ or smaller.

\subsection{Q=1}
\begin{figure}[h]
    \begin{center}
        \includegraphics[height=3cm]{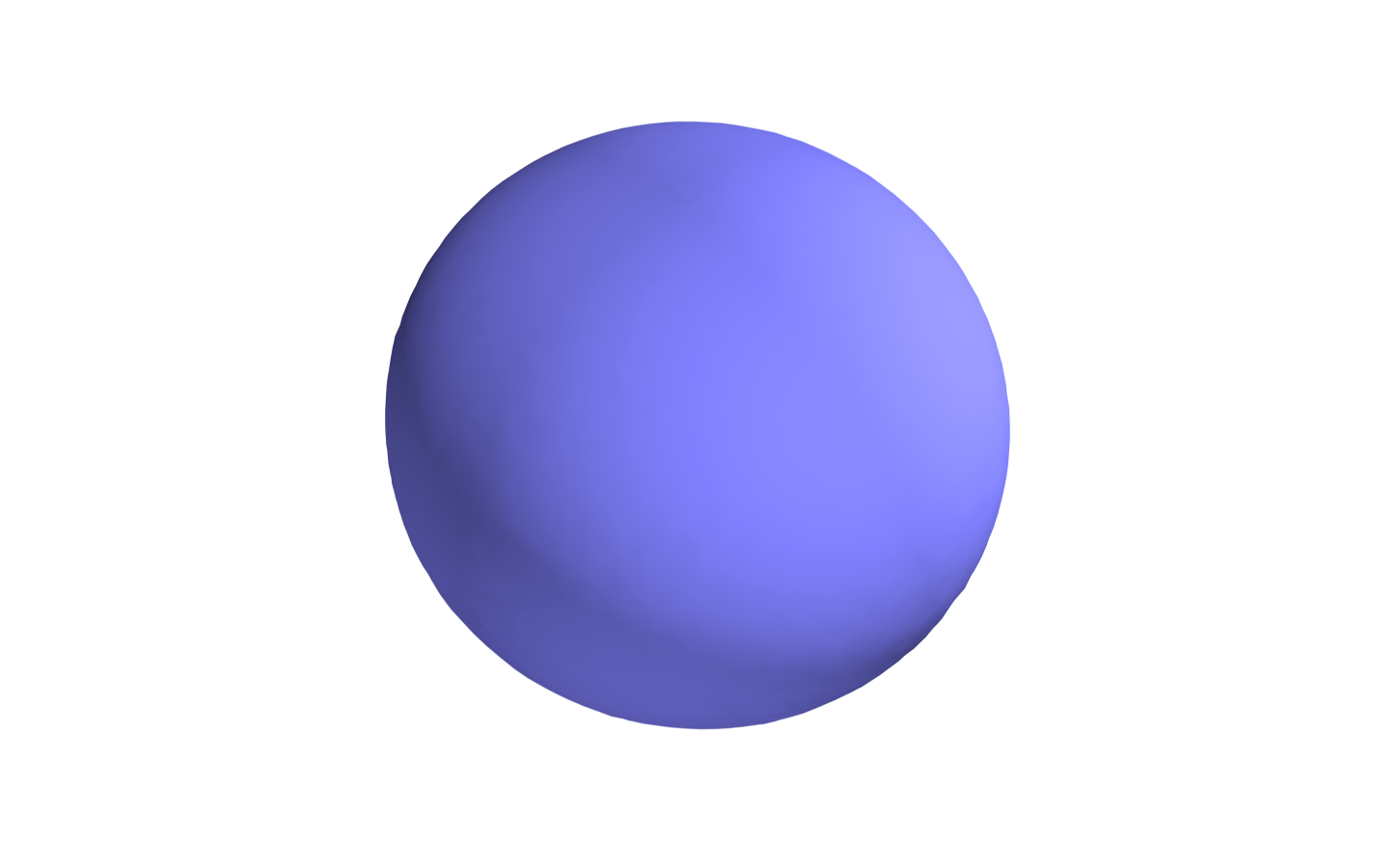}
        \includegraphics[height=3cm]{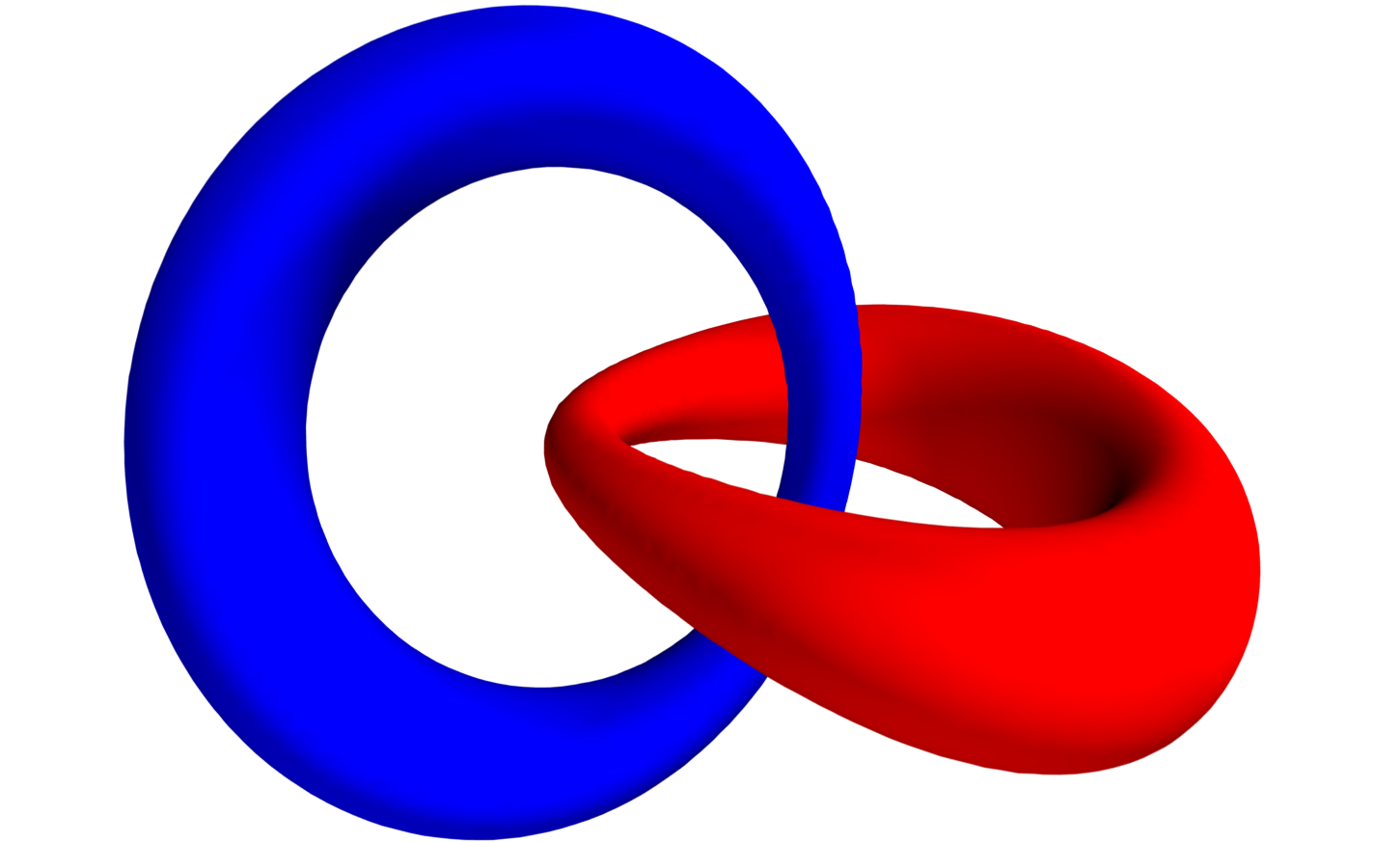}
        \includegraphics[height=3cm]{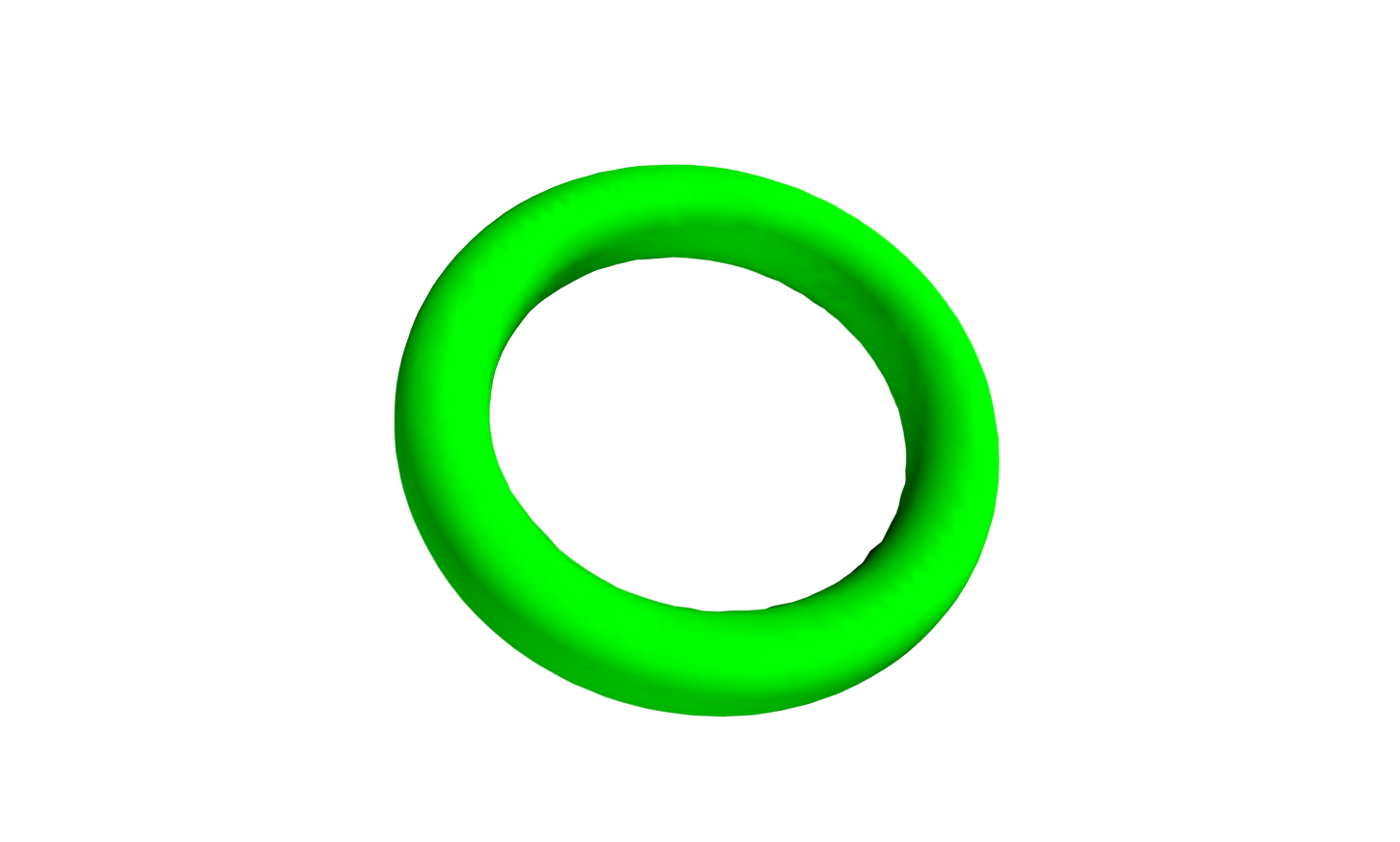}
    \end{center}
    \begin{center}
        \includegraphics[height=3cm]{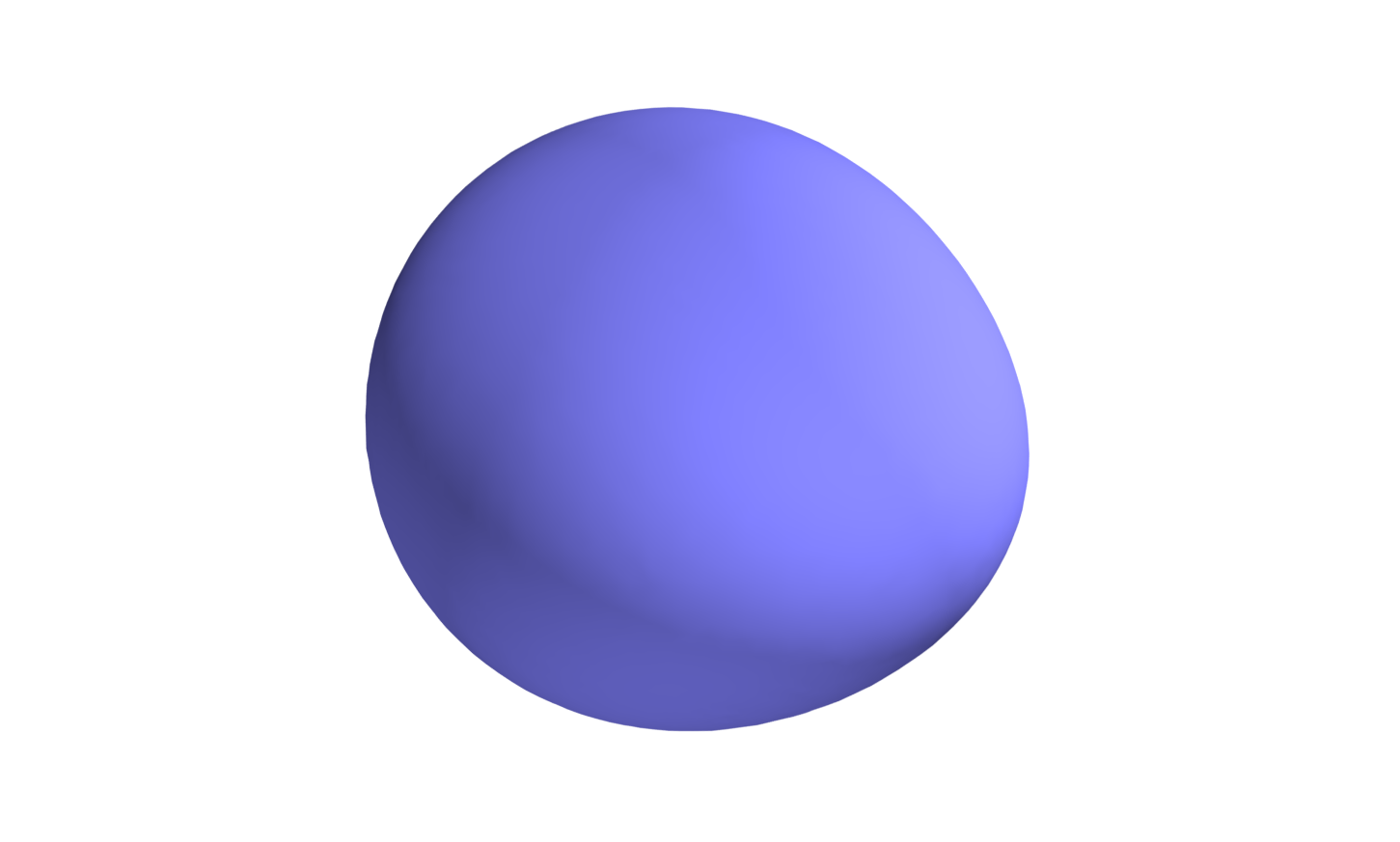}
        \includegraphics[height=3cm]{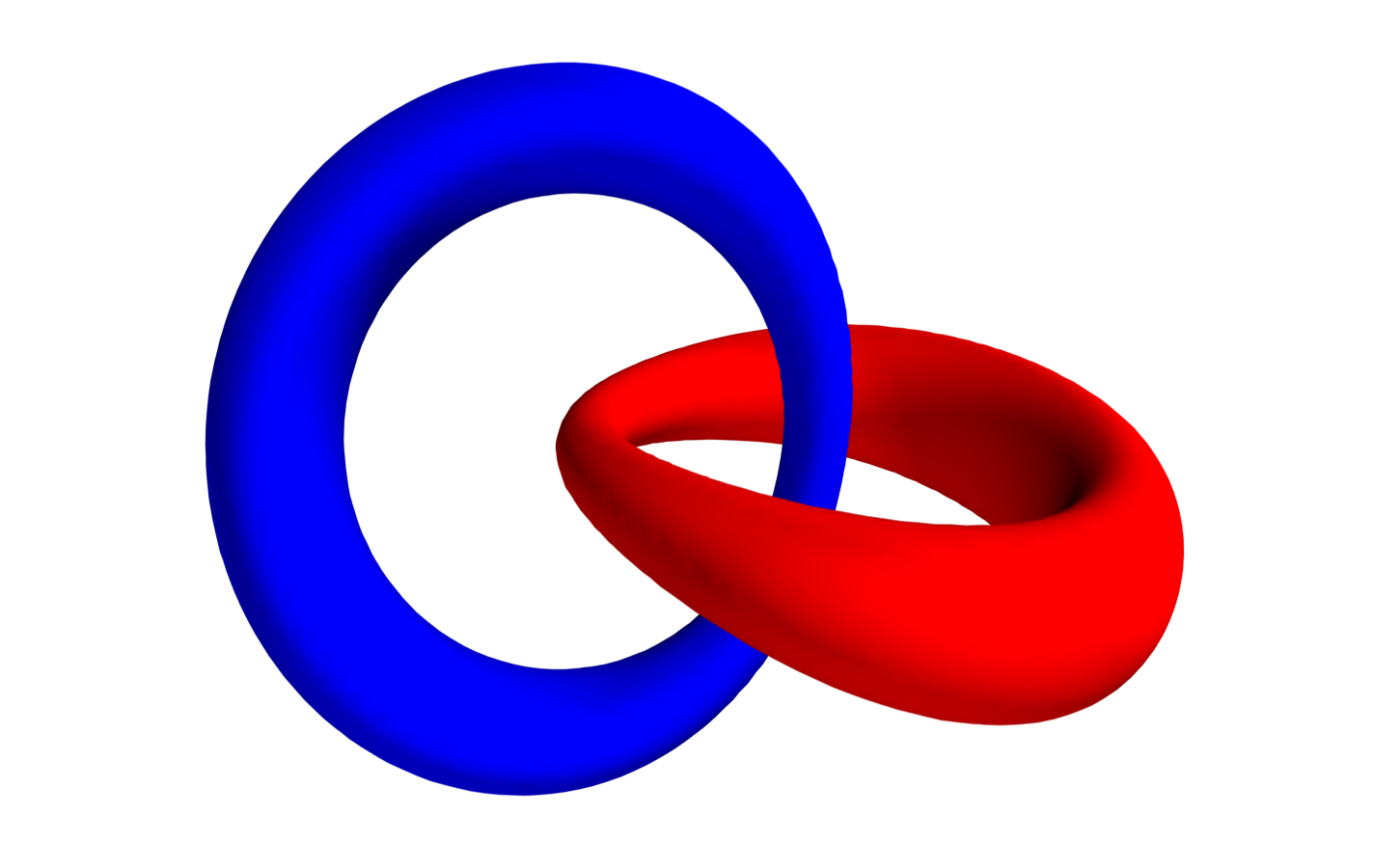}
        \includegraphics[height=3cm]{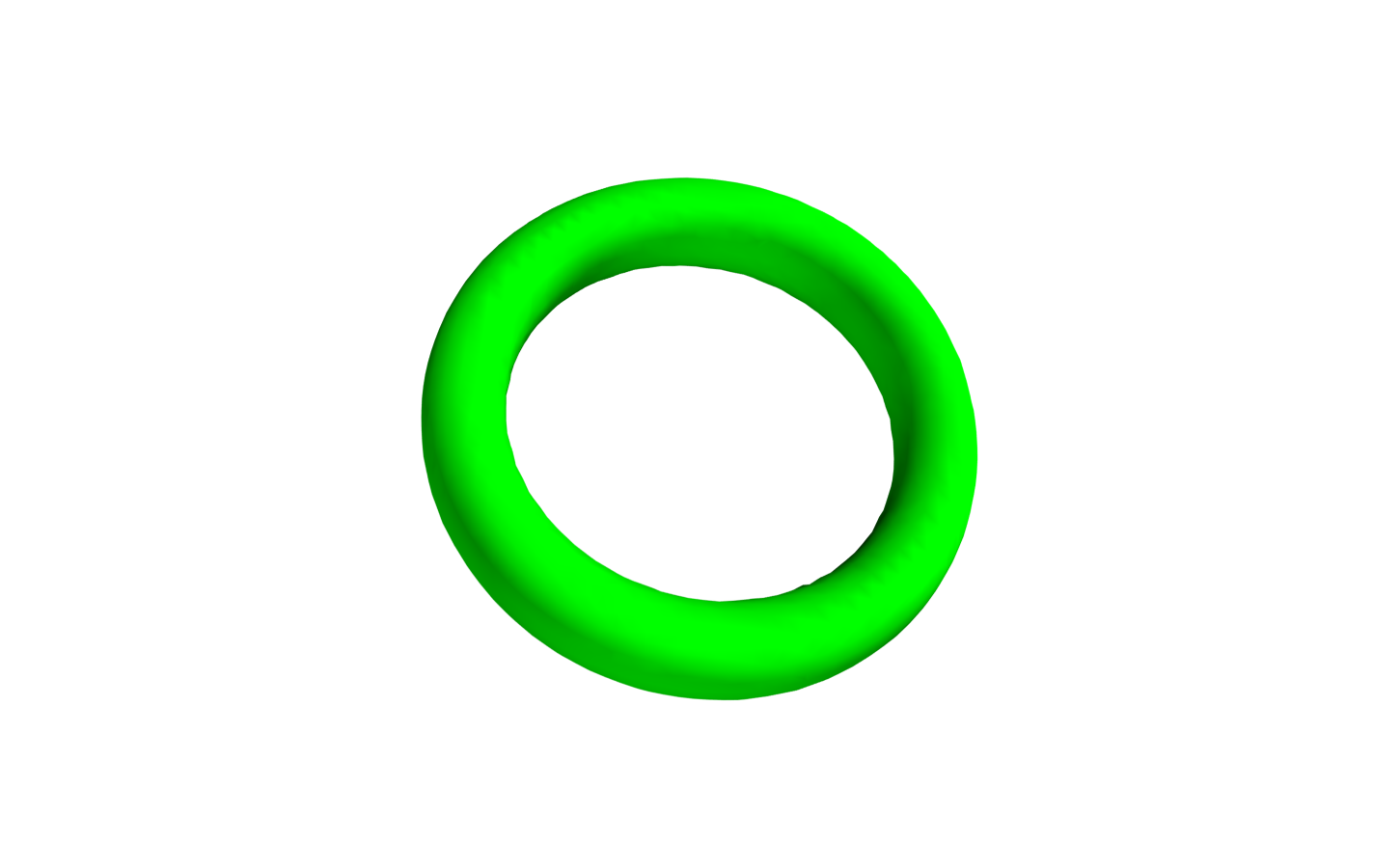}
    \end{center}
    \begin{center}
    \includegraphics[height=3cm]{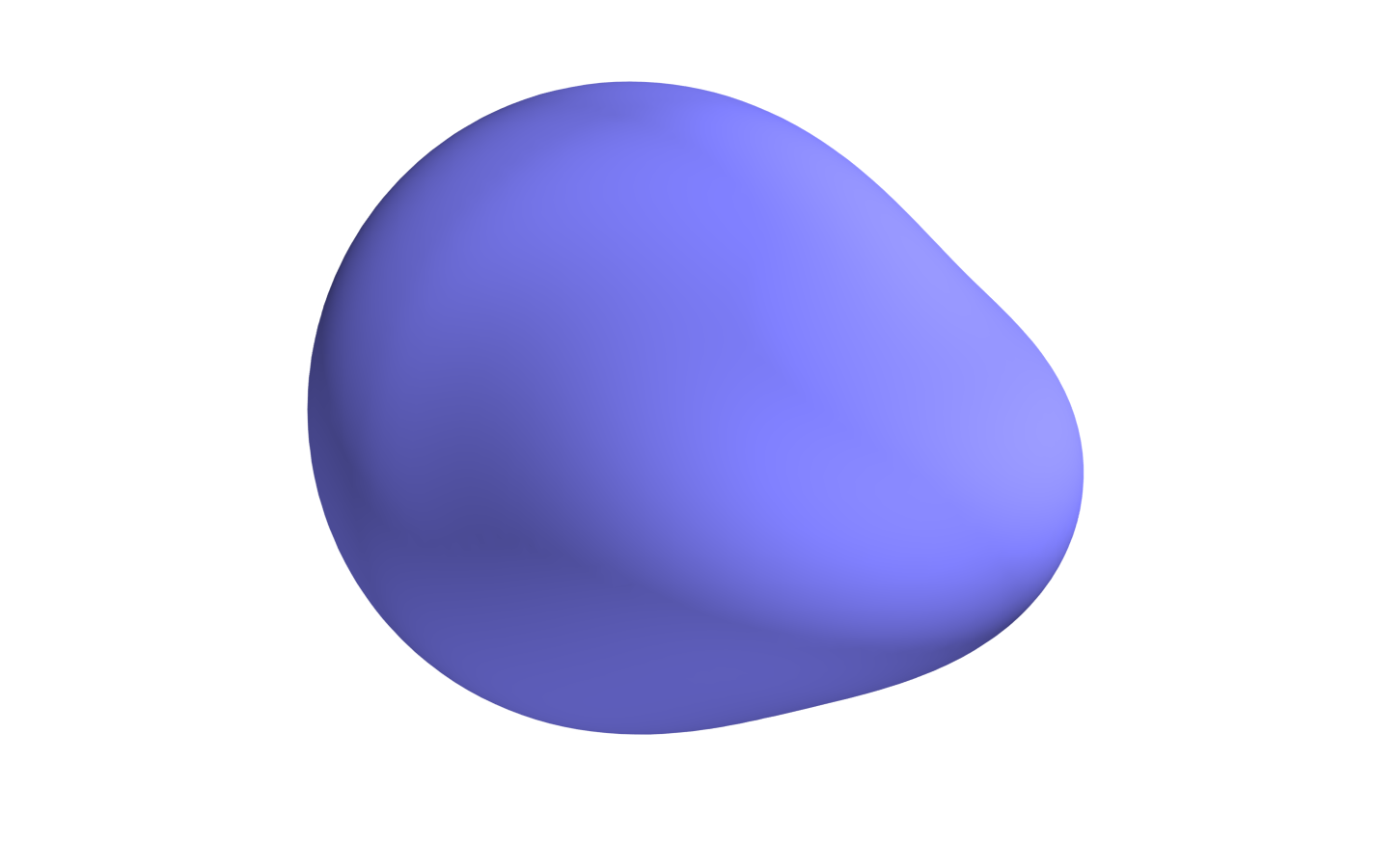}
    \includegraphics[height=3cm]{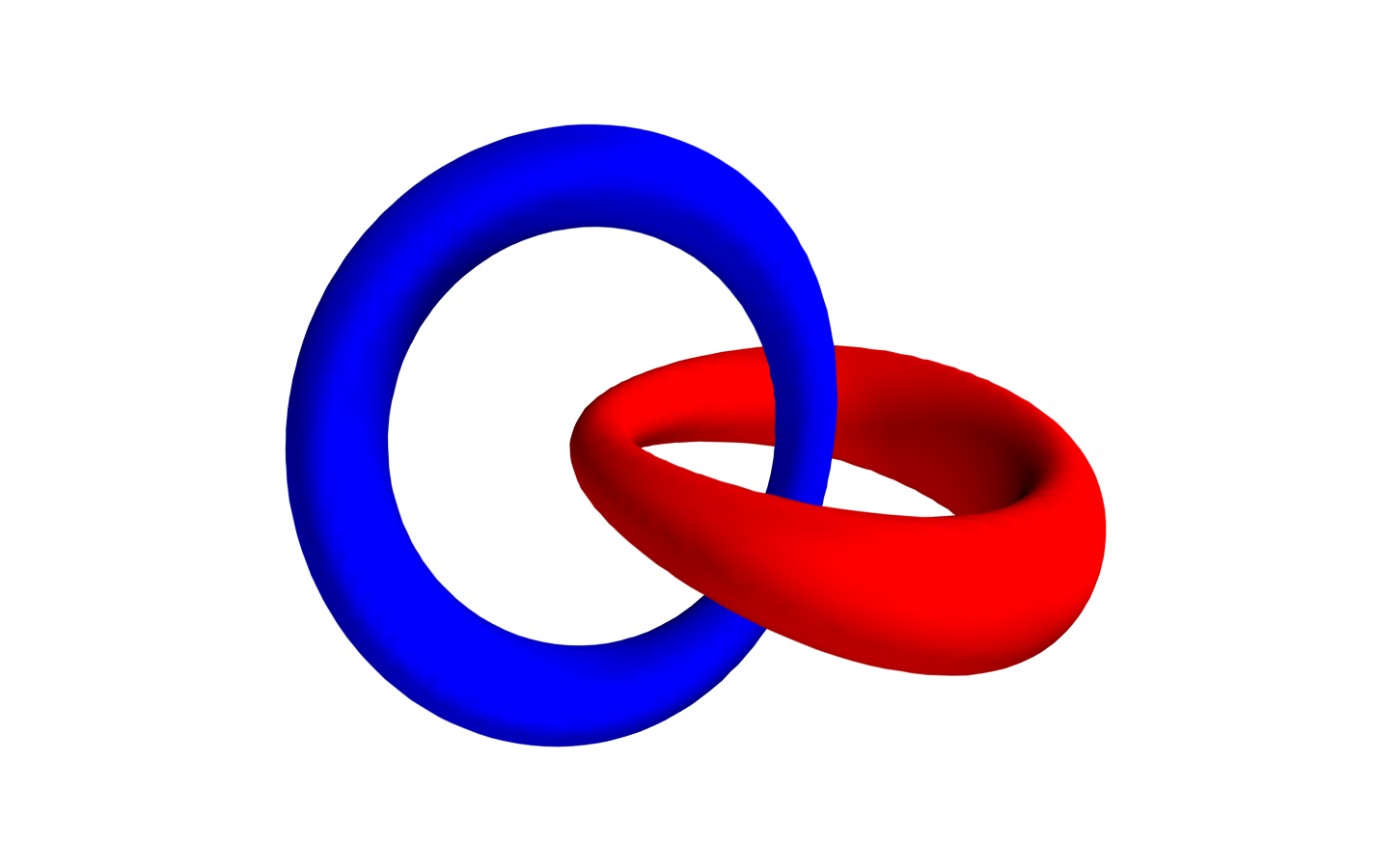}
    \includegraphics[height=3cm]{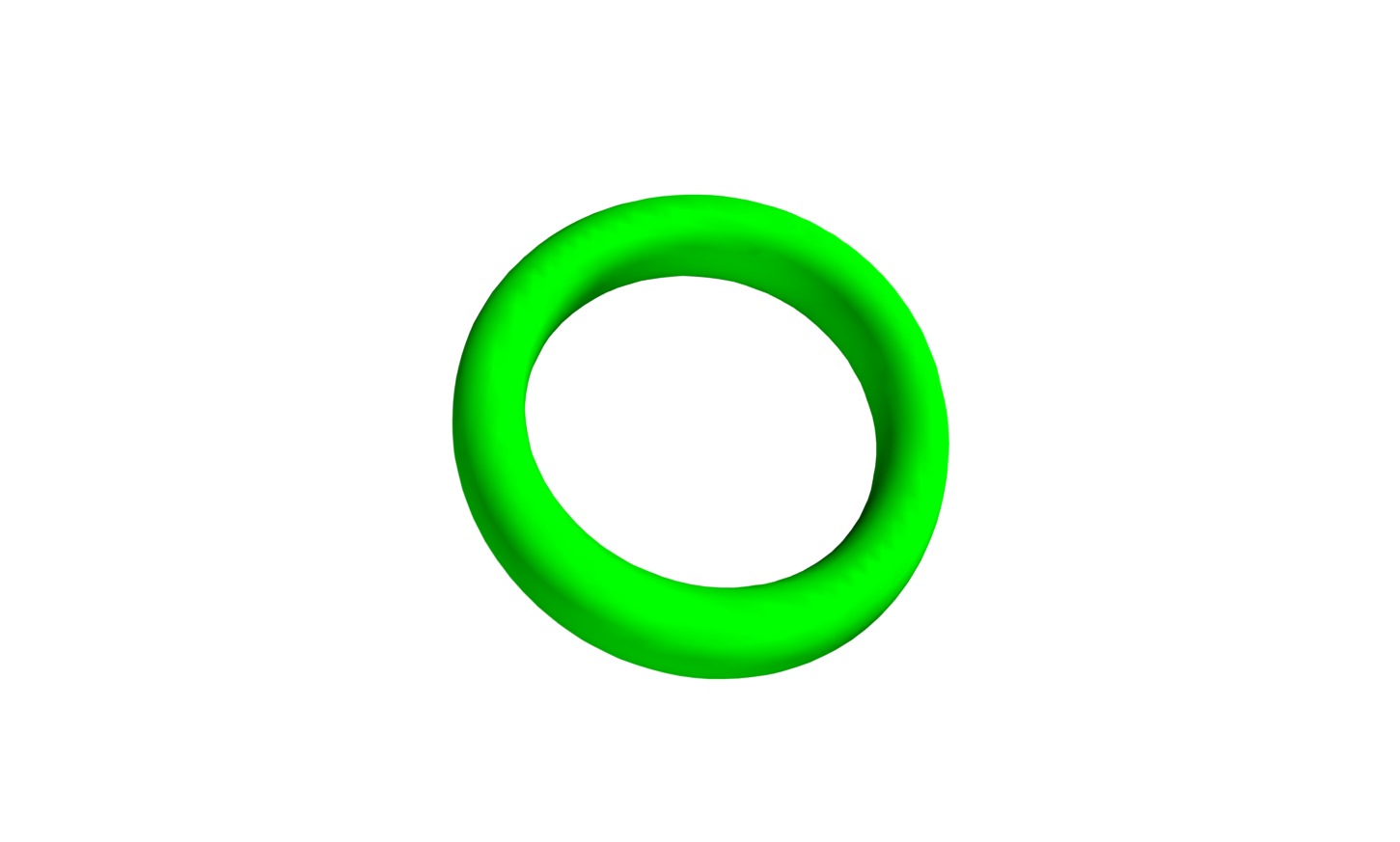}
\end{center}
    \begin{center}
    \includegraphics[height=3cm]{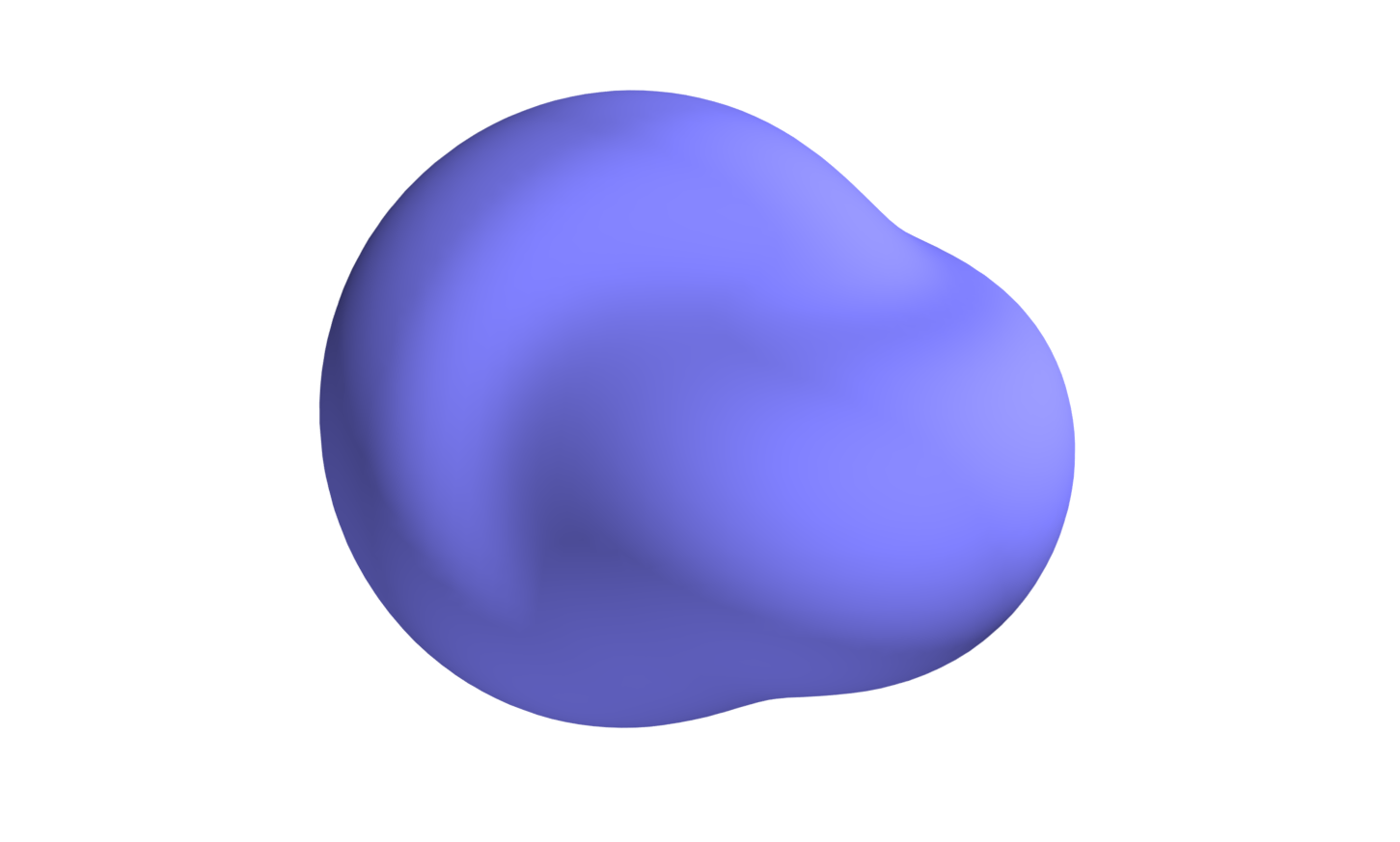}
    \includegraphics[height=3cm]{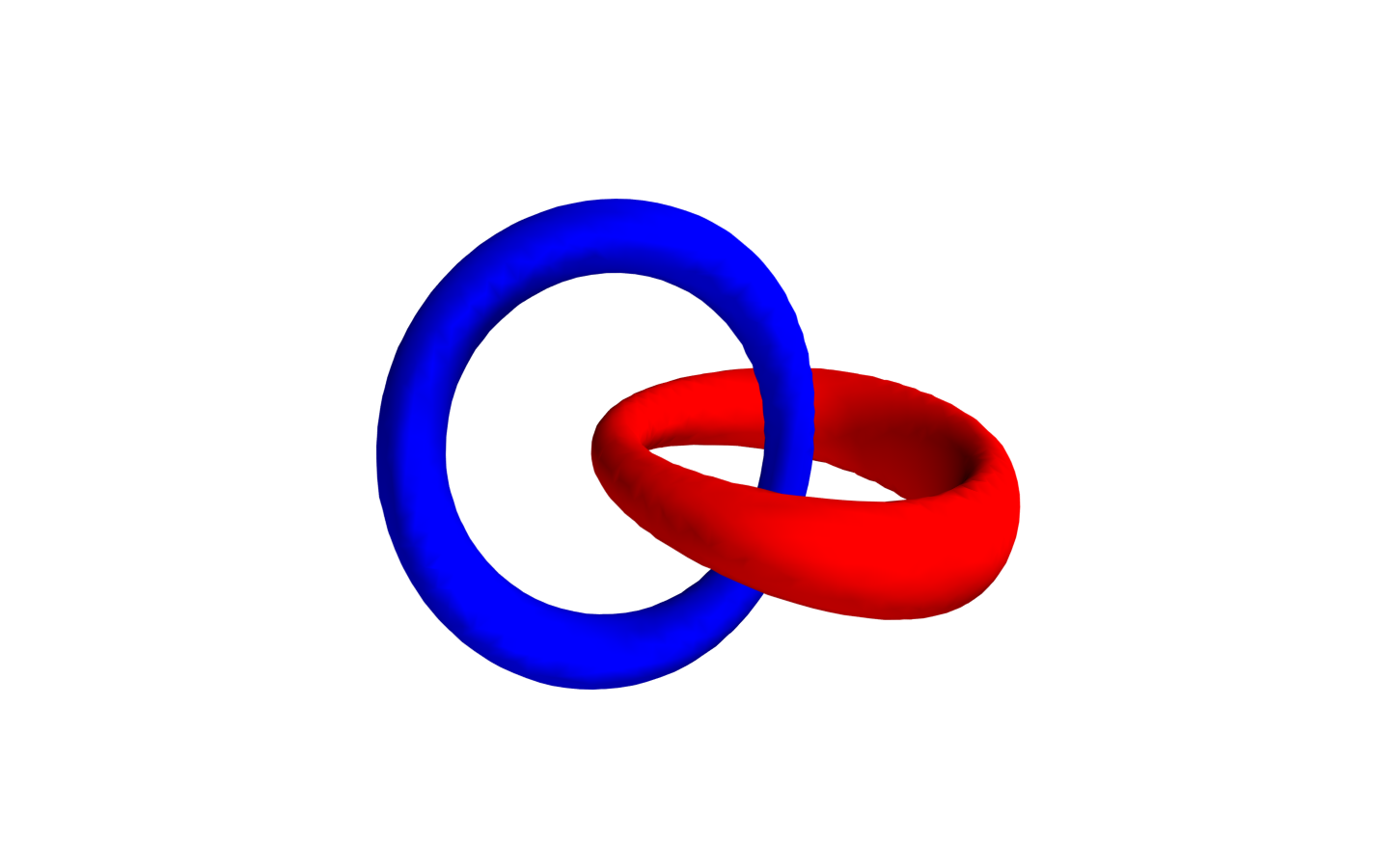}
    \includegraphics[height=3cm]{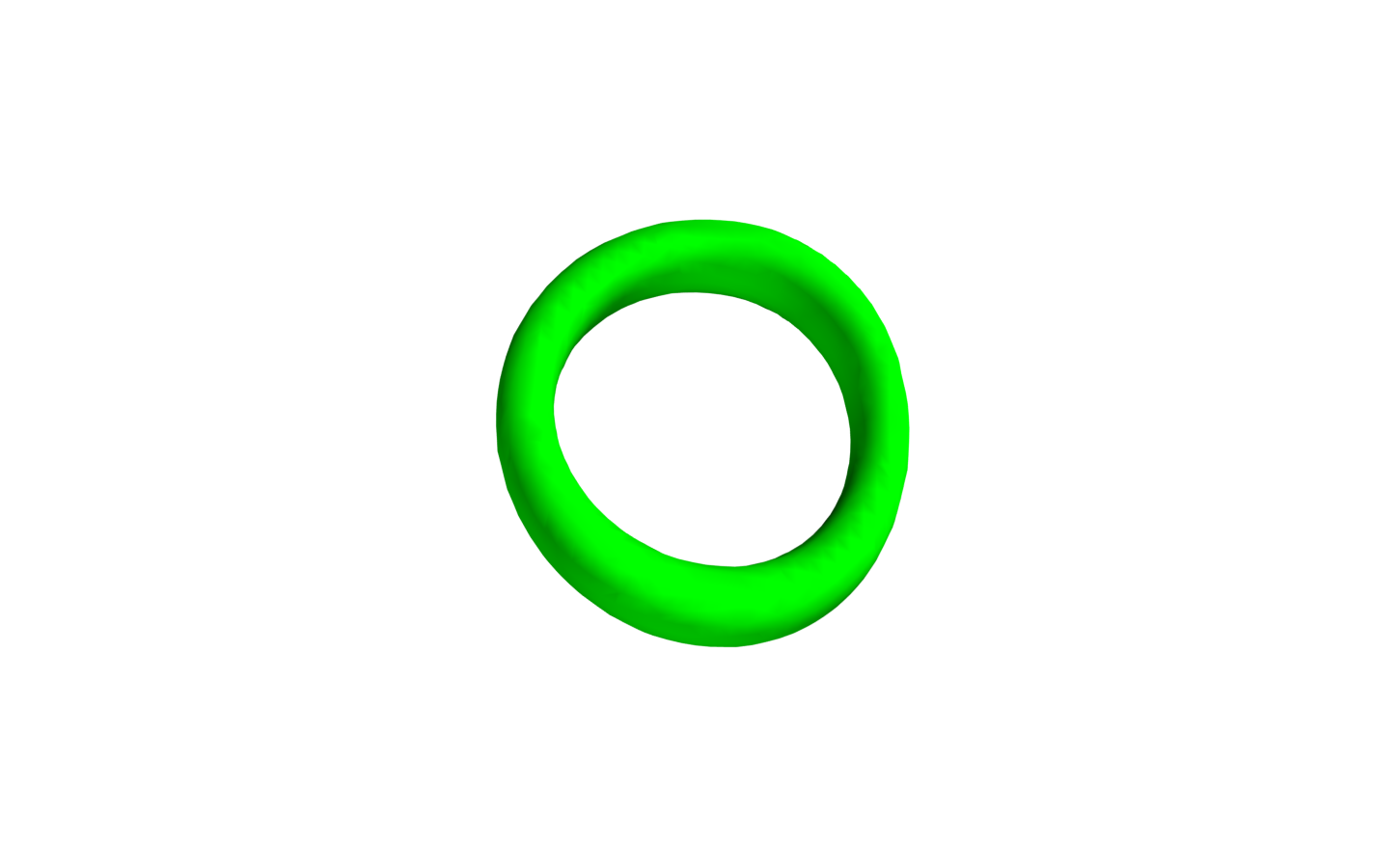}
\end{center}
    \caption{Isosurfaces showing the energy density distributions (first column), $\phi_1=\pm0.9$ (second column) and $\phi_3=-0.9$ of the $Q=1$ Hopfion solutions in the model \re{lagr} with potential $V=m^2\phi_1^2$ at  $m=0$ (first row),
    $m=1$ (second row), $m=2$ (third row), $m=4$ (fourth row) and $c_2=1$, $c_4=1$. Each plot corresponds to a
    value of the energy density $\mathcal{E}=15$ in unrescaled  units.}
    \label{q1enmass}
\end{figure}

First, we considered simplest unknot Hopfion in the sector of degree one. We set in \re{potential}
the angular parameter $\alpha=\pi/2$, thus this is the Heisenberg type potential which explicitly breaks the symmetry as $m\neq 0$.
In Fig.~\ref{q1enmass} we present isosurfaces of the energy density of the $Q=1$ solutions of the model
\re{lagr} with the potential $V=m^2\phi_1^2$ at $c_2=1$, $c_4=1$ for $m=0, 1, 2, 4$.
As $m=0$ we reproduce the usual axially symmetric $\mathcal{A}_{11}$ configuration. However, as the mass parameter $m$ starts to increase,
the configuration deforms taking the form of  two linked rings.

Clearly, the position curve $\phi_3=-1$,
which is a single loop, does not define the maximum of the energy
density distribution alone, as seen in  Fig.~\ref{q1enmass}.
Further, the increase of $m$ in the model with potential \re{potential} does not
localize the energy distribution, however it makes the field configuration more compact, see Fig.\ref{q1enmass}.
Note that, as the mass parameter
$m$ increases, the axial symmetry of the position curve is violated, see the plots in the third column in  Fig.~\ref{q1enmass}.
Thus, the contribution of the
potential term now becomes critical, the tube-like isosurfaces, which are the preimages of the vector
$\vec \phi = (\pm (1 - \mu), 0,0)$,
actually define the position of the maxima  of the energy density distribution of the configuration, as can be seen by comparing
the left and the middle columns in  Fig.~\ref{q1enmass}. Here we generally choose $\mu=0.1$.

\begin{figure}[h]
    \begin{center}
        \includegraphics[height=5.3cm]{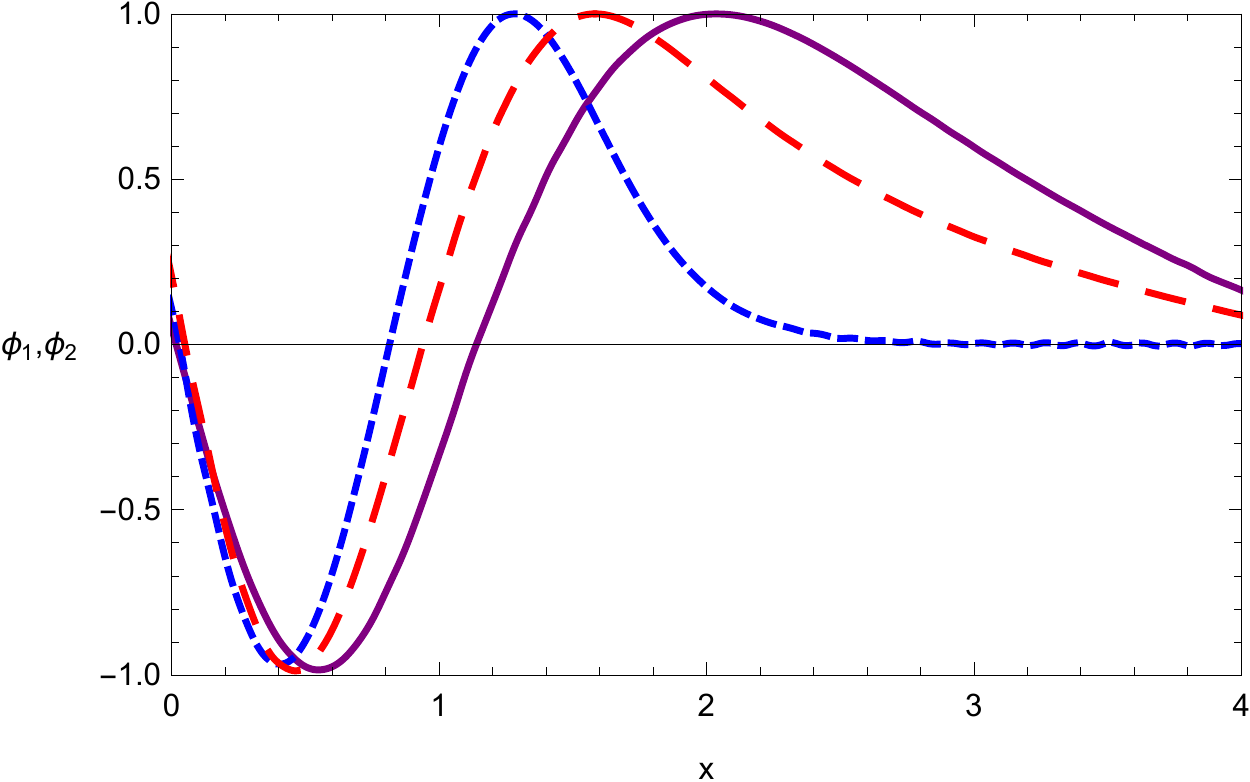}
        \includegraphics[height=5.3cm]{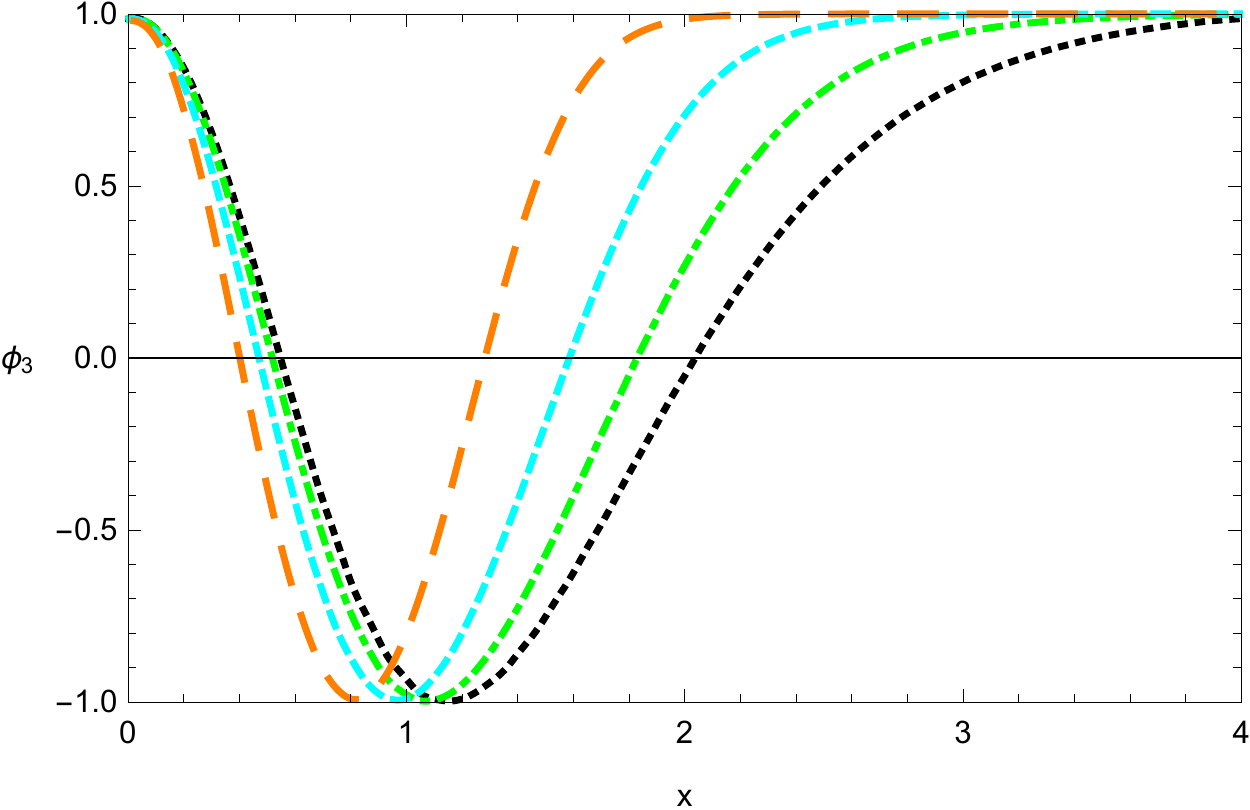}
    \end{center}
    \caption{Profiles of the field components of the $Q=1$ Hopfions
    in the model \re{lagr} with potential $V=m^2\phi_1^2$ and $c_2=1$, $c_4=1$.
    Left: Field components $\phi_1(x,0,0)$ at $m=0$ (purple solid), $\phi_1(x,0,0)$ (blue dashed) and
    $\phi_2(0,y,0)$ (red long-dashed)
    at $m=4$. Right: Field component $\phi_3(x,0,0)$ at $m=0$ (black dotted), $m=1$ (green dotted-dashed),
    $m=2$ (cyan dashed) and $m=4$ (orange long-dashed).}
    \label{f1f2f3}
\end{figure}

It is seen in the left plot of Fig.~\ref{f1f2f3}, which displays profiles of the components
$\phi_1(x,0,0)$ and $\phi_2(0,y,0)$ for some set of
values of mass parameter $m $, that, as $m$ increases, both components become more localized. However
the asymptotic behavior of these components is different, while the field $\phi_1$ decays exponentially, the component $\phi_2$
remains massless.

Indeed, let us consider the asymptotic form of the fields  in the limit $r\rightarrow\infty$, when
$\vec{\phi}\simeq(\phi_1,\phi_2,1-\frac{1}{2}(\phi_1^2+\phi_2^2))$ as $\phi_1,\phi_2\rightarrow0$.
Then the energy density takes the form
$$
\mathcal{E}\simeq c_2(\pa_i\phi_1)^2+c_2(\pa_i\phi_2)^2+m^2\left(\phi_1\sin\alpha\right)^2+\mathcal{O}((\pa_i\phi_a)^4,\phi_a^3)
$$
where we take into account the explicit form of the potential \re{potential}. Thus, the linearized Euler-Lagrange equations on the
spacial asymptotic are
$$(\Delta-\frac{(m\sin\alpha)^2}{c_2})\phi_1=0$$
$$\Delta\phi_2=0 $$

Hence, as $r\rightarrow\infty$,
\be
\phi_1\sim\frac{\exp^{-r \frac{m\sin\alpha}{\sqrt{c_2}}}}{r}; \qquad
\phi_2\sim\frac{1}{r}
\ee
Note that the field component $\phi_3$ also possesses the asymptotic Yukawa massive interaction tail, see the right plot in
Fig.~\ref{f1f2f3}, increase of the mass parameter $m$ also effectively decreases the characteristic size of this component.

\begin{figure}[h]
    \begin{center}
        \includegraphics[height=5cm]{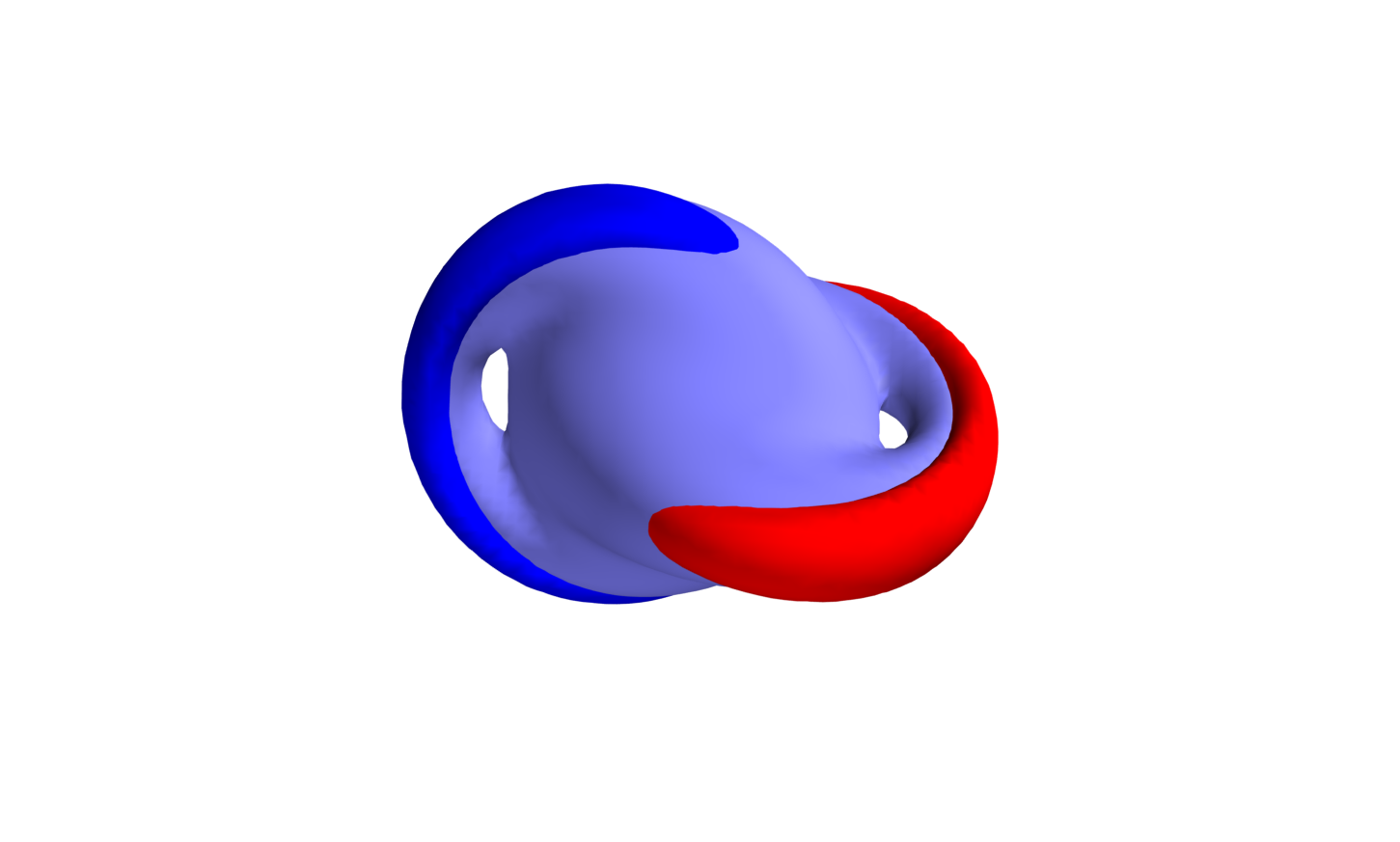}
        \includegraphics[height=5cm]{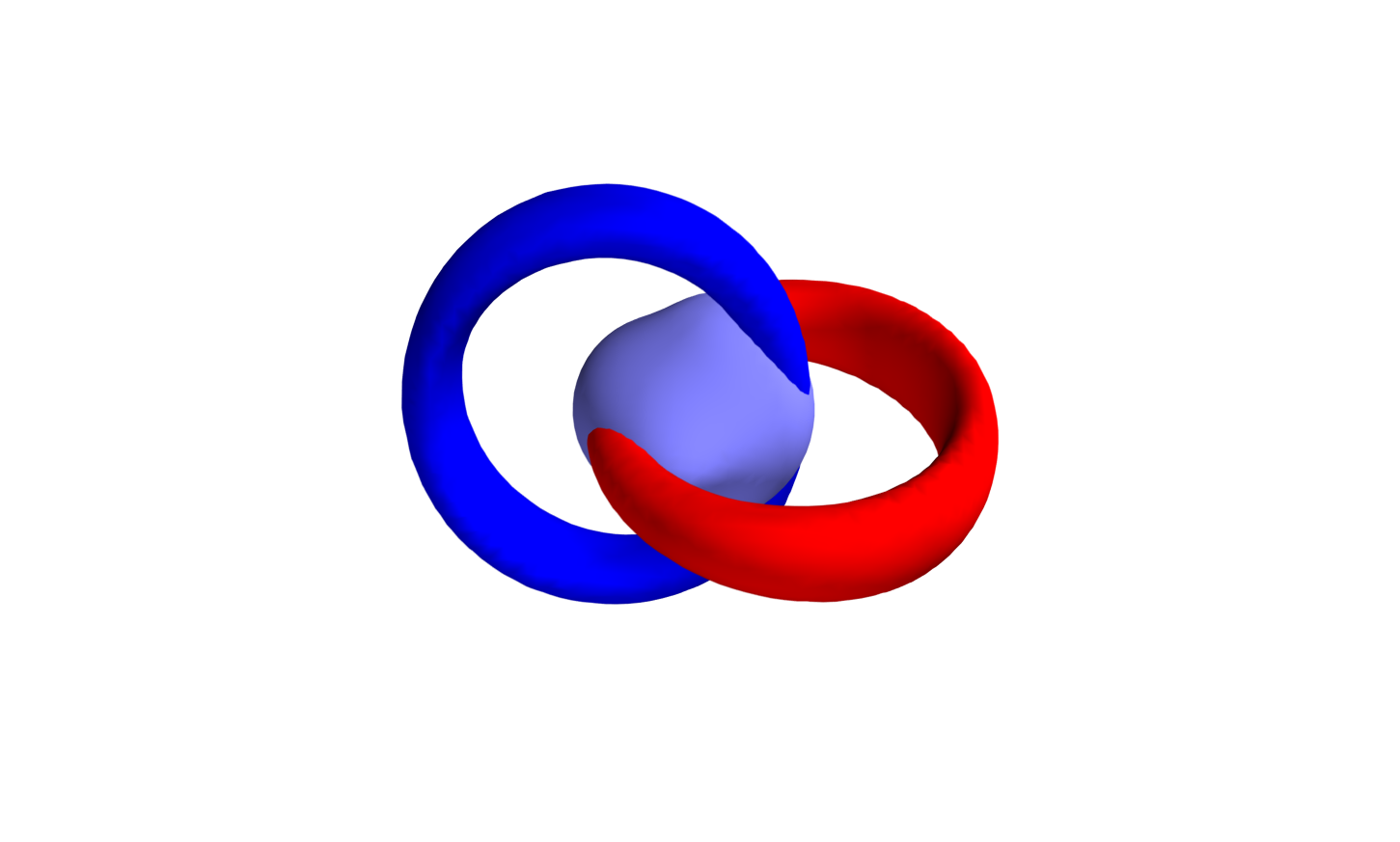}
    \end{center}
    \begin{center}
        \includegraphics[height=5cm]{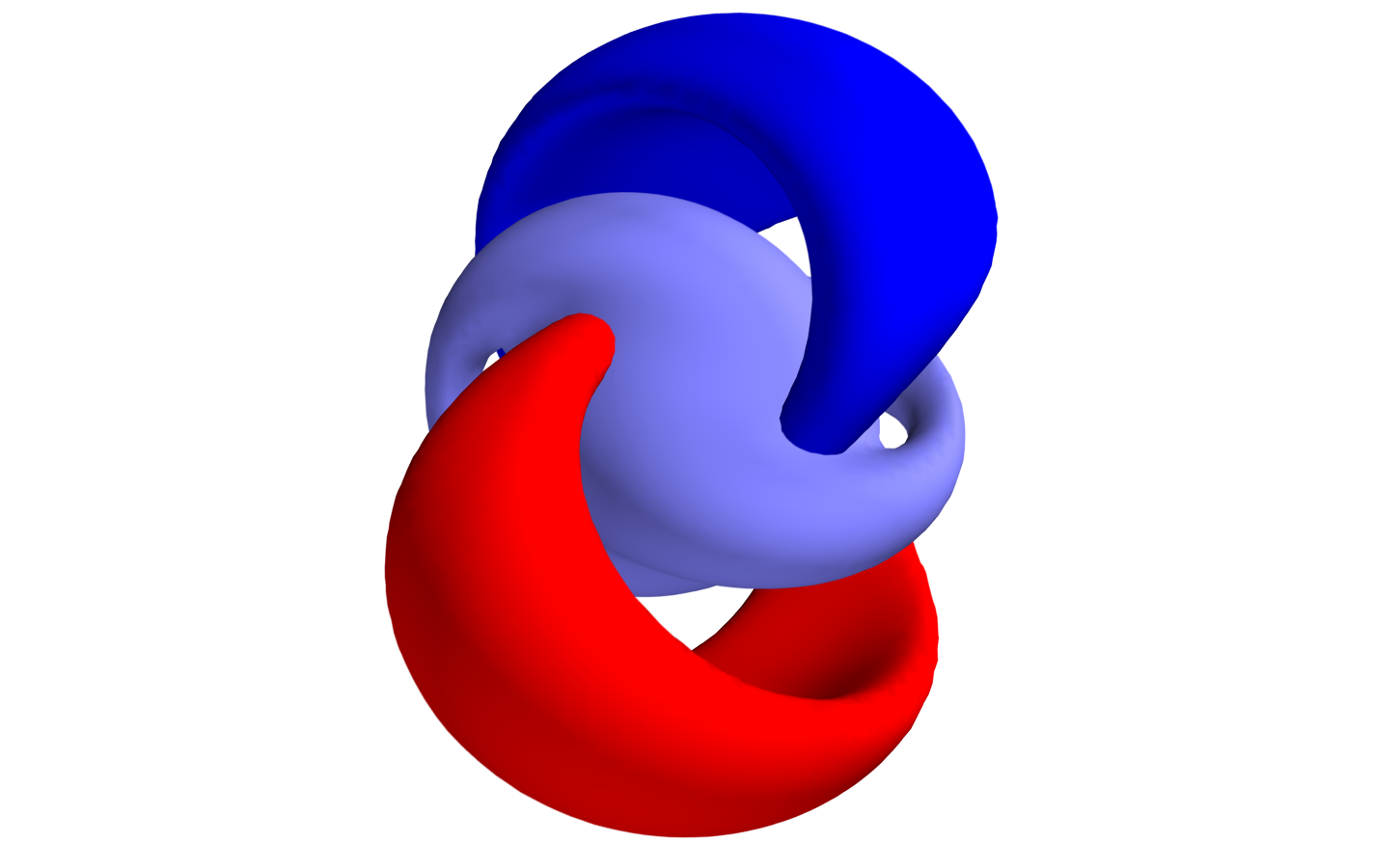}
        \includegraphics[height=5cm]{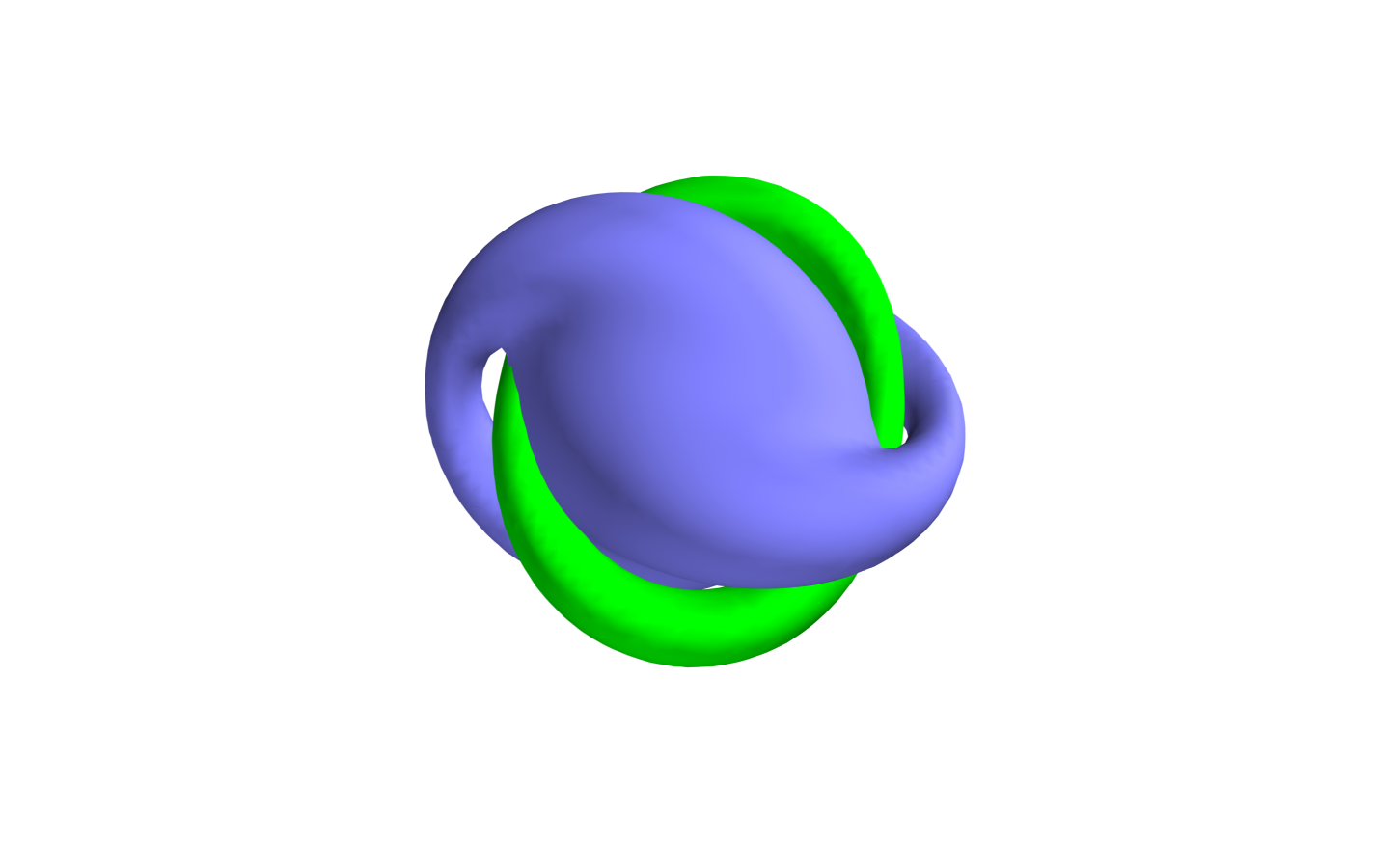}
    \end{center}
    \caption{$Q=1$ Hopfion solution at $c_2=0.5, c_4=1, m=4$. Top row: location curves
    of the preimages of the points $\vec \phi=(\pm 0.9, 0,0)$ together
with the energy isosurface at $\mathcal{E}=45$ (left) and $\mathcal{E}=100$ (right). Bottom row:
location curves of the vector $\vec \phi=(0,\pm 0.9,0)$ (left) and $\vec \phi=(0, 0, -0.9)$ together
with the energy isosurface at $\mathcal{E}=45$.
}
    \label{q1combine}
\end{figure}

It is instructive to plot the energy isosurface of the $Q=1$ Hopfion together with the location curves of the field
components, see Fig.~\ref{q1combine}. Clearly, the shape of the energy density distribution is similar to the
location curve, which correspond to the isosurfaces of the preimages of the vectors  $\vec \phi=(\pm 1, 0,0)$, it is
visibly distinguished from the curve, which corresponds to the isosurface of
$\vec \phi^{-1}(0,0,-1)$. Further, it is seen in the right upper plot in Fig.~\ref{q1combine}, which displays
the location curves of the preimages of the points $\vec \phi=(\pm 1, 0,0)$ together
with the energy isosurface, the maximum of the energy density distribution corresponds to the
domain, where these curve interlace. Indeed, as the value of the mass parameter $m$ increases, the contribution of the
potential \re{potential} into the total energy becomes more significant. However, in the limit $m\rightarrow\infty$
the virial condition for the Hopfions becomes
$\CE_4 = 3V$, where $\CE_4$ is the energy contribution of the  quartic in derivatives term in \re{hopfenergy}. Thus, it does not
allow us to neglect the corresponding contribution of the gradients of the fields.

Let us now consider the energy functional \re{hopfenergy} with the potential\footnote{Note that the vacuum
boundary conditions on the field components $\vec \phi_\infty$ are now different. However, an appropriate rotation of the components
$$
\phi_1 \to \phi_1\sin\alpha+\phi_3\cos\alpha
$$
with $\cos\alpha = 1/3$, allows us to recover
the general form of the potential \re{potential} with the usual boundary conditions on the field $\vec \phi_\infty = (0,0,1)$.}
\be
V=m^2\left(\phi_1-1/3\right)^2
\label{pot-Nitta}
\ee
This potential represents an intermediate form between the limiting cases of the Heisenberg type potential,
and the usual pion mass potential.
Note that potential term of similar form was suggested in the usual Skyrme model in 3+1 dimensions
to construct charge one Skyrmion solution,
which consists of two components with fractional topological charges $1/3$ and $2/3$, respectively \cite{Gudnason:2015nxa}.

\begin{figure}[h]
    \begin{center}
        \includegraphics[height=5cm]{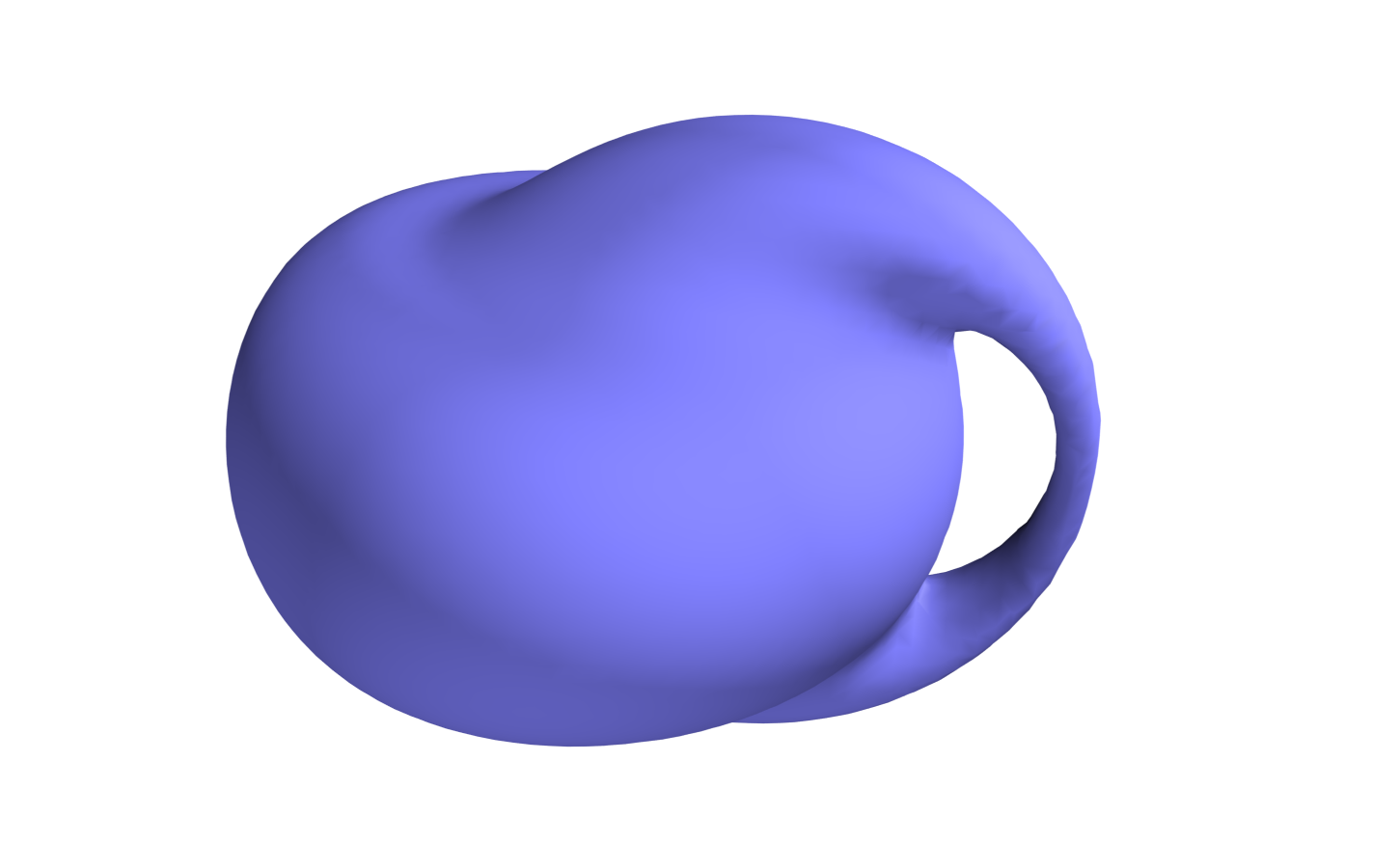}
        \includegraphics[height=5cm]{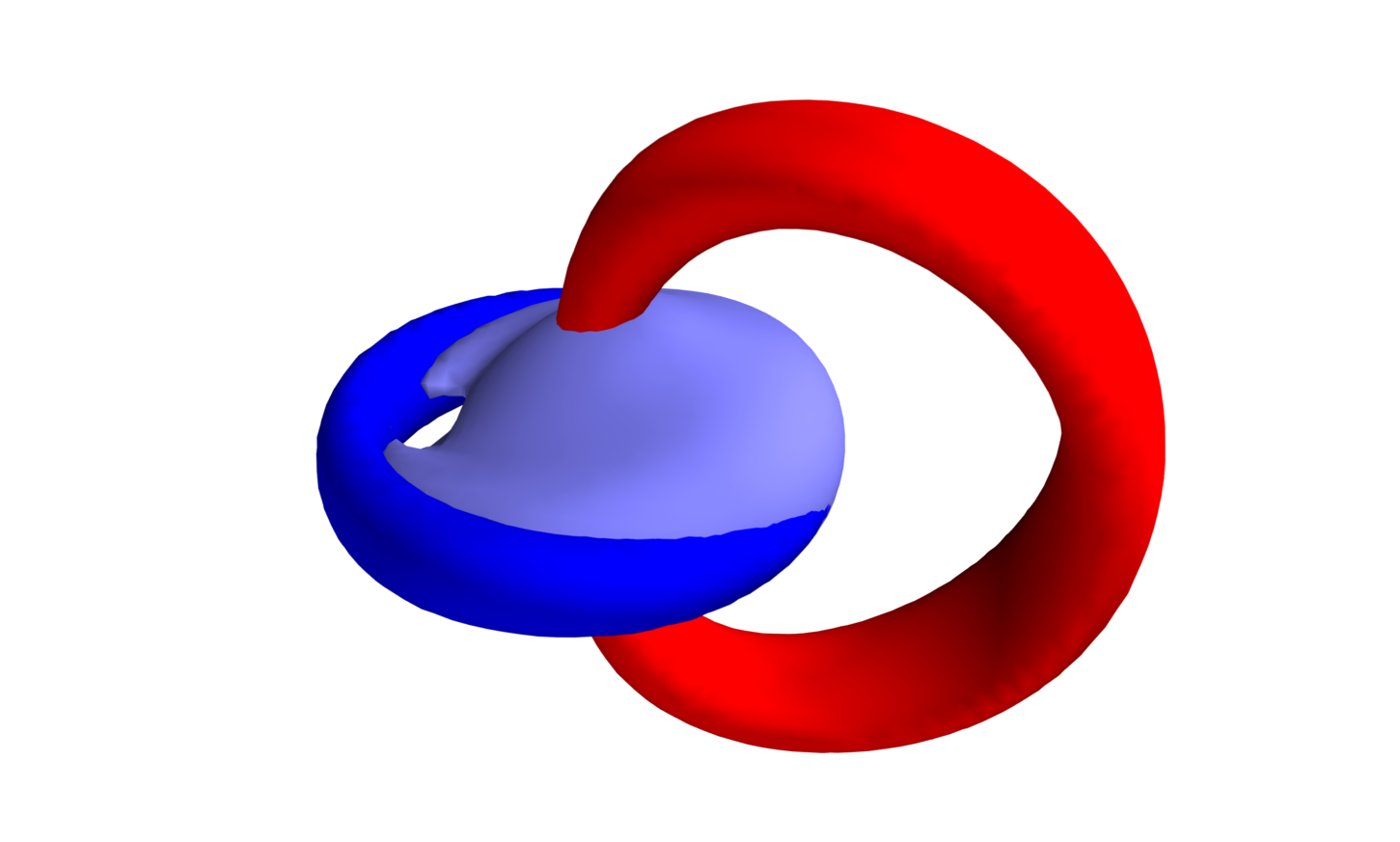}
    \end{center}
    \caption{Isosurfaces showing the  energy density of the   $(\CA_1\between \CA_1)_{\CA_{1,1}}$  Hopfion of degree one
        in the Faddeev-Skyrme model \re{hopfenergy}
        with the potential $V=m^2\left(\phi_1-\frac13\right)^2$ at $c_2=0.5, c_4=1, m=4$ for $\mathcal{E}=20$
        (left) and $\mathcal{E}=100$ together
        with the location curves of the preimages of the points $\vec \phi=(\pm 0.9, 0,0)$ (right).
    }
    \label{q1al1o23}
\end{figure}

It is seen in Fig.~\ref{q1al1o23}, which displays energy isosurfaces of the corresponding $Q=1$ configuration, that
similar to our consideration above, the Hopfion is composed of two linked loops. However, one of these tubes is
much thinner and has a smaller size than the other, see left plot on Fig.~\ref{q1al1o23}. Right plot of Fig.~\ref{q1al1o23} presents location curves of the preimages of the points $\vec \phi=(\pm (1 - \mu), 0,0)$ together
with the corresponding energy isosurface (cf. Fig.~\ref{q1combine}). Note that the blue curve, which corresponds to the
preimage of the point $\vec \phi=(-1 + \mu, 0,0)$ has bigger size and it is a bit thicker than the red curve, which
corresponds to the preimage of the point $\vec \phi=(1 - \mu, 0,0)$.

This is consistent with our observation above that in the model \re{lagr} the
shape of the energy density distribution
follows the maxima of the potential although the virial relation holds as before.

\subsection{Q=2}
Let us consider now the Hopfion solutions  in the topological sector of degree two.

\begin{figure}[h]
    \begin{center}
        \includegraphics[height=5cm]{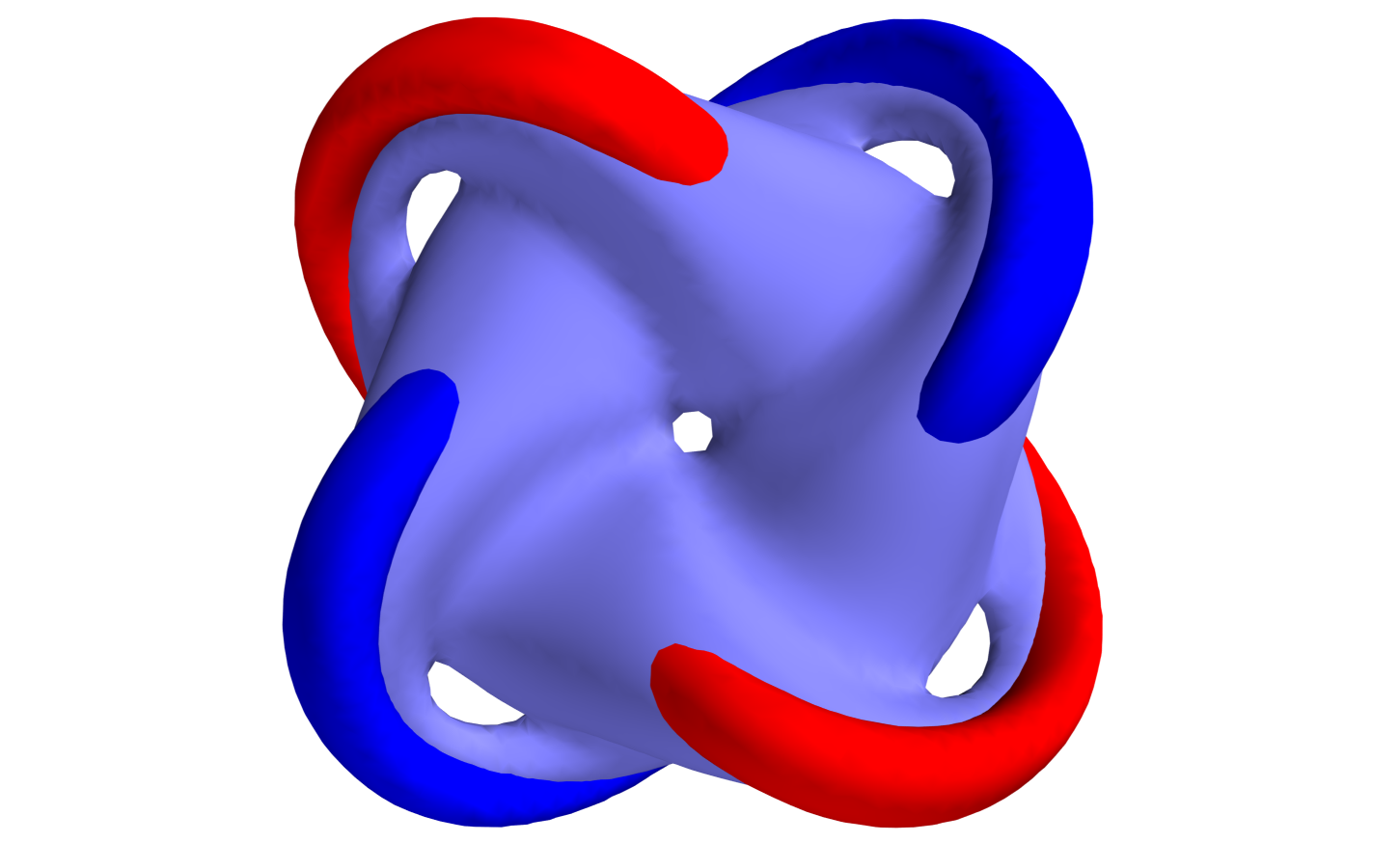}
        \includegraphics[height=5cm]{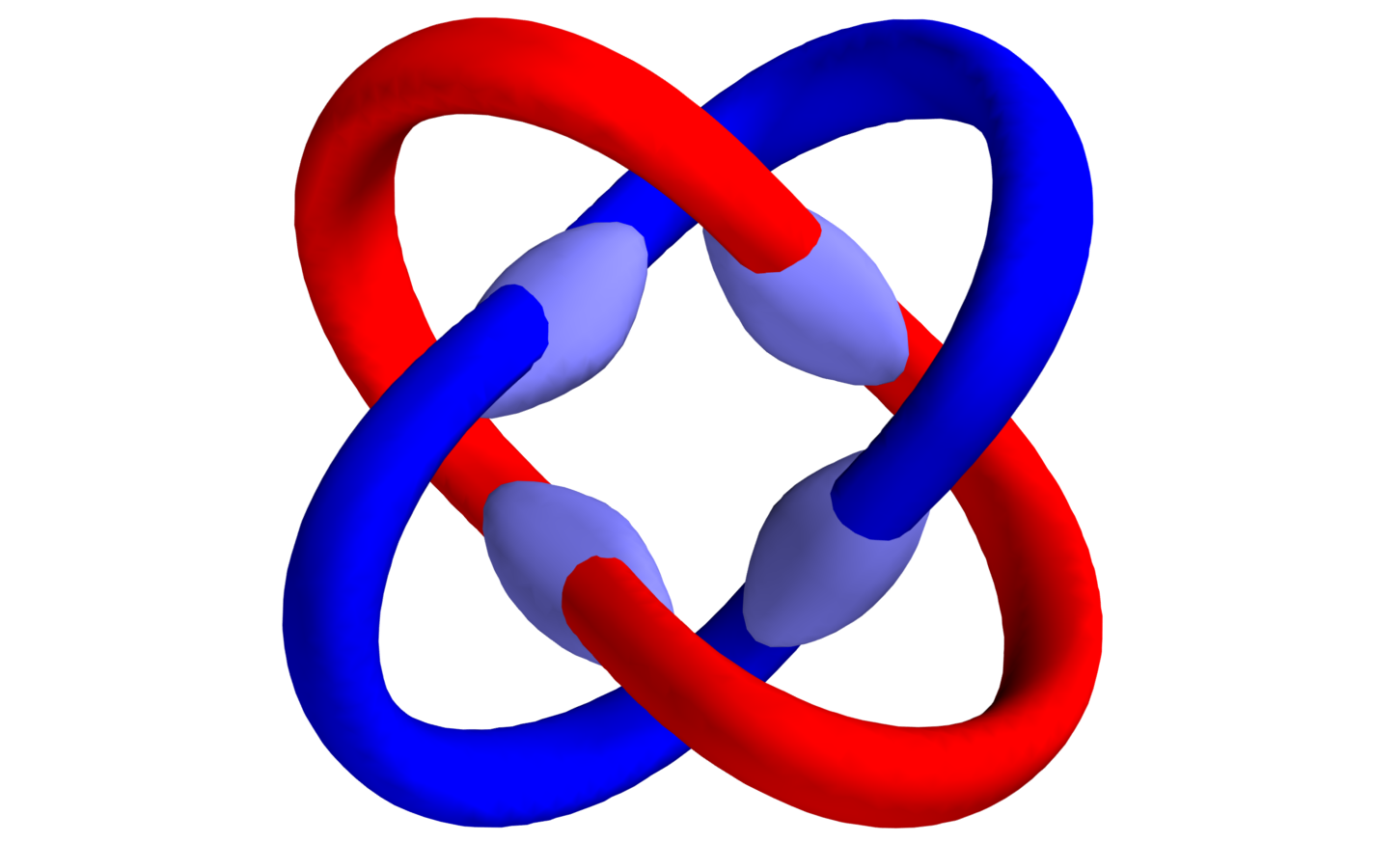}
    \end{center}
    \caption{Isosurfaces showing the energy density of the $2(\widetilde \CA_1\between \widetilde \CA_1)_{\CA_{2,1}}$ Hopfion in the Faddeev-Skyrme model \re{hopfenergy}   with the potential $V=m^2{\phi_1}^2$ at $c_2=0.5, c_4=1, m=4$ for $\mathcal{E}=52$ (left) and $\mathcal{E}=90$ (right) together with the location curves of the preimages of the points $\vec \phi=(\pm 0.9, 0,0)$.}
    \label{q2combine}
\end{figure}

We can see that, as in previously considered case of the $Q=1$ Hopfion,
for relatively large value of the mass parameter $m$
the spacial distribution of the energy density
follows the location curves $\mathcal{C}_1 = \vec \phi^{-1}(1,0,0)$ and $\mathcal{C}_{-1} = \vec \phi^{-1}(-1,0,0)$.
Indeed, in Fig.~\ref{q2combine}, left plot, we display the energy density isosurfaces of the $Q=2$ configuration together with the isosurfaces of the preimages of the vector $\vec \phi=(\pm 0.9, 0,0)$.
Clearly, for higher values of the energy density, left plot, the energy density isosurface
has a form of twisted bead necklace with four beads. This structure is similar to the energy density isosurface of the
corresponding four-component solution of the Skyrme model with Heisenberg type potential \cite{Gudnason:2015nxa}, which however,
is not twisted.

\begin{figure}[h]
    \begin{center}
        \includegraphics[height=4cm]{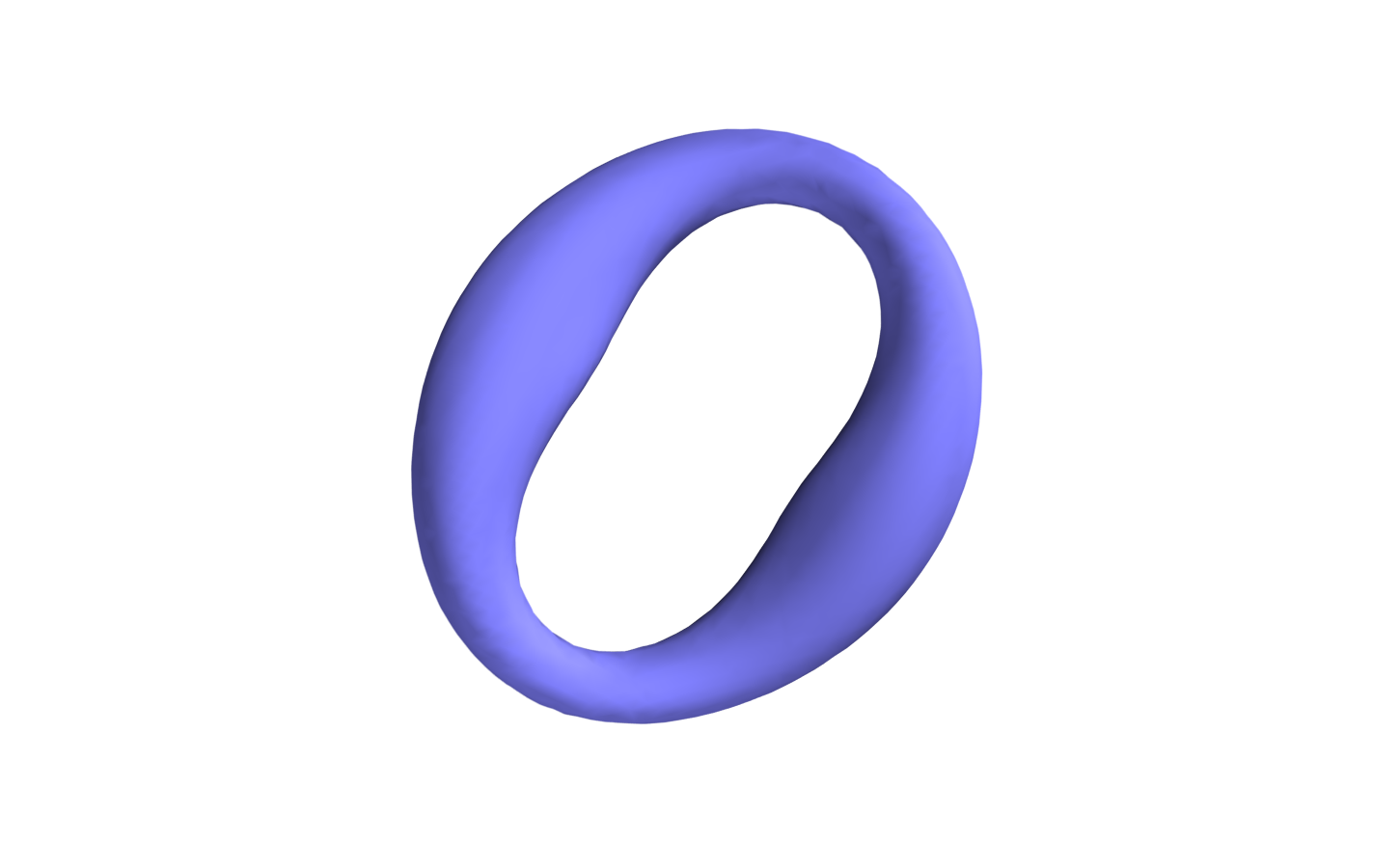}
        \includegraphics[height=4cm]{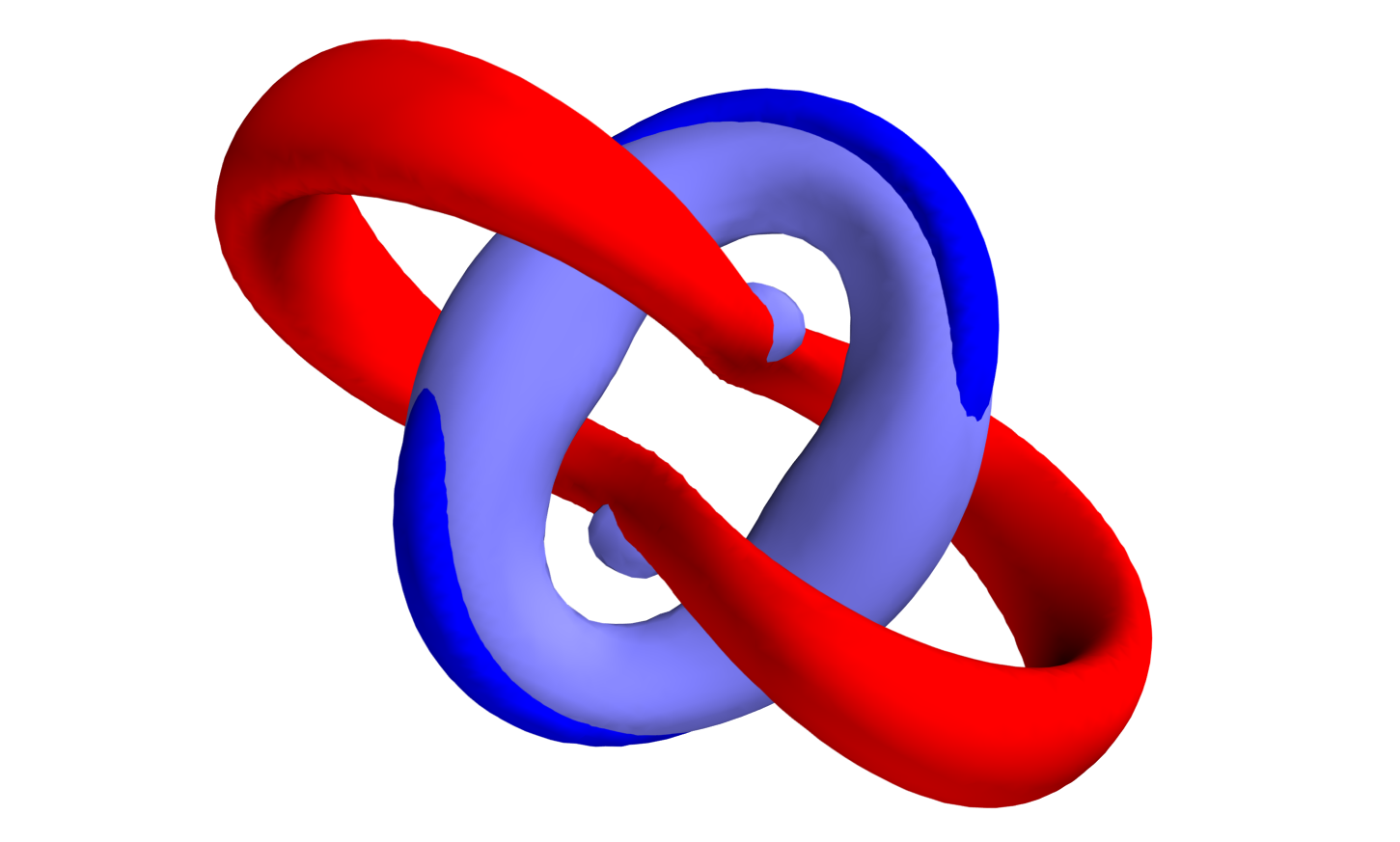}
    \end{center}
    \caption{Isosurfaces showing the  energy density of the $Q=2$ Hopfion in the Faddeev-Skyrme model \re{hopfenergy}
        with the potential $V=m^2\left(\phi_1-\frac13\right)^2$ at $c_2=0.5, c_4=1, m=4$ and $\mathcal{E}=100$ (left plot). The energy isosurfaces at $\mathcal{E}=90$ together
        with the location curves of the preimages of the points $\vec \phi=(\pm 0.9, 0,0)$ (right plot).
    }
    \label{q2al1o23}
\end{figure}

We can consider now the $Q=2$ Hopfion solution in the model  \re{lagr} with the potential \re{pot-Nitta}.
As it is seen in Fig.~\ref{q2al1o23}, in this case the energy density isosurfaces also represent two buckled tubular loops
twisted two times
around each other. However, similar to the corresponding $Q=1$ solution above, cf. Fig.~\ref{q1al1o23}, the energy is not equally distributed
among these loops (see table \ref{hopftable2}), one of them is thinner than another, it
disappears as the value of the energy density increases, see left plot in Fig.~\ref{q2al1o23}. Interestingly, for higher values
of the energy density the corresponding tube is not decomposed into four isolated beads, as it happens in the model with Heisenberg type
potential above.

To classify solutions of the Faddeev-Skyrme model with symmetry breaking potential we introduce new notations, which generalize the
above-mentioned  classification by Sutcliffe \cite{Sutcliffe:2007ui}. Since the spacial position of maxima of the energy distribution
corresponds to the location curves $\mathcal{C}_1$ and $\mathcal{C}_{-1}$,
which could be of different types, the  configuration with linking number $Q$ can be denoted as
\be
Q\left(\mathcal{C}_1 \between \mathcal{C}_{-1}\right)_{\mathcal{C}}
\ee
where the subindex corresponds to the type of the loop $\mathcal{C} = \vec \phi^{-1}(0,0,-1)$,  which is the antipodal
to the vacuum. In this notations, for example,
the configuration of degree two is denoted as $2(\widetilde \CA_1\between \widetilde \CA_1)_{\CA_{2,1}}$.
Further remark is that for larger values of the mass parameter $m$, the loop $\mathcal{C}$ in the sector of degree $Q$
becomes symmetric with respect to the dihedral group $D_{Q}$. This is a symmetry of the truncated $c=0$ model.

Note that in our numerical simulation we find another solution of the Faddeev-Skyrme model with potential $m^2{\phi_1}^2$ in this sector,
$2(\widetilde \CA_1\between \widetilde \CA_1)_{\widetilde \CA_{1,2}}$, see Table \ref{hopftable}.
This configuration represents a local minimum, its energy is
about $16 \%$ higher than the energy of the global minimum,
see Table \ref{energytable}. However, as the parameter $\alpha$ in \re{potential} varies,
this configuration becomes a saddle point, our algorithm is unable to find this solution in the model with potential \re{pot-Nitta}
(see Table \ref{energytable}). Indeed, it was noticed that the Hopfion of that type, which in the usual $m=0$
Faddeev-Skyrme model represents an unstable configuration of two charge one solitons
stacked one above the other \cite{Hietarinta:1998kt,Ward:1998pj}, it
has the energy around $ 13\%$ above the global minimum \cite{Sutcliffe:2007ui}.

\subsection{Multisoliton configurations}

\subsubsection{Degrees 3 and 4}
A peculiar feature of the Faddeev-Skyrme model is that
for a given degree $Q$, there are usually several different stable static soliton solutions of
similar energy, furthermore, the number of solutions grows with $Q$.
We considered both the configurations with minimal energies  and the local minima in a given sector.
Results of our numerical simulations are summarized in
Table \ref{hopftable}.

\begin{figure}[h]
    \begin{center}
        \includegraphics[height=5cm]{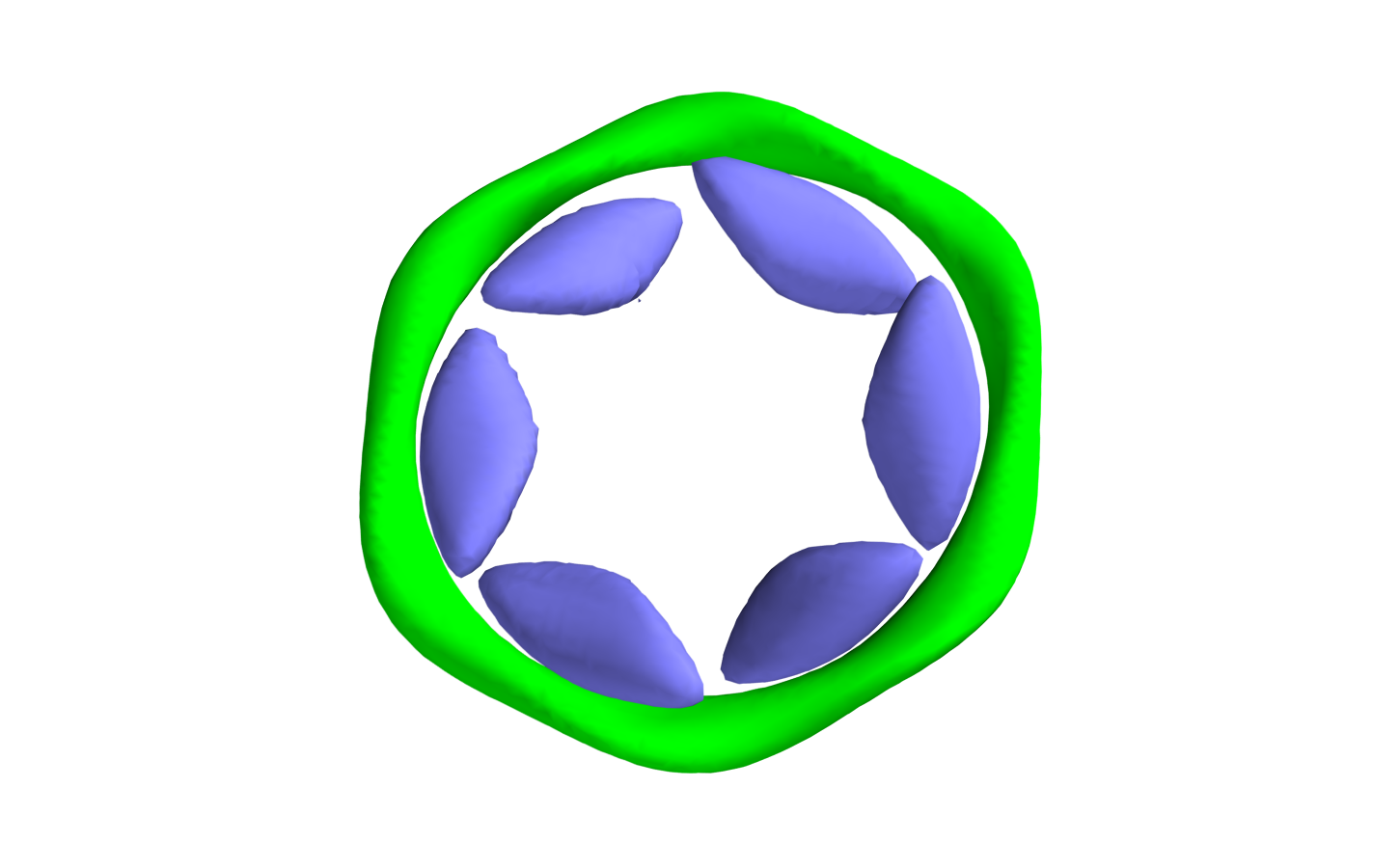}
    \end{center}
    \begin{center}
        \includegraphics[height=5cm]{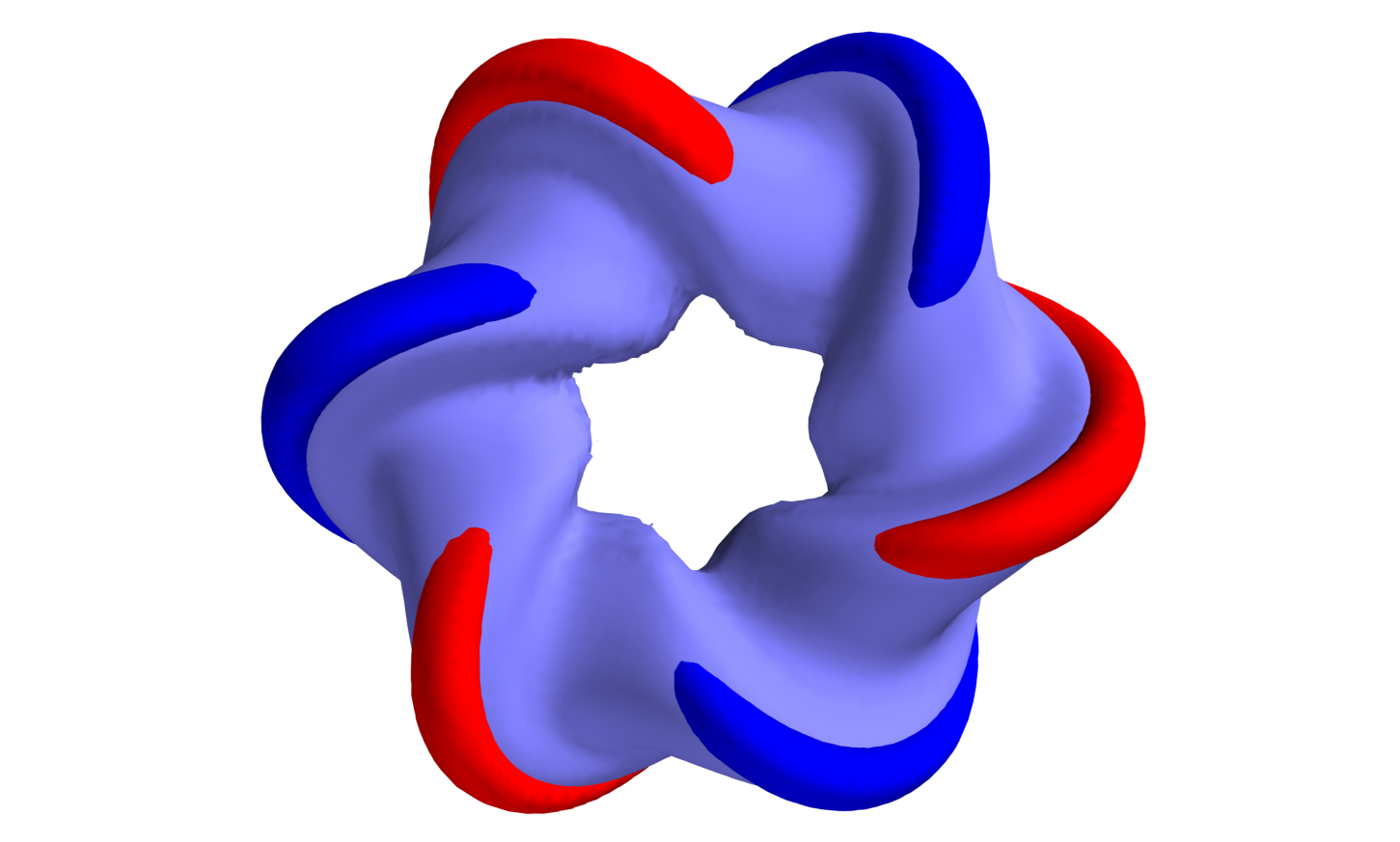}
        \includegraphics[height=5cm]{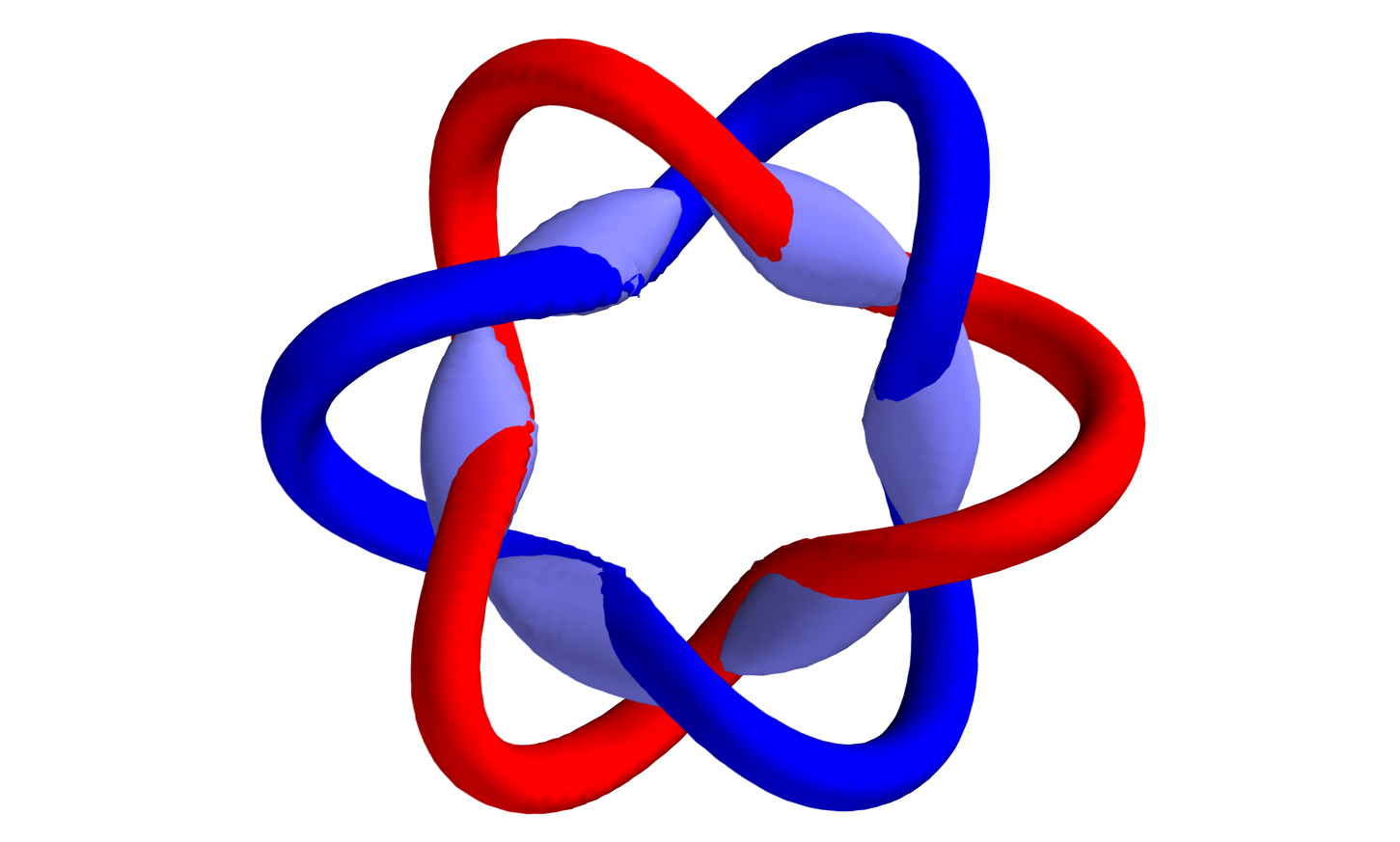}
    \end{center}
    \caption{Isosurface of the energy density of the $3(\widetilde \CA_1\between \widetilde \CA_1)_{\CA_1}$
            Hopfion in the Faddeev-Skyrme model \re{hopfenergy}
            with the potential $V=m^2{\phi_1}^2$ at $c_2=0.5, c_4=1, m=4$ and $\mathcal{E}=90$ together with
            the location curve of the preimage of the point $\vec \phi=(0, 0, -0.9)$ (upper plot).
            Bottom row displays these isosutfaces together
            with the isosurfaces of the preimages of the points $\vec \phi=(\pm 0.9, 0,0)$ for $\mathcal{E}=50$ (left) and
            $\mathcal{E}=90$ (right).
    }
    \label{q3combine}
\end{figure}

For Hopf degree $Q=3$ the minimum-energy configuration in the model with the symmetry breaking potential
$m^2{\phi_1}^2$ is presented in Fig.~\ref{q3combine}. Similar to the corresponding solutions in the sectors of degrees one and two,
cf. Figs.~\ref{q2combine}, \ref{q1combine}, we displayed there the
energy isosurface of the $Q=3$ configuration $3(\widetilde \CA_1\between \widetilde \CA_1)_{\CA_1}$
together with the location curves of the field components. Clearly, the energy density distribution follows the location curves
$\mathcal{C}_1 = \vec \phi^{-1}(1,0,0)$ and $\mathcal{C}_{-1} = \vec \phi^{-1}(-1,0,0)$, as above.
Note that for the higher values of the energy, the energy density isosurface has a form of a bead necklace with six beads.
The antipodal to the vacuum curve $\mathcal{C} = \vec \phi^{-1}(0,0,-1)$, which is bounding these beads, is planar but it is
not exactly circular. Interestingly it has the hexagonal shape possessing dihedral symmetry $D_3$, see
Fig.~\ref{q3combine}, bottom-left plot.

\begin{table}
\begin{center}
\begin{TAB}[1pt]{|c|c|c|c|}{|c|c|c|c|c|c|c|c|c|}
Configuration & $\rho_E$ &  $\phi_1$ &  $\phi_3$  \\
$1(\CA_1\between\CA_1)_{\CA_{1,1}}$  & \includegraphics[height=2cm]{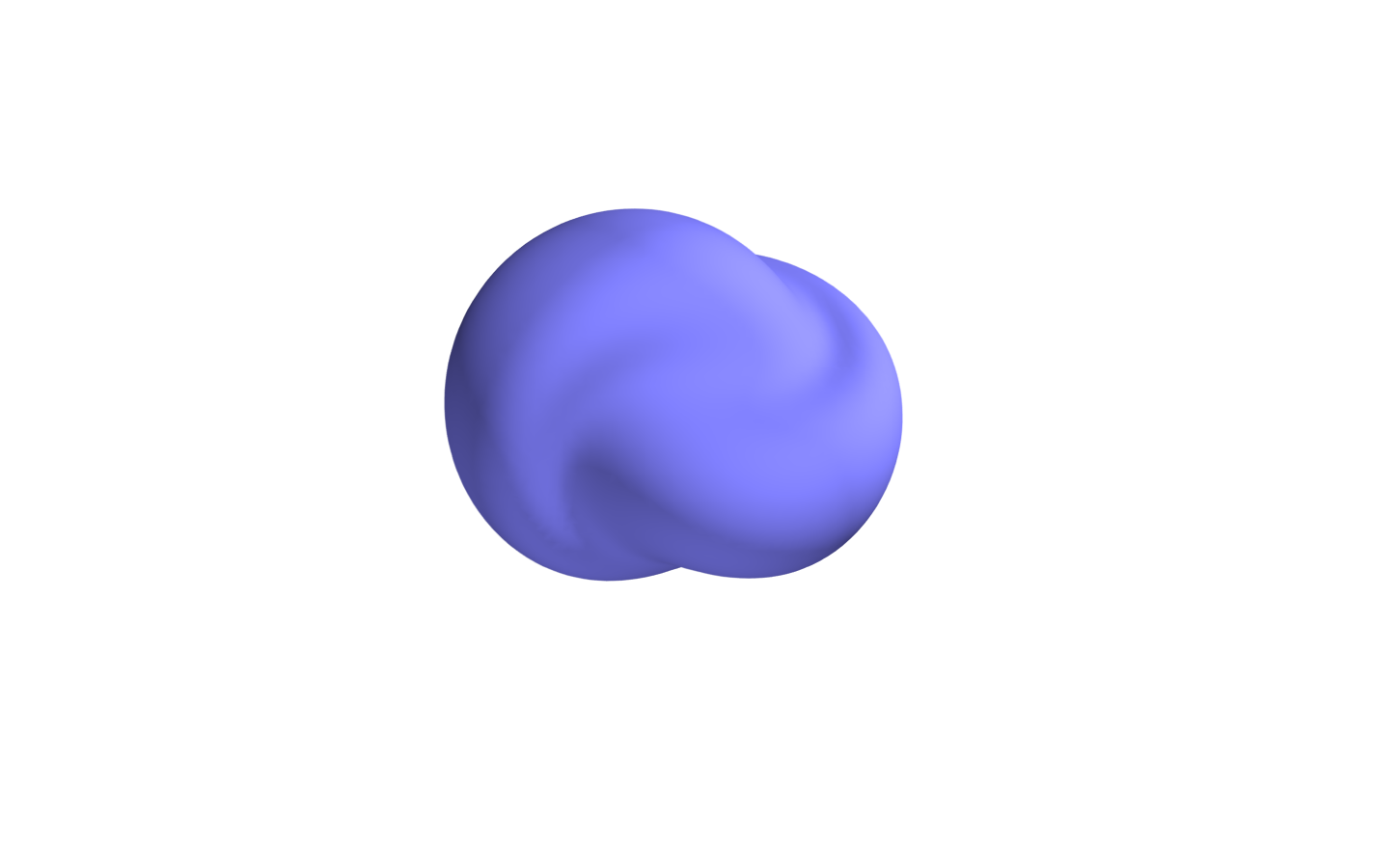} & \includegraphics[height=2cm]{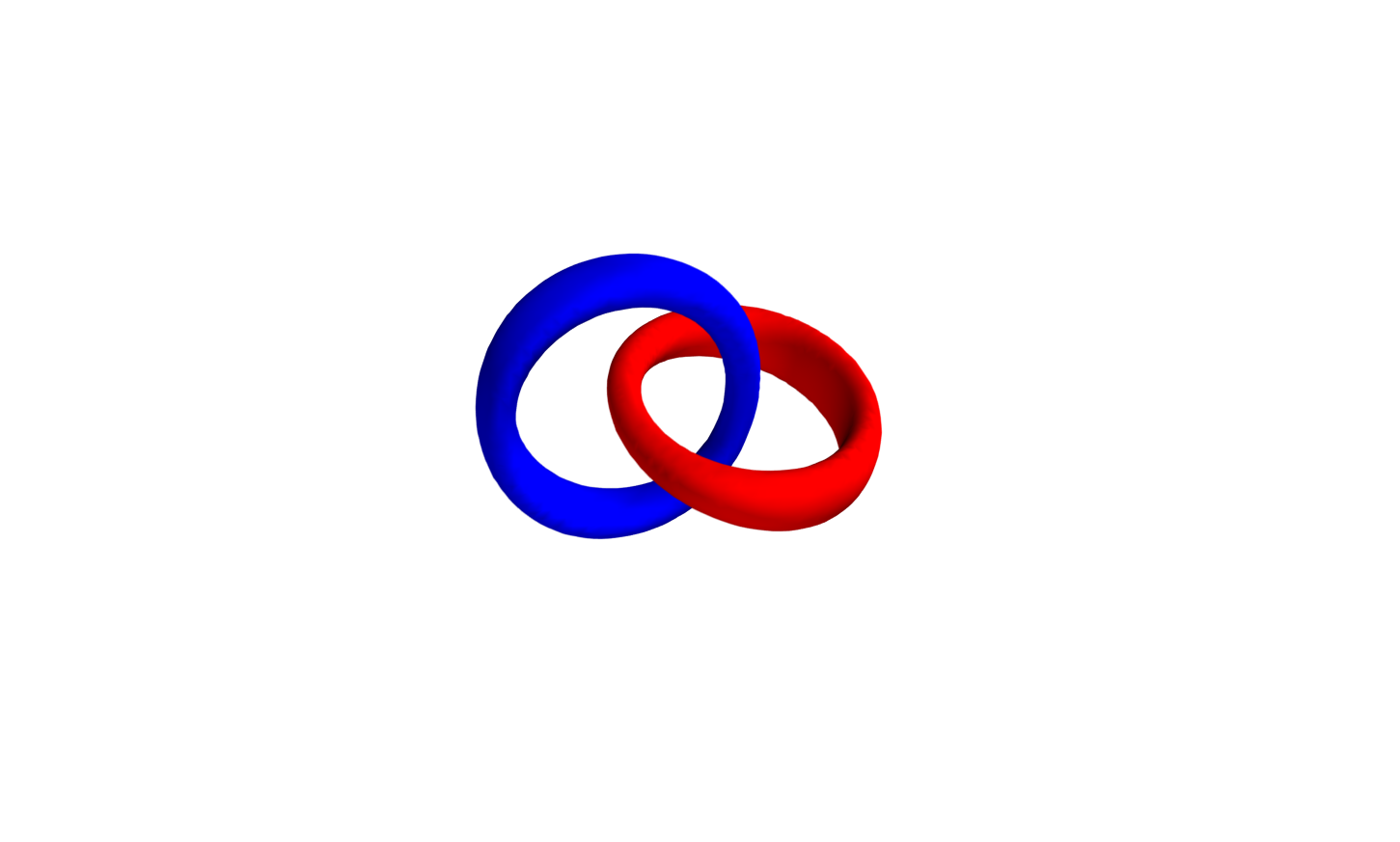} &
\includegraphics[height=2cm]{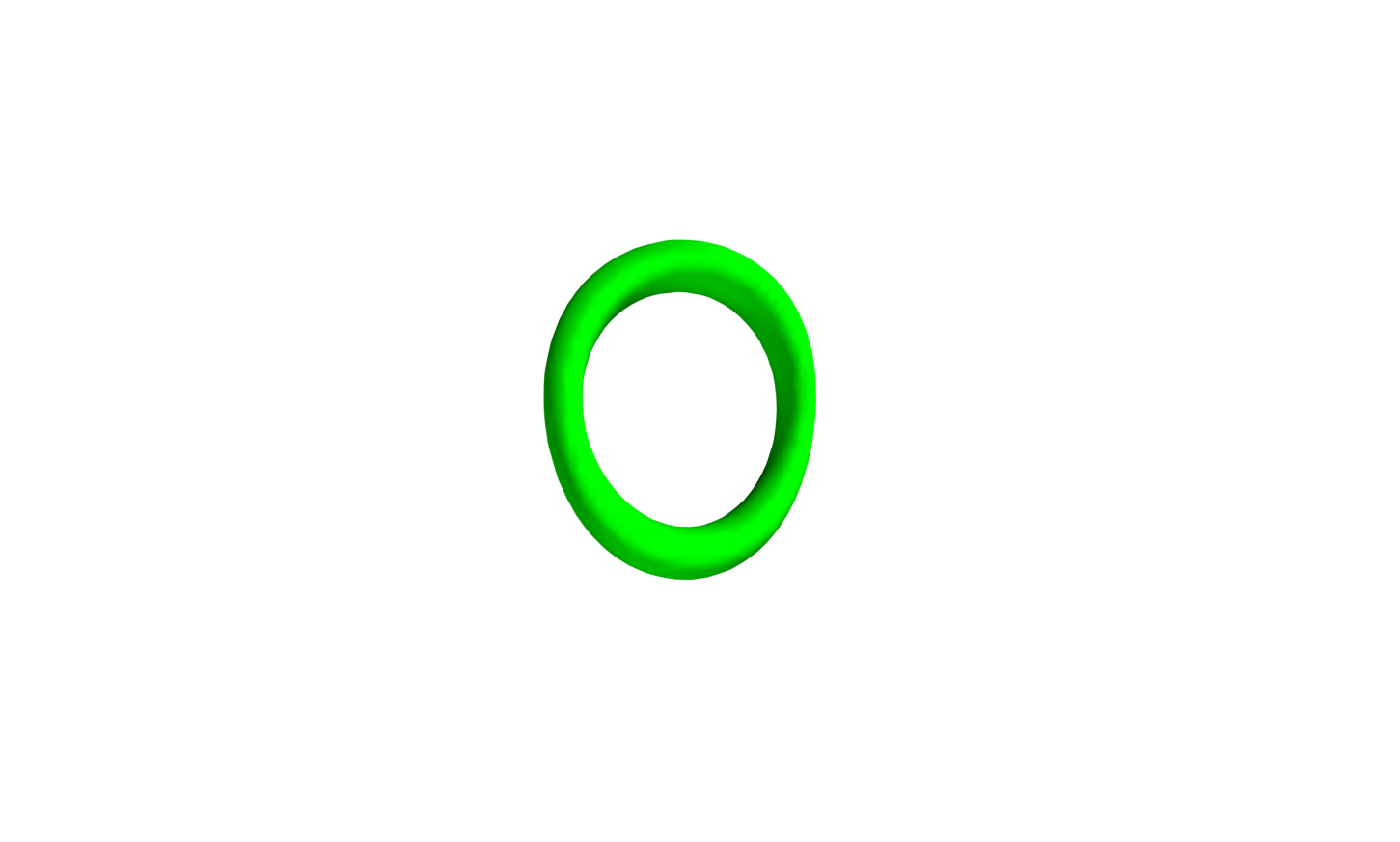}\\
$2(\CA_1\between\CA_1)_{\CA_{2,1}}$  & \includegraphics[height=2cm]{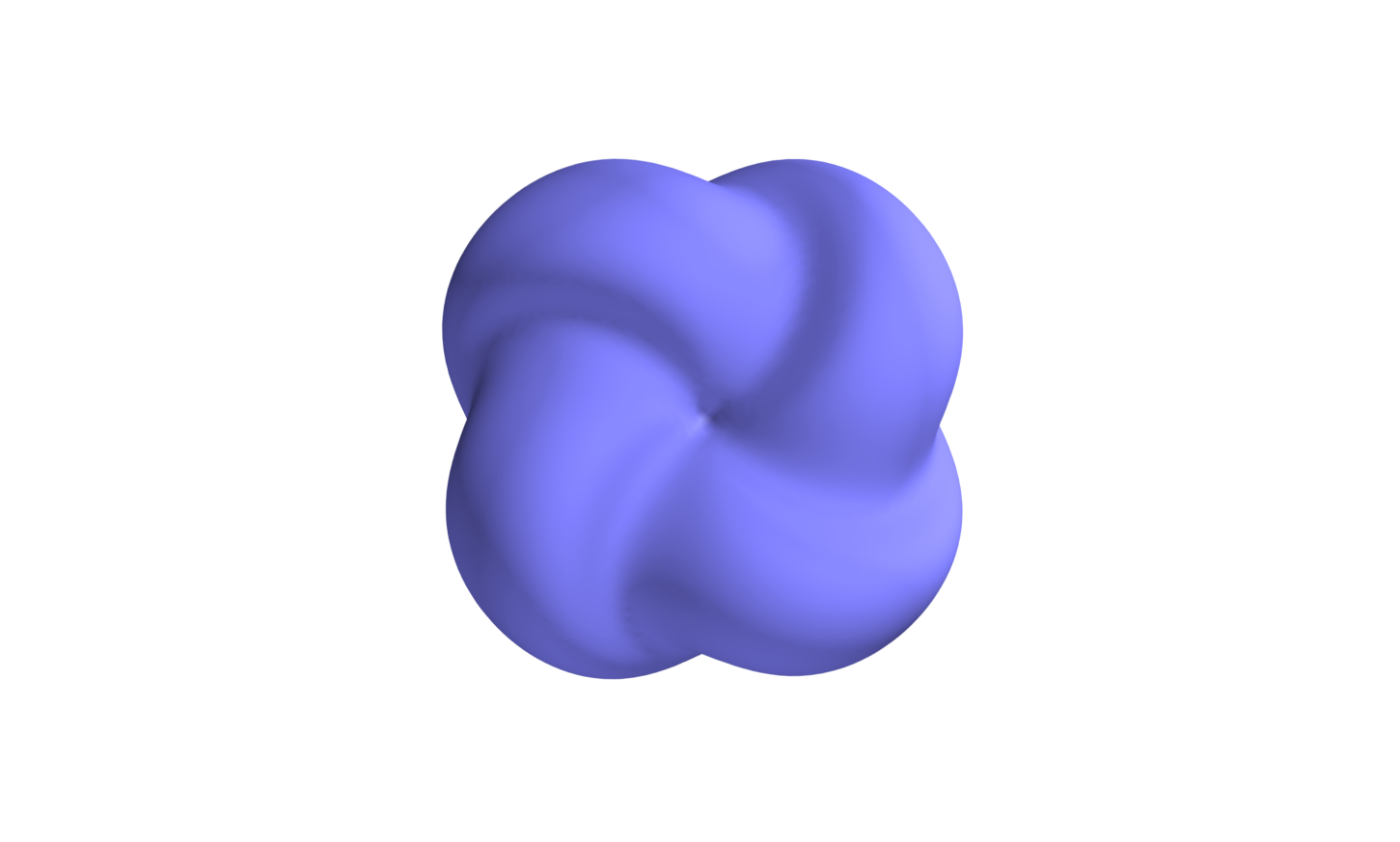} & \includegraphics[height=2cm]{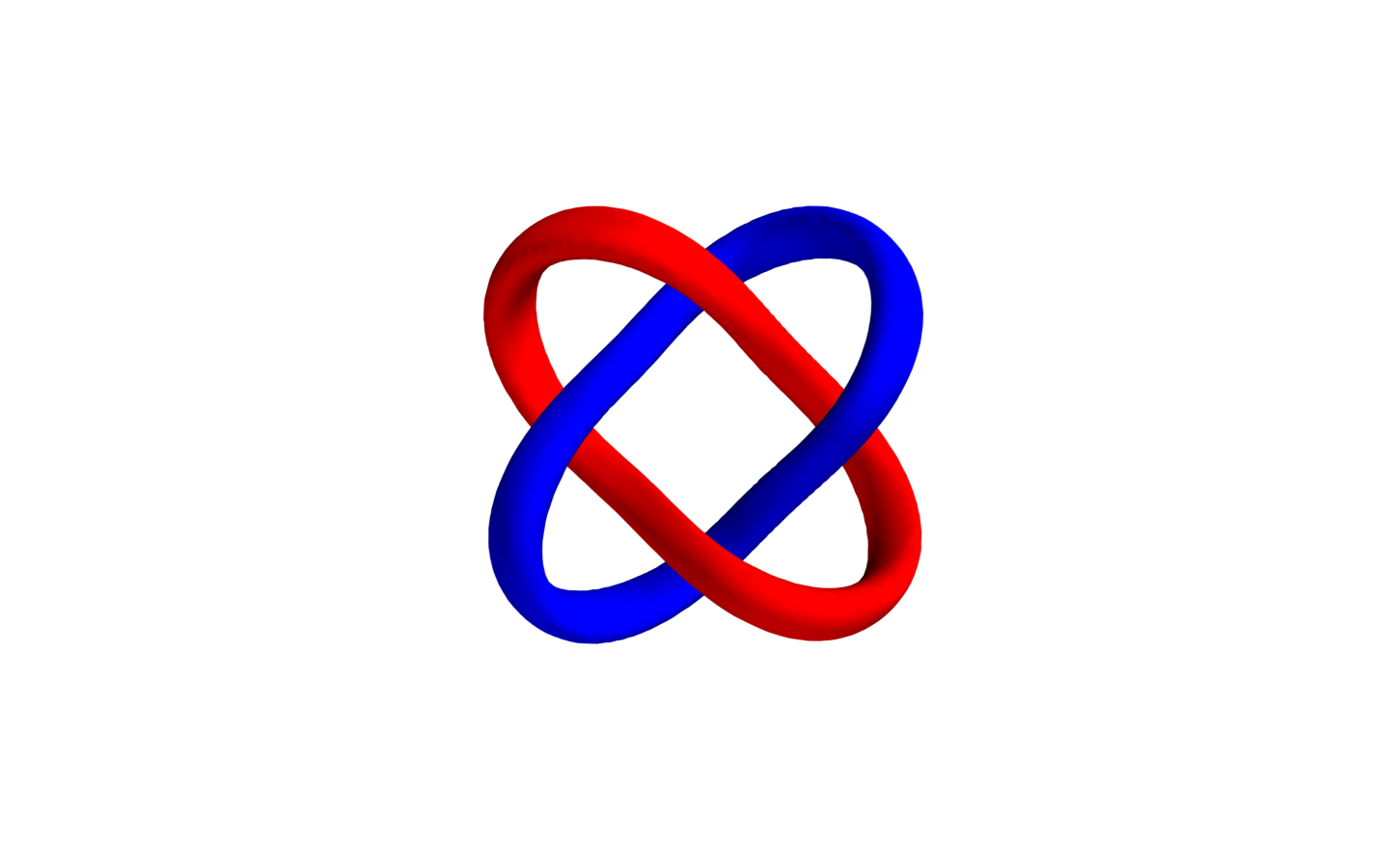} & \includegraphics[height=2cm]{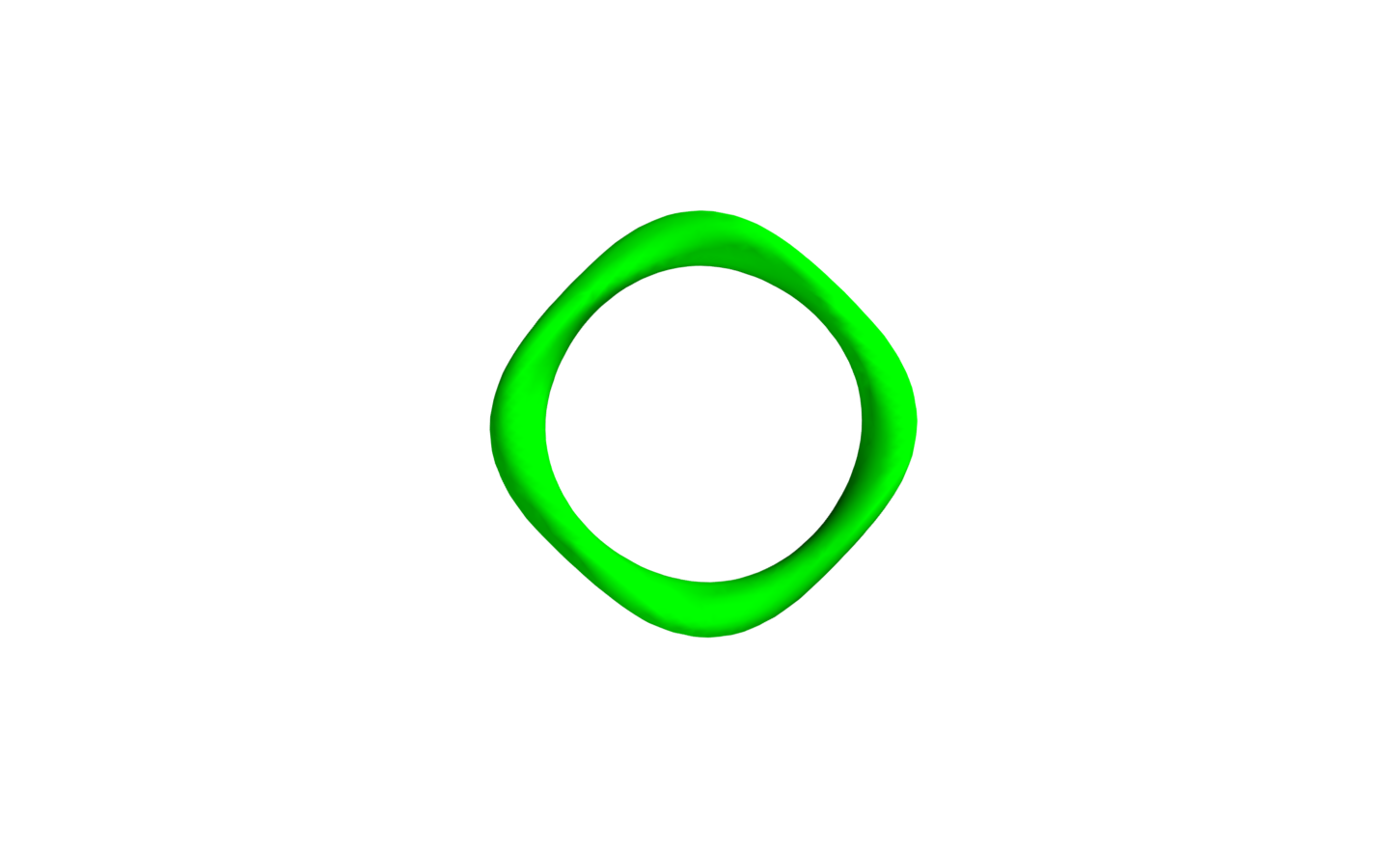} \\
$2(\widetilde\CA_1\between\widetilde\CA_1)_{\CA_{1,2}}$ &  \includegraphics[height=2cm]{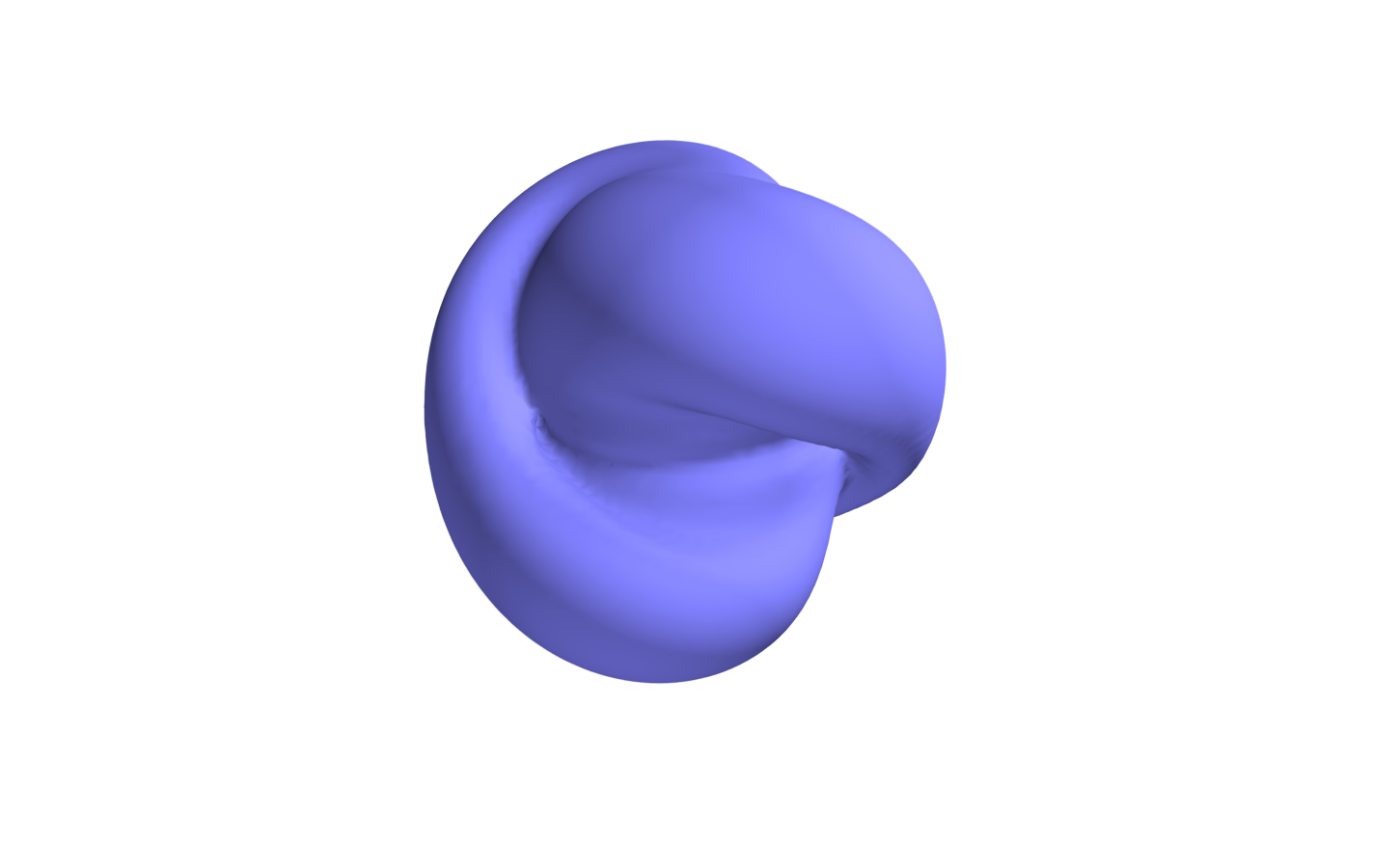} & \includegraphics[height=2cm]{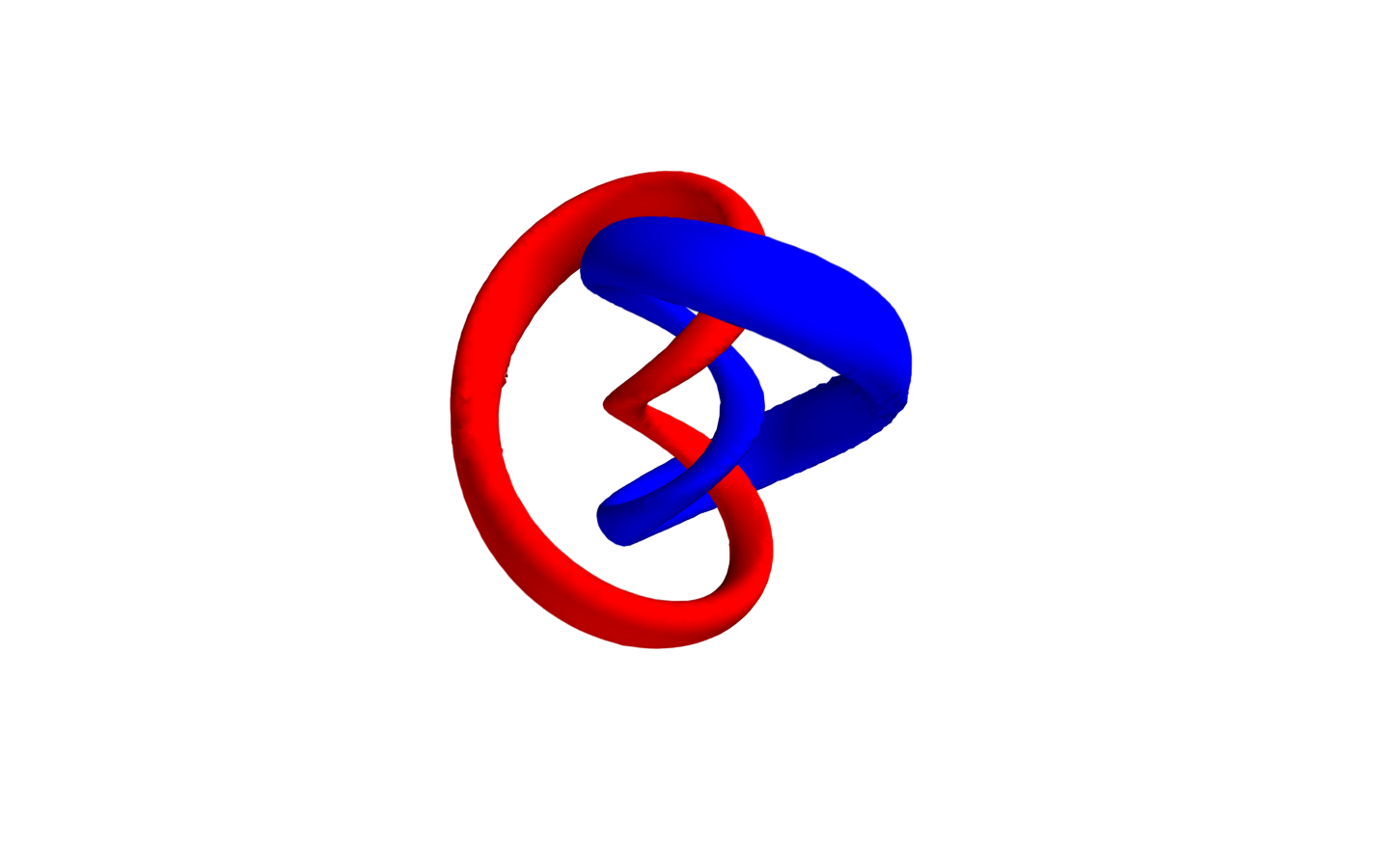} & \includegraphics[height=2cm]{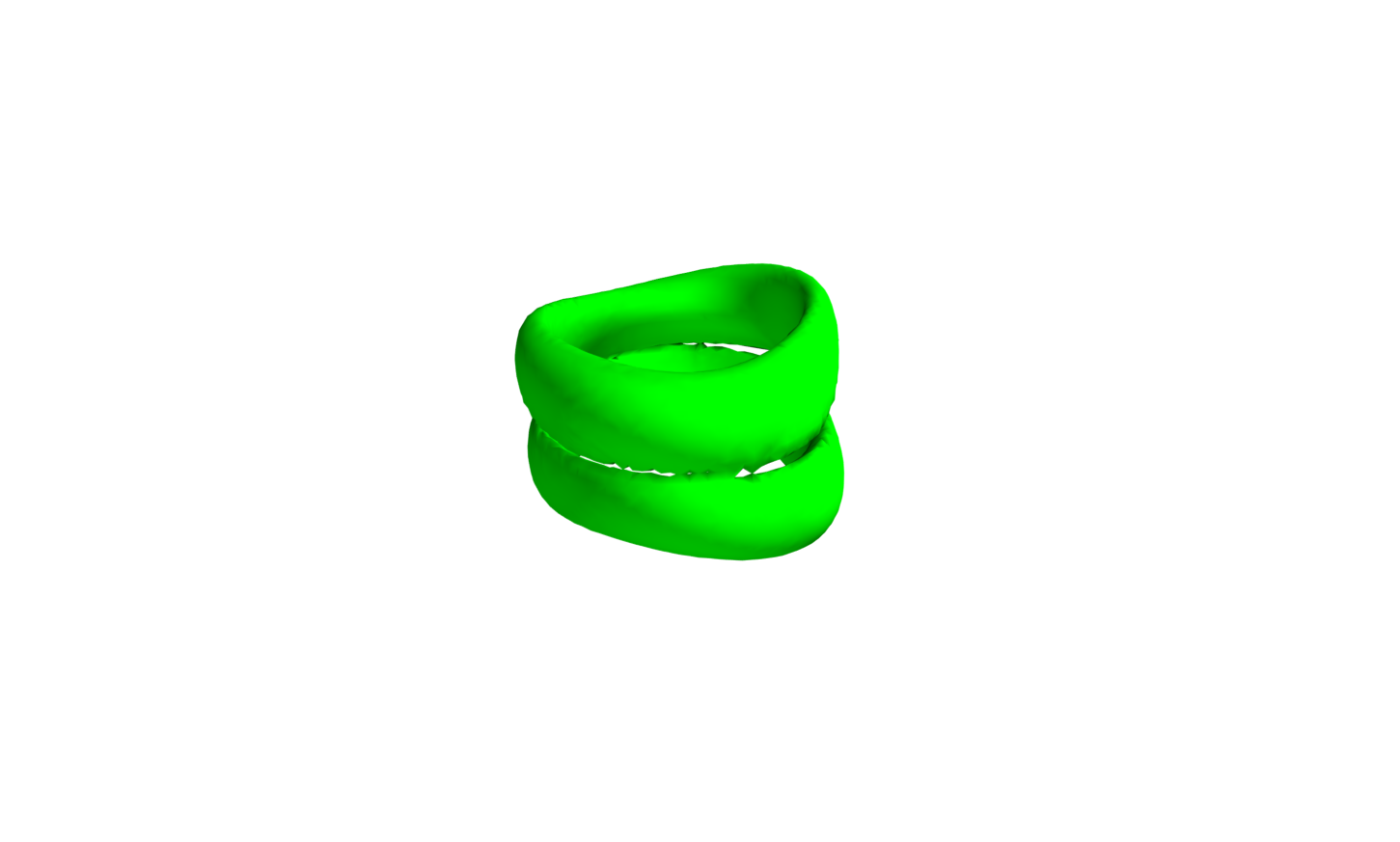} \\
$3(\CA_1\between\CA_1)_{\CA_{3,1}}$ &  \includegraphics[height=2cm]{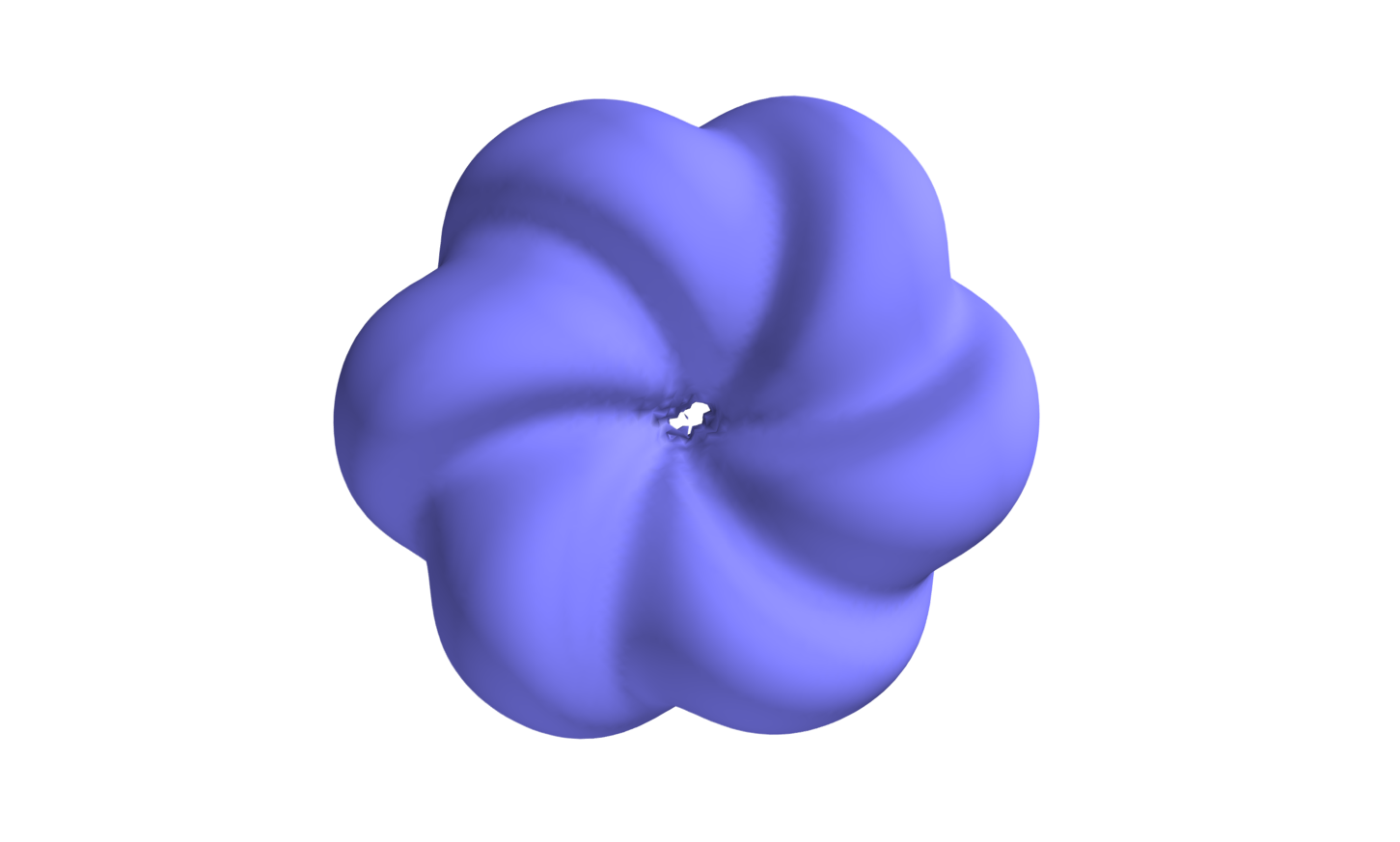} & \includegraphics[height=2cm]{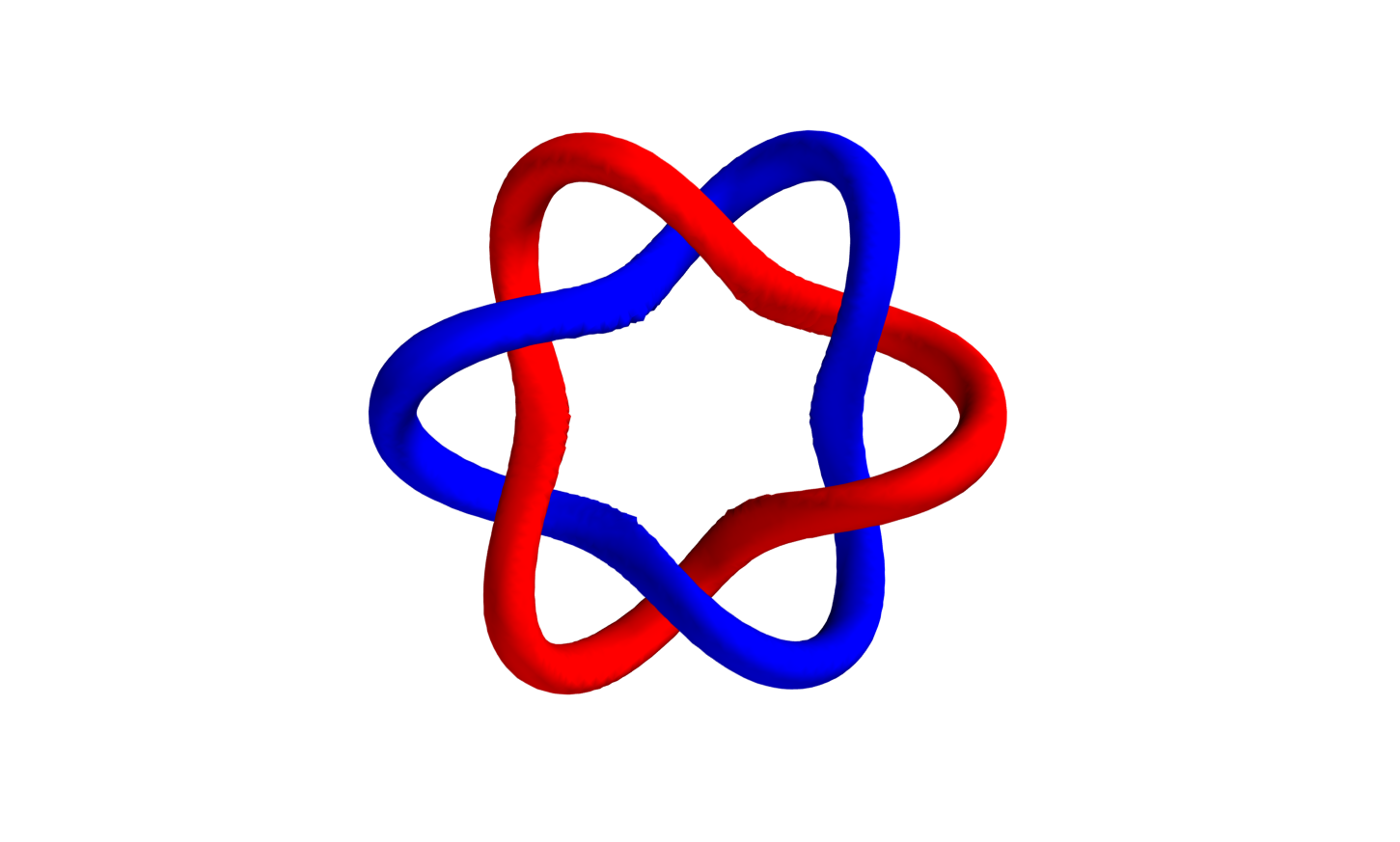} & \includegraphics[height=2cm]{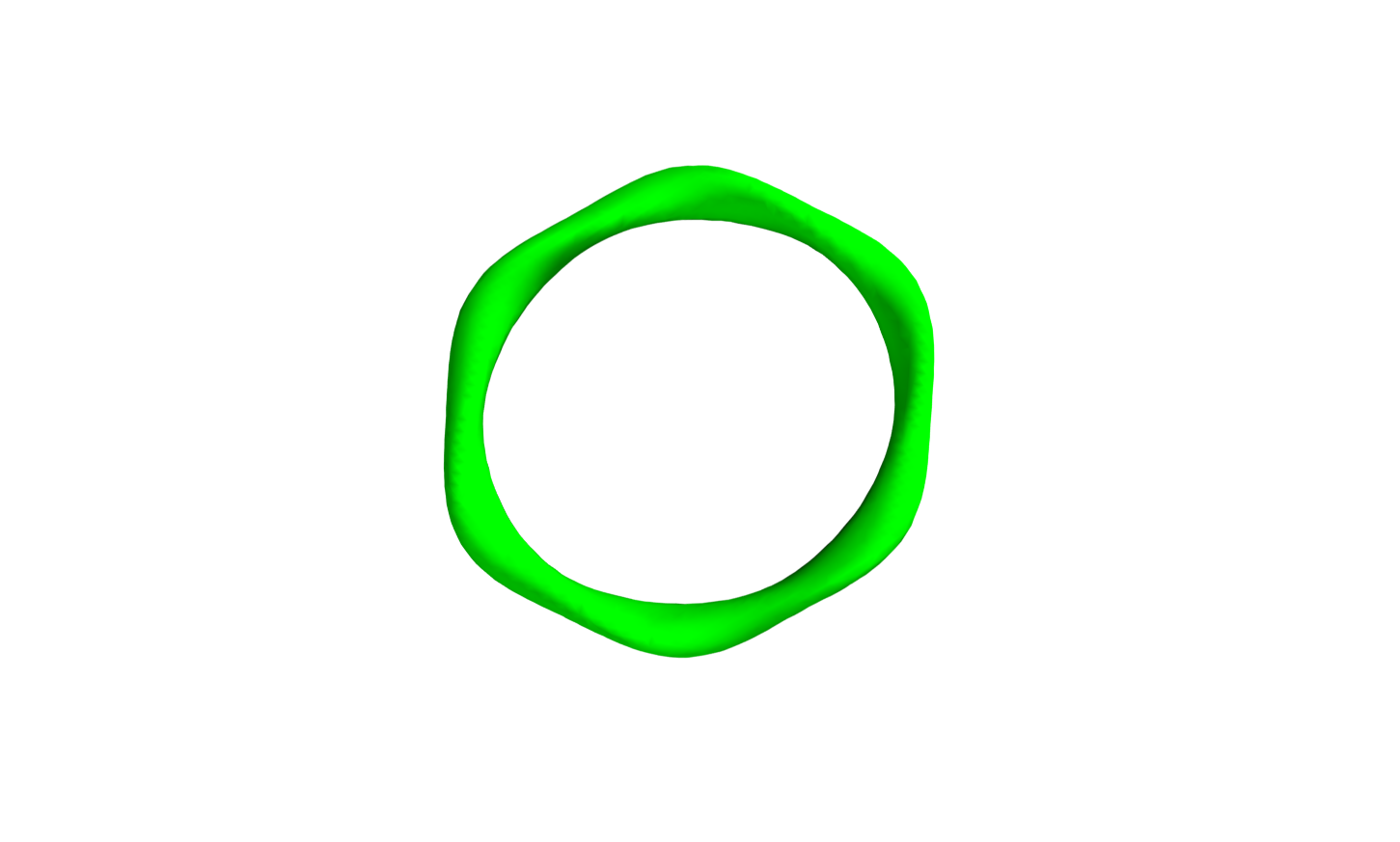}\\
$3(\widetilde{\CA}_1\between\widetilde{\CA}_1)_{\widetilde\CA_{3,1}}$ & \includegraphics[height=2cm]{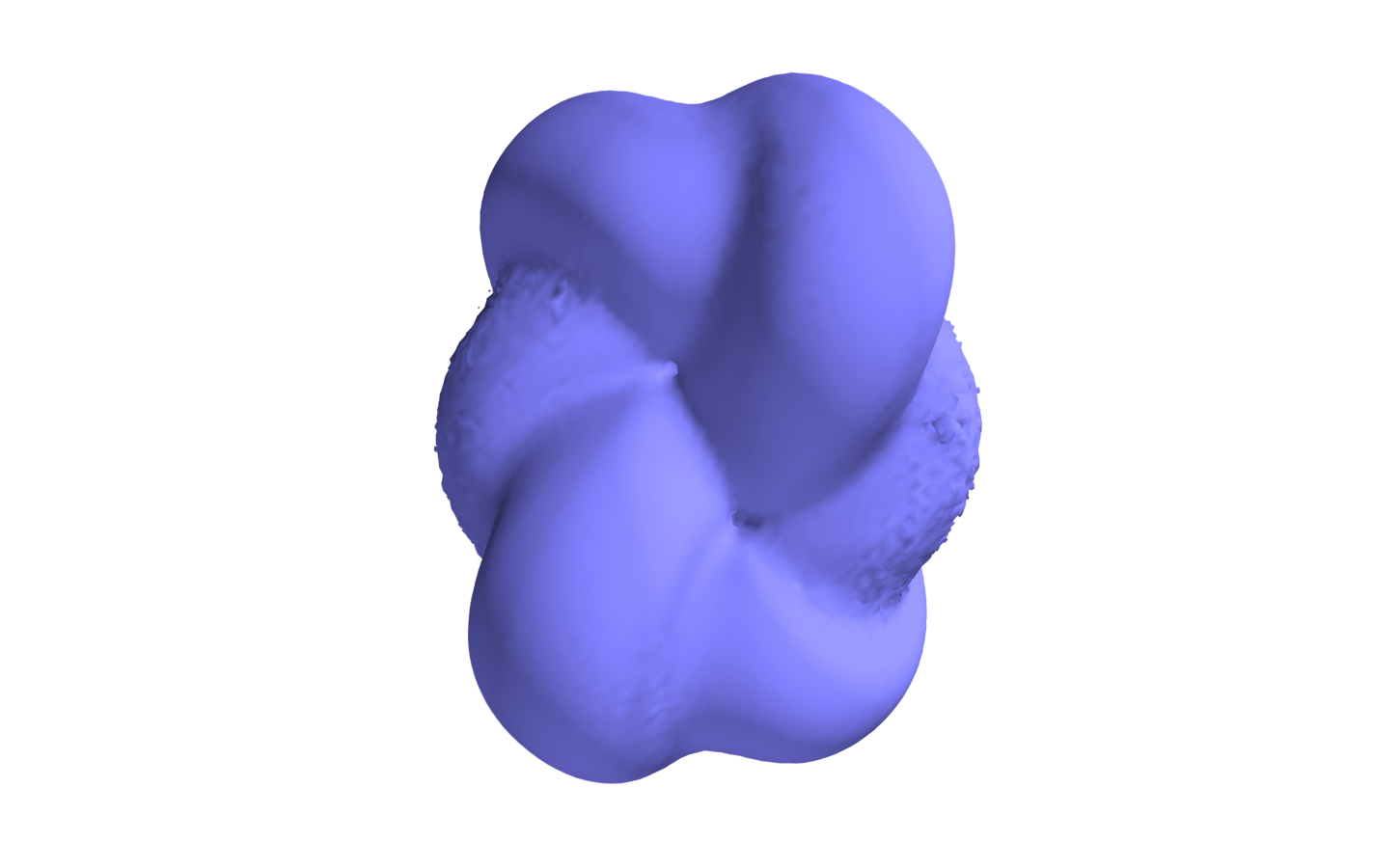} & \includegraphics[height=2cm]{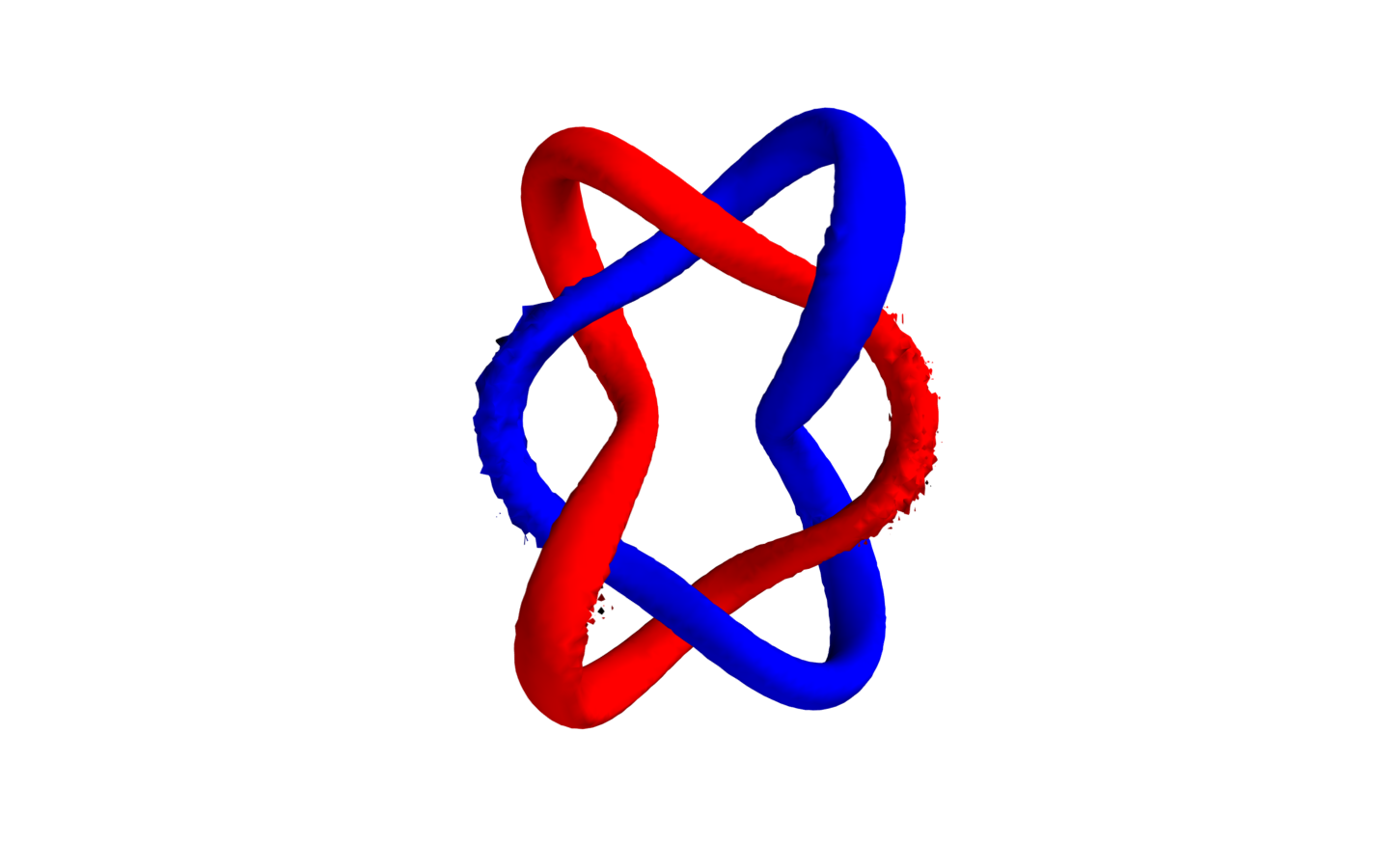} & \includegraphics[height=2cm]{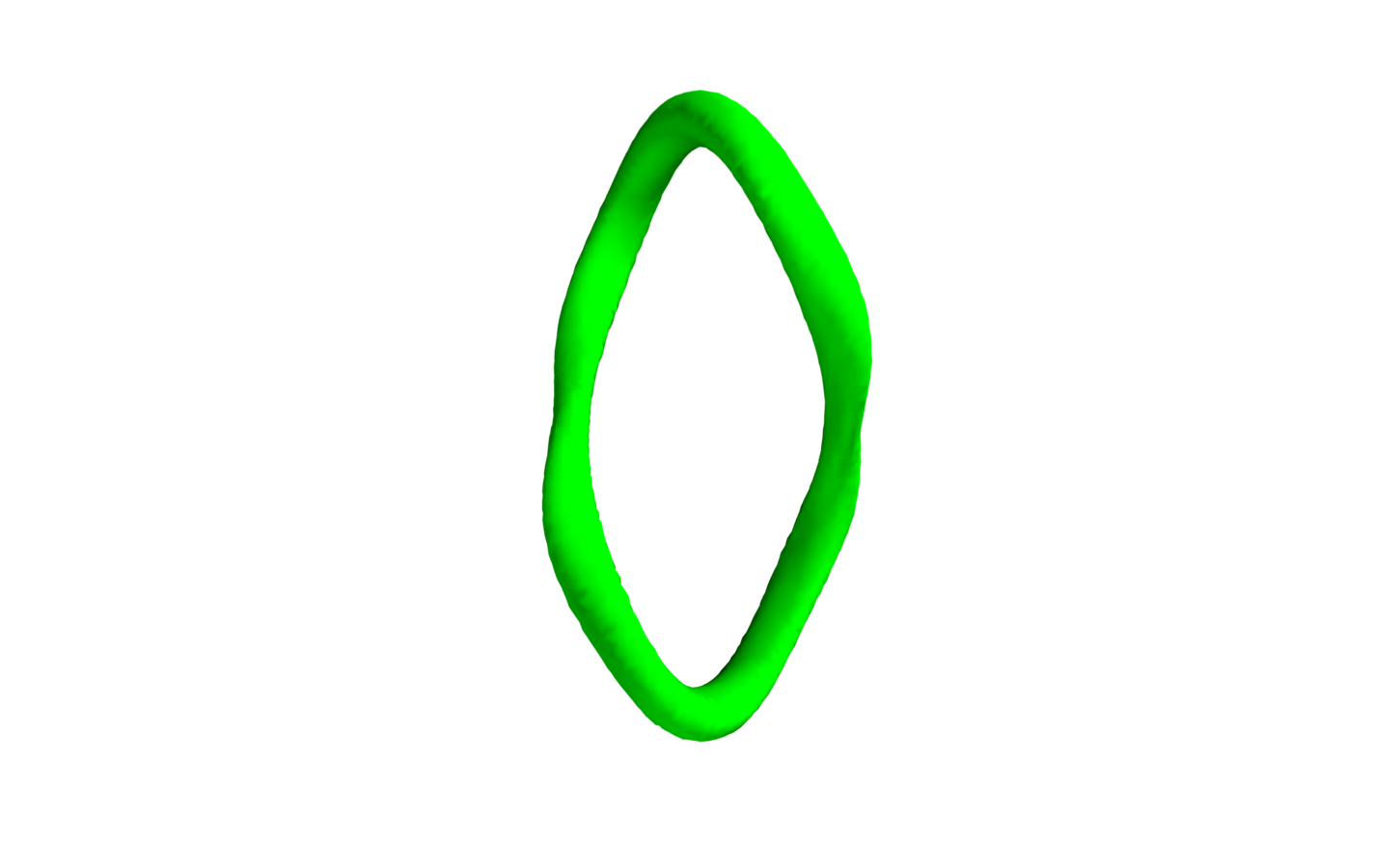} \\
$4(\CL_{1,1}\between\CL_{1,1})_{\CA_{2,2}}$ & \includegraphics[height=2cm]{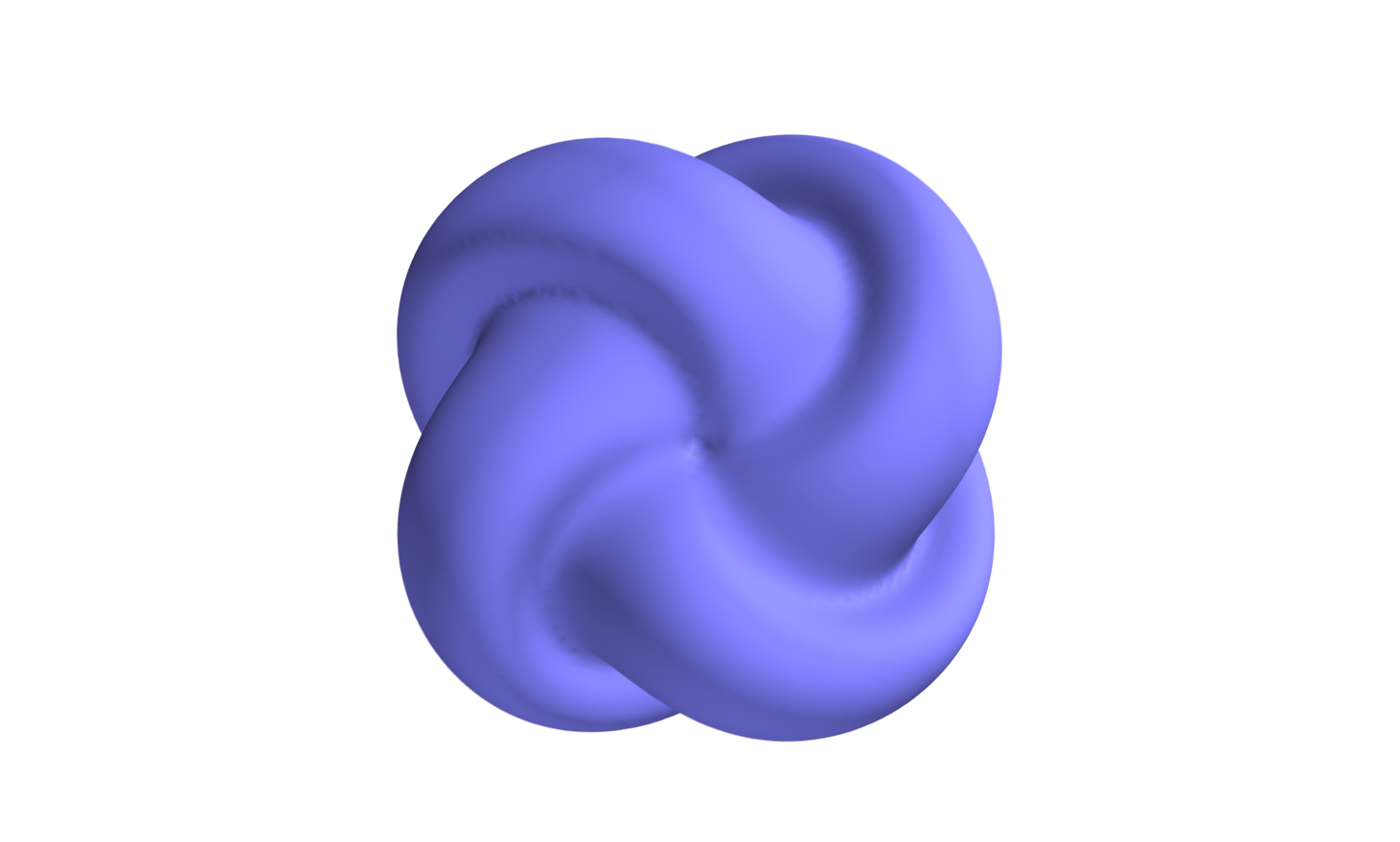} & \includegraphics[height=2cm]{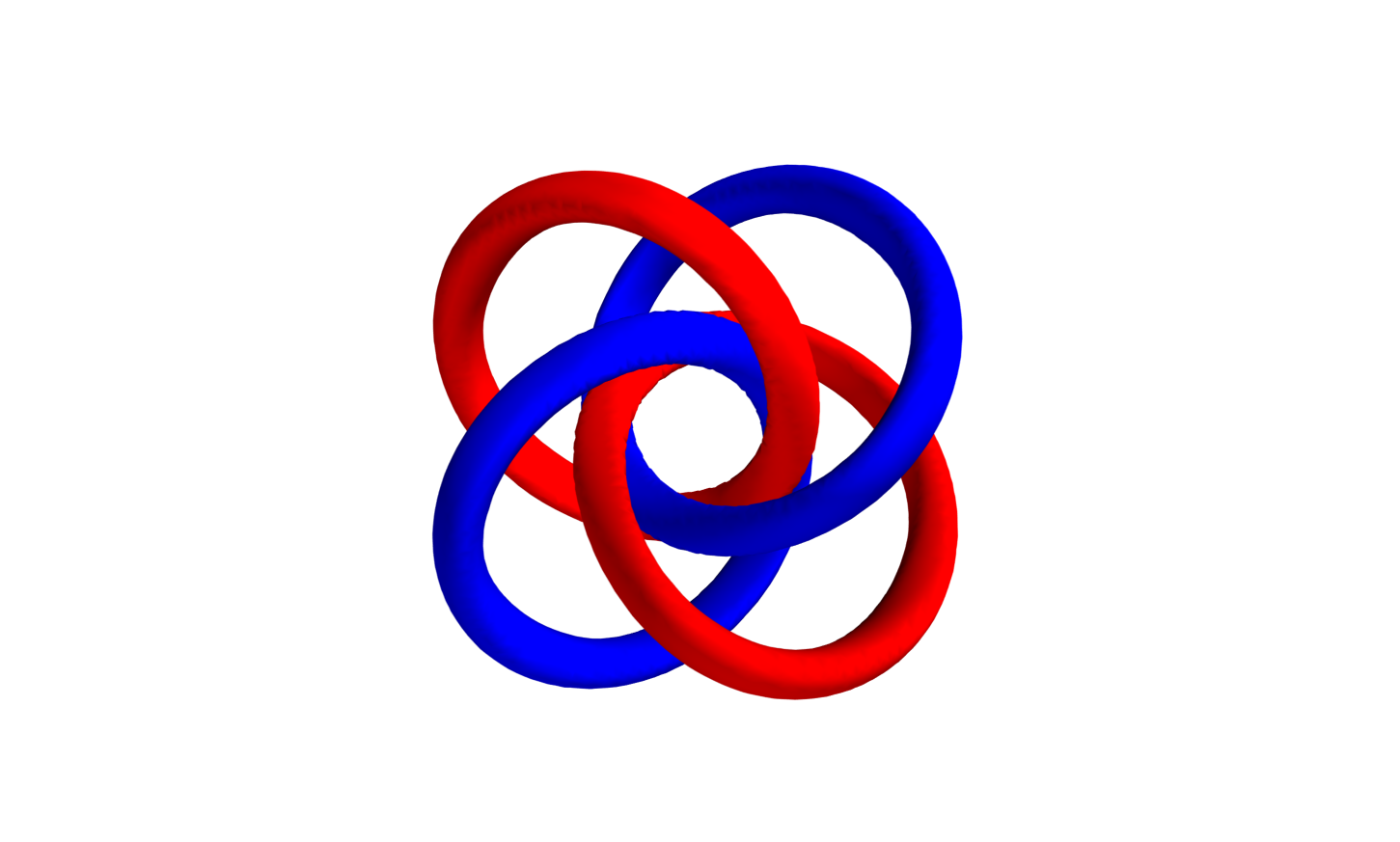} &
\includegraphics[height=2cm]{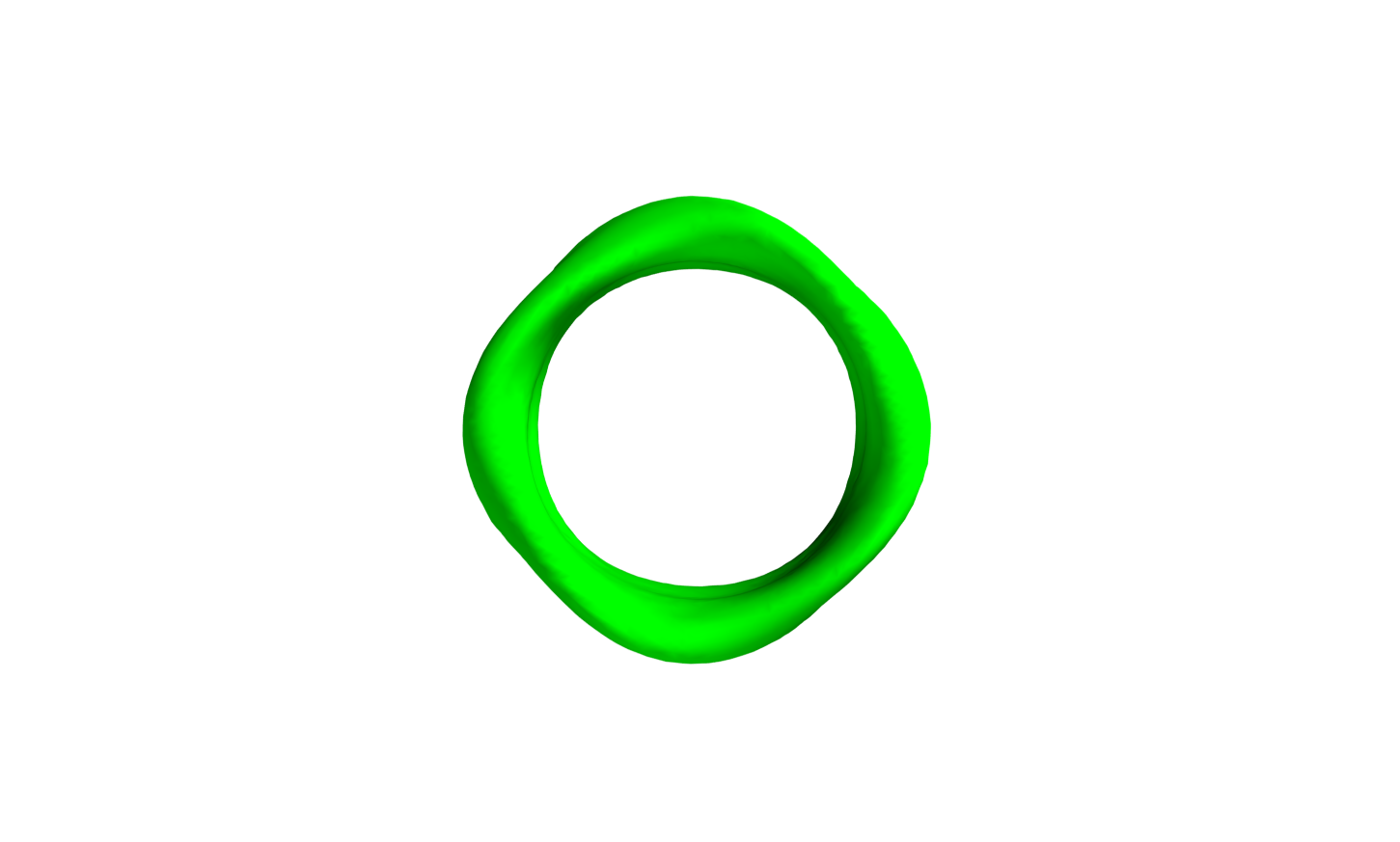}\\
$4(\widetilde\CA_1\between\widetilde\CA_1)_{\widetilde\CA_{4,1}}$ & \includegraphics[height=2cm]{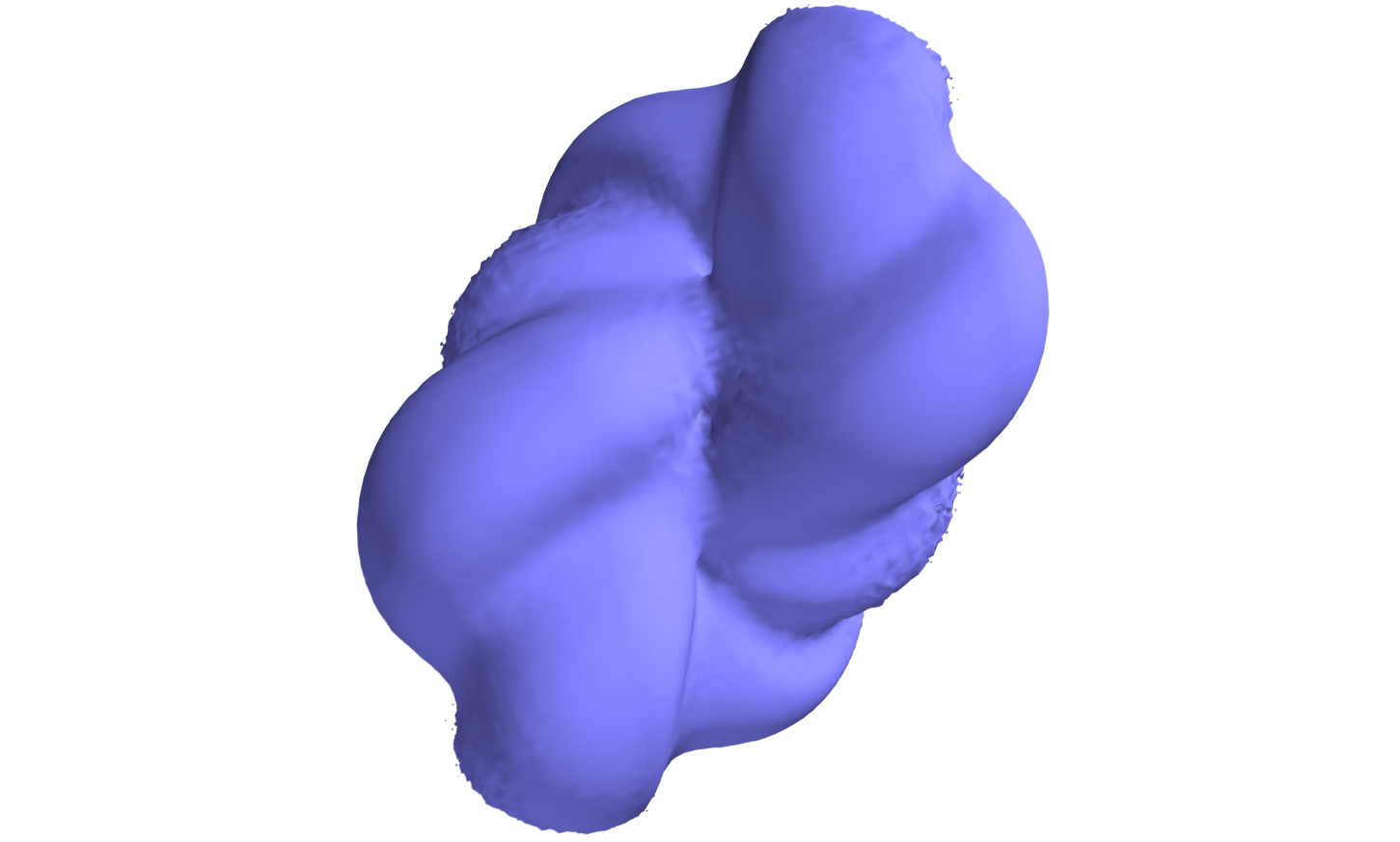} & \includegraphics[height=2cm]{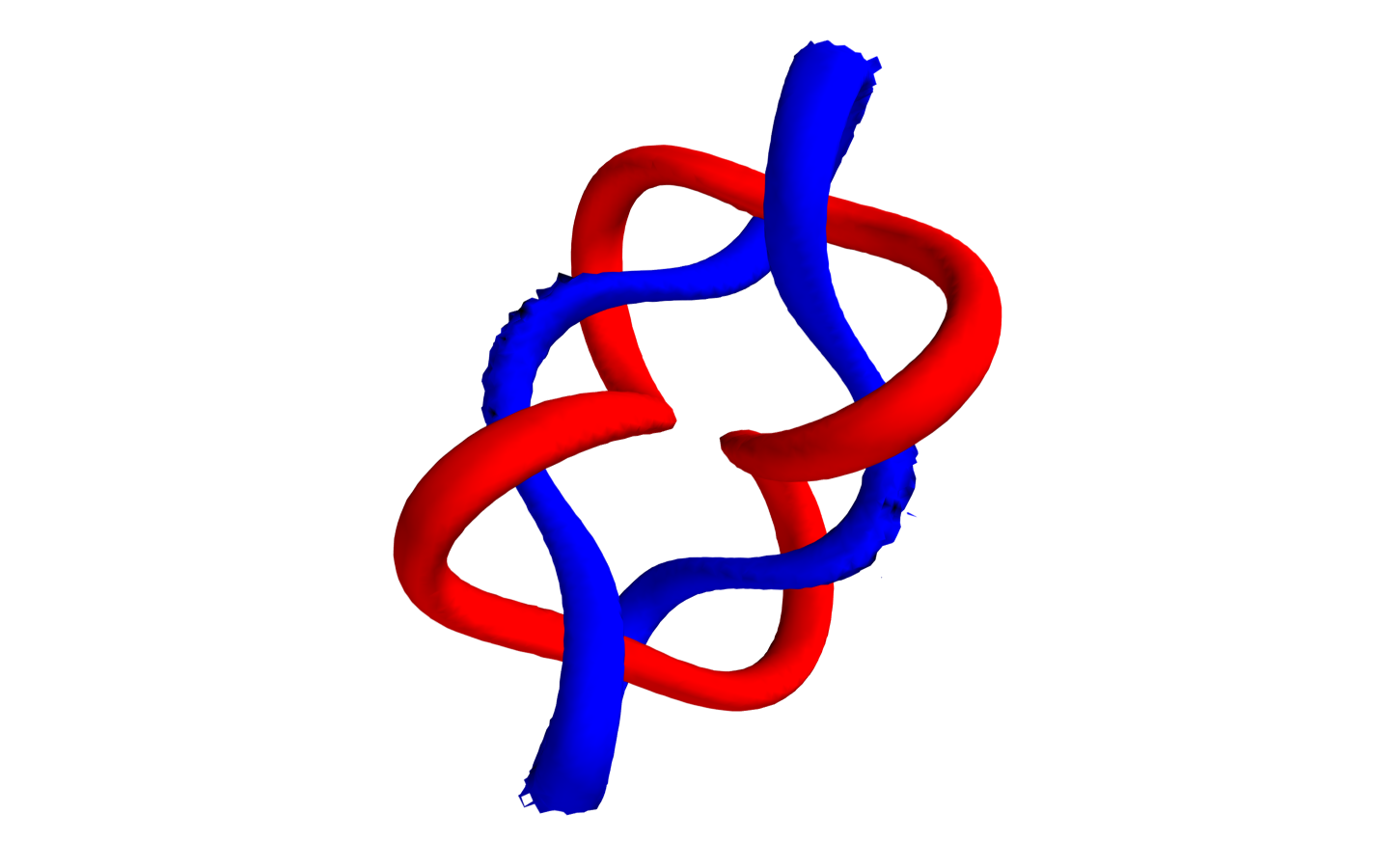} & \includegraphics[height=2cm]{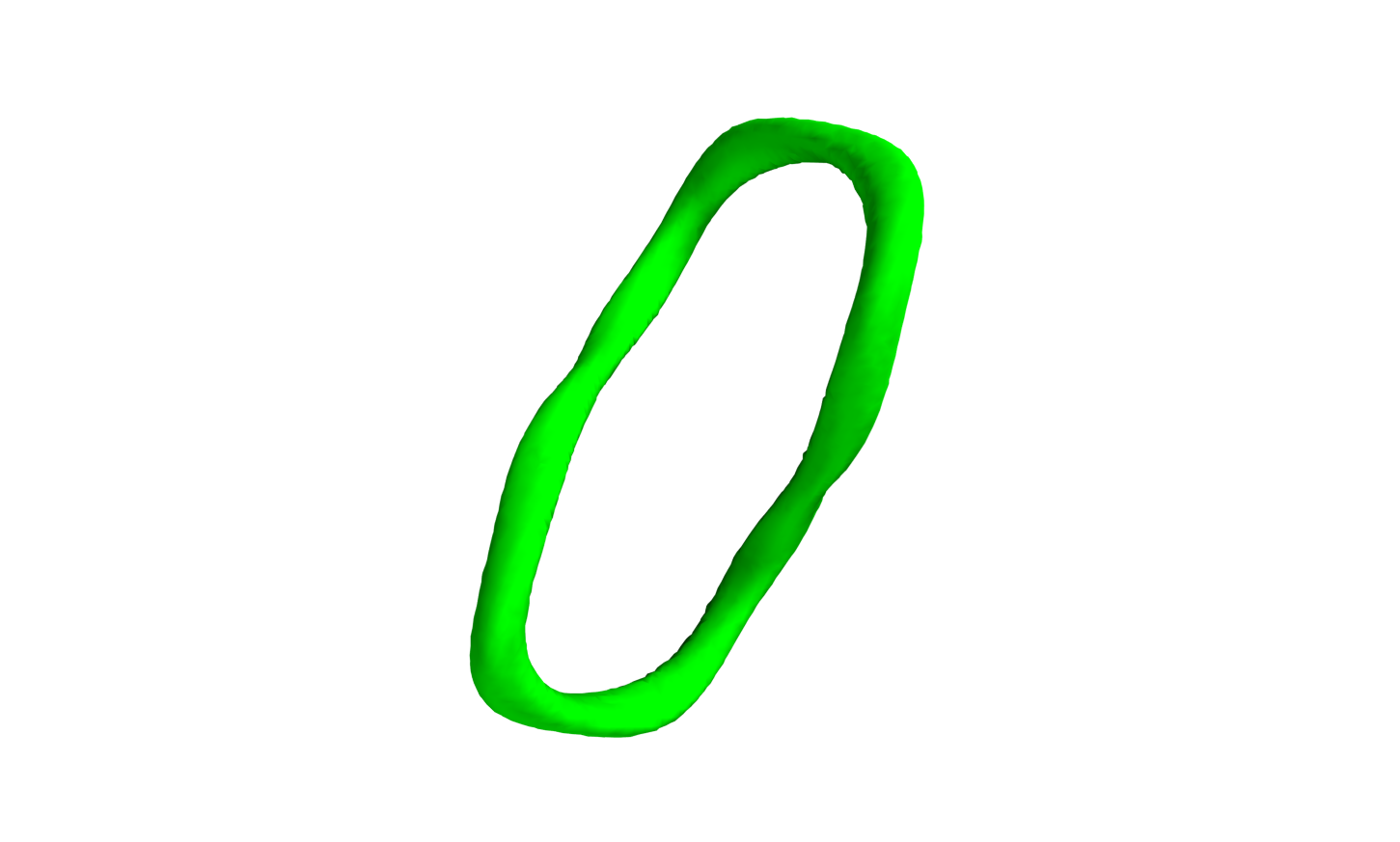} \\
$4(\CA_1\between\CA_1)_{\CA_{4,1}}$ & \includegraphics[height=2cm]{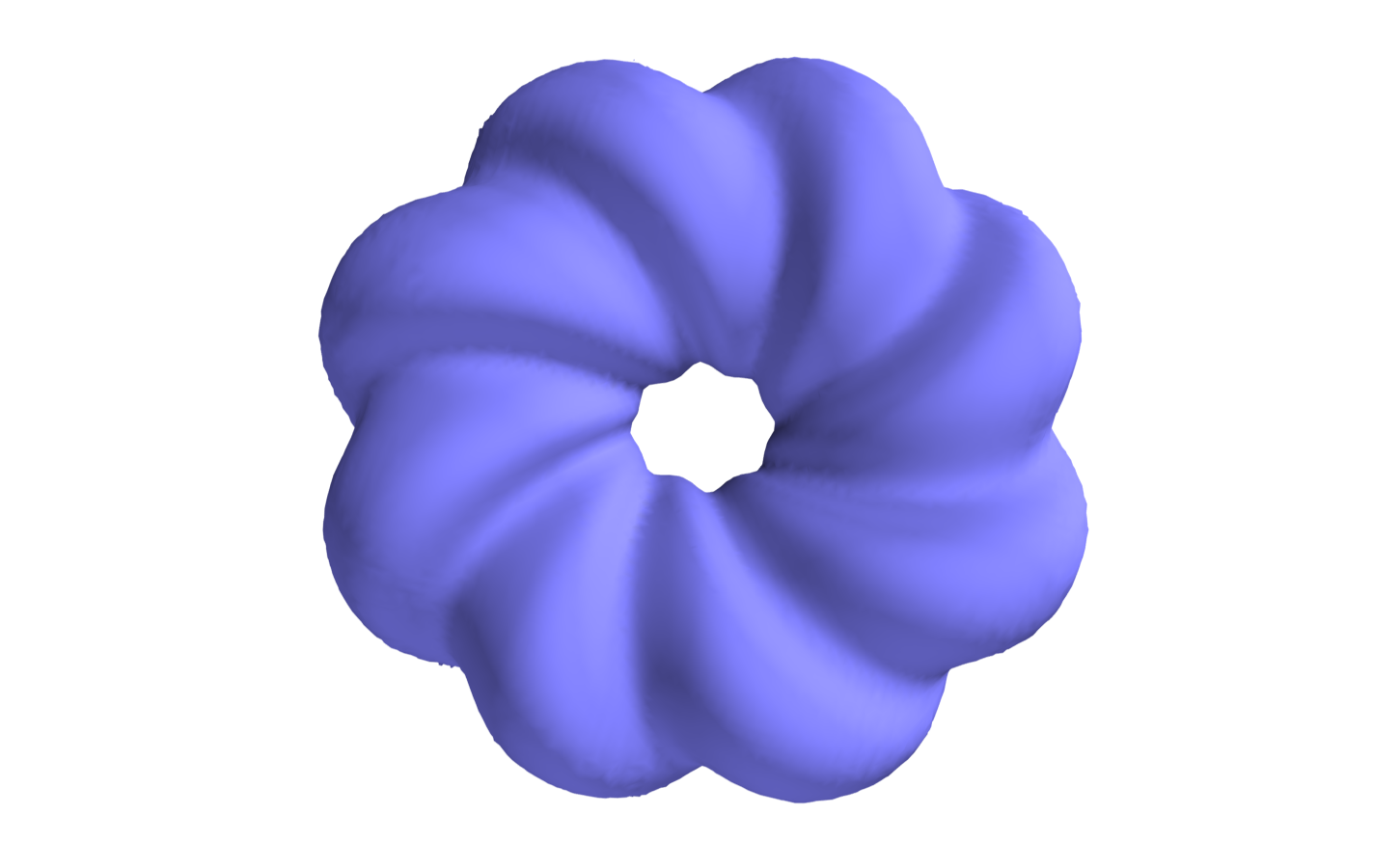} & \includegraphics[height=2cm]{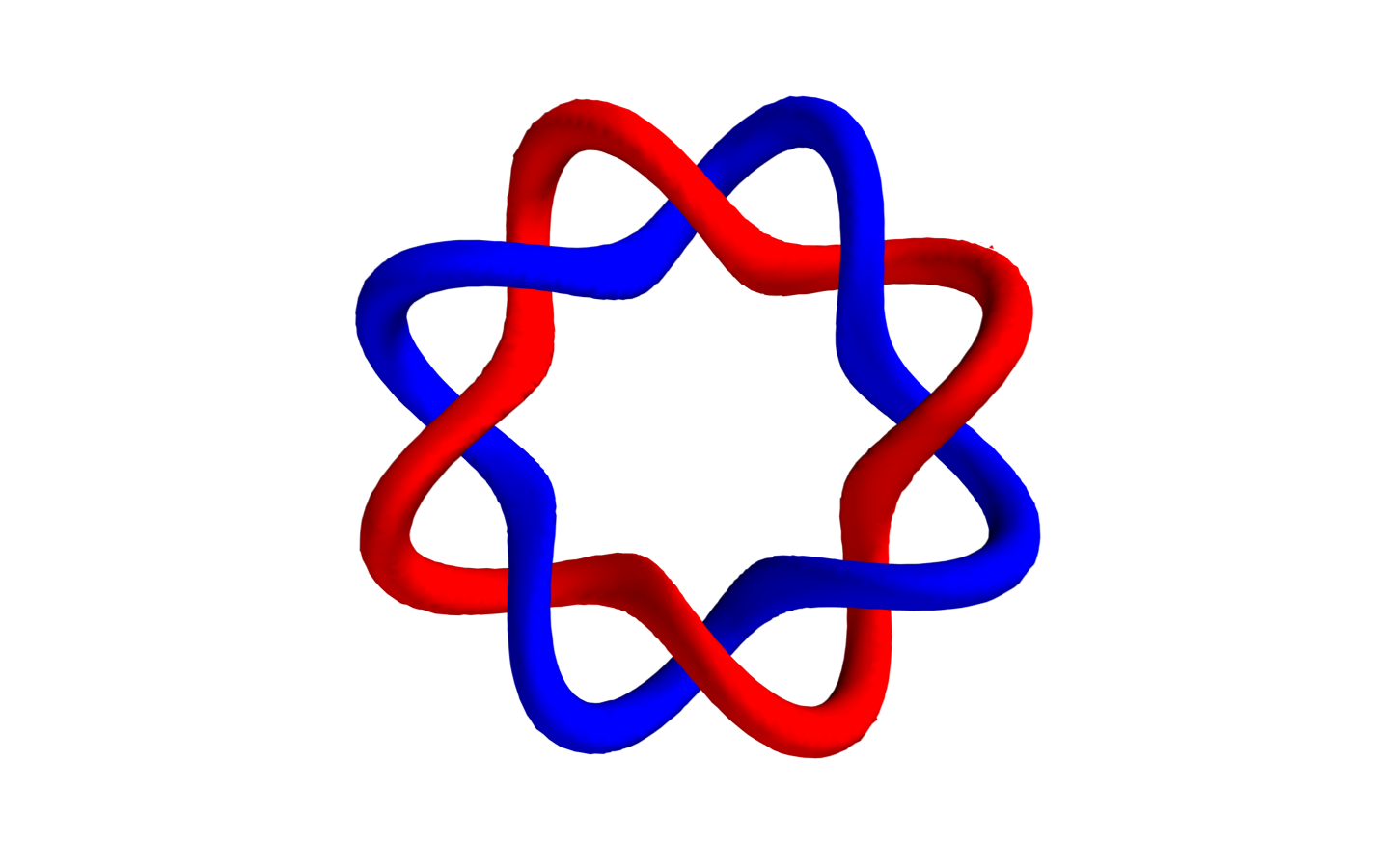} & \includegraphics[height=2cm]{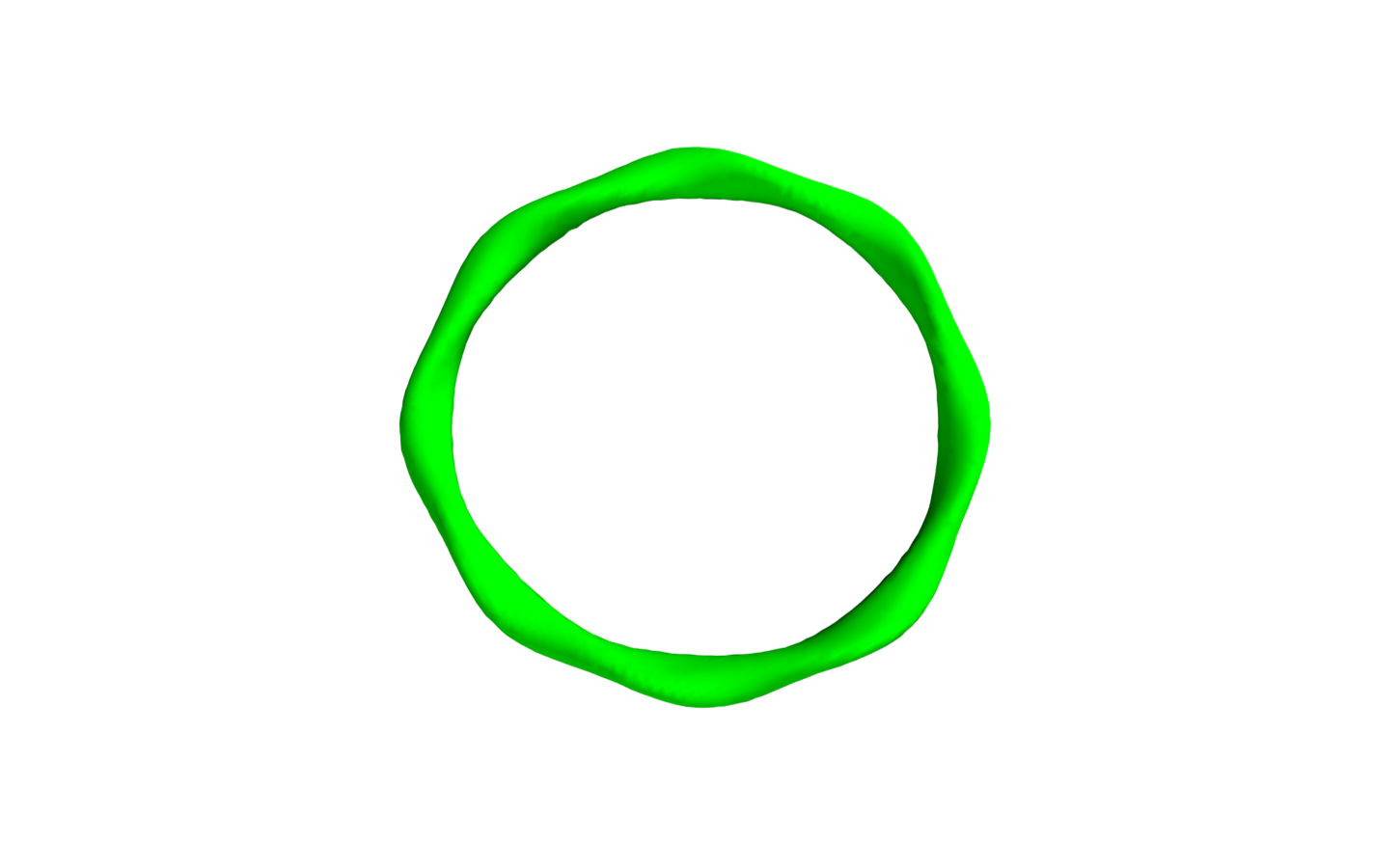} \\
\end{TAB}
\end{center}
\end{table}

\begin{table}
\begin{center}
\begin{TAB}[1pt]{|c|c|c|c|}{|c|c|c|c|c|c|c|c|c|}
Configuration & $\rho_E$ &  $\phi_1$ &  $\phi_3$  \\
$5(\CL_{1,2}\between\CL_{1,2})_{\CL_{1,2}}$  & \includegraphics[height=2cm]{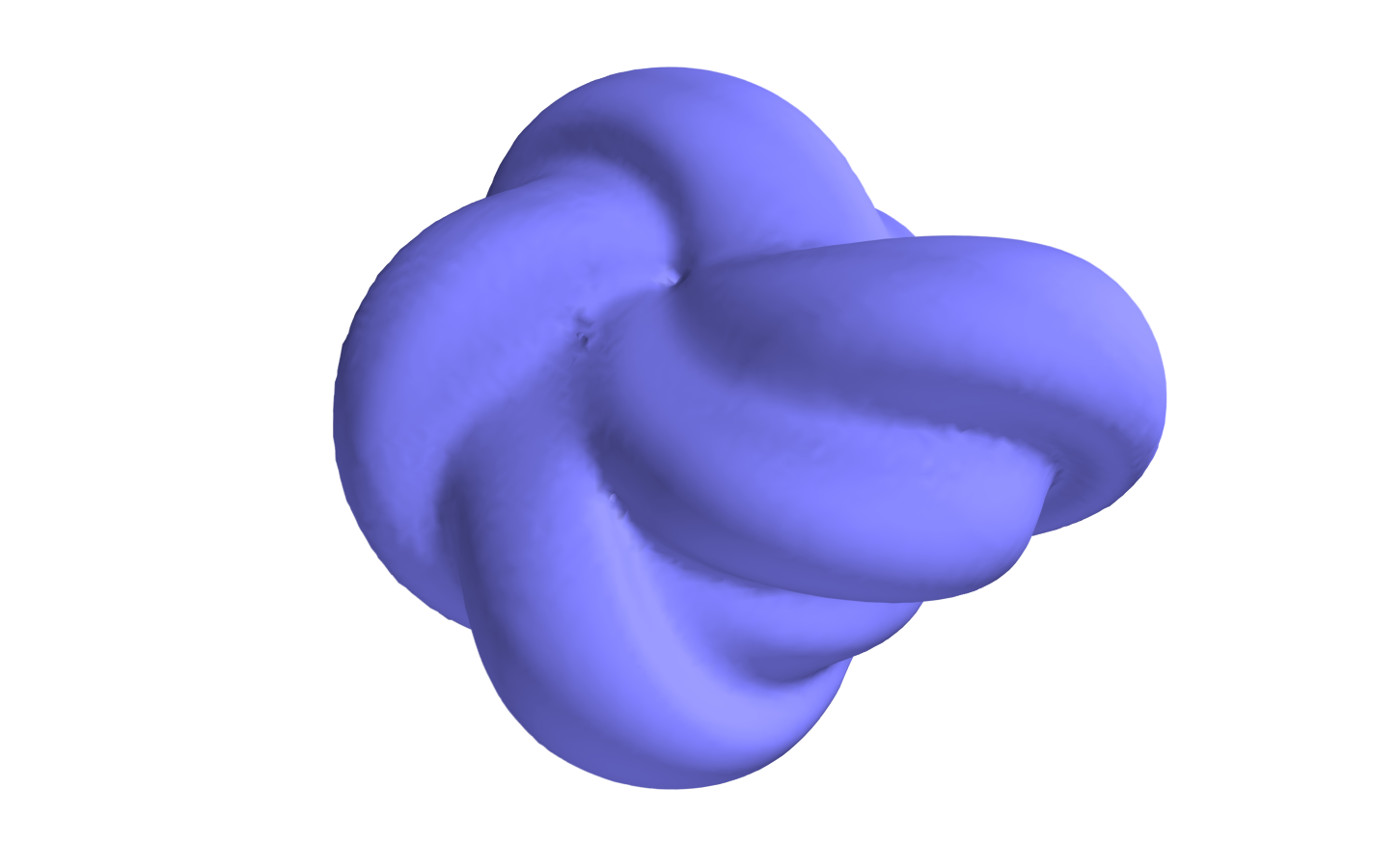} &  \includegraphics[height=2cm]{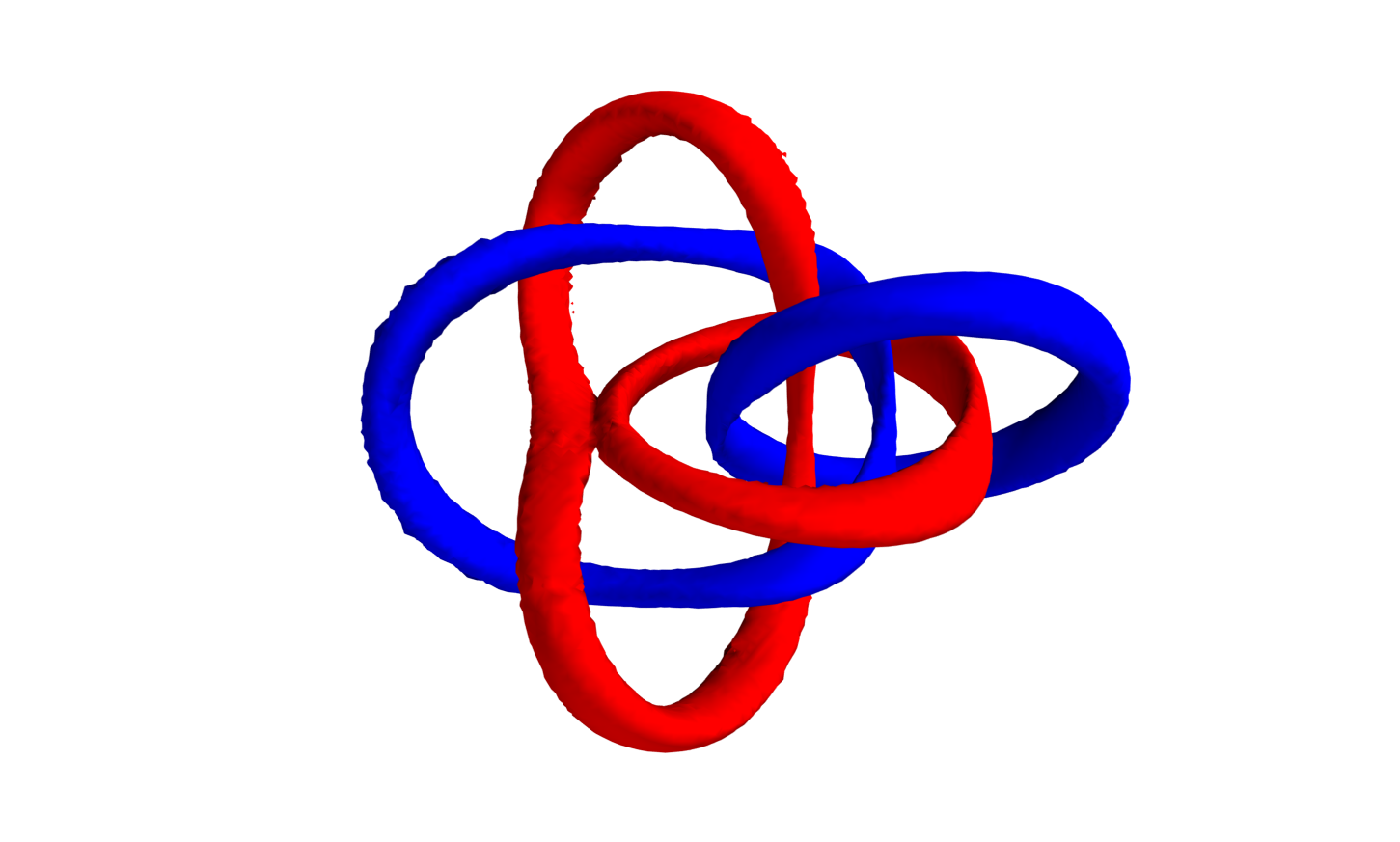} & \includegraphics[height=2cm]{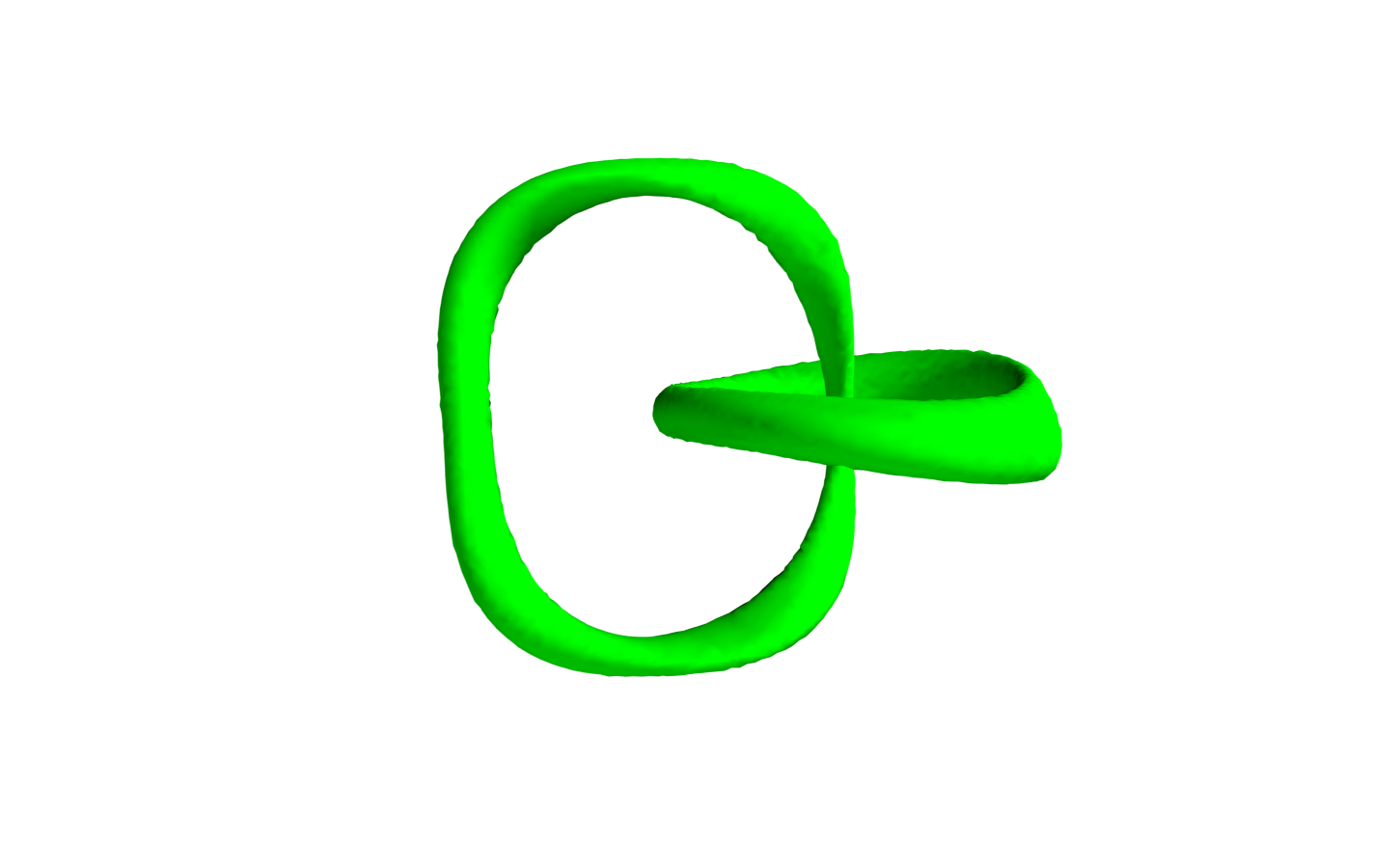}\\
$5(\CA_1\between\CA_1)_{\widetilde\CA_{5,1}}$  & \includegraphics[height=2cm]{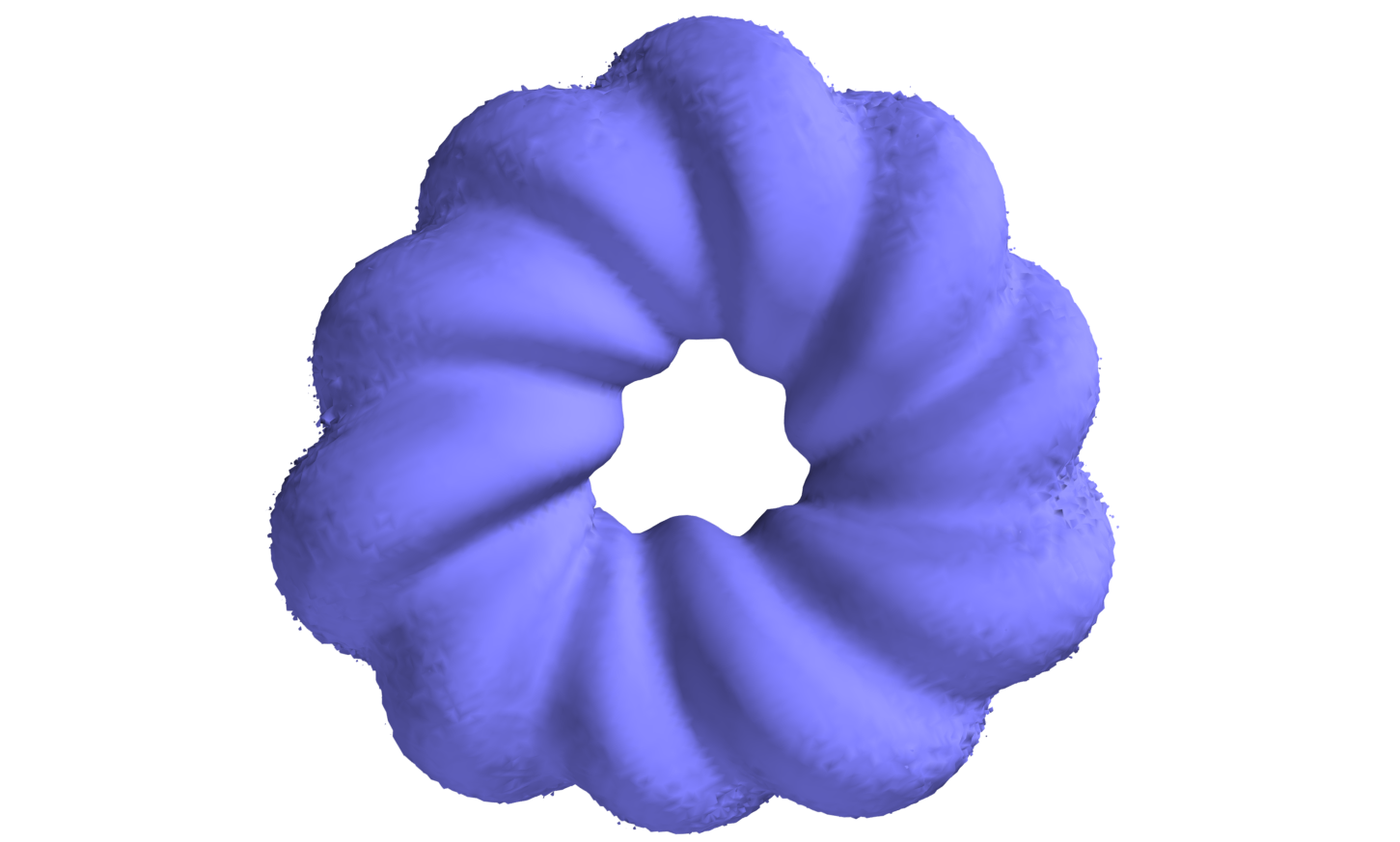} & \includegraphics[height=2cm]{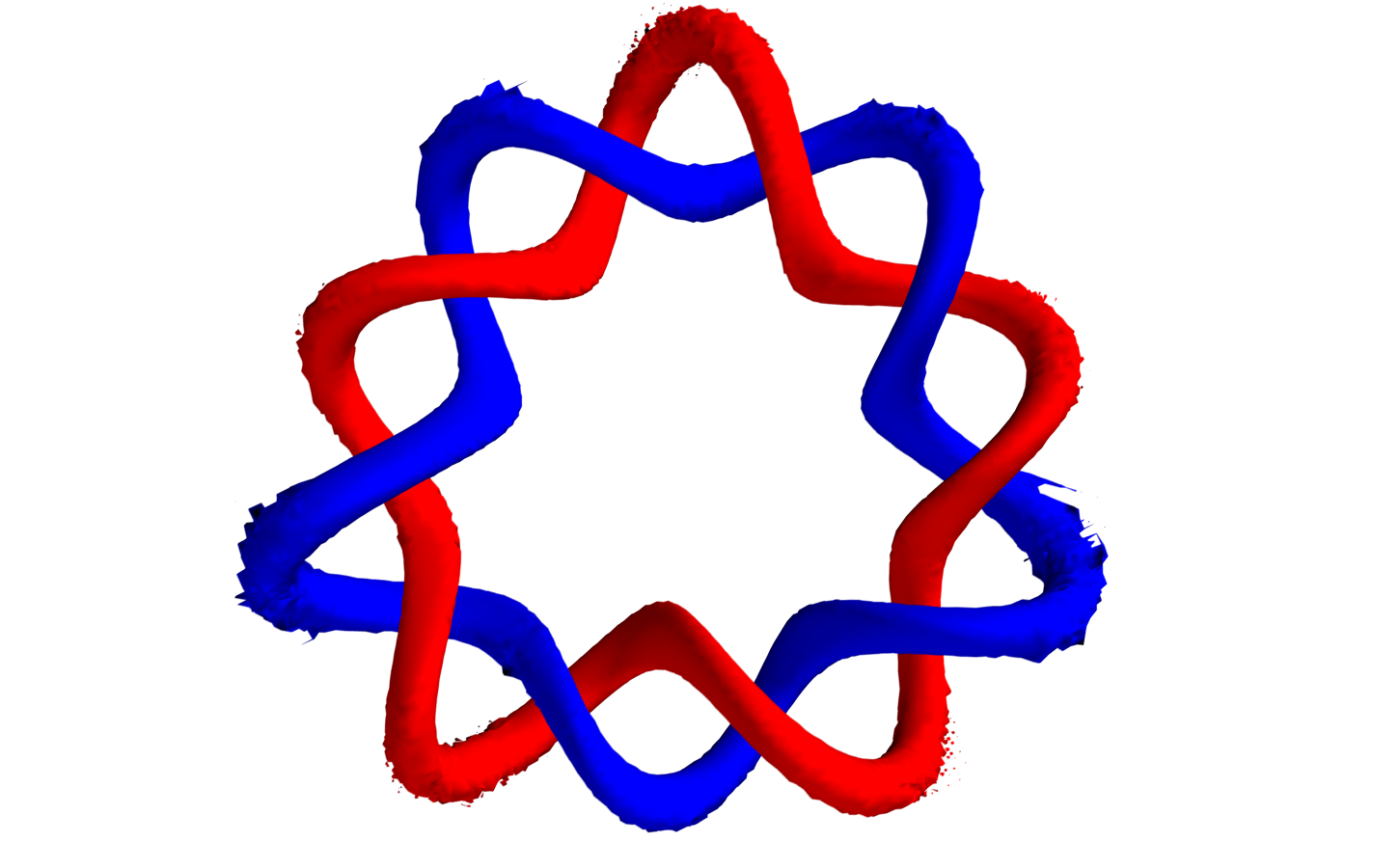} & \includegraphics[height=2cm]{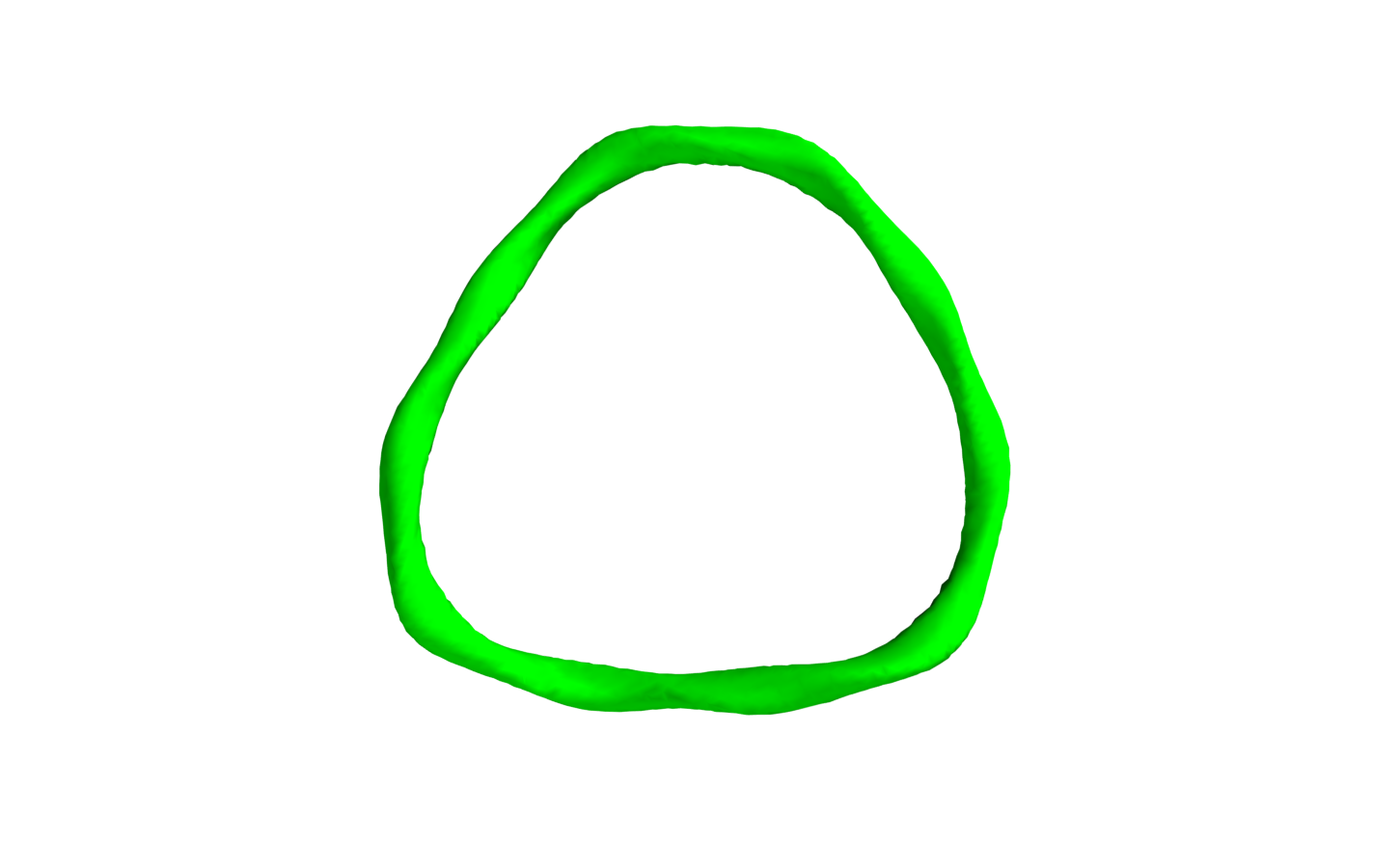} \\
$6(\CK_{3,2}\between\CK_{3,2})_{A_{3,2}}$ &  \includegraphics[height=2cm]{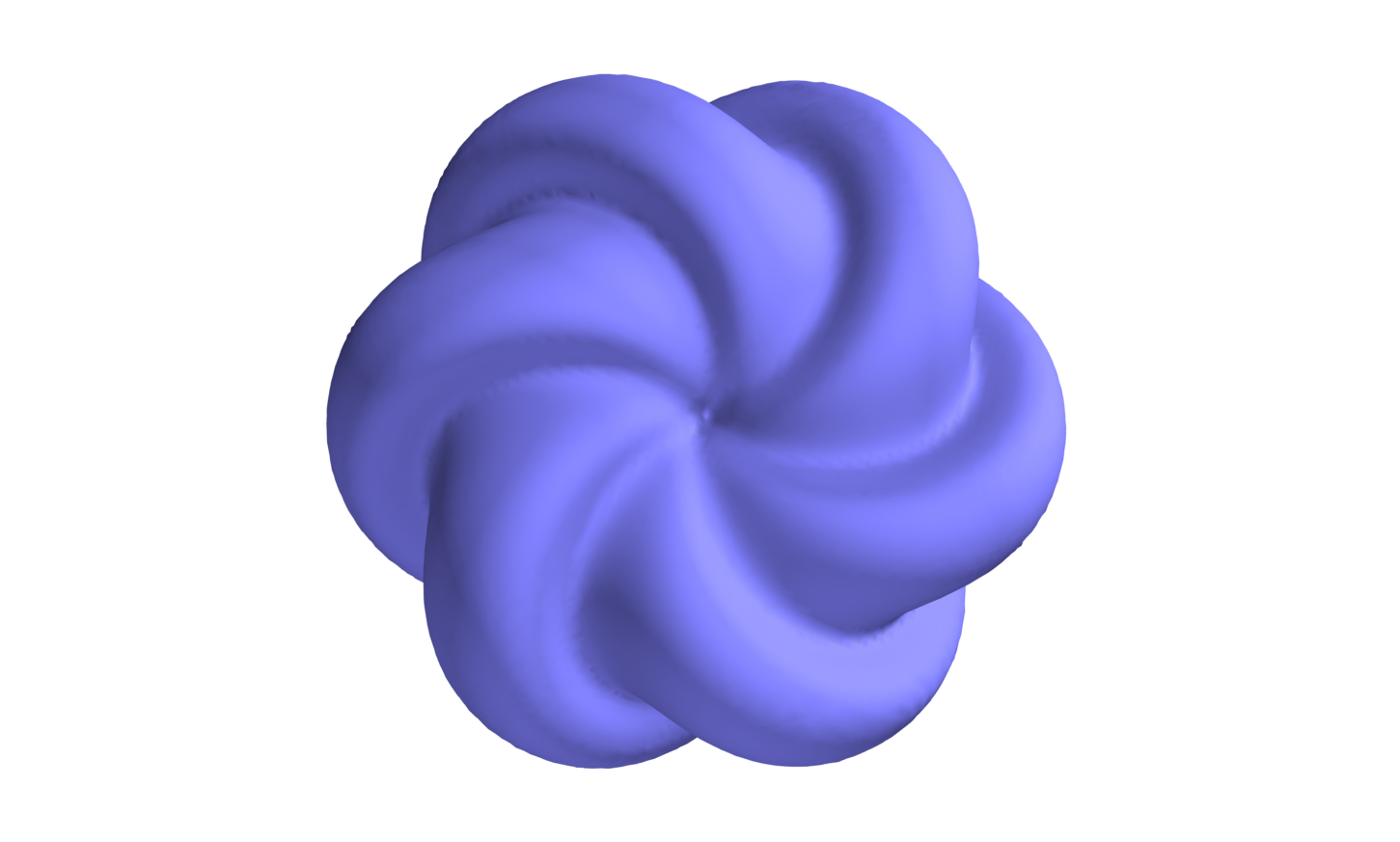} & \includegraphics[height=2cm]{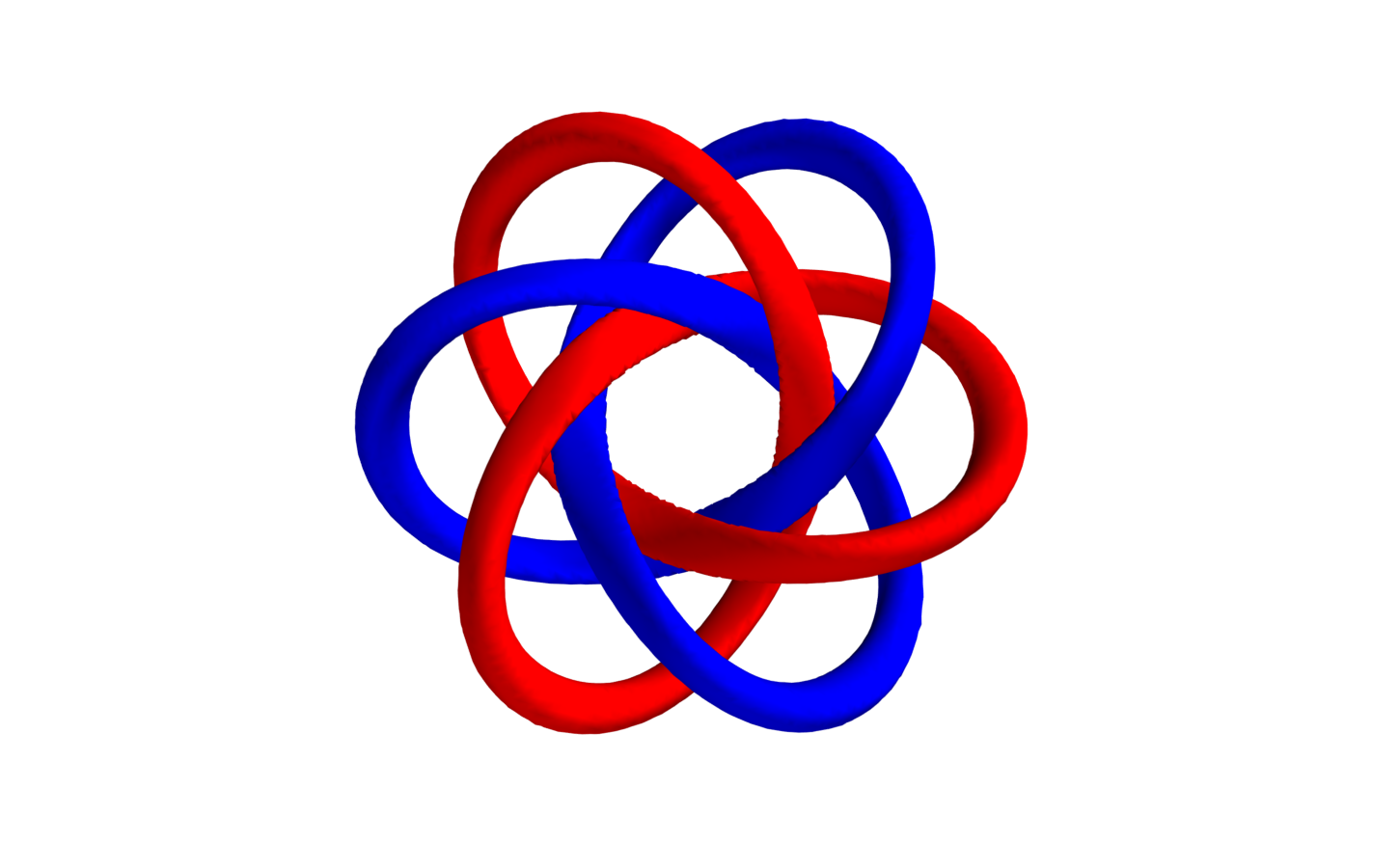} & \includegraphics[height=2cm]{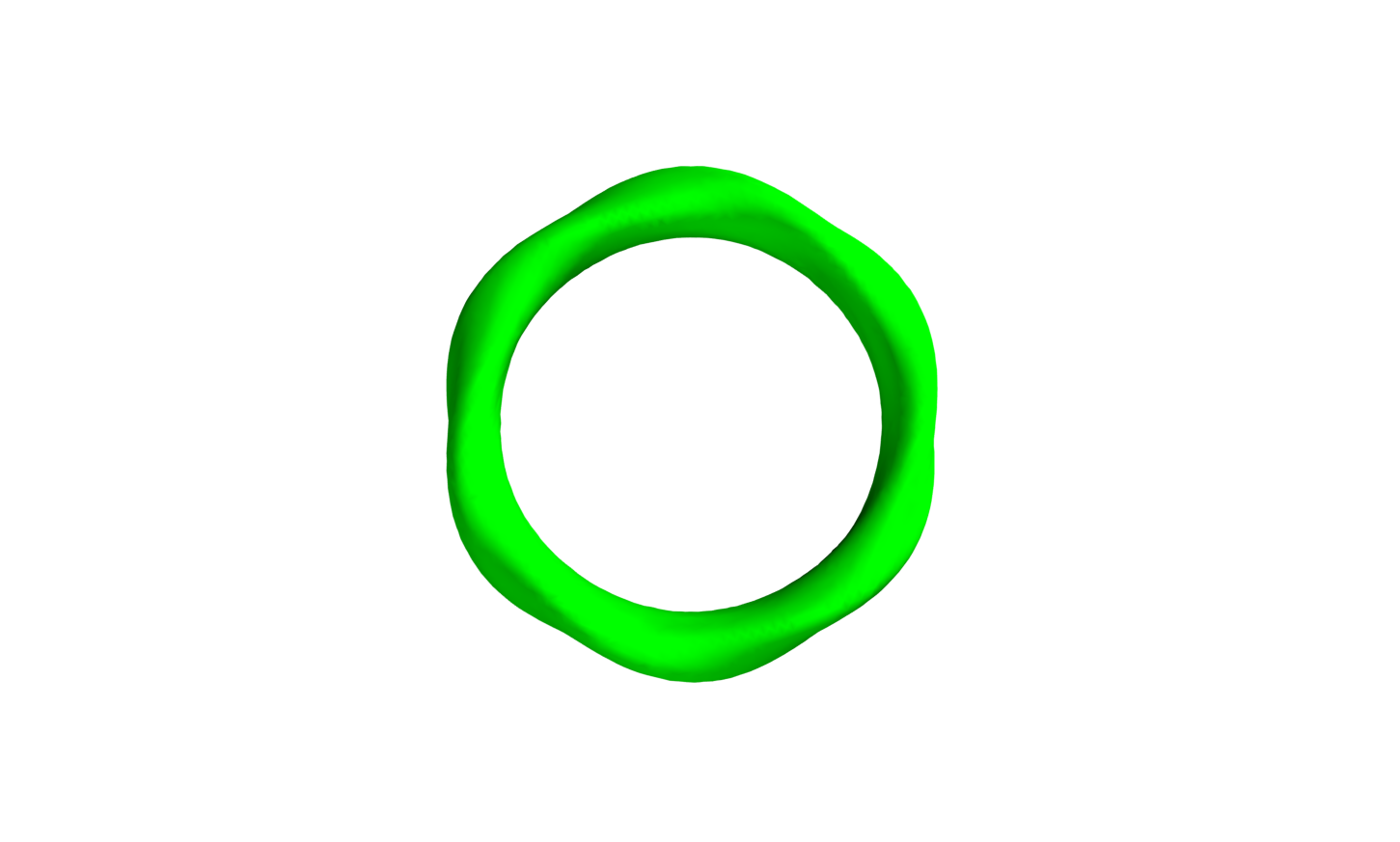} \\
$6(\CL_{1,3}\between\CK_{3,2})_{\CL_{1,3}^{1,1}}$ &  \includegraphics[height=2cm]{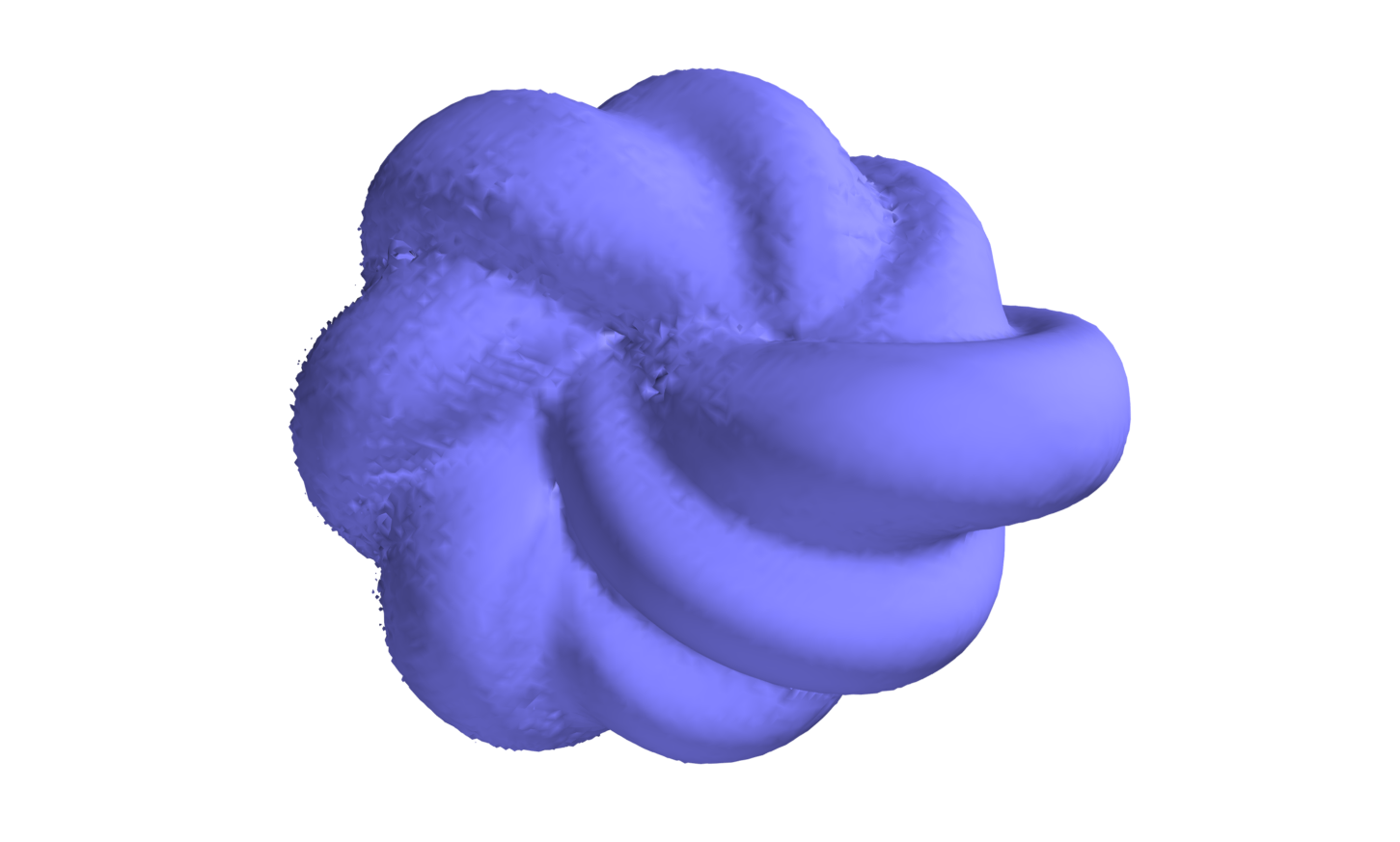} & \includegraphics[height=2cm]{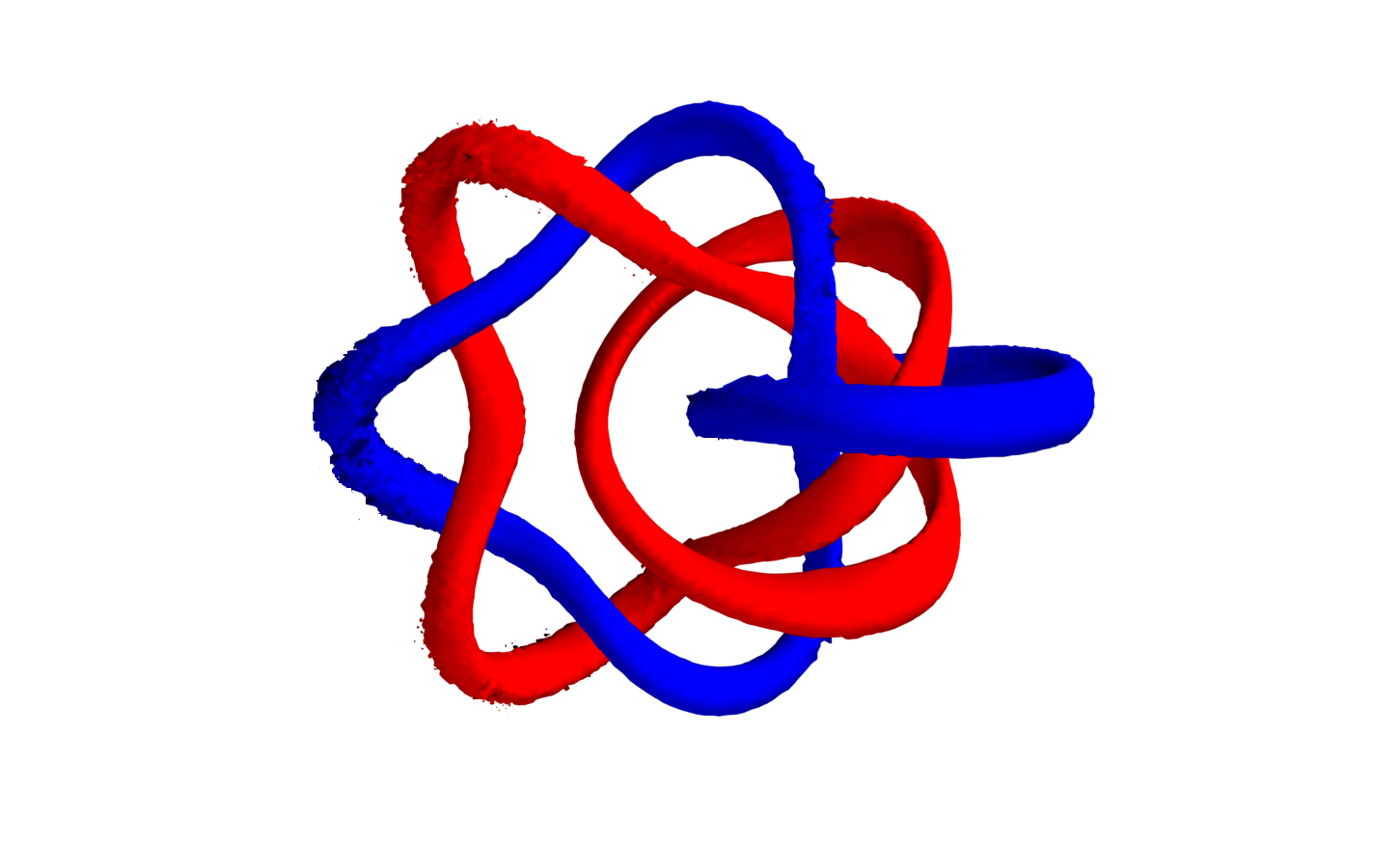} & \includegraphics[height=2cm]{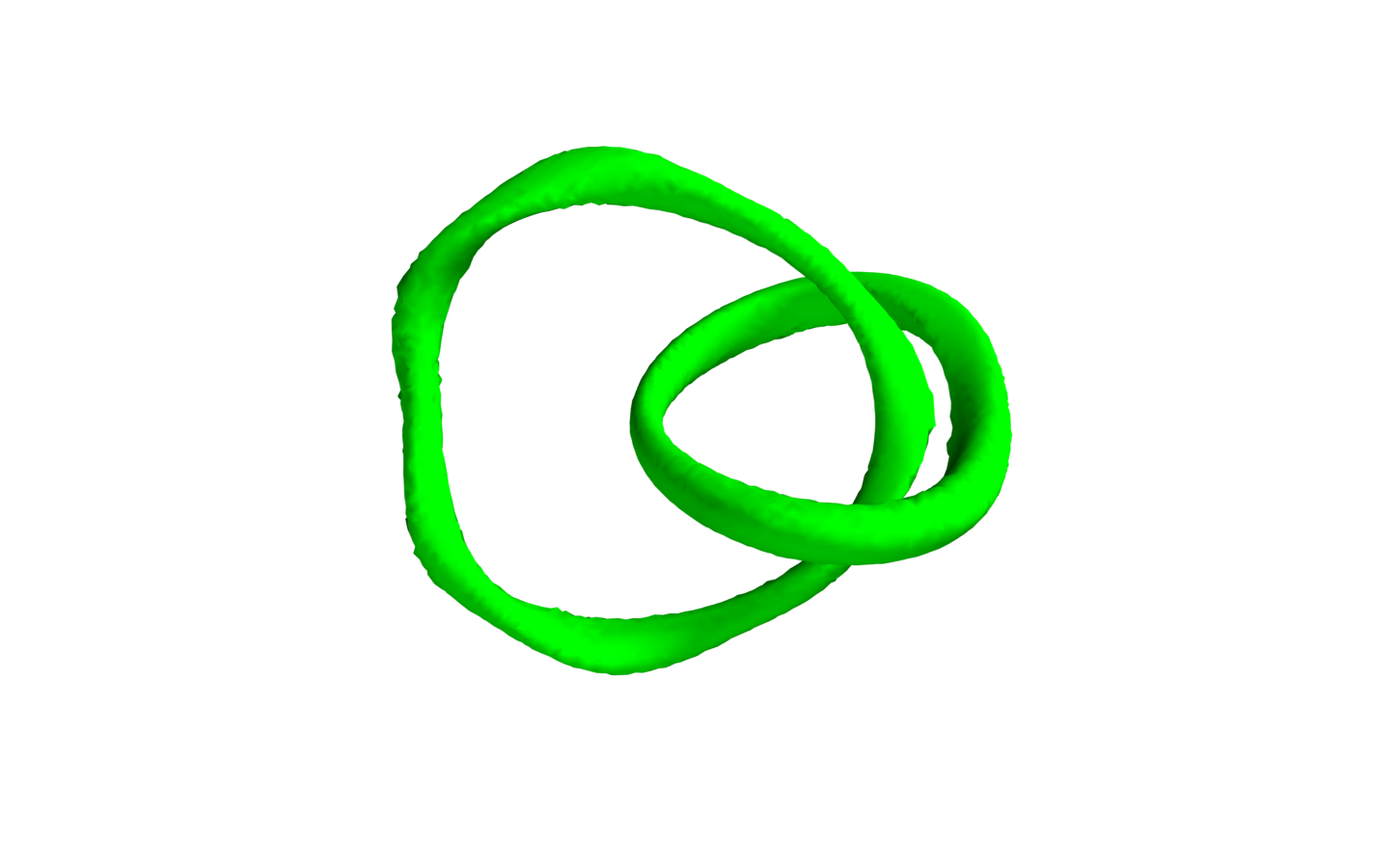}\\
$6(\widetilde\CA_1\between\widetilde\CA_1)_{\widetilde\CA_{6,1}}$ & \includegraphics[height=2cm]{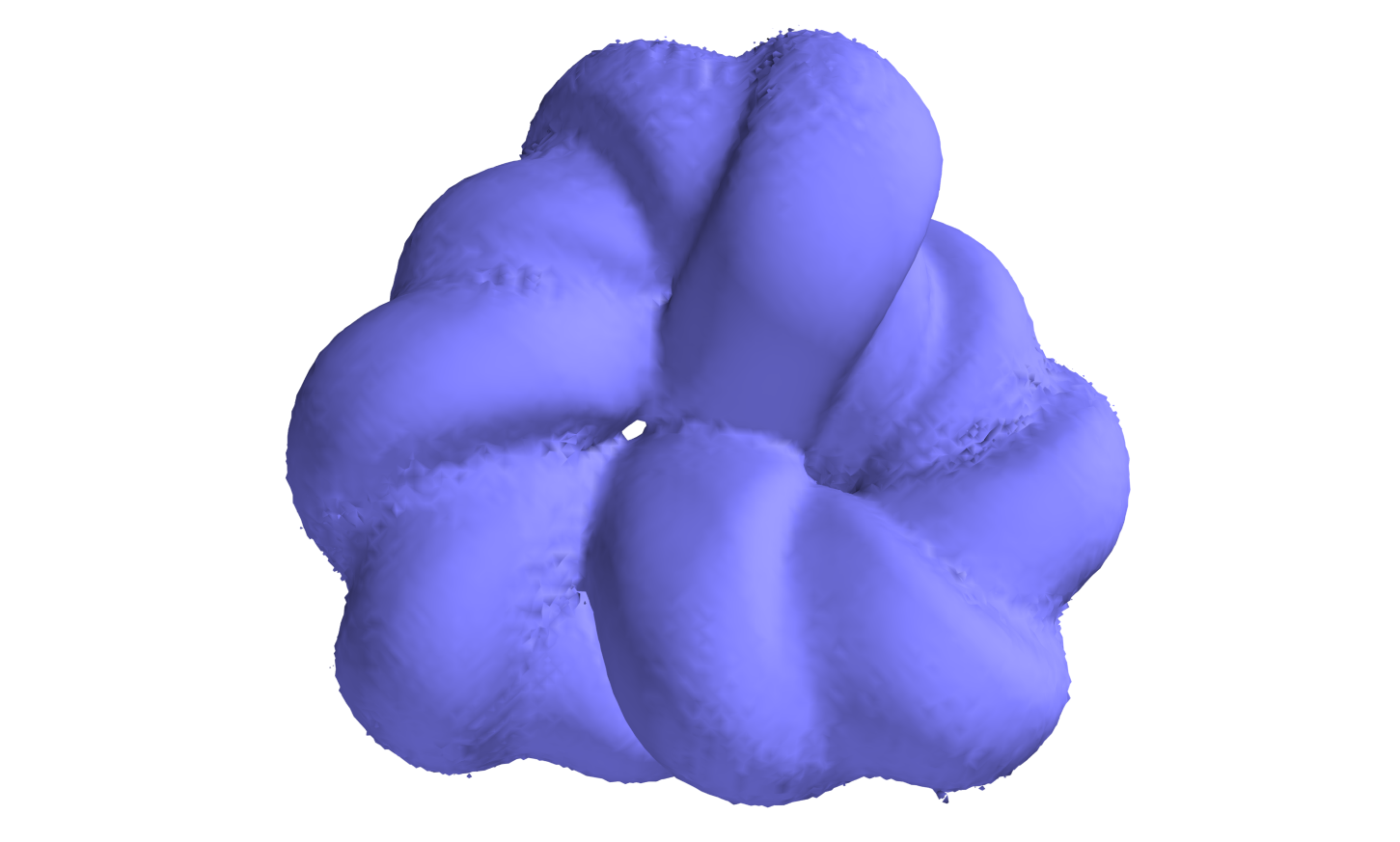} & \includegraphics[height=2cm]{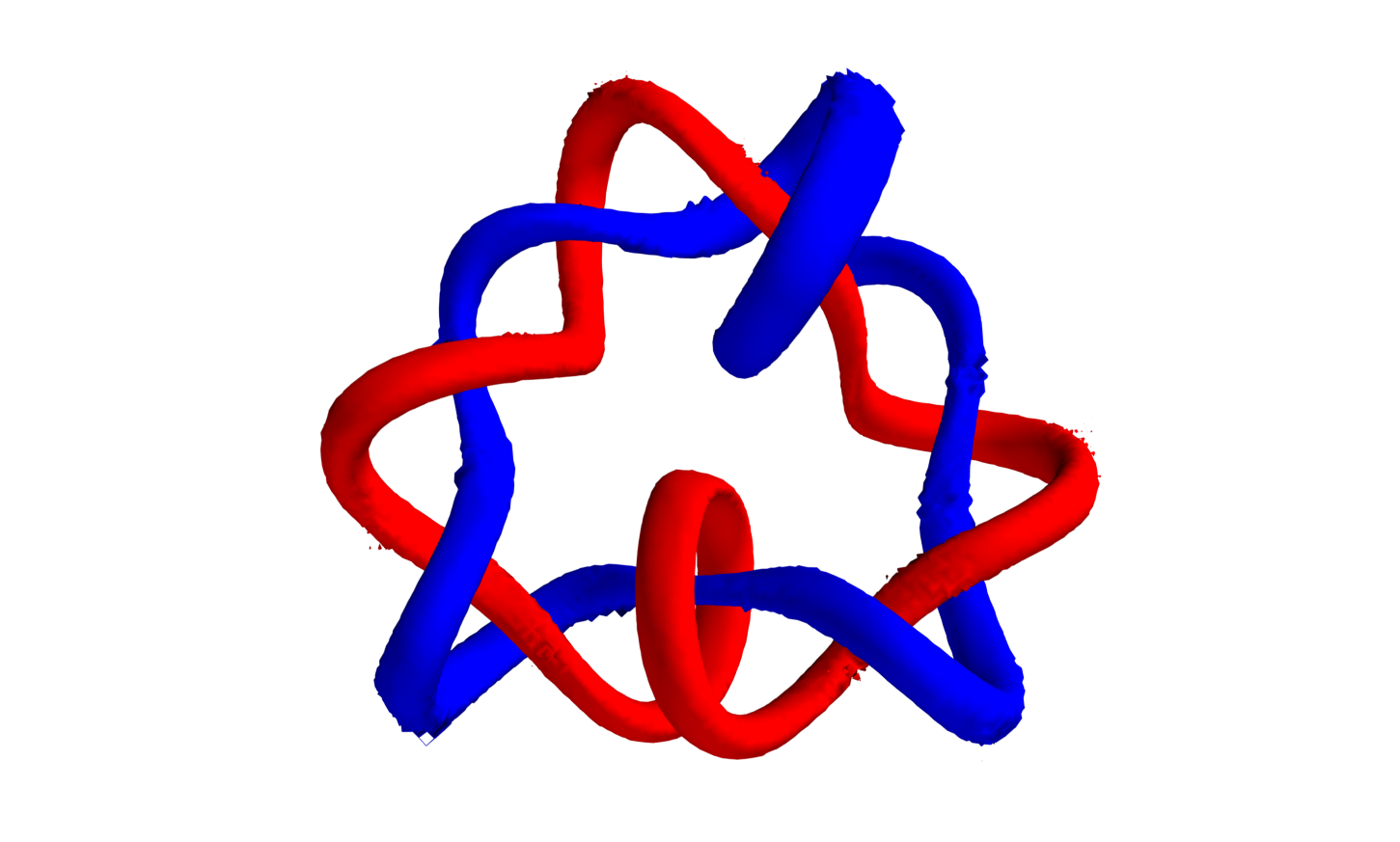} & \includegraphics[height=2cm]{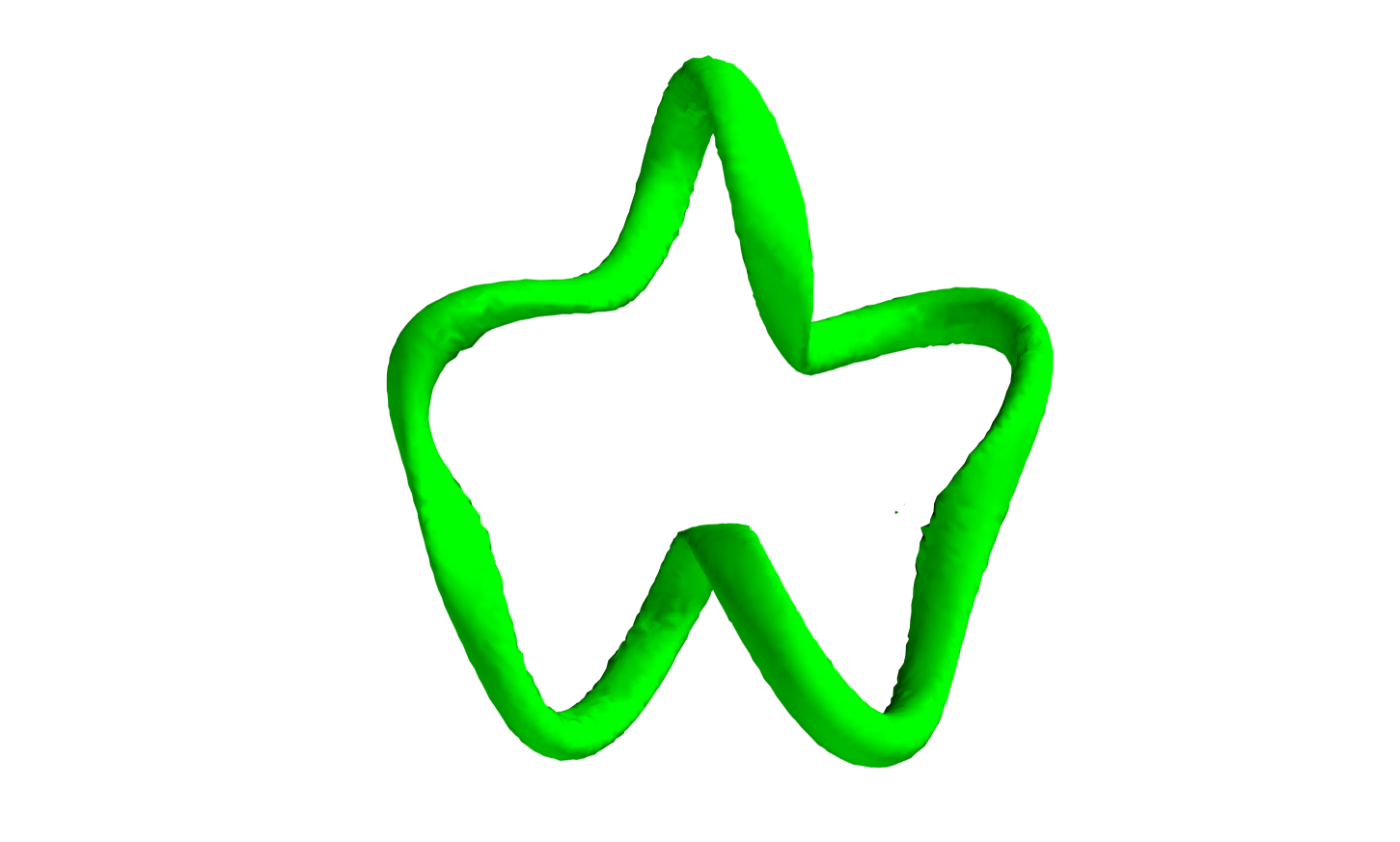} \\
$7(\CL_{1,2}^{2,2}\between\CK_{3,2})_{\CK_{3,2}}$ & \includegraphics[height=2cm]{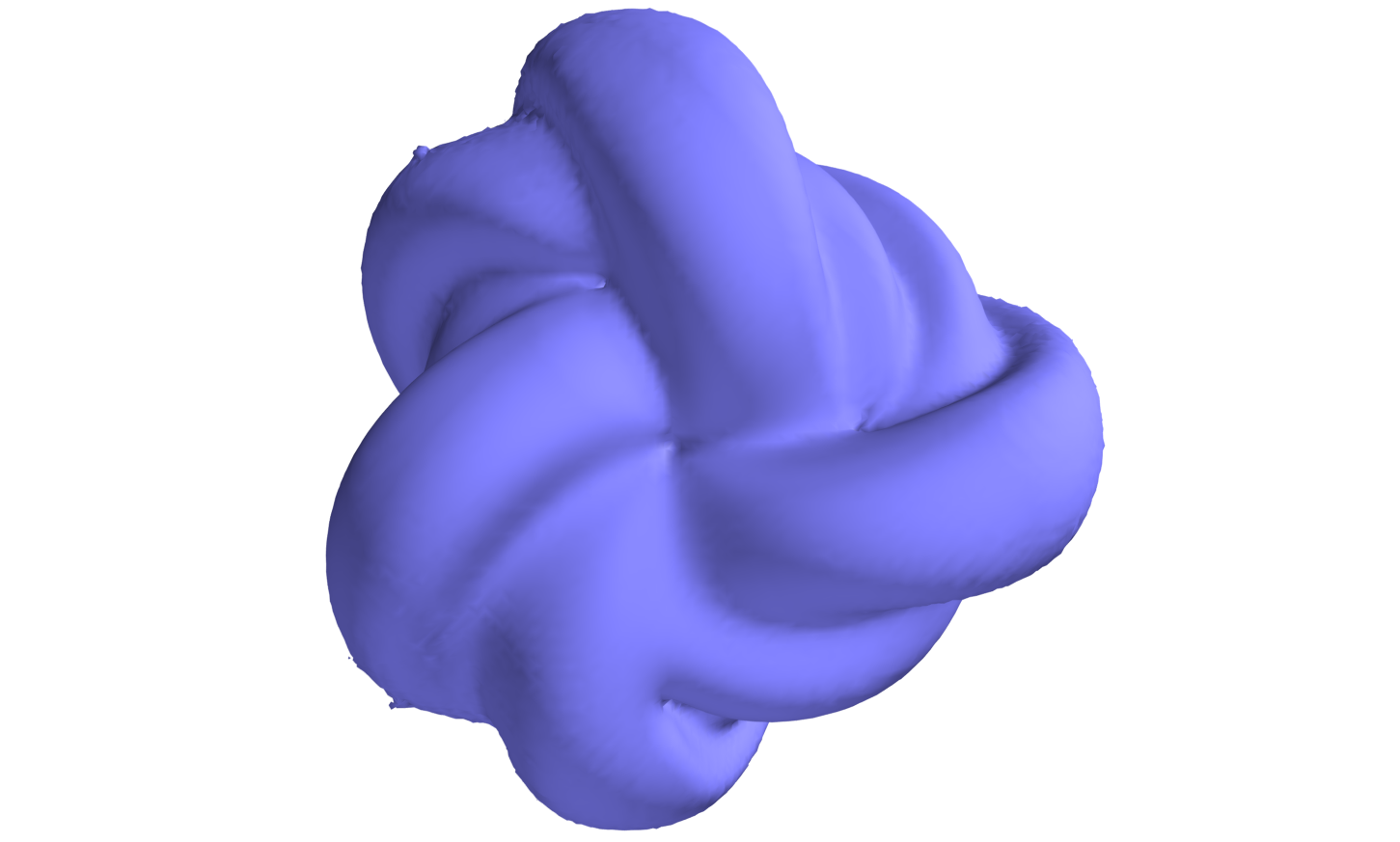} & \includegraphics[height=2cm]{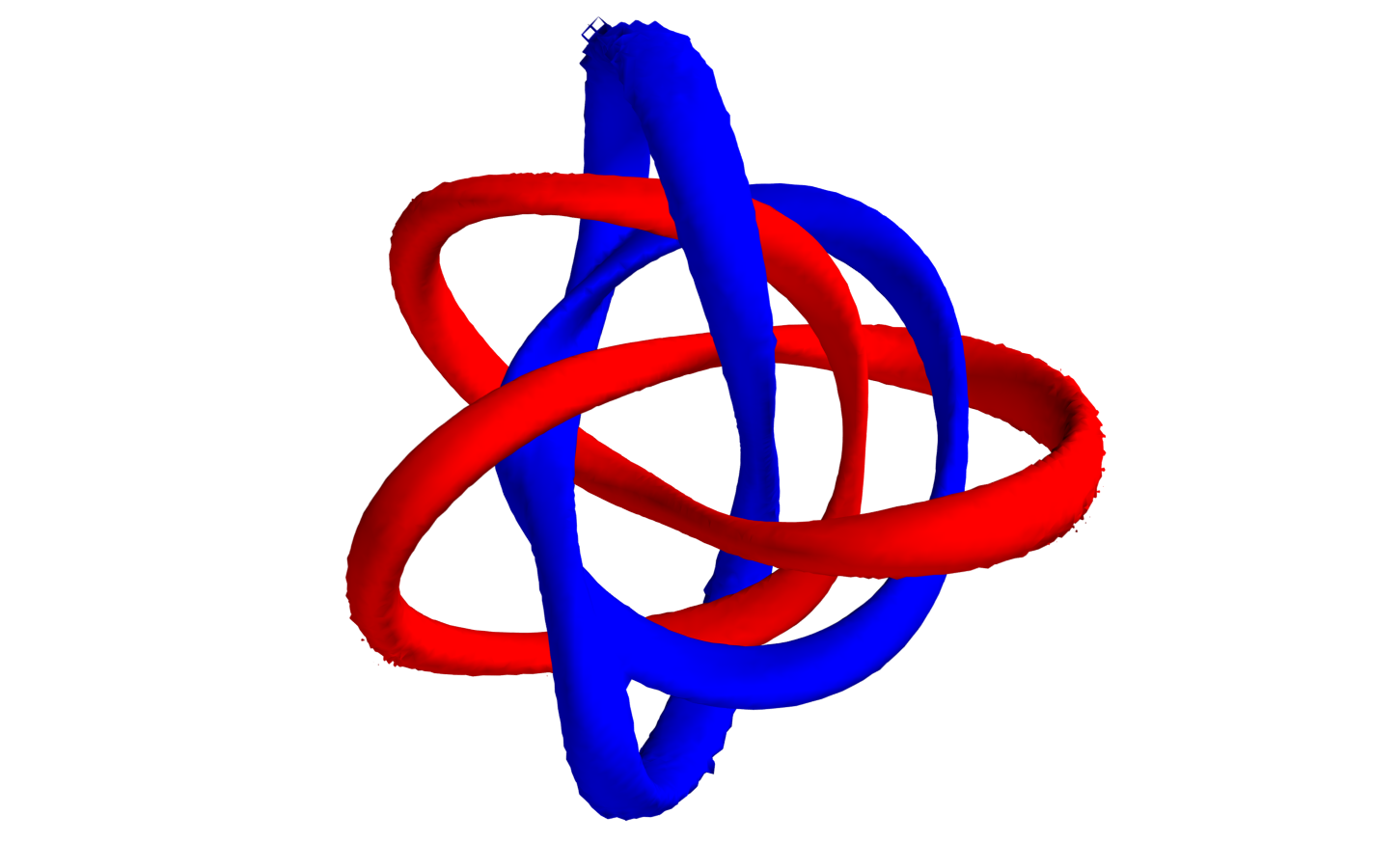} &
\includegraphics[height=2cm]{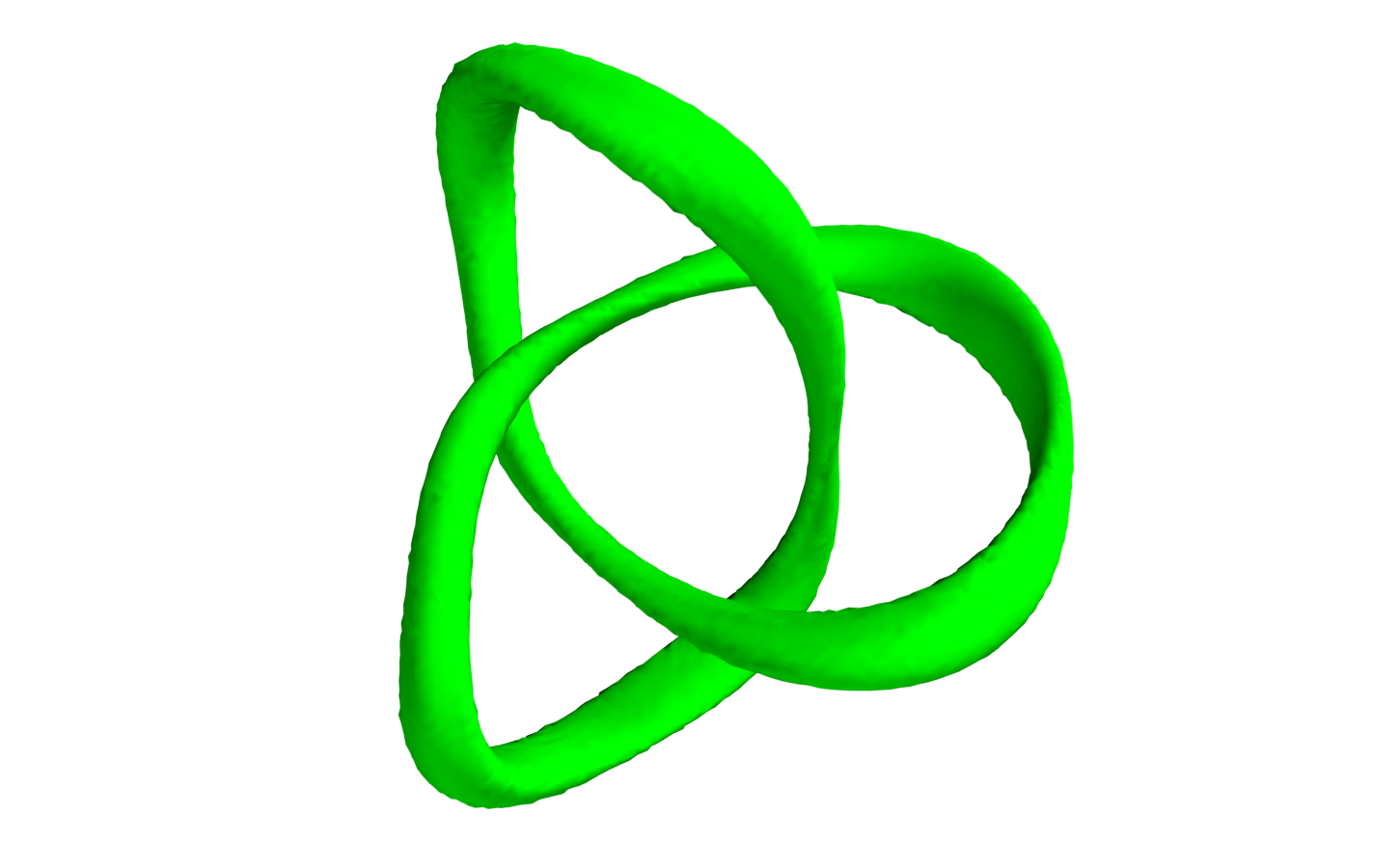}\\
$7(\CK_{2,3}\between\CK_{2,3})_{\CK_{2,3}}$ & \includegraphics[height=2cm]{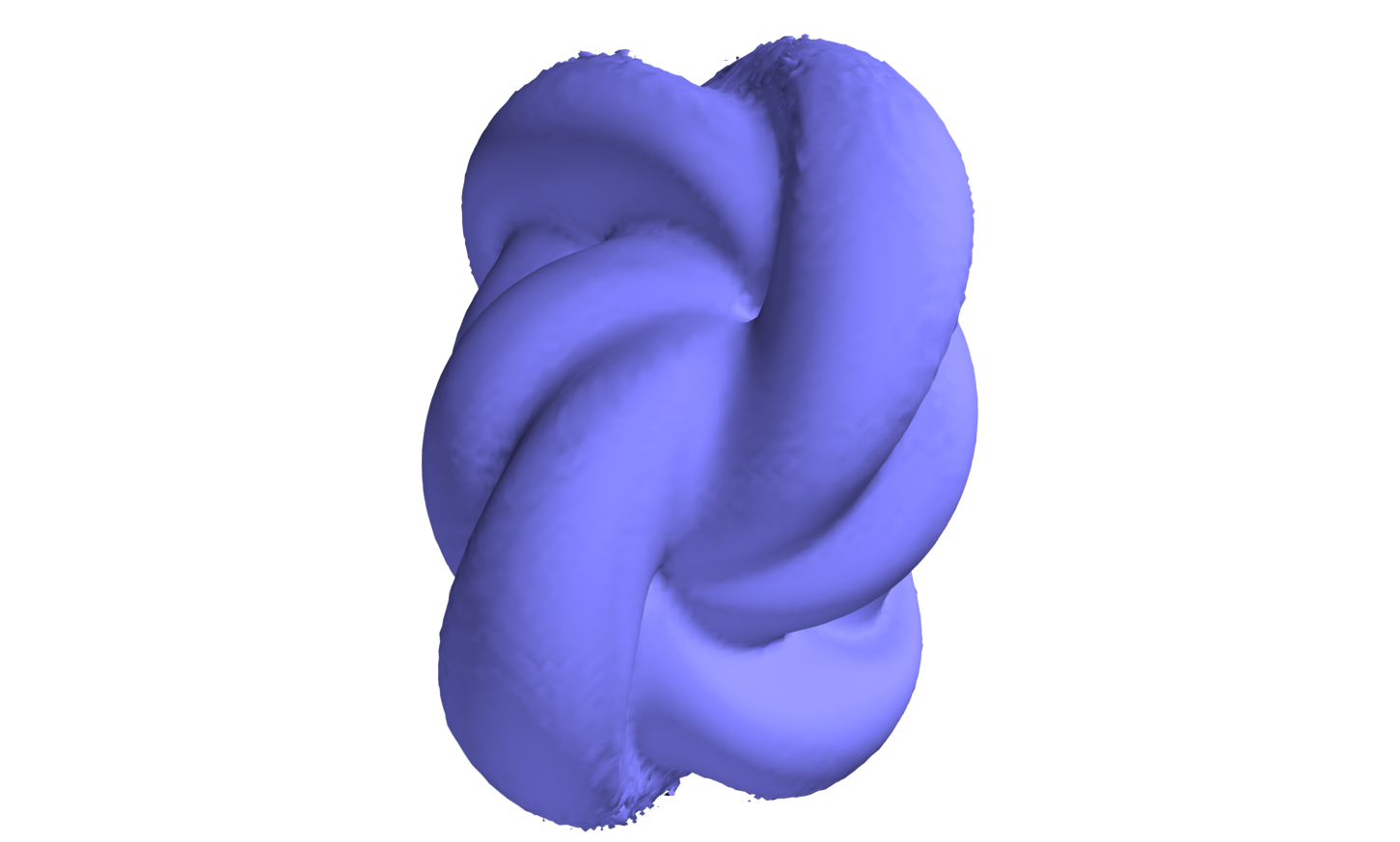} & \includegraphics[height=2cm]{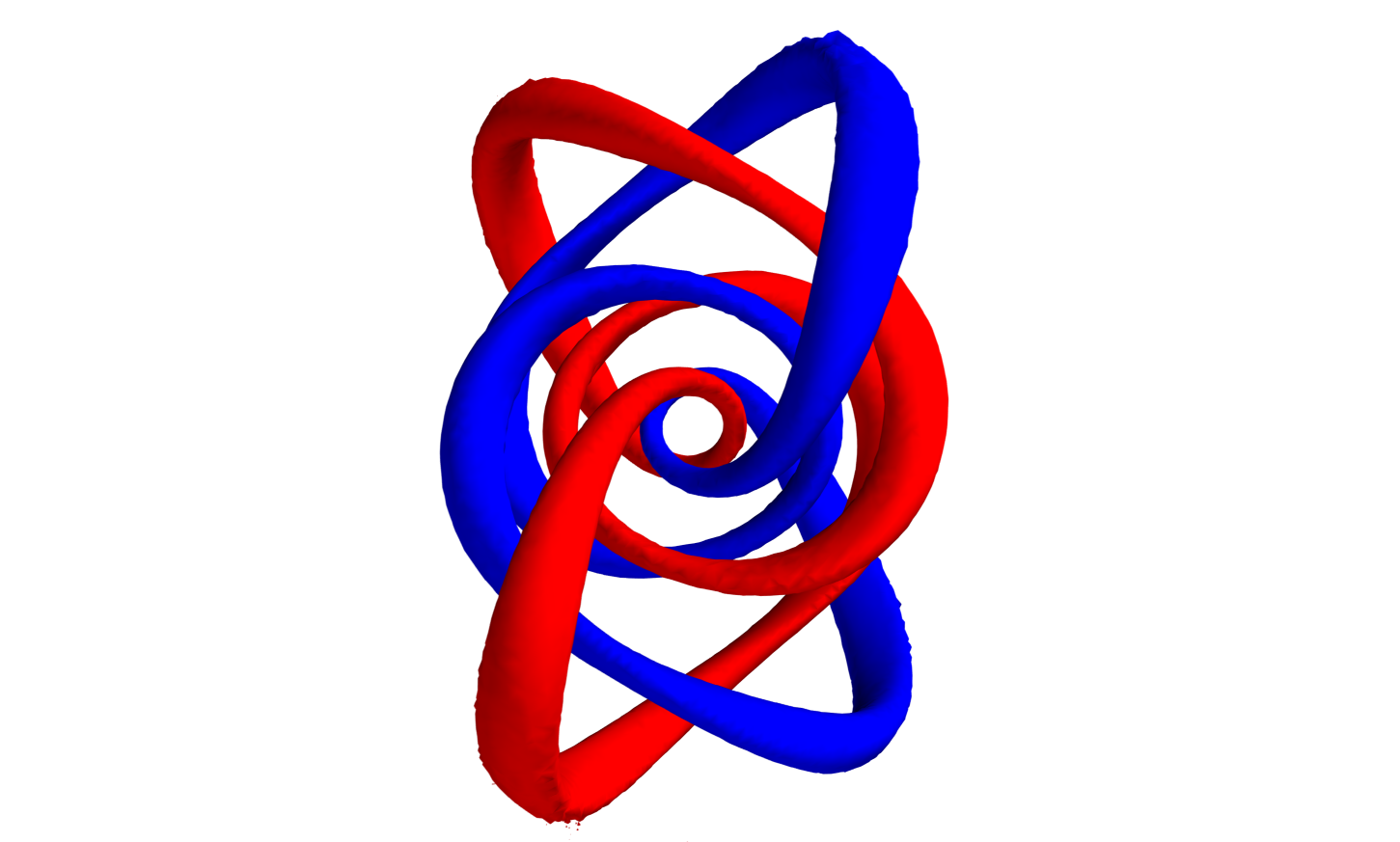} & \includegraphics[height=2cm]{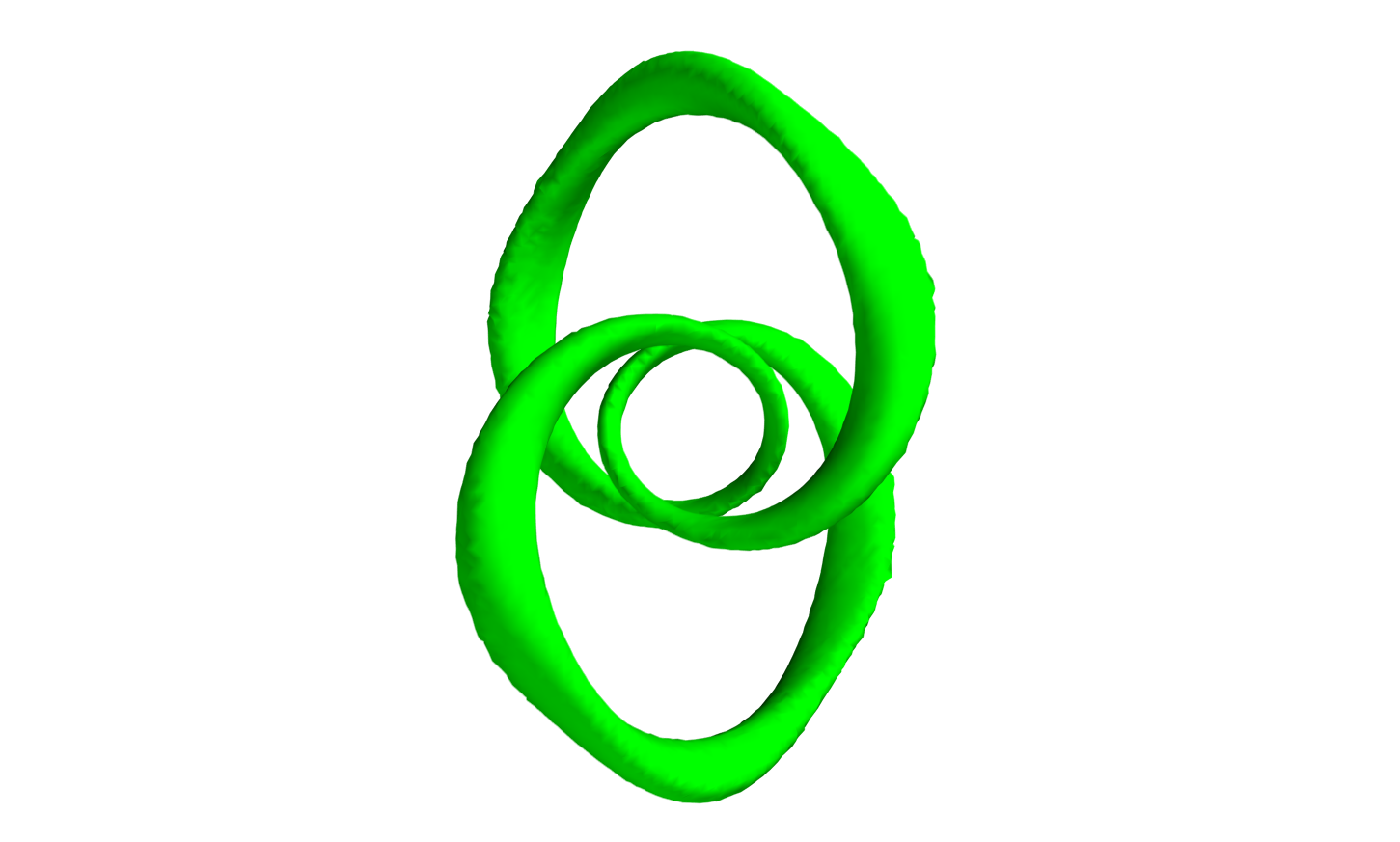} \\
$7(\widetilde{\CA}_1\between\widetilde{\CA}_1)_{\CA_{7,1}}$ & \includegraphics[height=2cm]{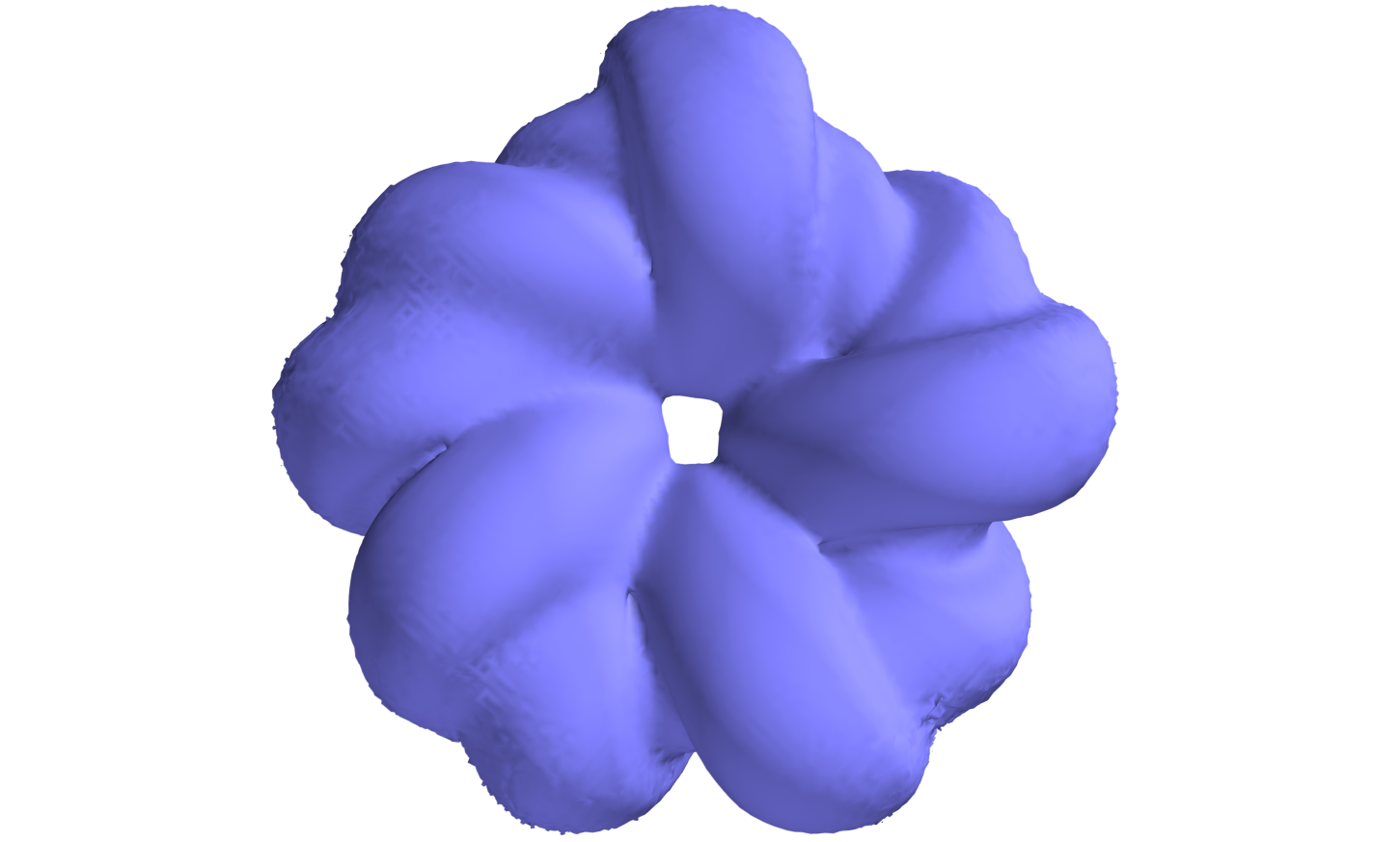} & \includegraphics[height=2cm]{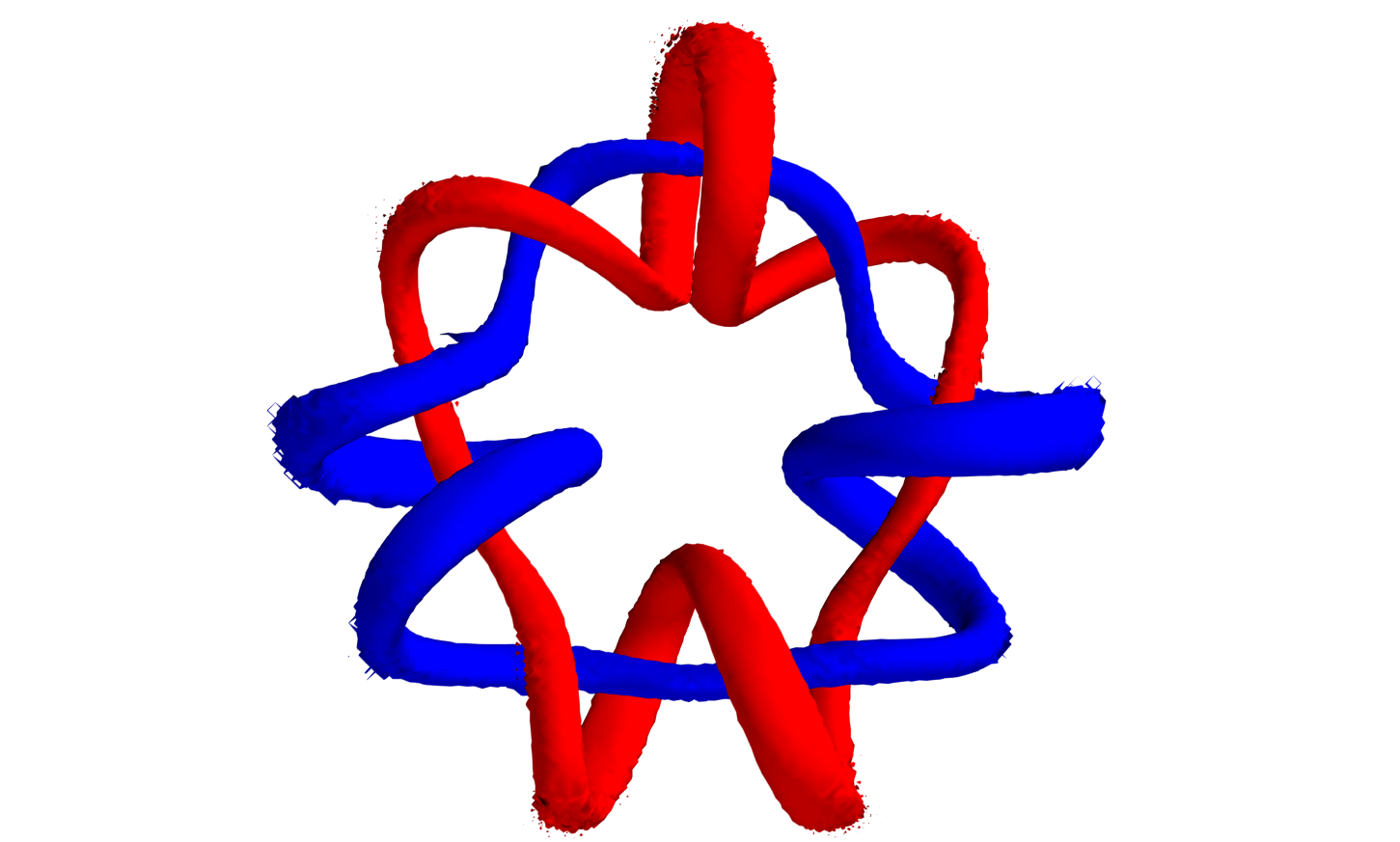} & \includegraphics[height=2cm]{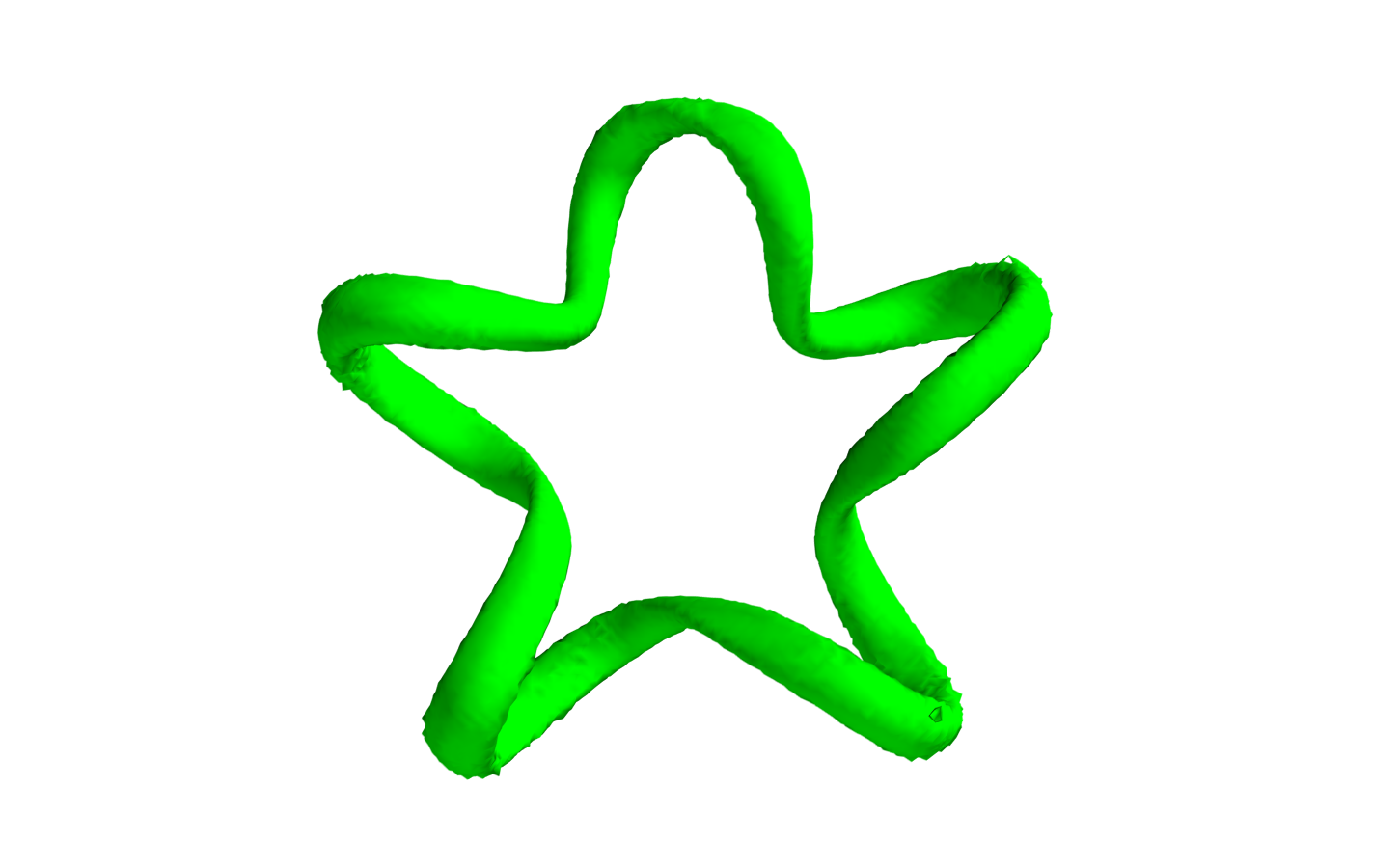} \\
\end{TAB}
\end{center}
\caption{$\rho_E=15$ isosurface, $\phi_1=\pm0.9$ and $\phi_3=-0.9$ for $Q=1-7$ Hopfions in model with $c_2=0.5,
c_4=1, m=4$ and potential $V=m^2\phi_1^2$}
\label{hopftable}
\end{table}

\begin{table}
\begin{center}
\begin{TAB}[1pt]{|c|c|c|c|}{|c|c|c|c|c|c|c|c|c|c|}
Configuration & $\rho_E$ &  $\phi_1$ &  $\phi_3$  \\
$1(\CA_1\between\CA_1)_{\CA_{1,1}}$  & \includegraphics[height=2cm]{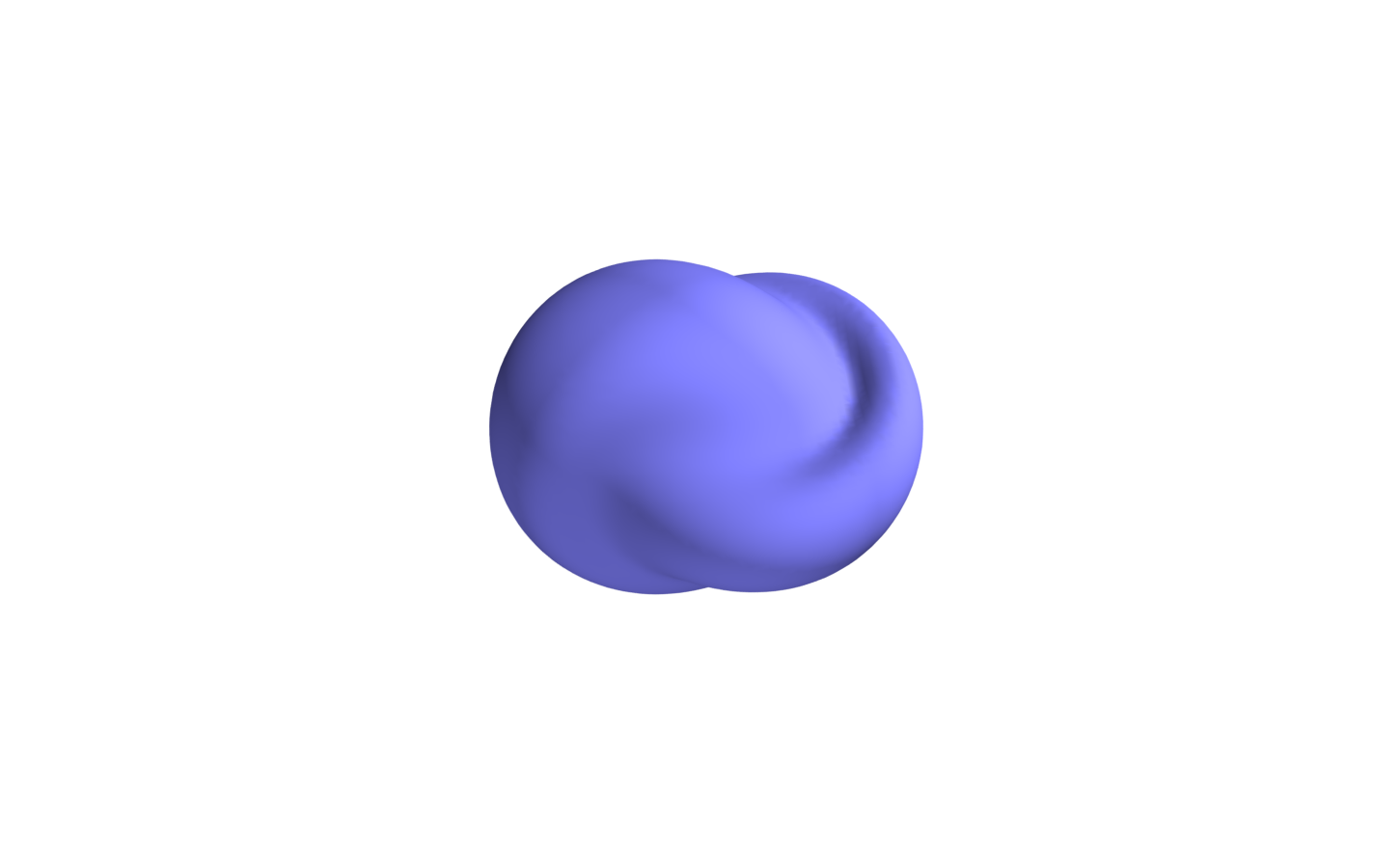} & \includegraphics[height=2cm]{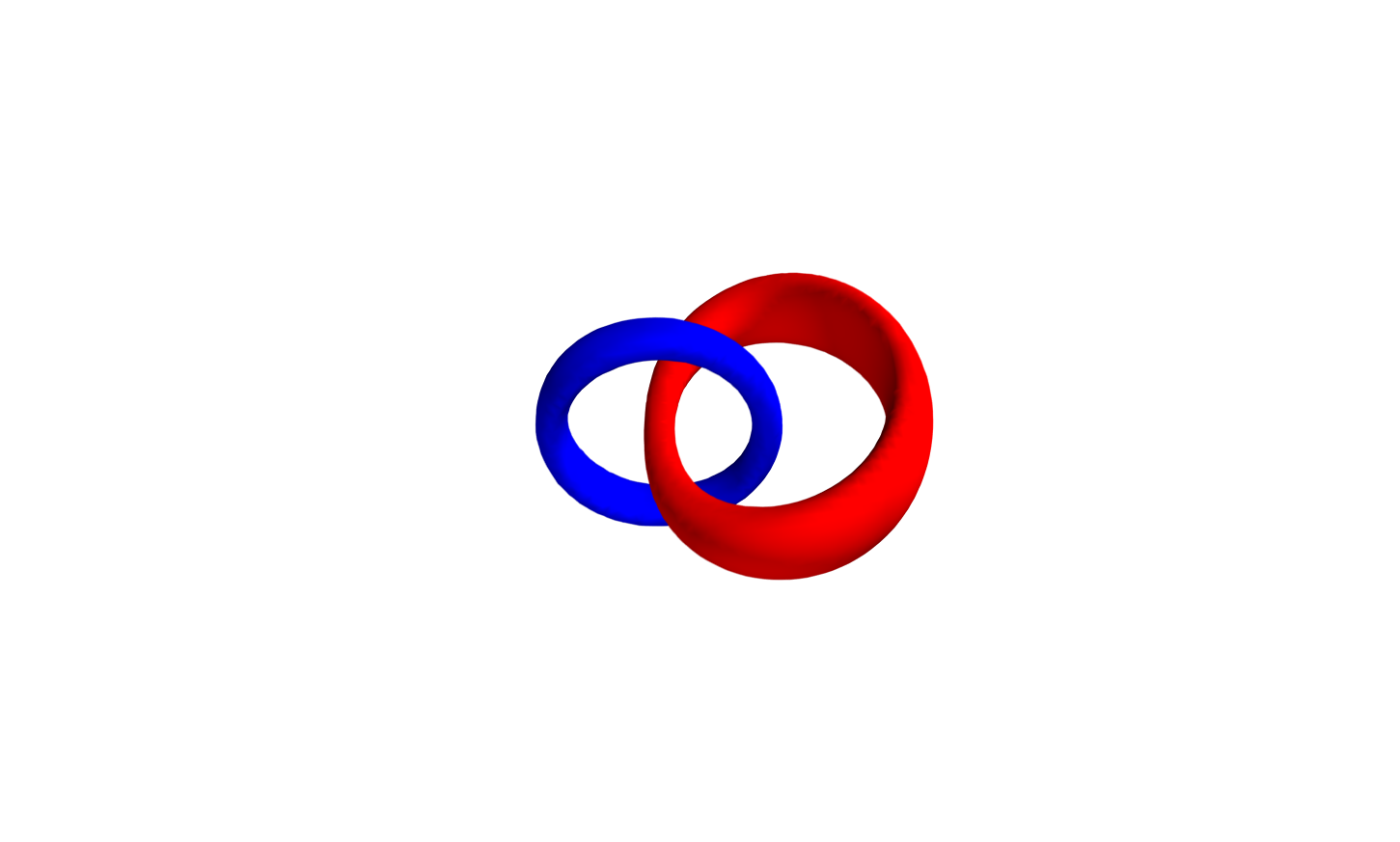} &
\includegraphics[height=2cm]{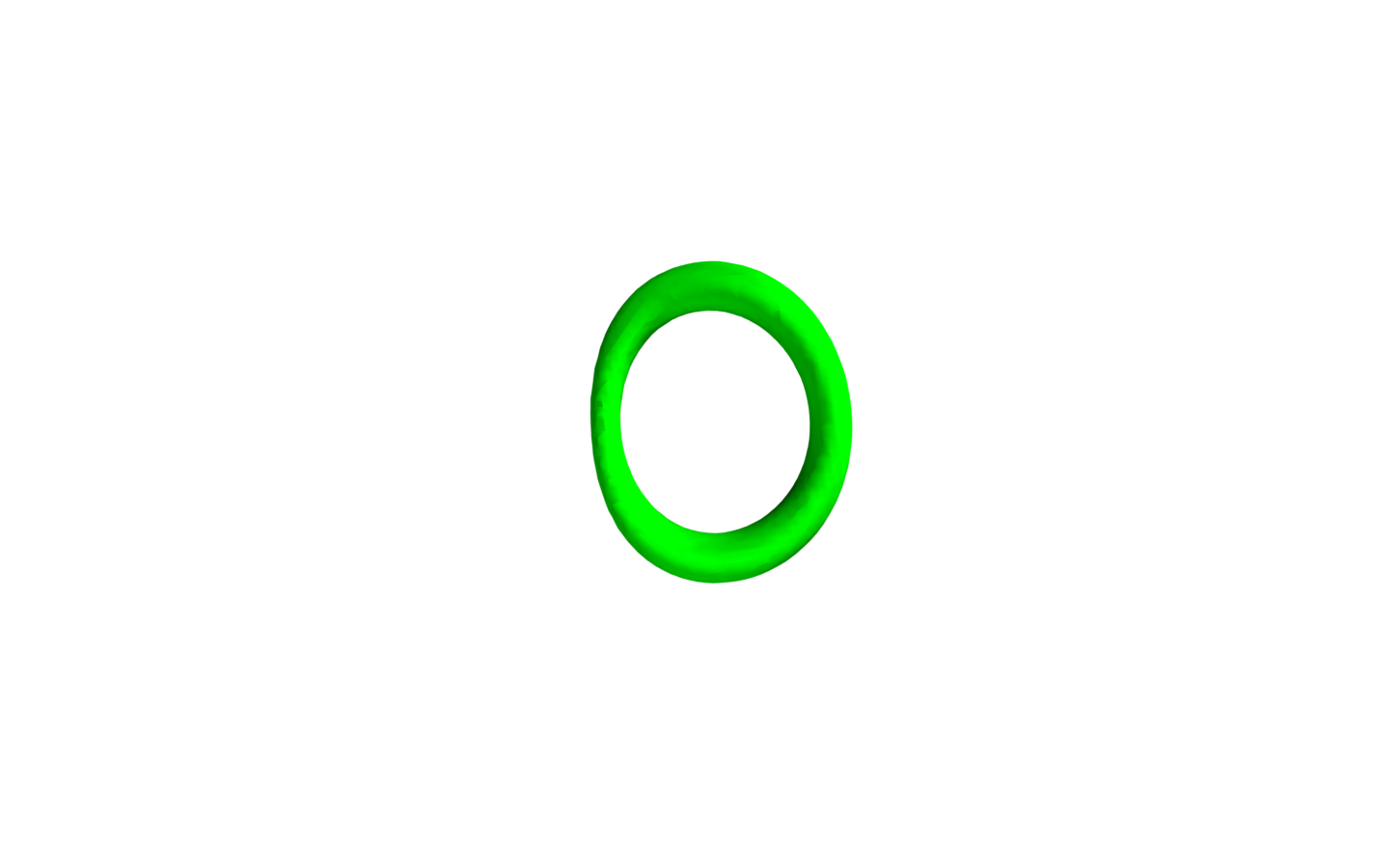}\\
$2(\CA_1\between\CA_1)_{\CA_{2,1}}$  & \includegraphics[height=2cm]{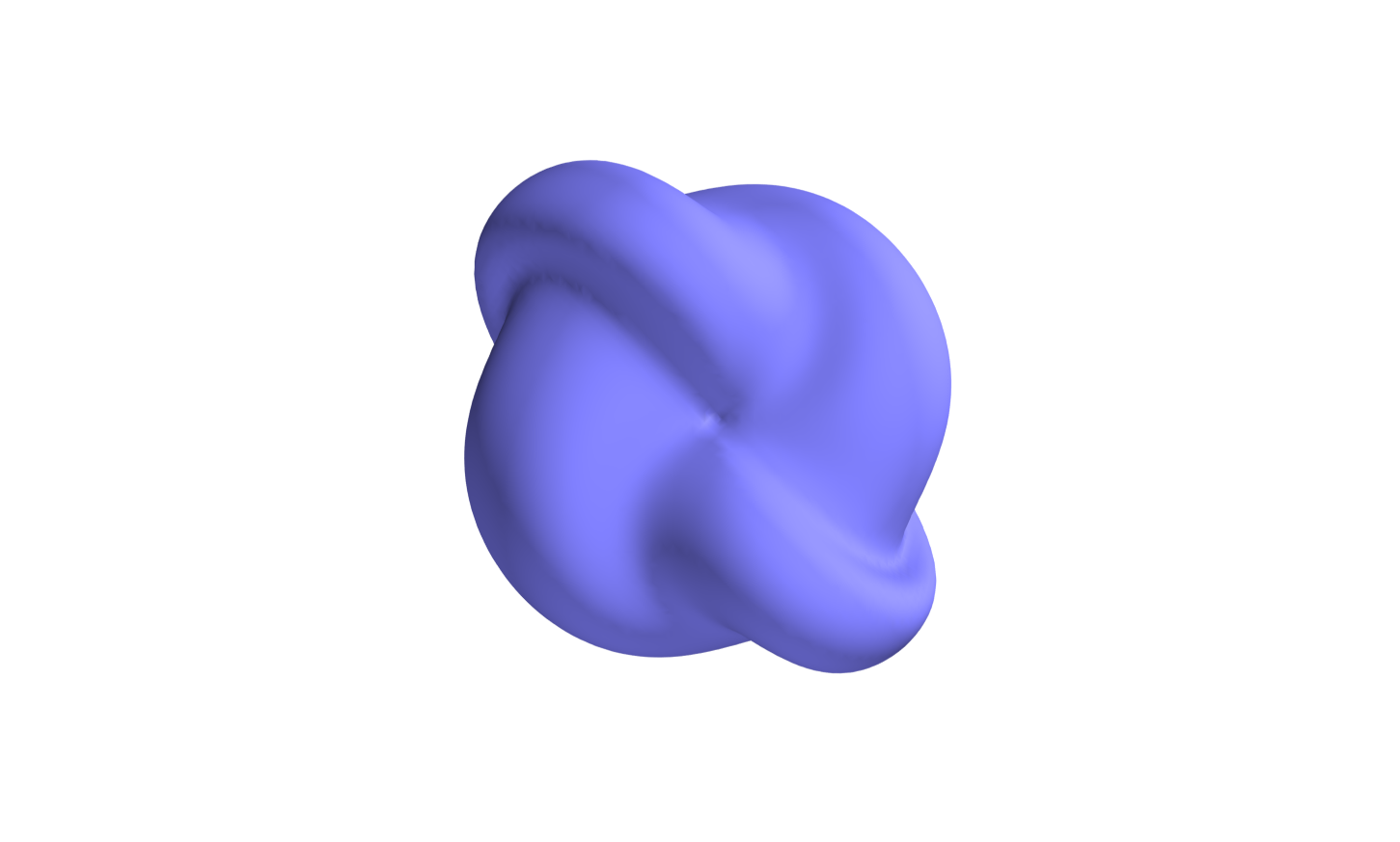} & \includegraphics[height=2cm]{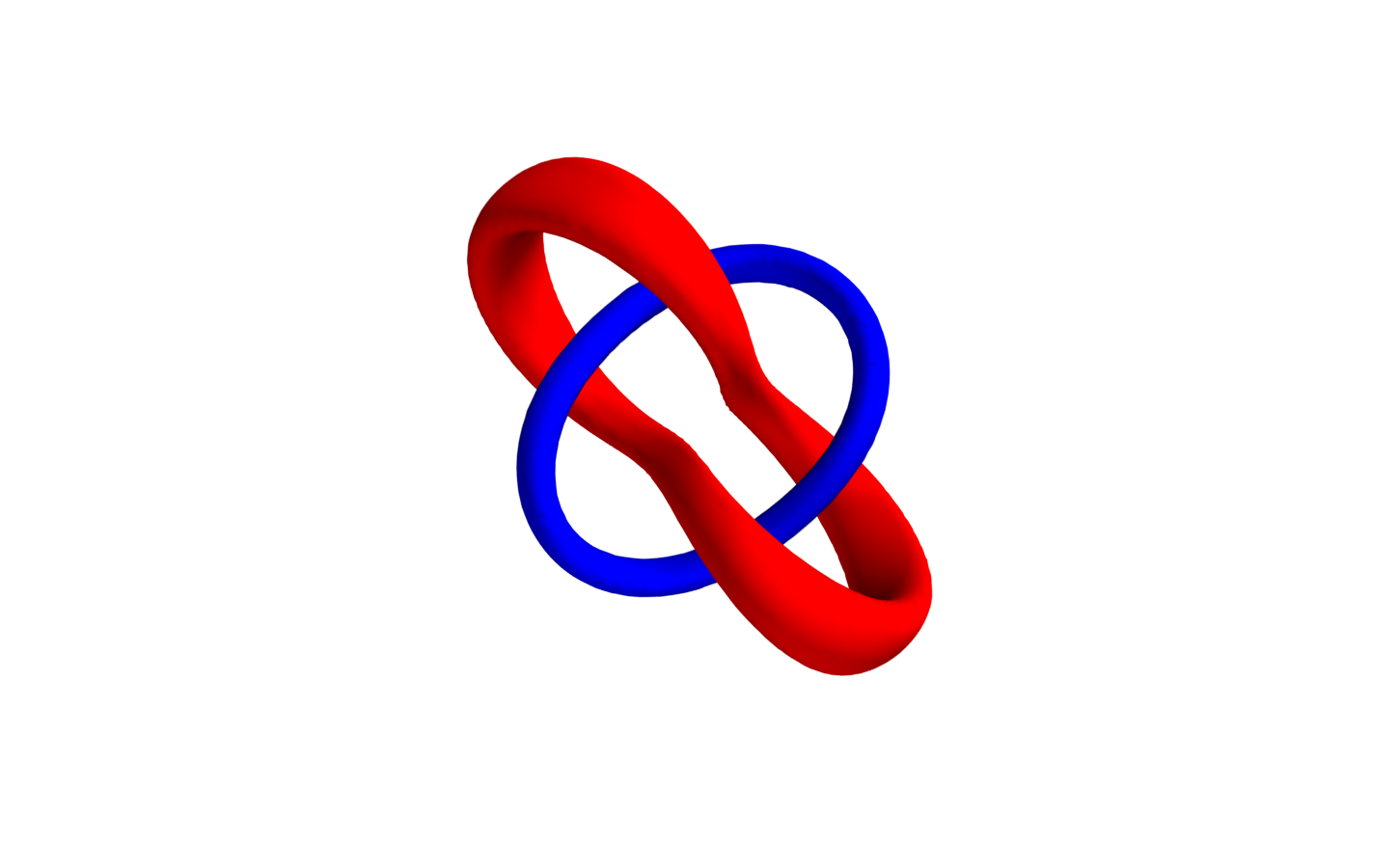} & \includegraphics[height=2cm]{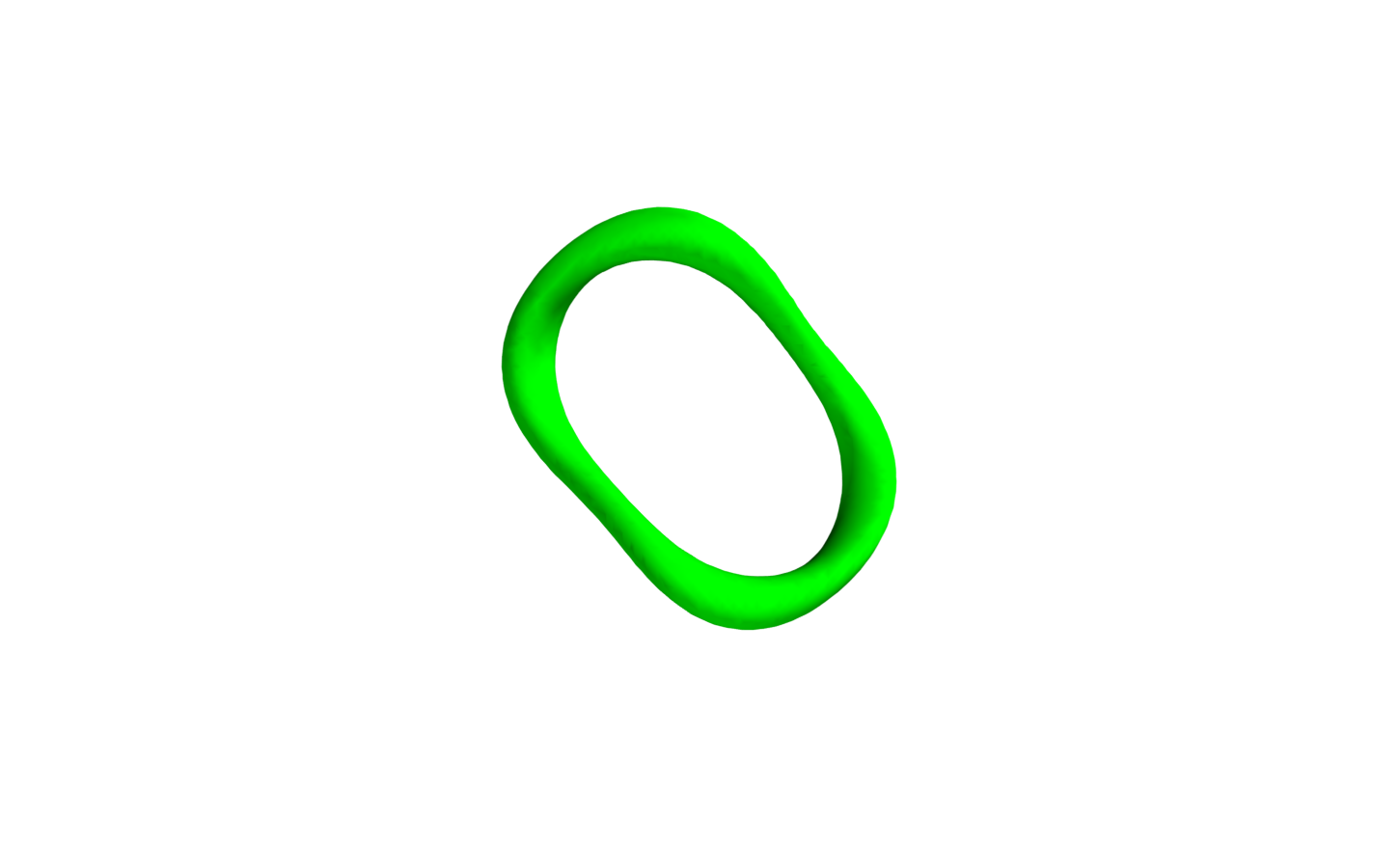} \\
$3(\CA_1\between\CA_1)_{\CA_{3,1}}$ &  \includegraphics[height=2cm]{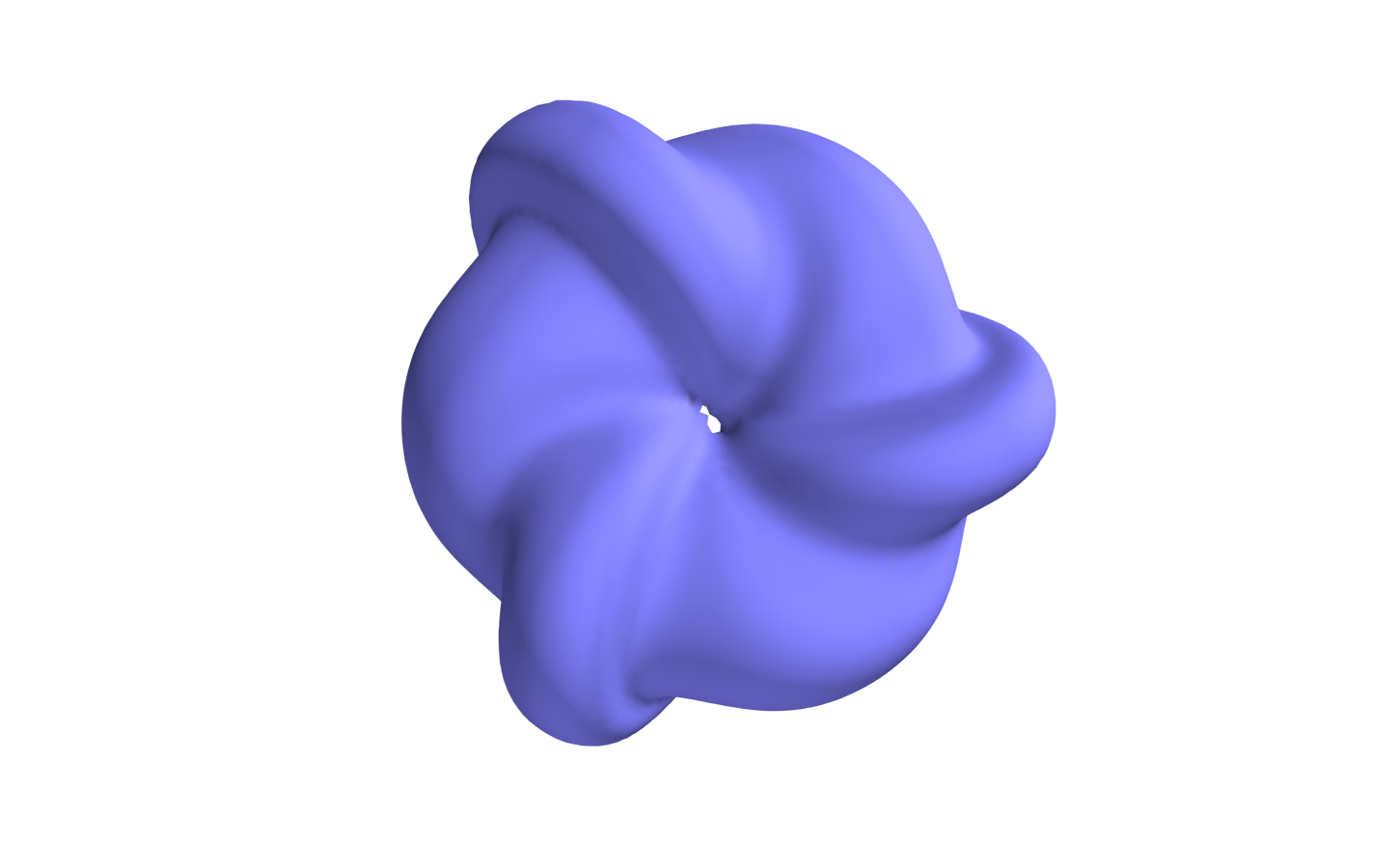} & \includegraphics[height=2cm]{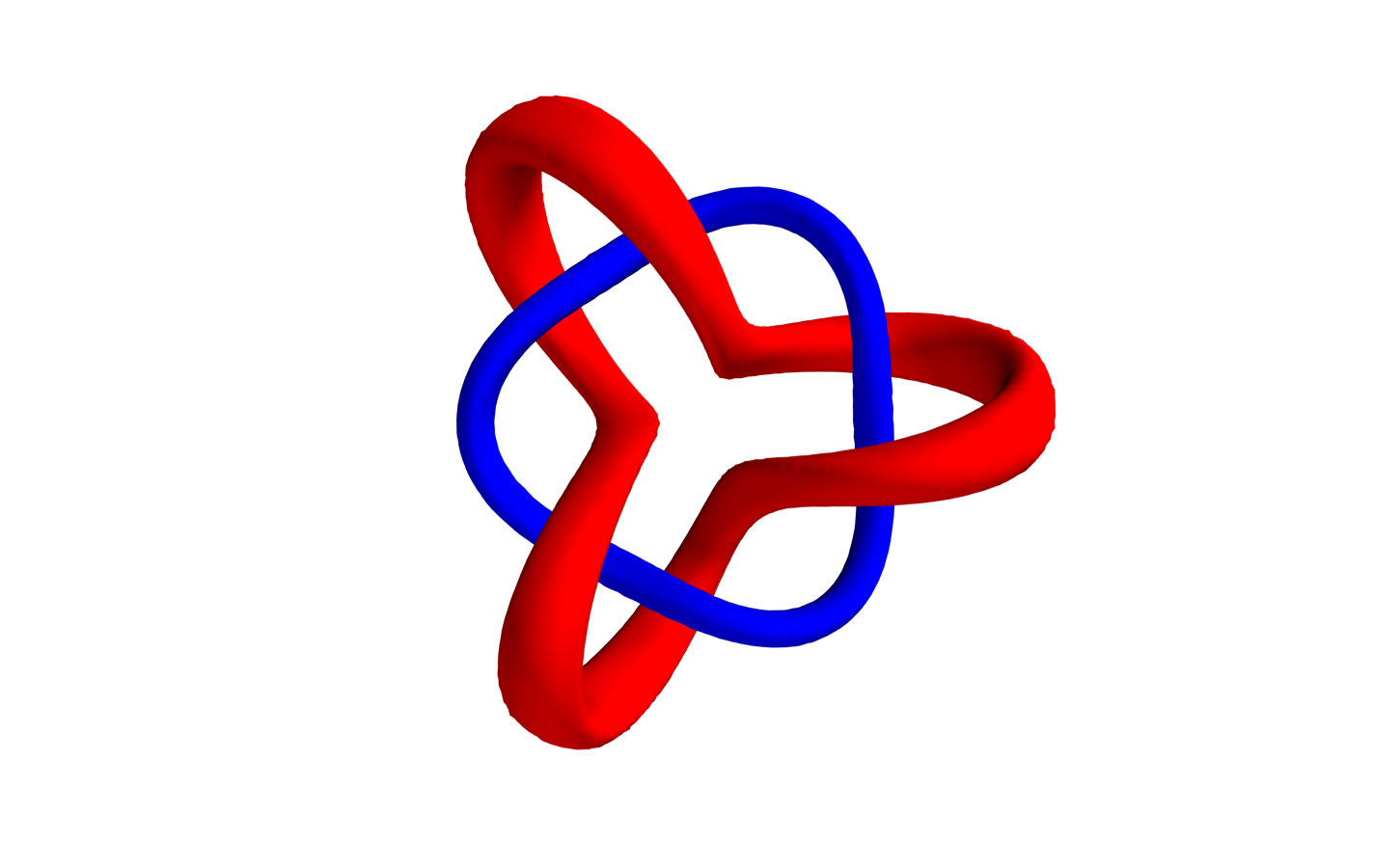} & \includegraphics[height=2cm]{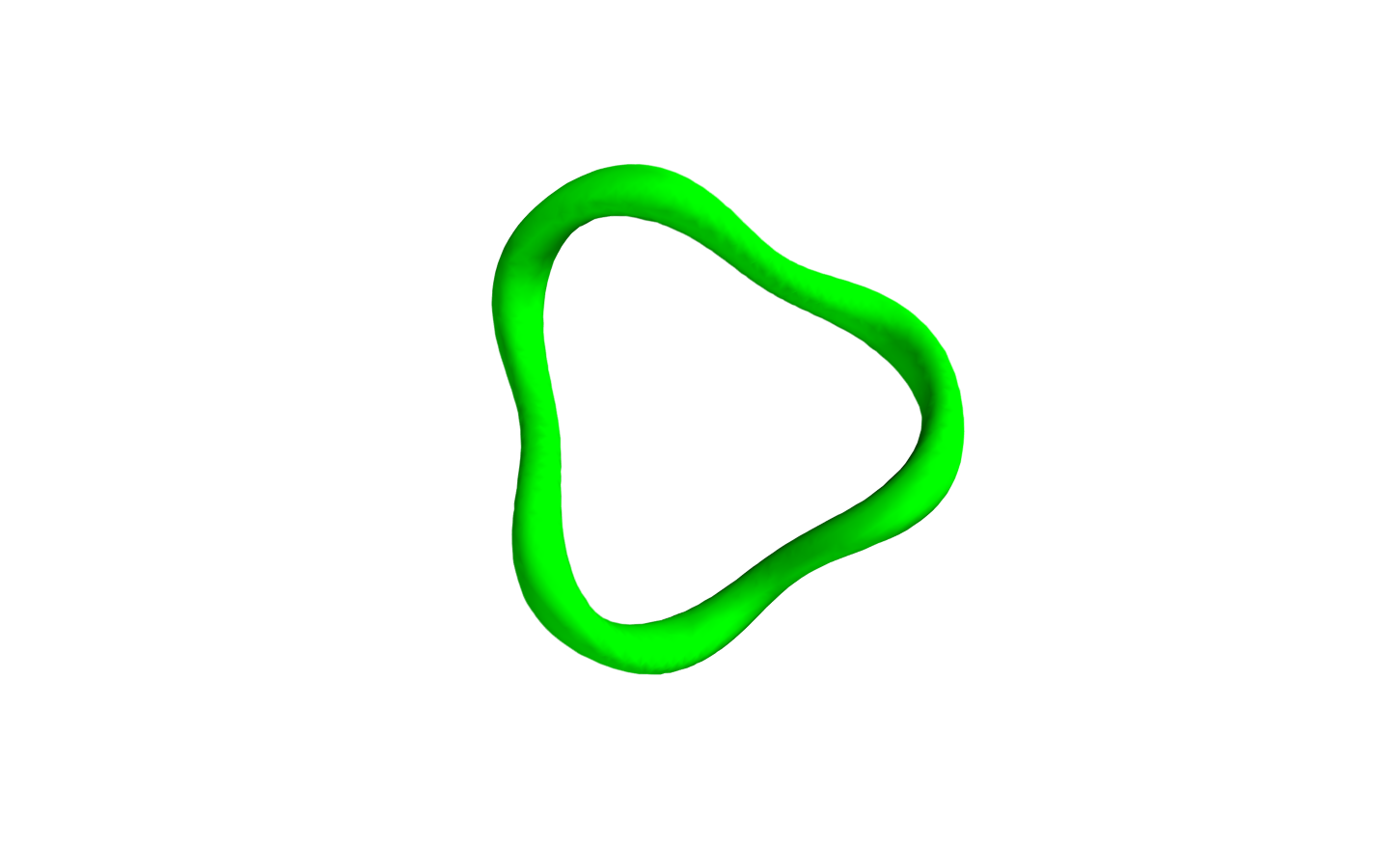}\\
$3(\widetilde{\CA}_1\between\widetilde{\CA}_1)_{\widetilde\CA_{3,1}}$ & \includegraphics[height=2cm]{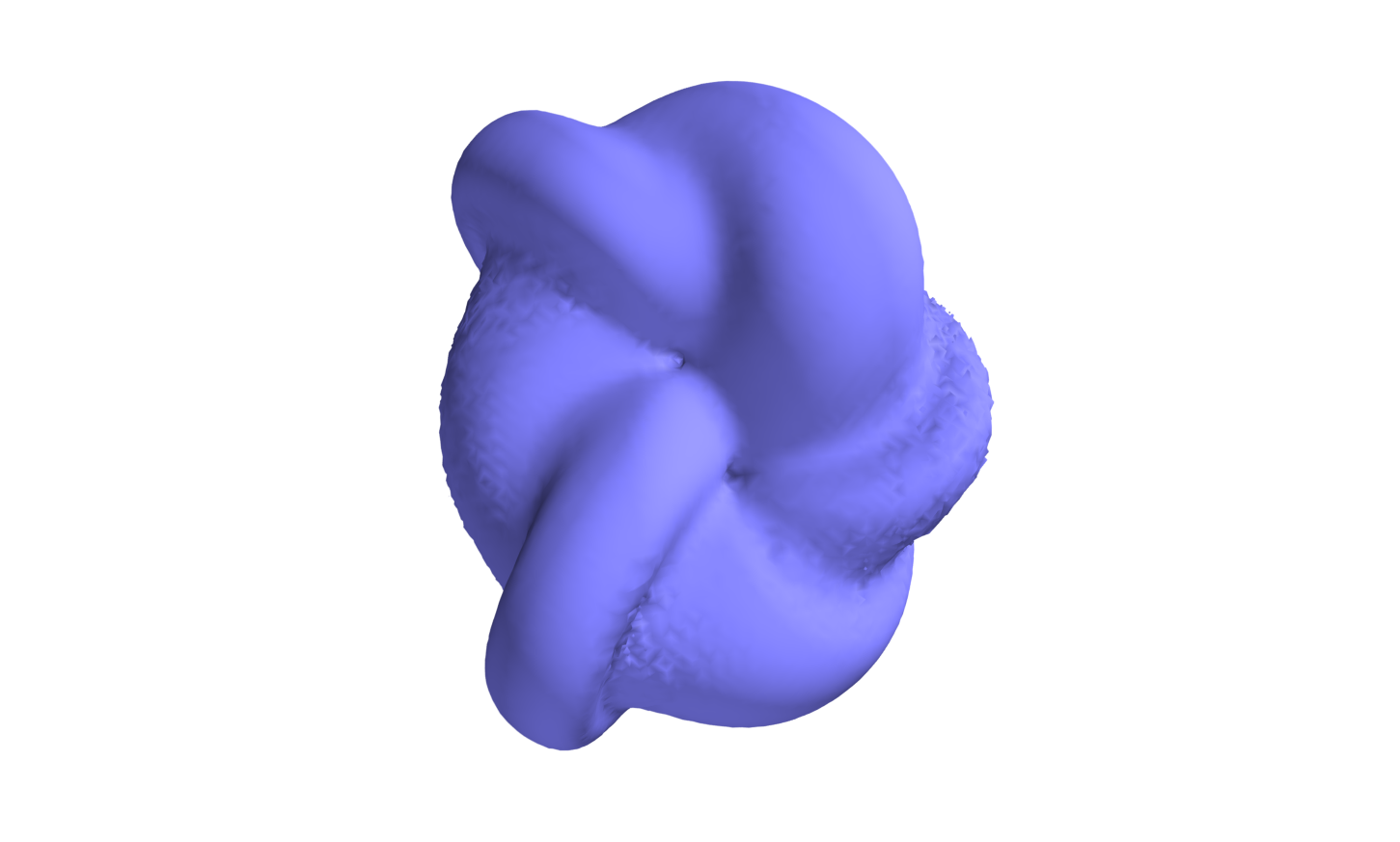} & \includegraphics[height=2cm]{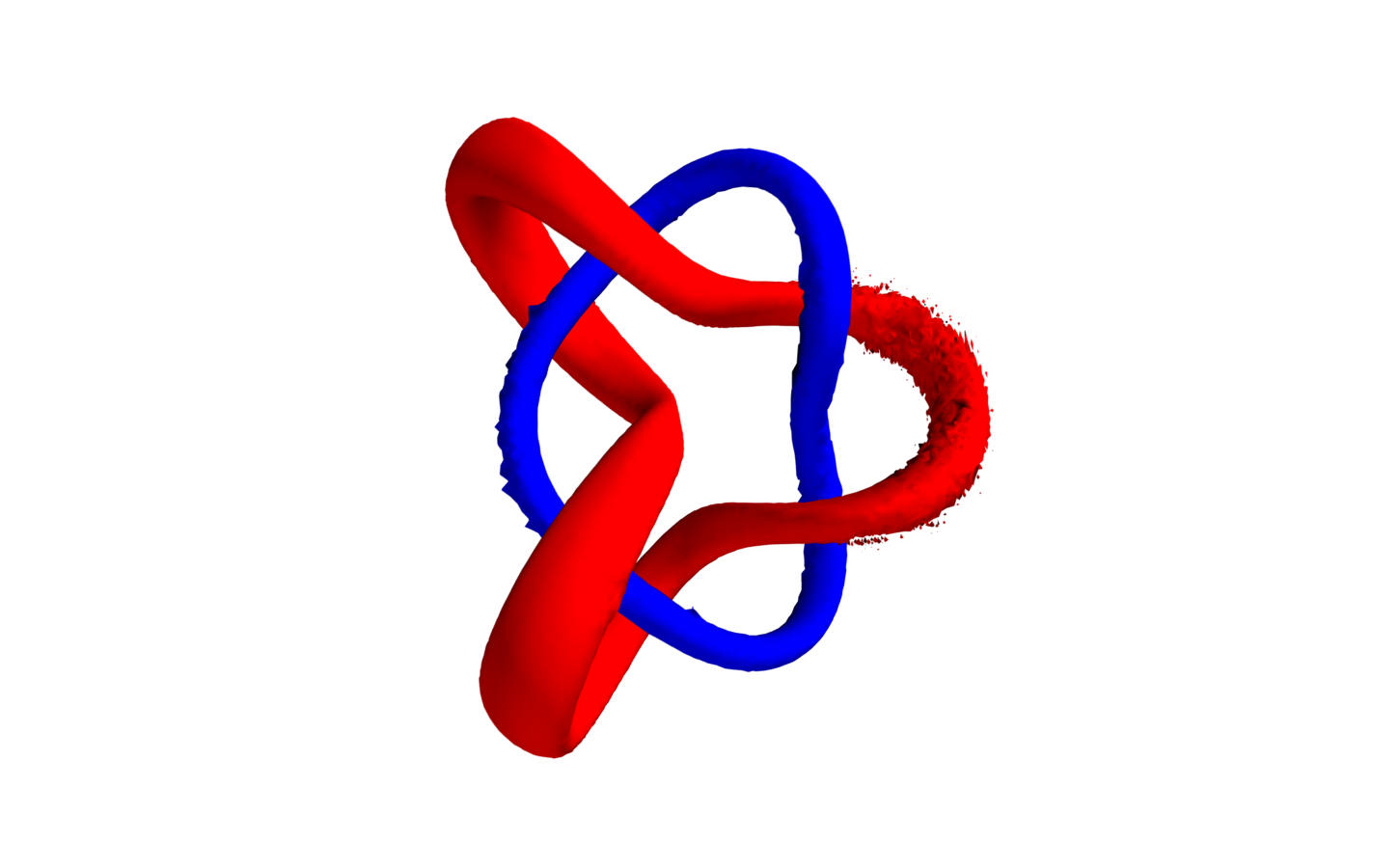} & \includegraphics[height=2cm]{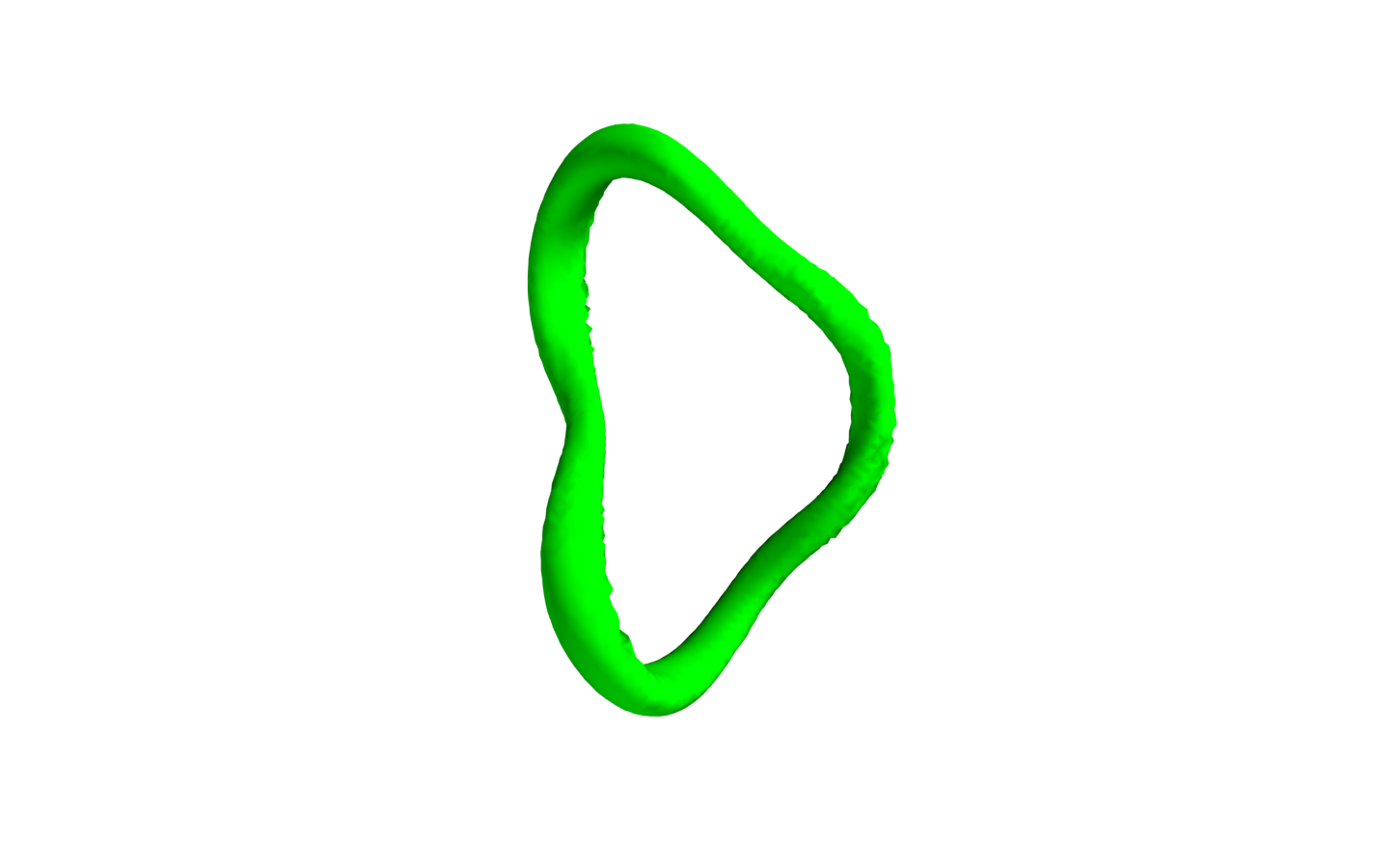} \\
$4(\widetilde\CA_1\between\CL_{1,1})_{\widetilde\CA_{4,1}}$ & \includegraphics[height=2cm]{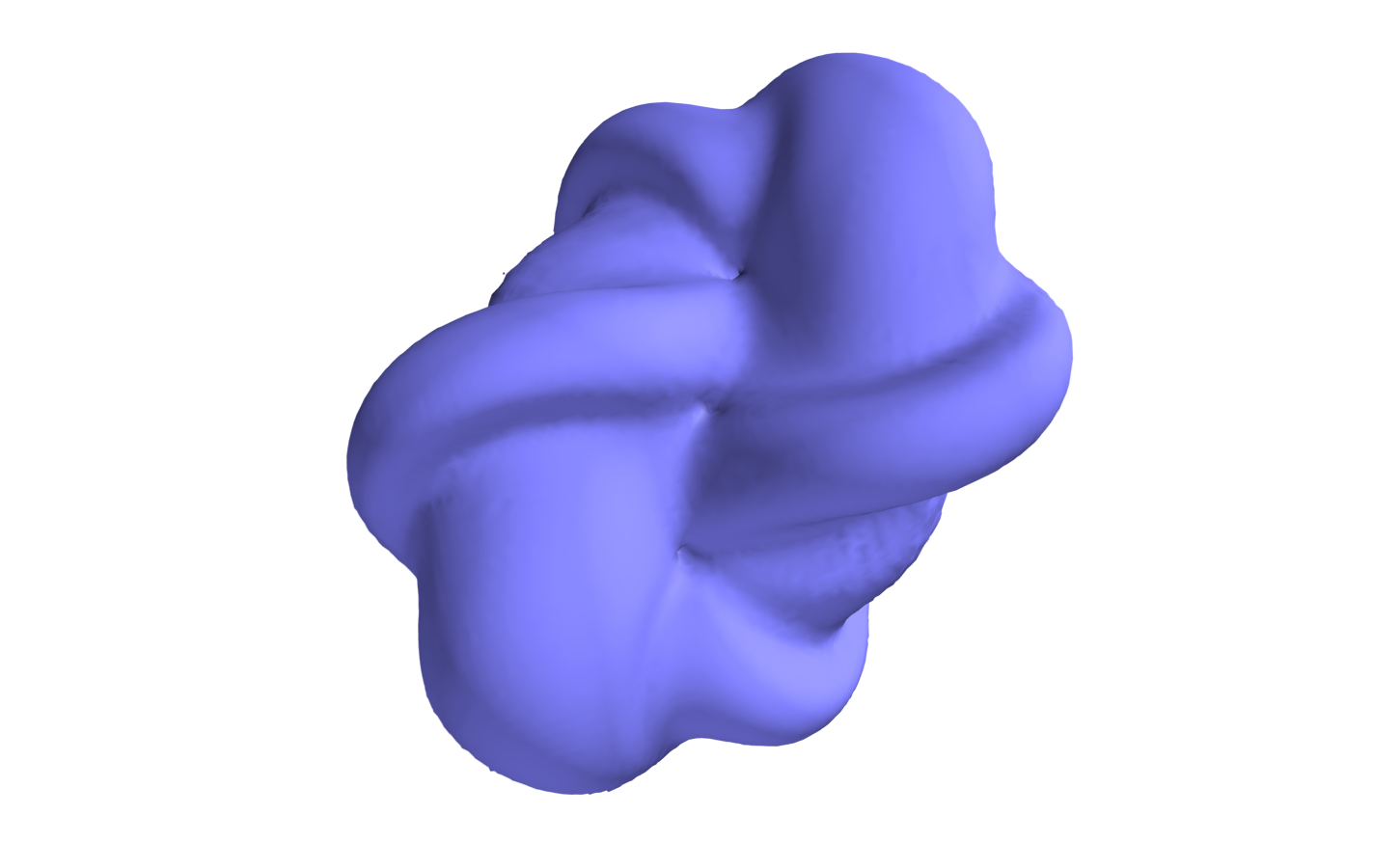} & \includegraphics[height=2cm]{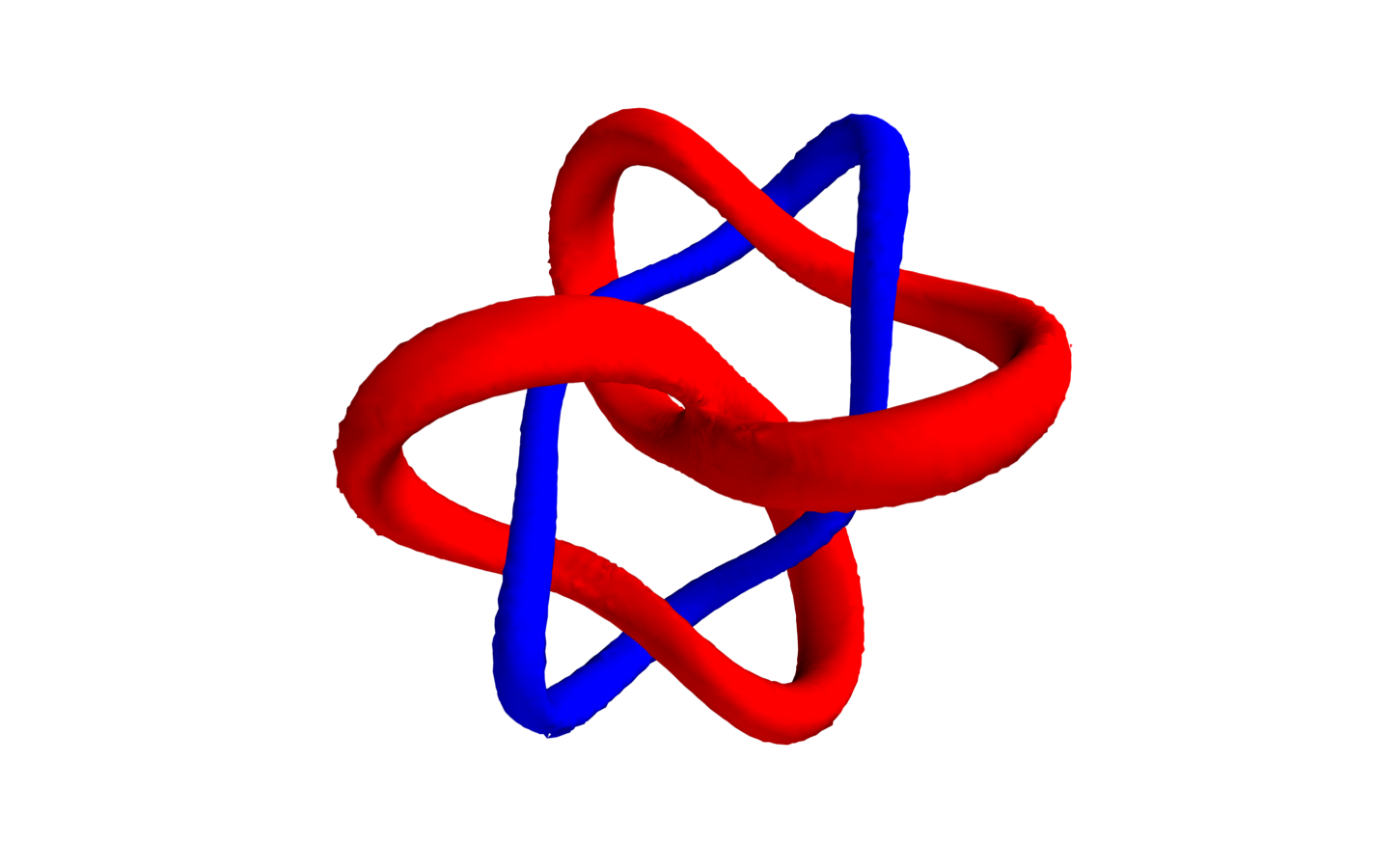} & \includegraphics[height=2cm]{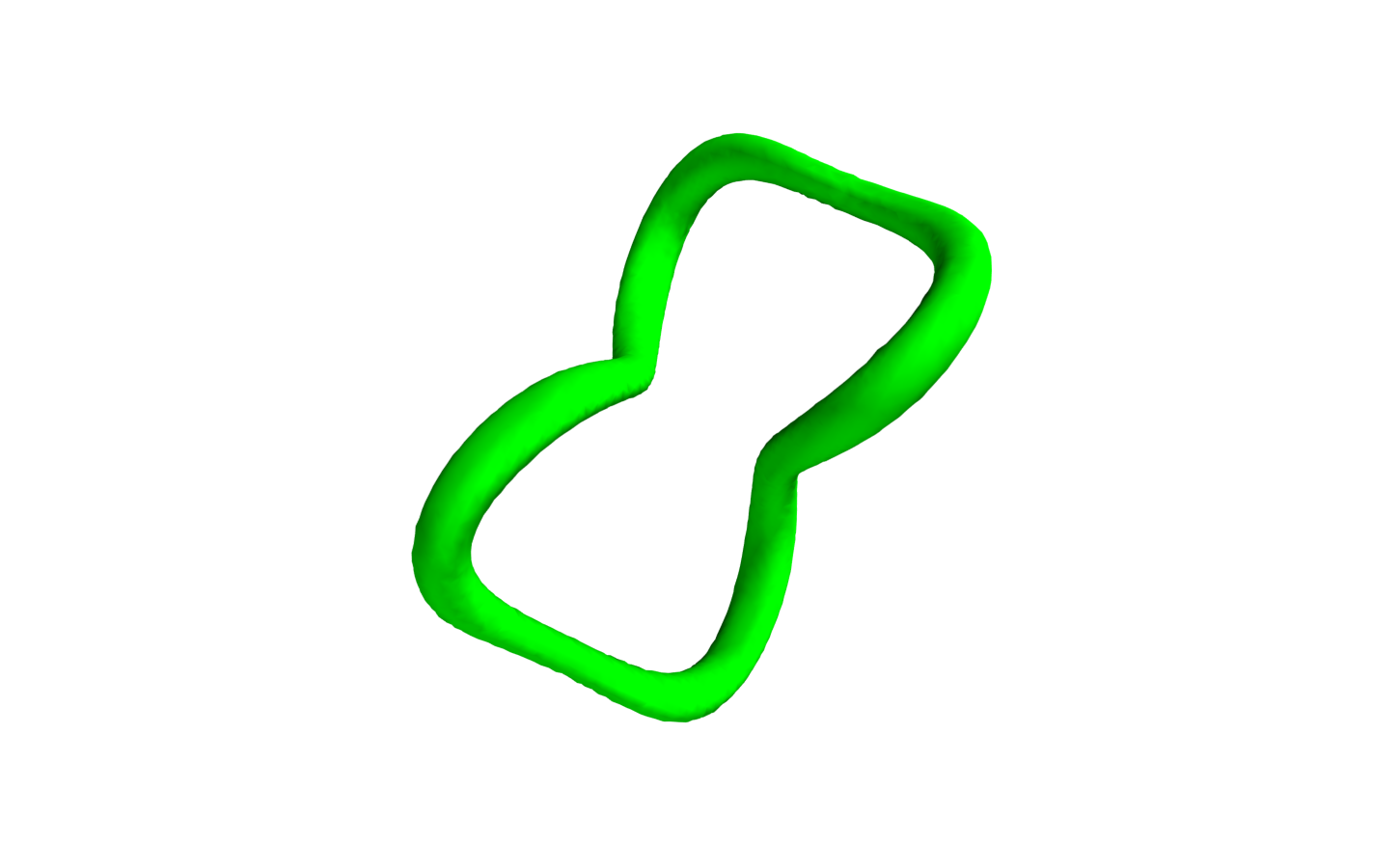} \\
$4(\CL_{1,1}\between\CL_{1,1})_{\CA_{2,2}}$ & \includegraphics[height=2cm]{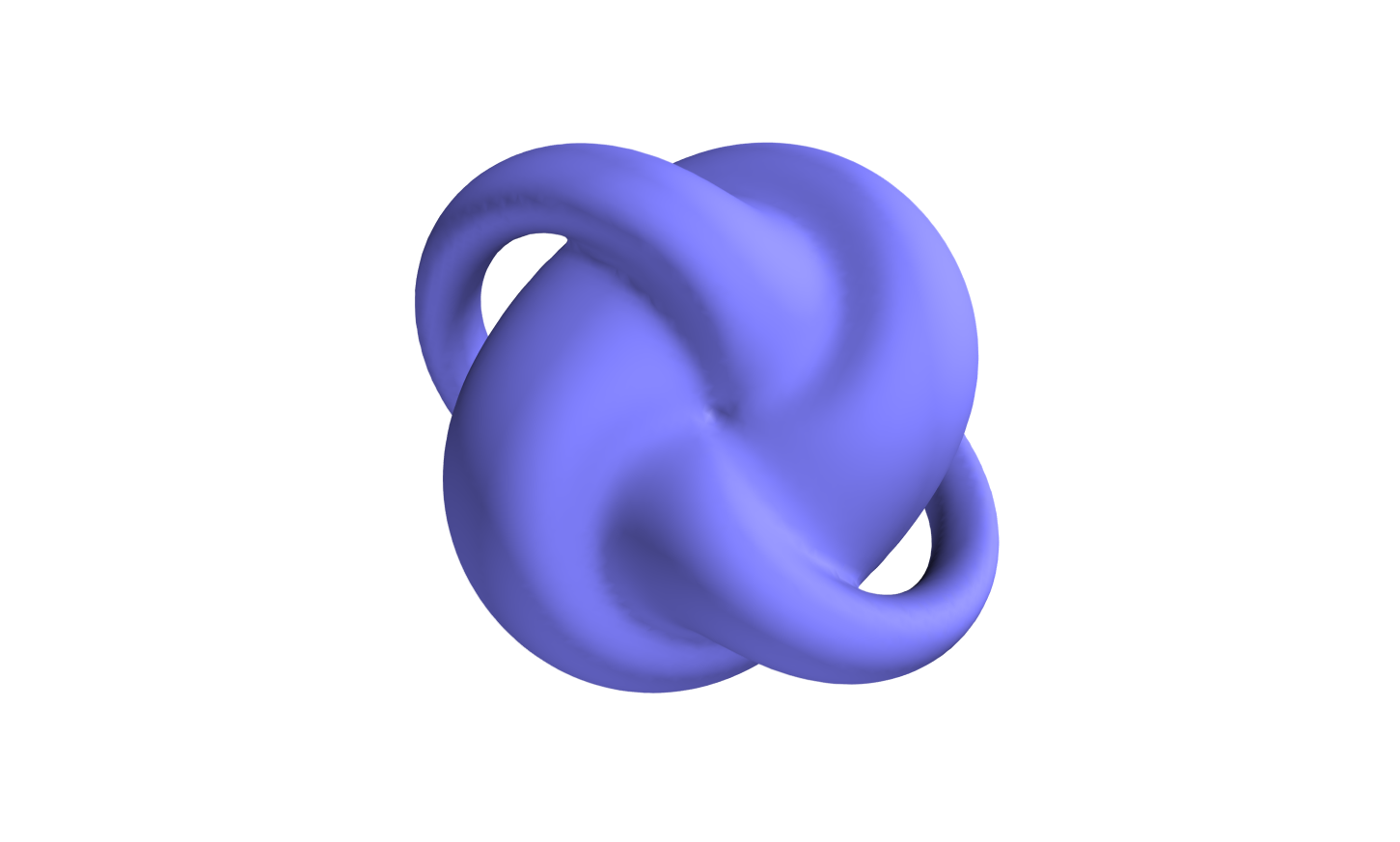} & \includegraphics[height=2cm]{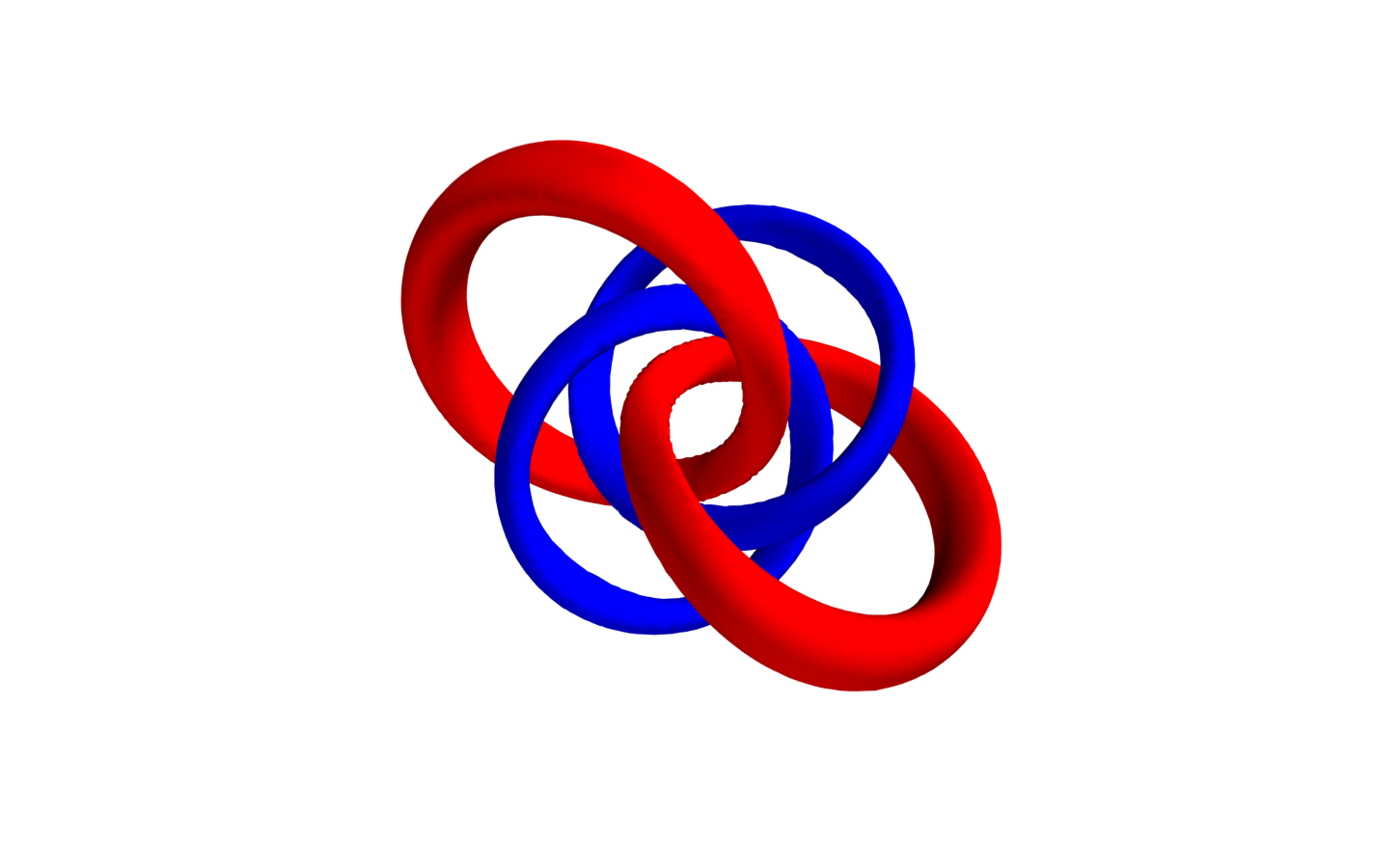} &
\includegraphics[height=2cm]{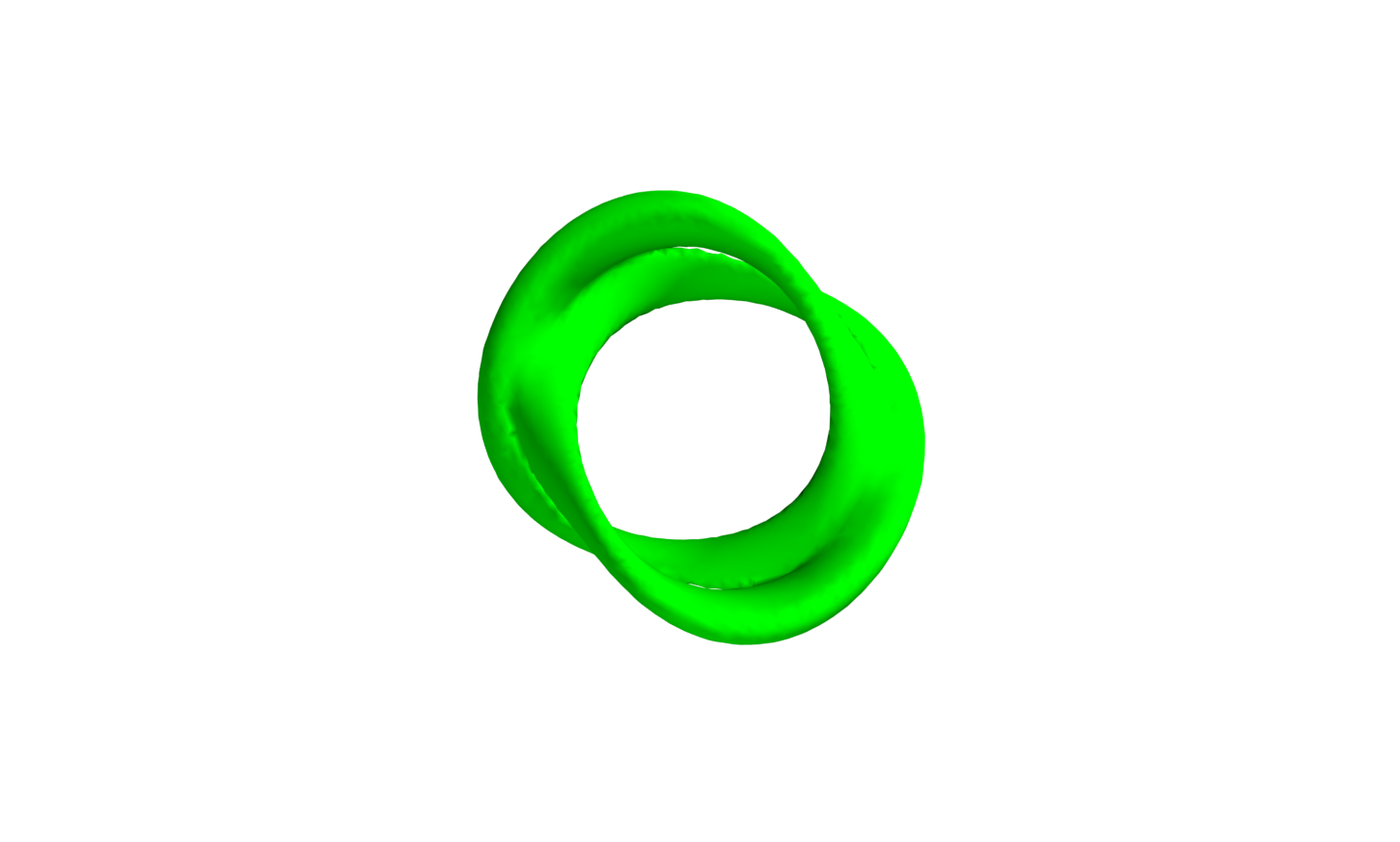}\\
$4(\CA_1\between\CA_1)_{\CA_{4,1}}$ & \includegraphics[height=2cm]{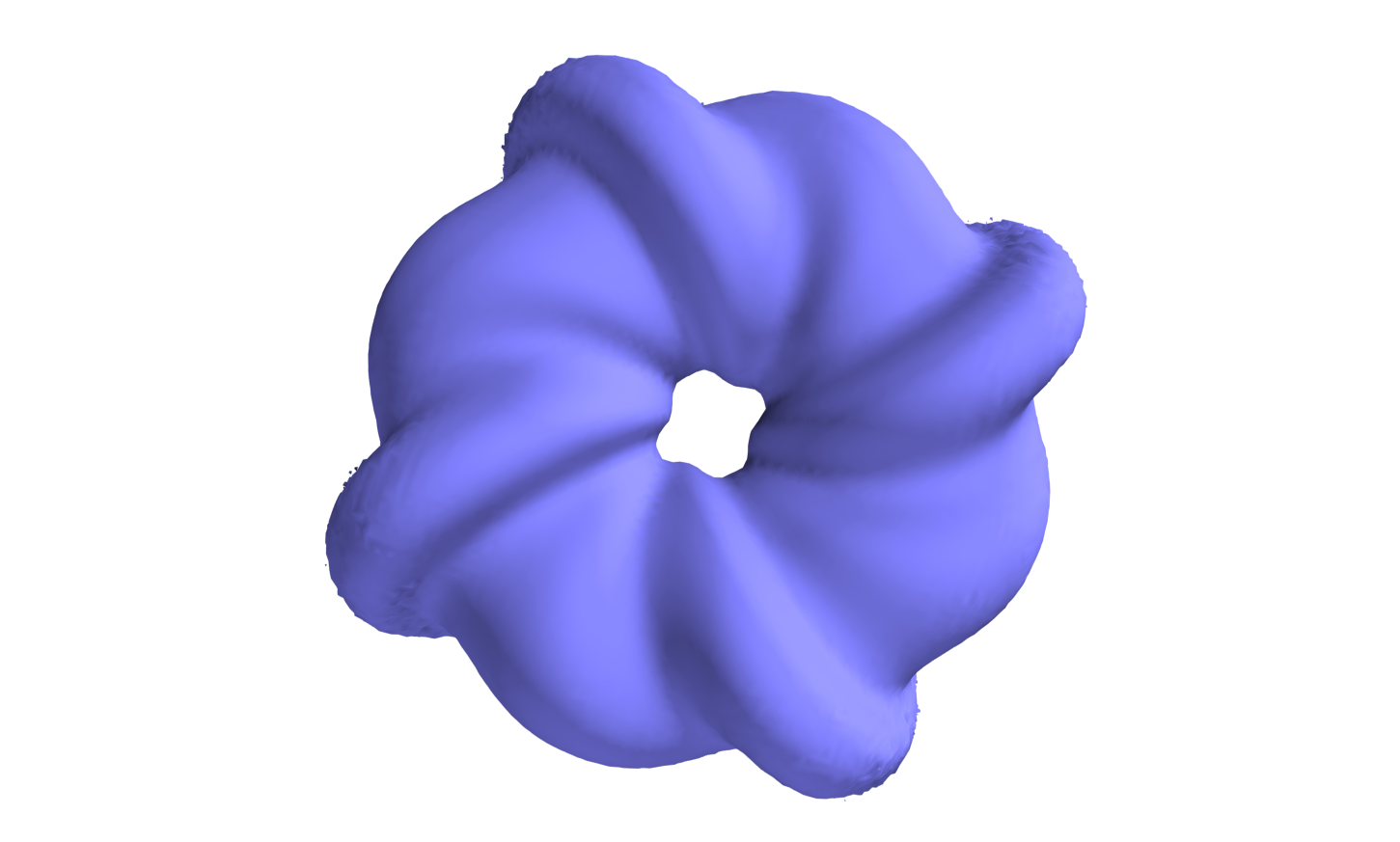} & \includegraphics[height=2cm]{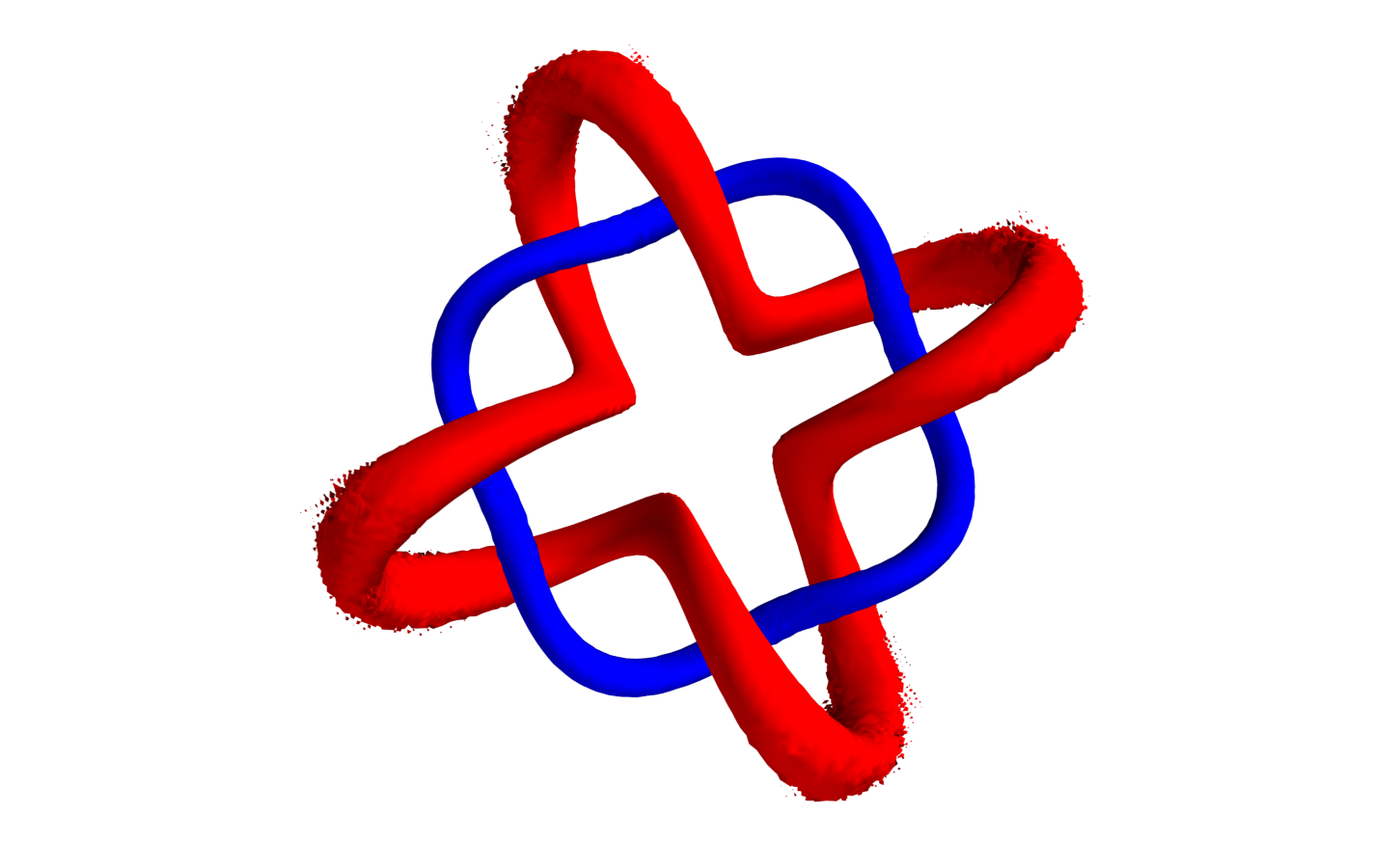} & \includegraphics[height=2cm]{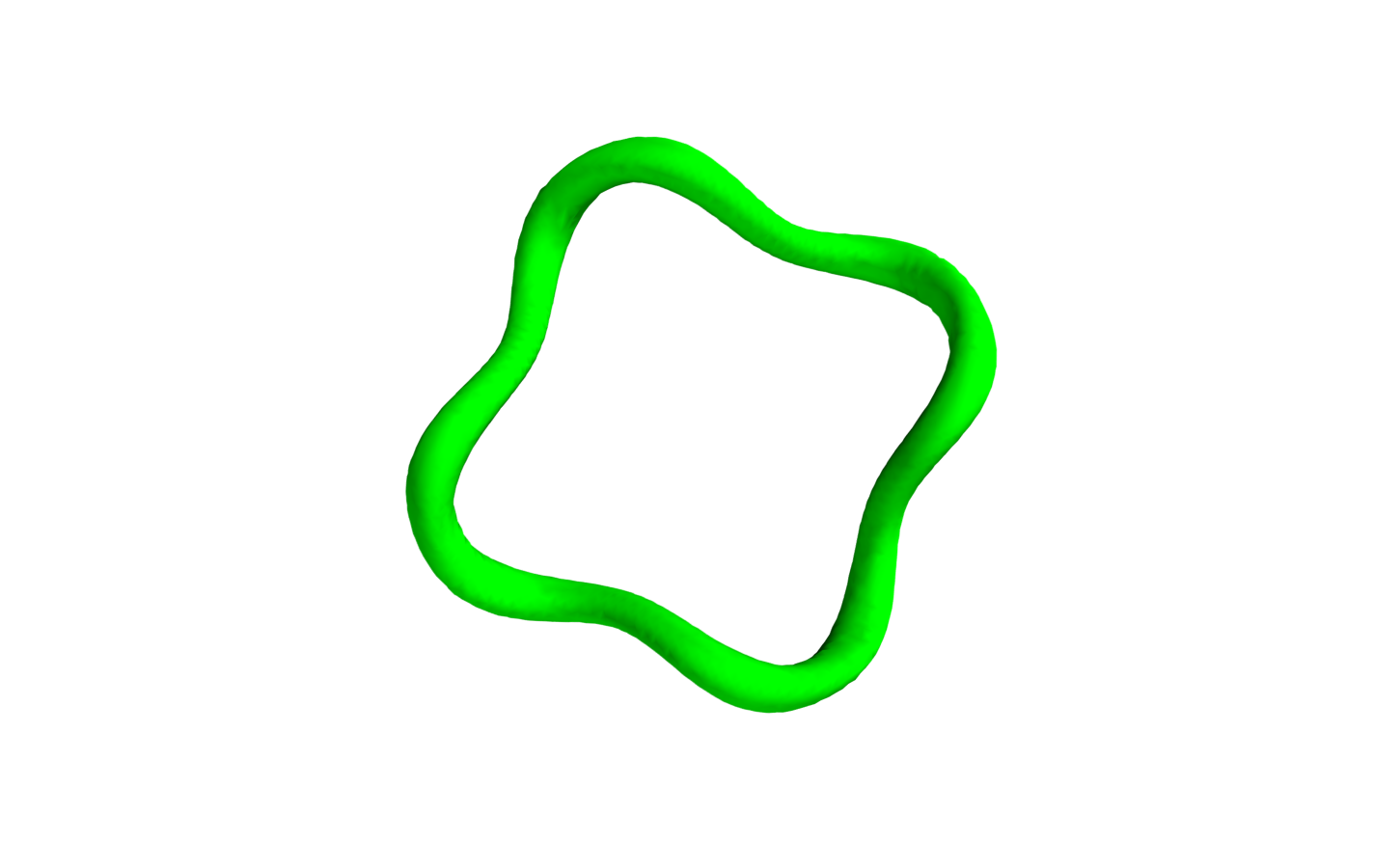} \\
$5(\CL_{1,2}\between\CK_{3,2})_{\CL_{1,2}}$  & \includegraphics[height=2cm]{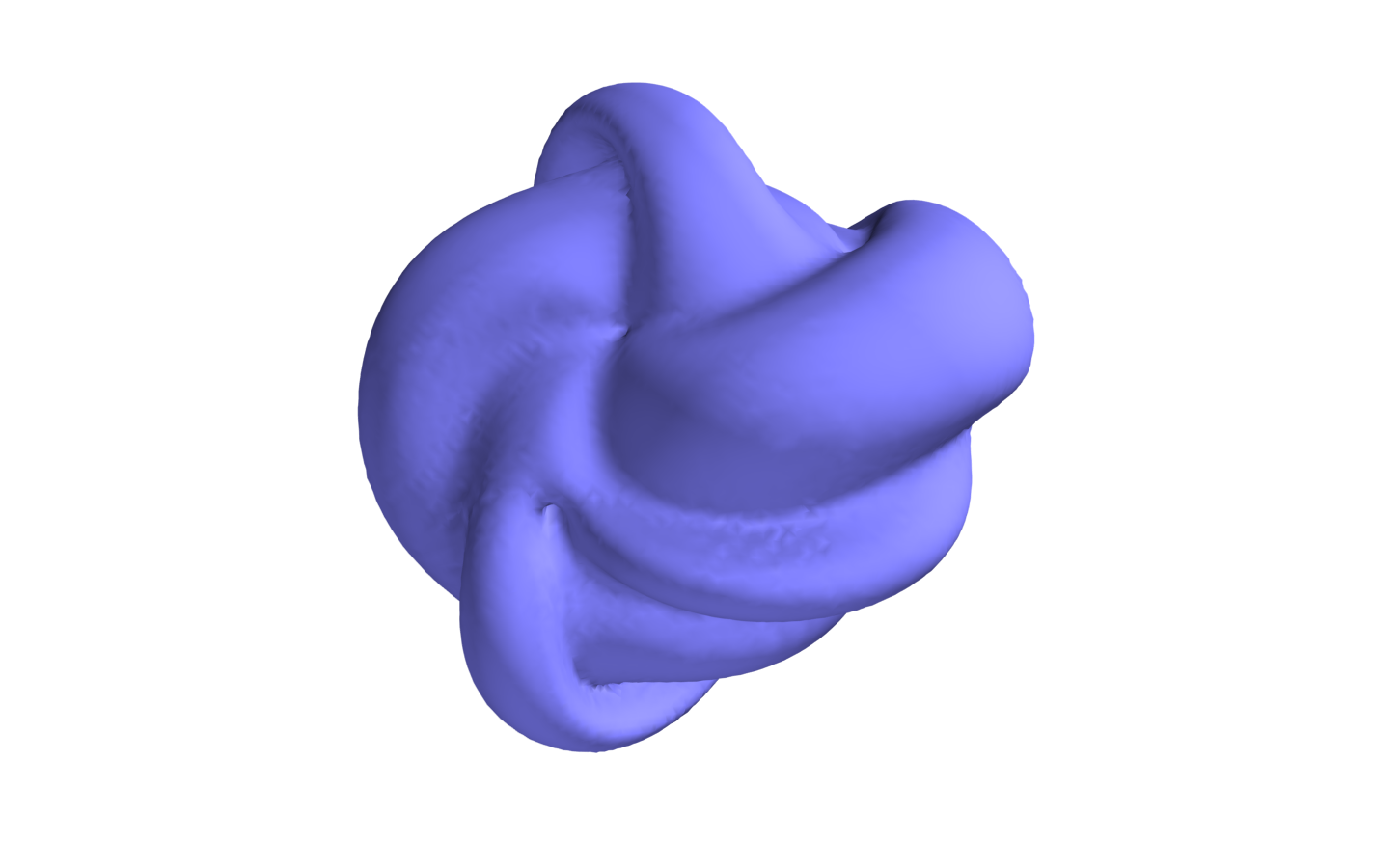} &  \includegraphics[height=2cm]{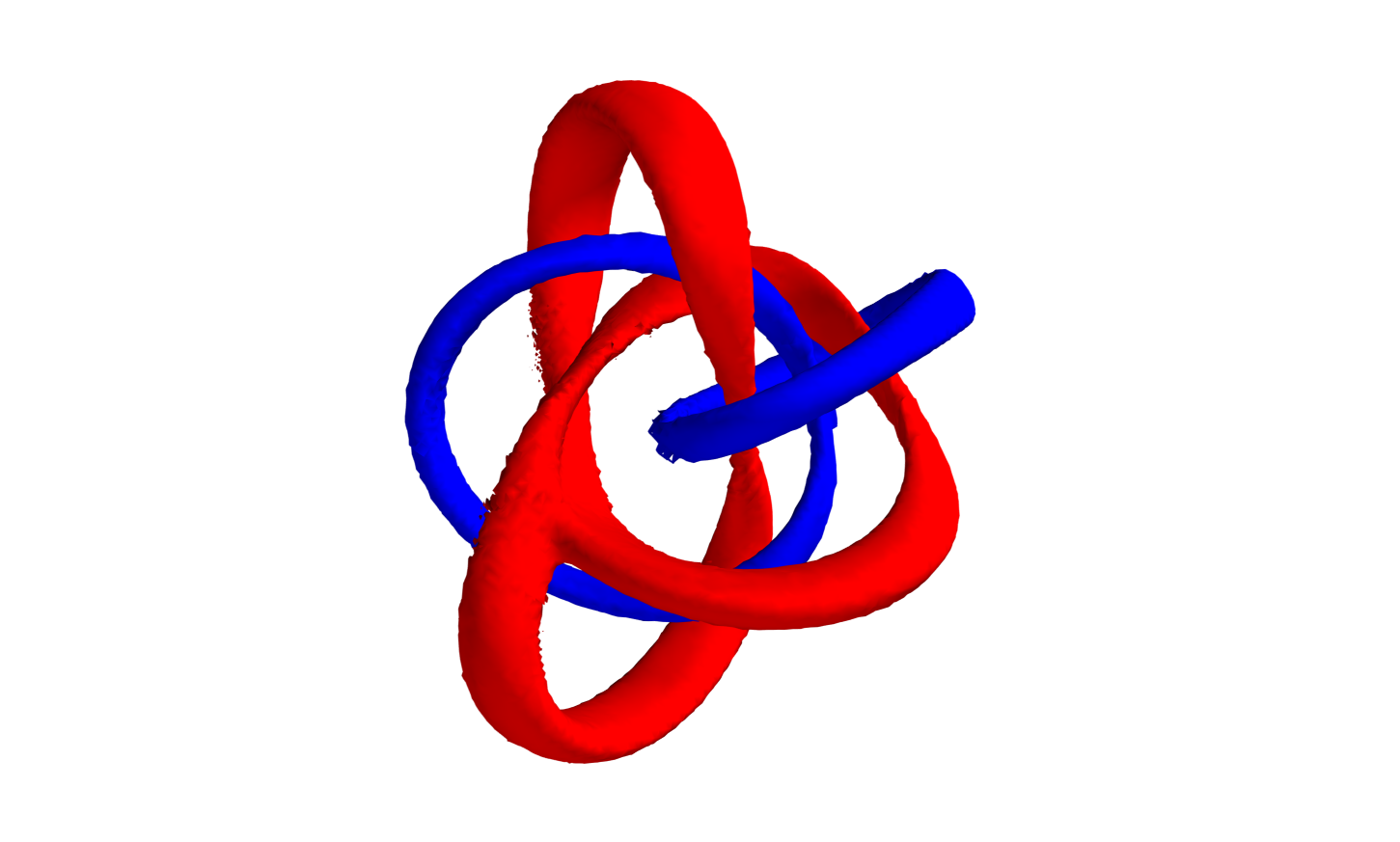} & \includegraphics[height=2cm]{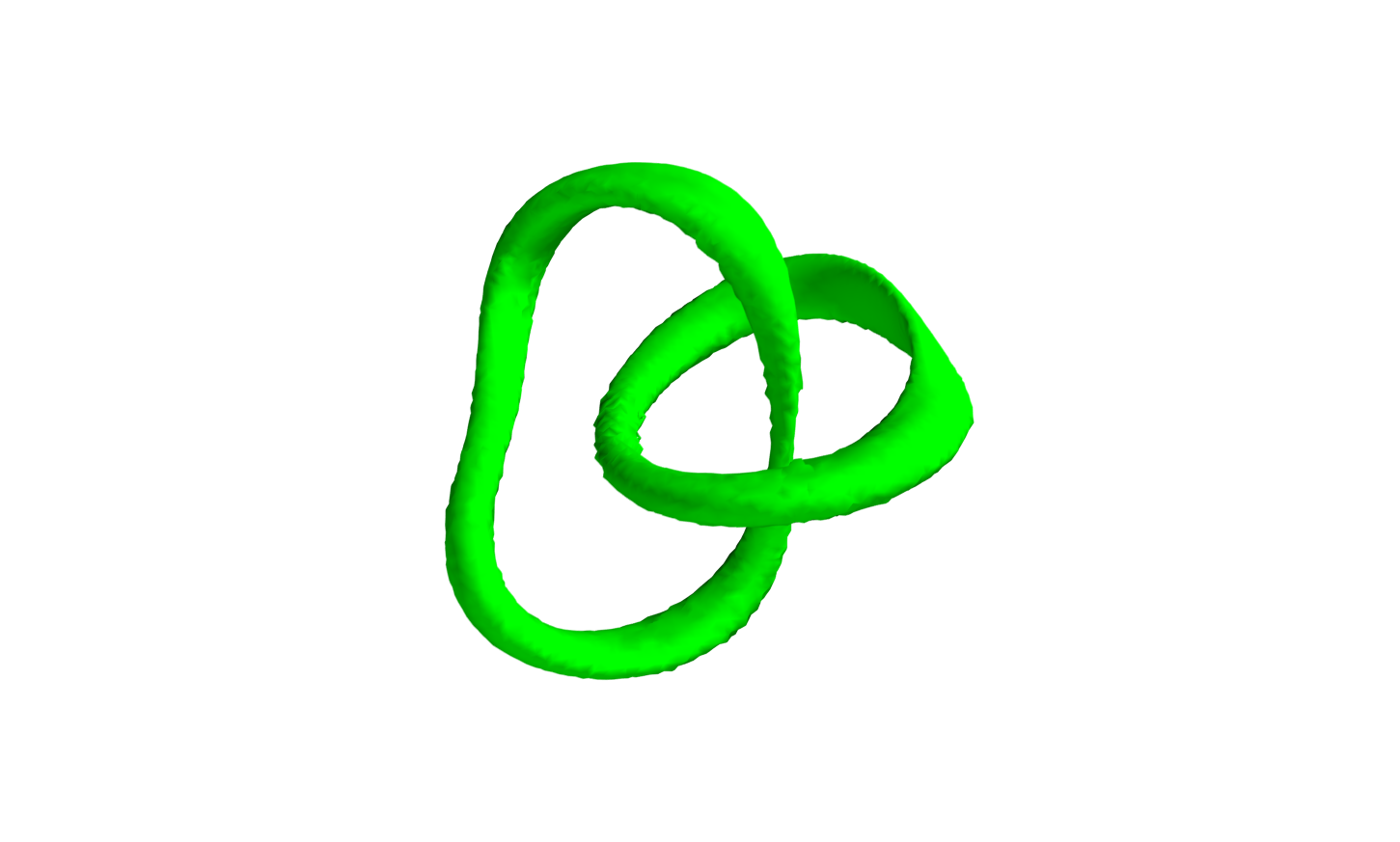}\\
$5(\CA_1\between\CA_1)_{\widetilde\CA_{5,1}}$  & \includegraphics[height=2cm]{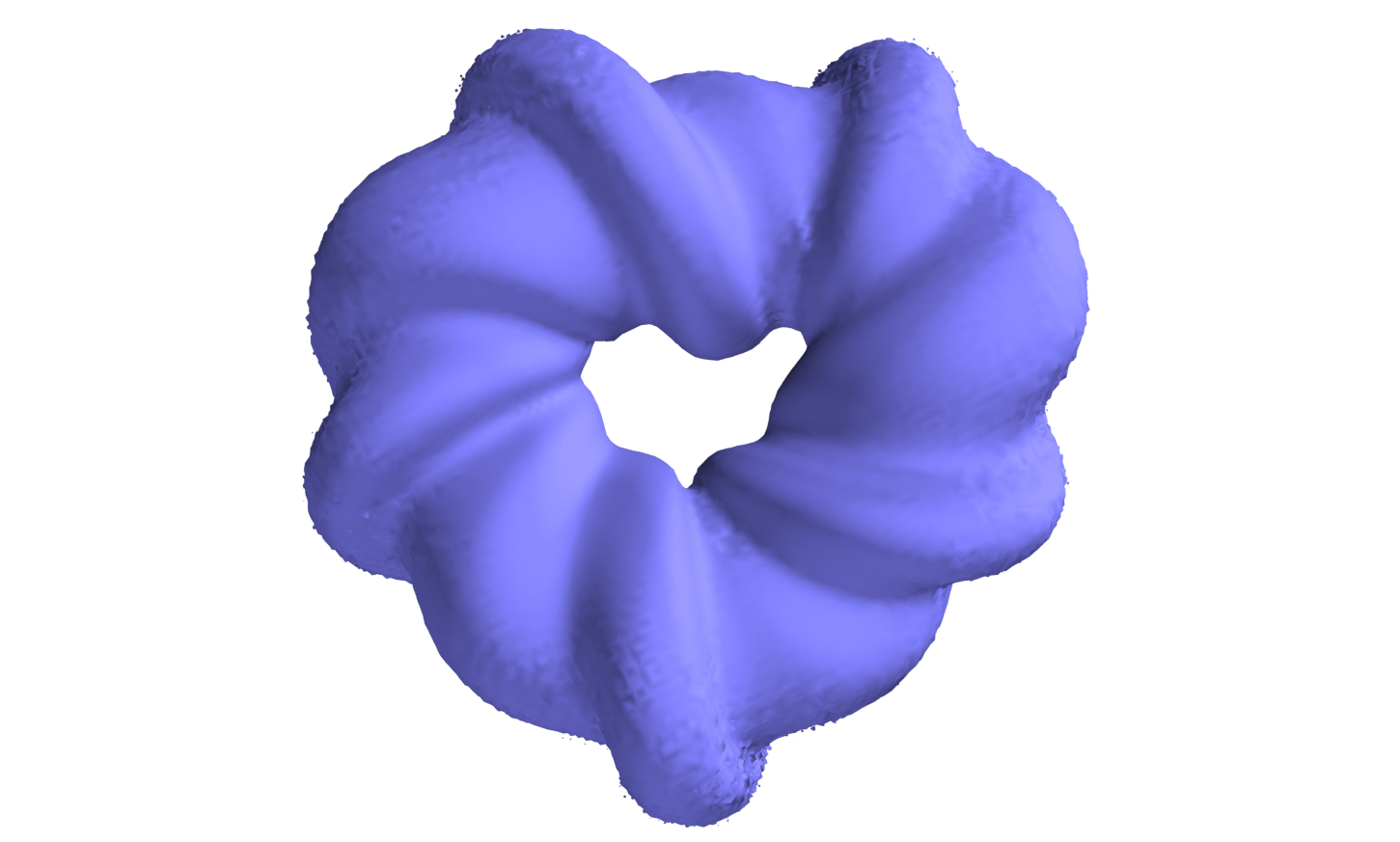} & \includegraphics[height=2cm]{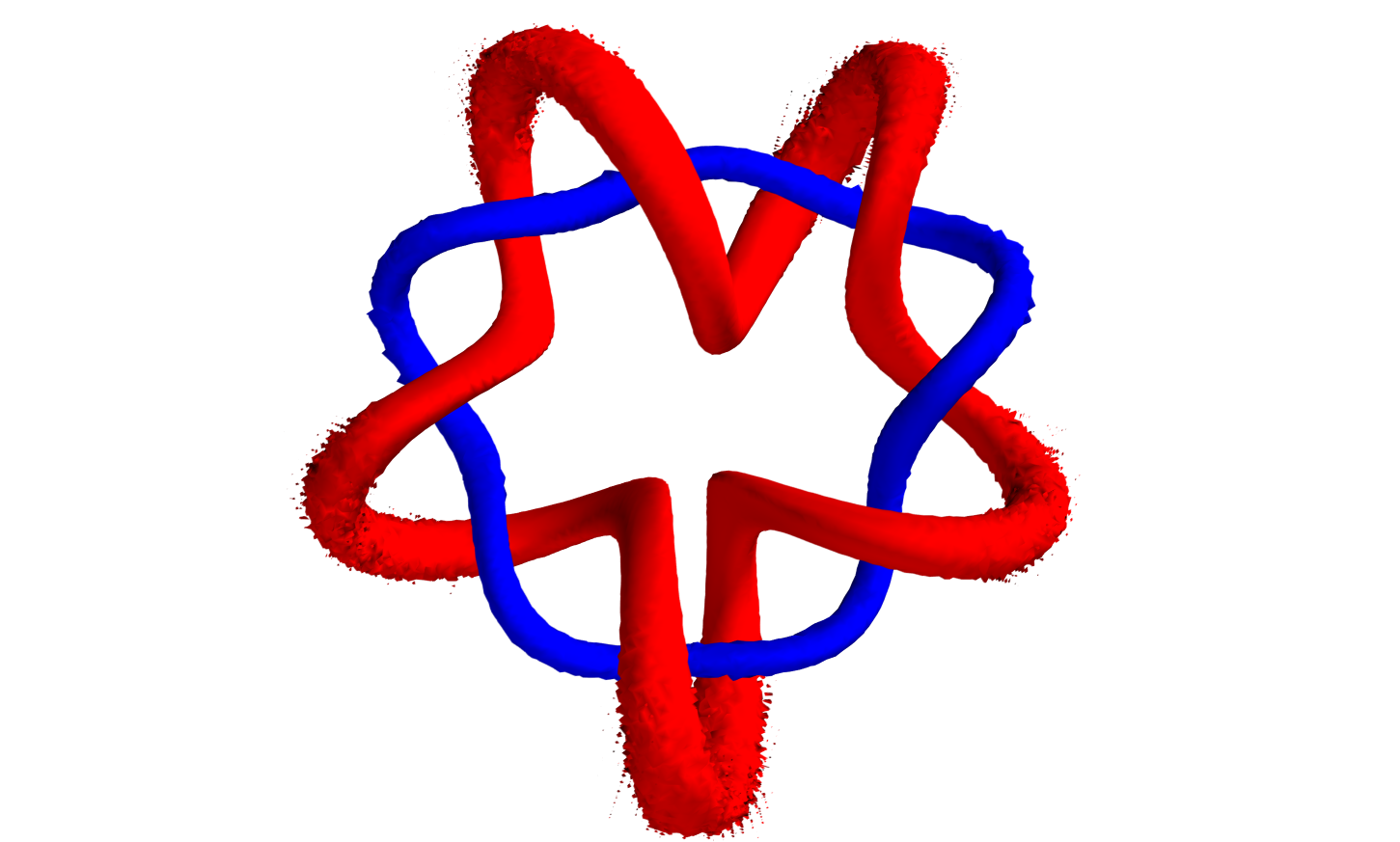} & \includegraphics[height=2cm]{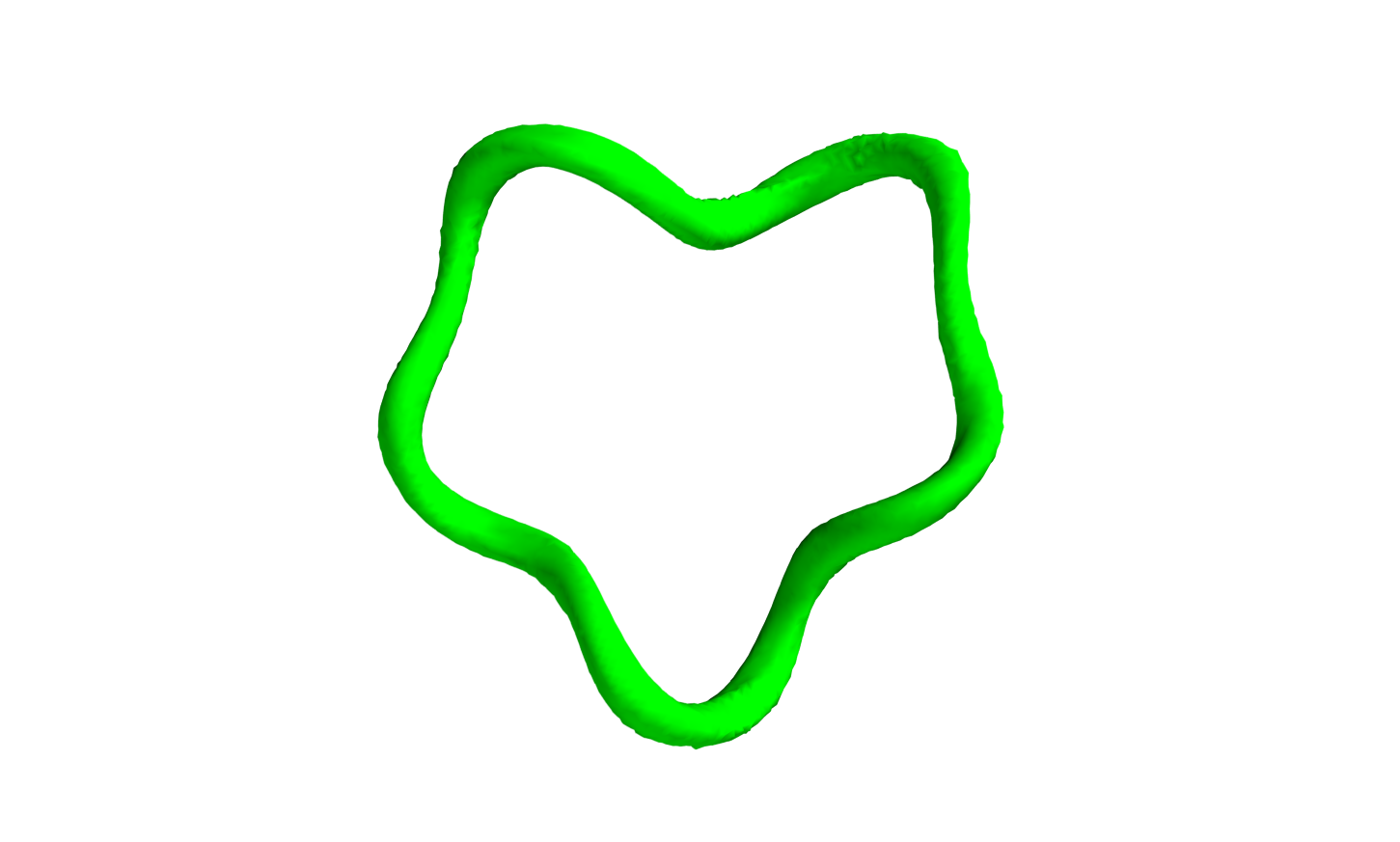} \\
\end{TAB}
\end{center}
\end{table}

\begin{table}
    \begin{center}
        \begin{TAB}[1pt]{|c|c|c|c|}{|c|c|c|c|c|c|c|}
            Configuration & $\rho_E$ &  $\phi_1$ &  $\phi_3$  \\
            $6(\CK_{3,2}\between\CK_{3,2})_{\CK_{3,2}}$  &  \includegraphics[height=2cm]{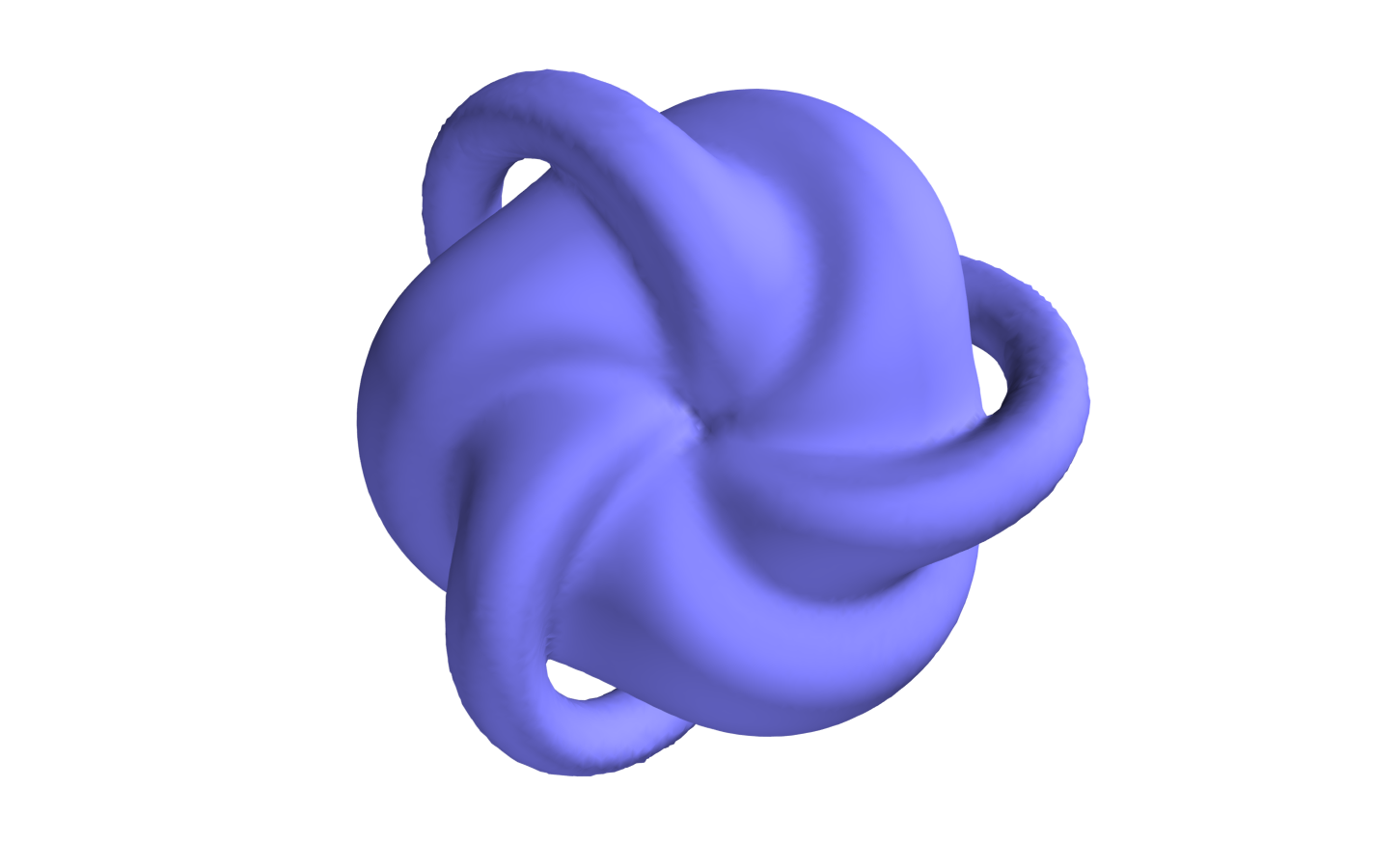} & \includegraphics[height=2cm]{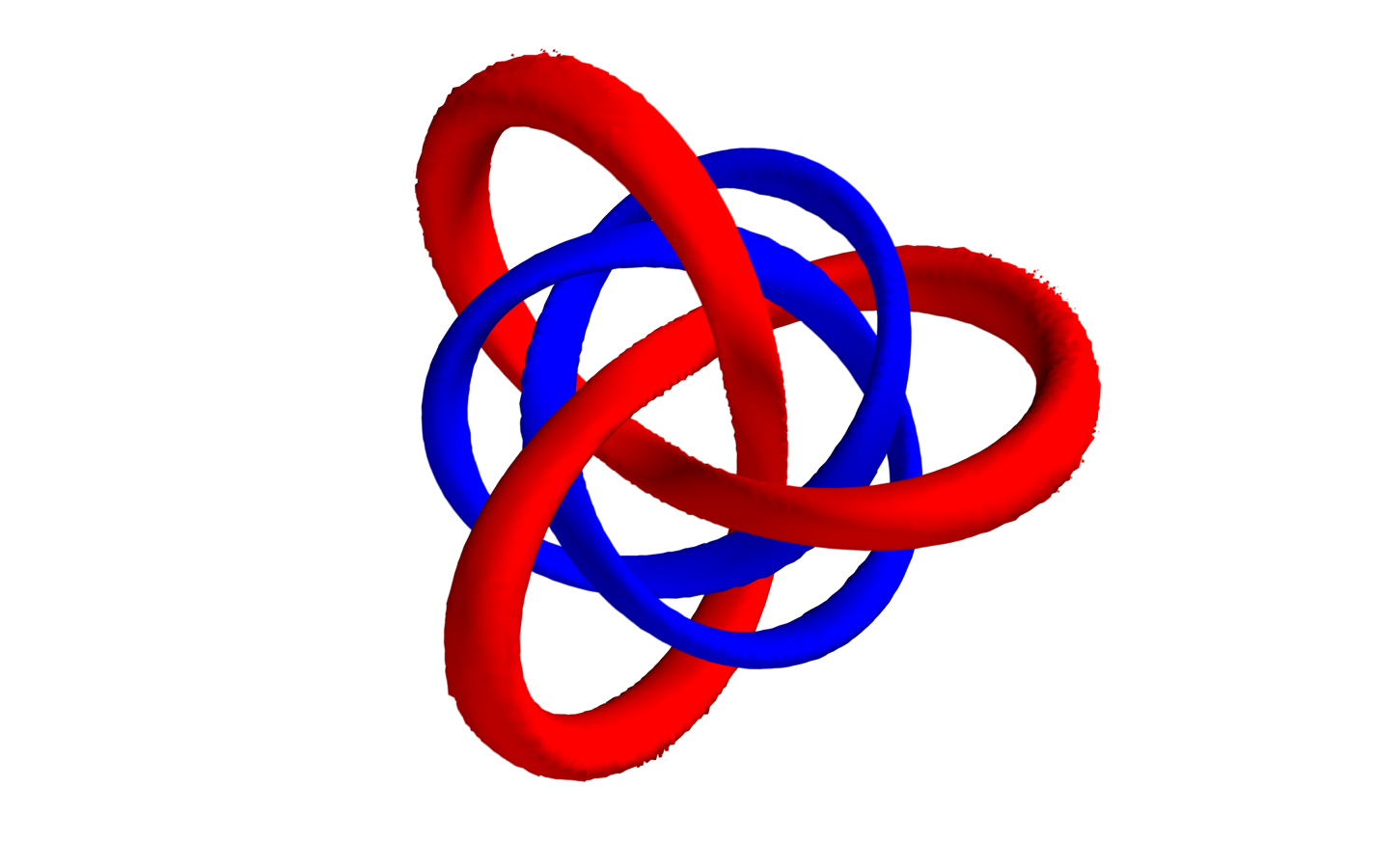} &\includegraphics[height=2cm]{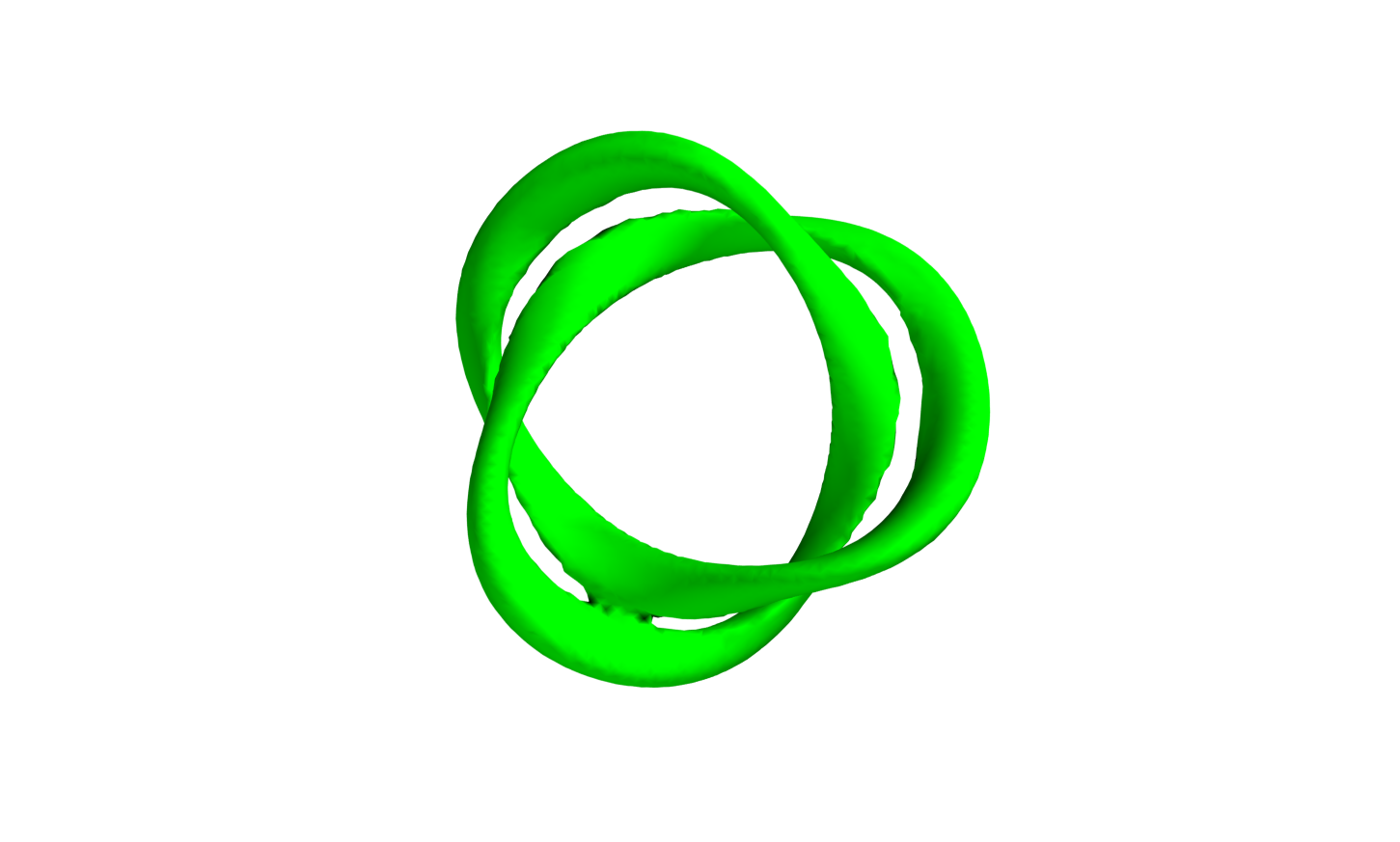}\\
            $6(\CL_{1,3}\between\CK_{3,2})_{\CK_{3,2}}$  &  \includegraphics[height=2cm]{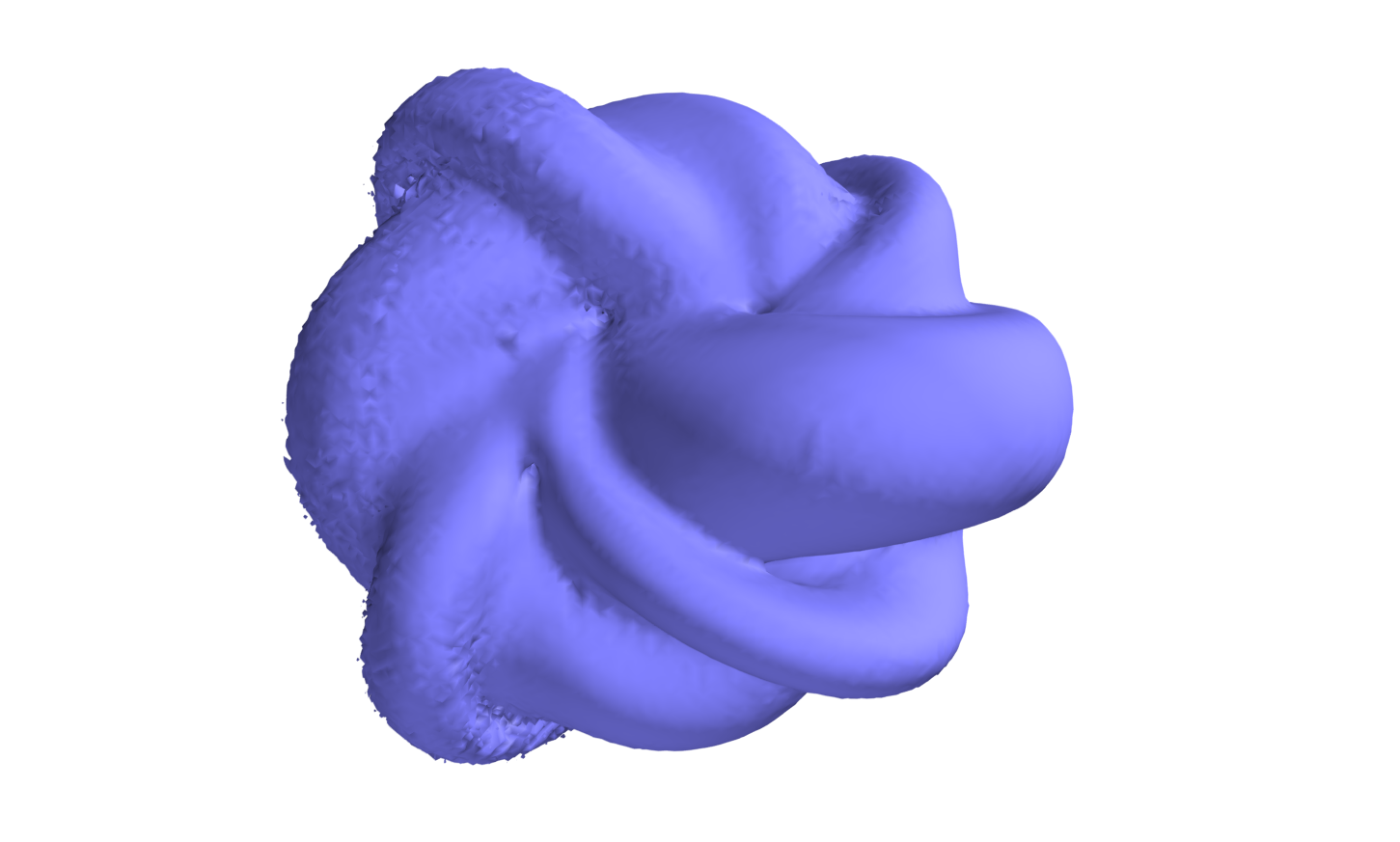} & \includegraphics[height=2cm]{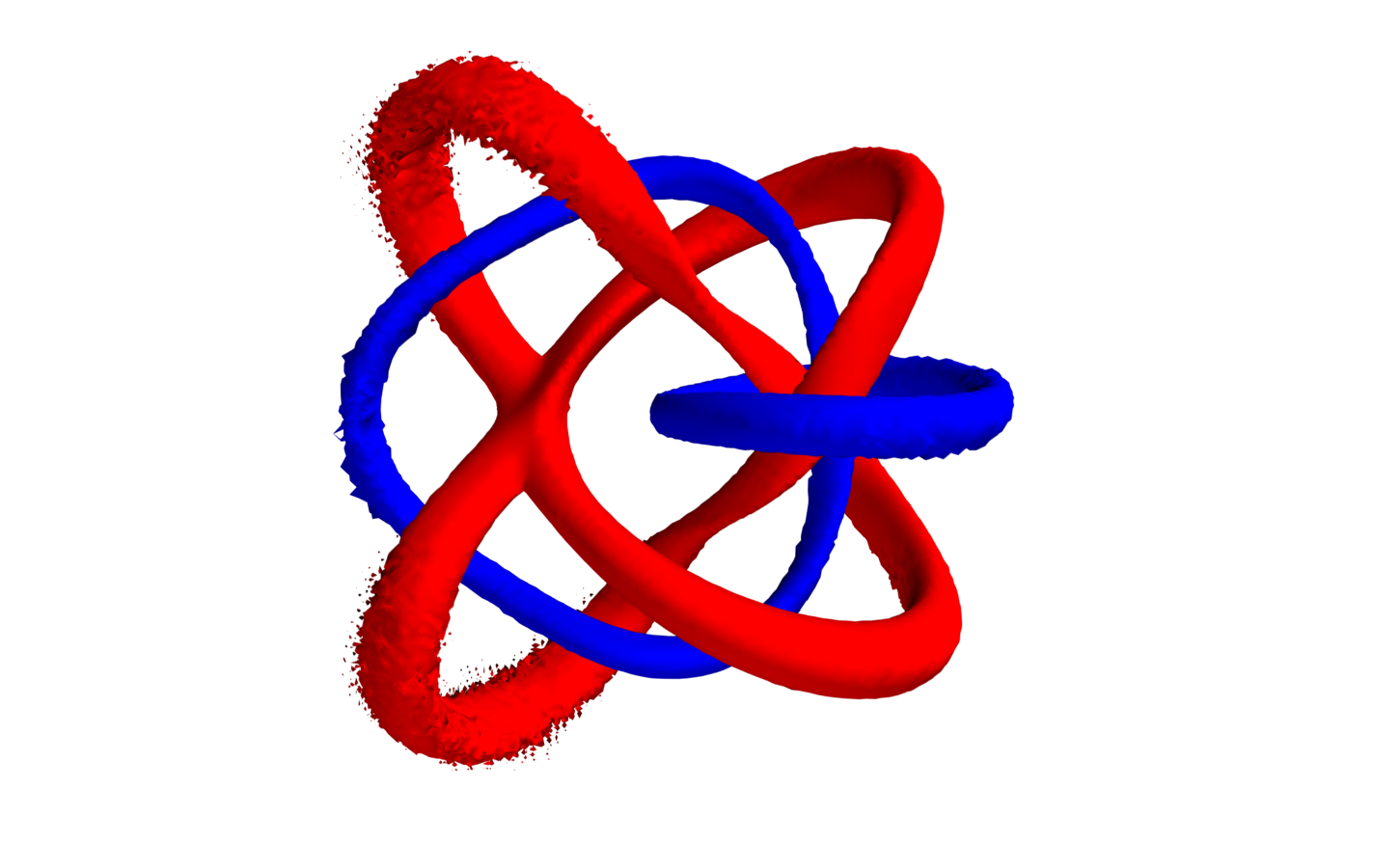}&\includegraphics[height=2cm]{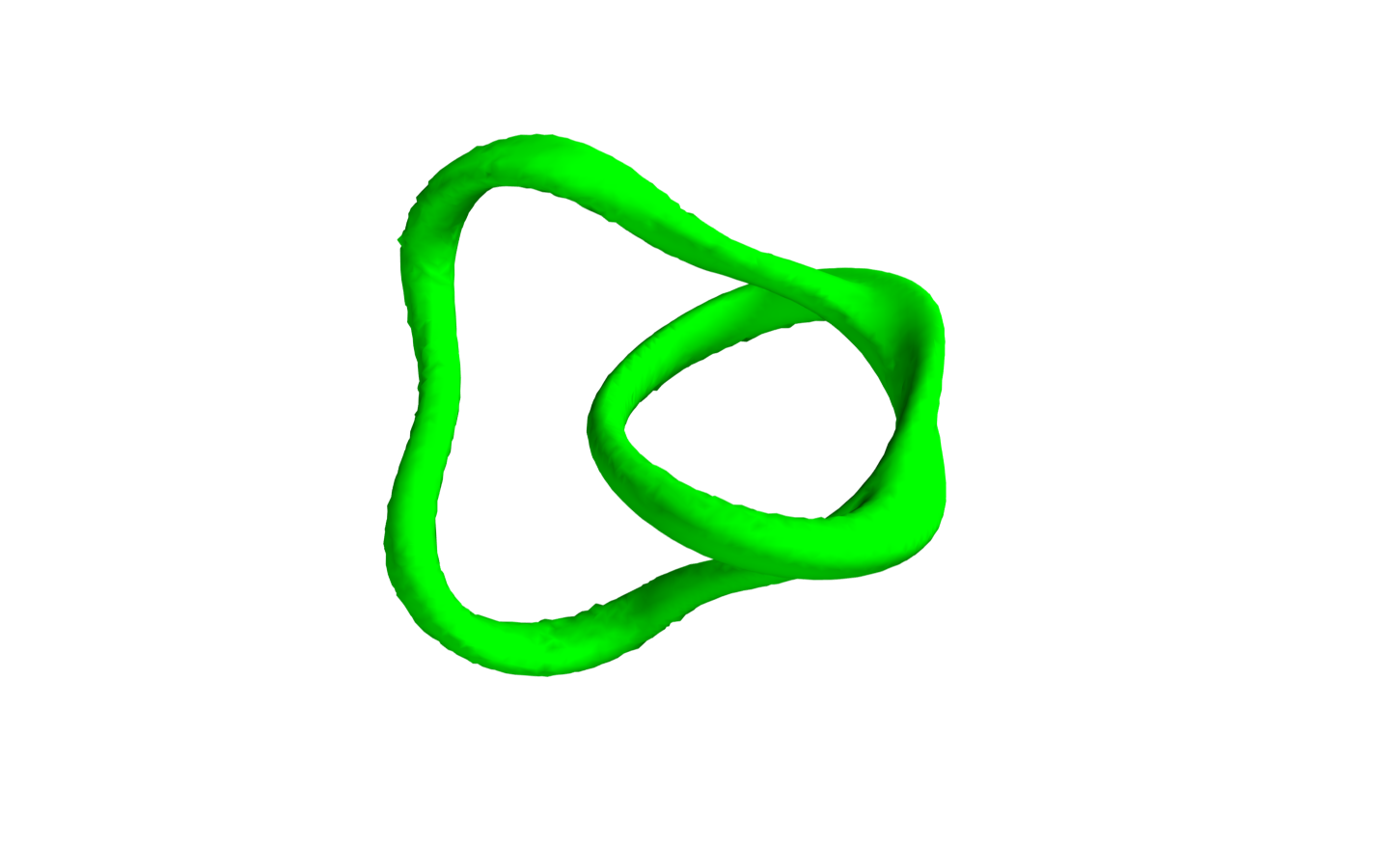} \\
            $6(\widetilde\CA_1\between\widetilde\CA_1)_{\widetilde\CA_{6,1}}$  &  \includegraphics[height=2cm]{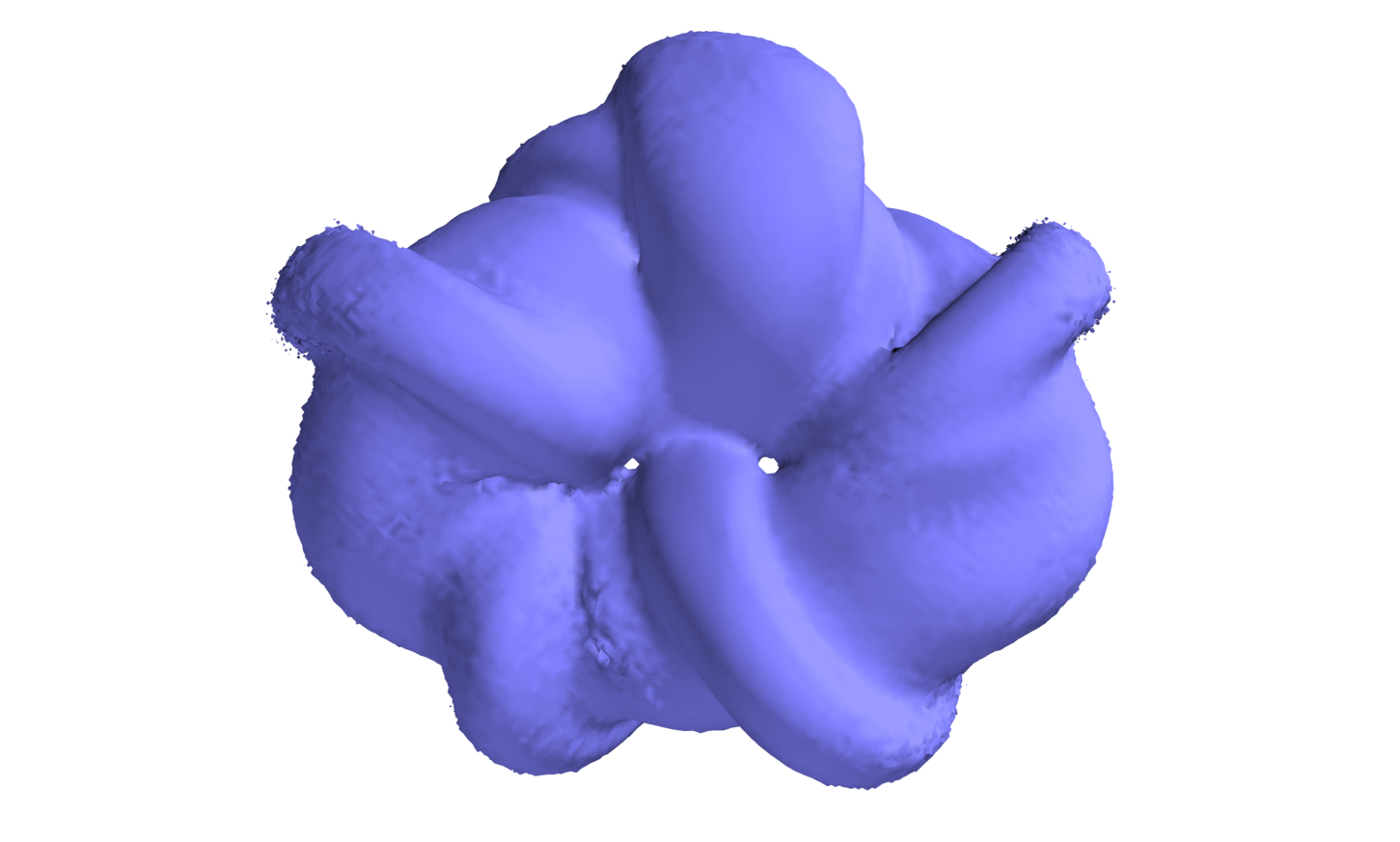} & \includegraphics[height=2cm]{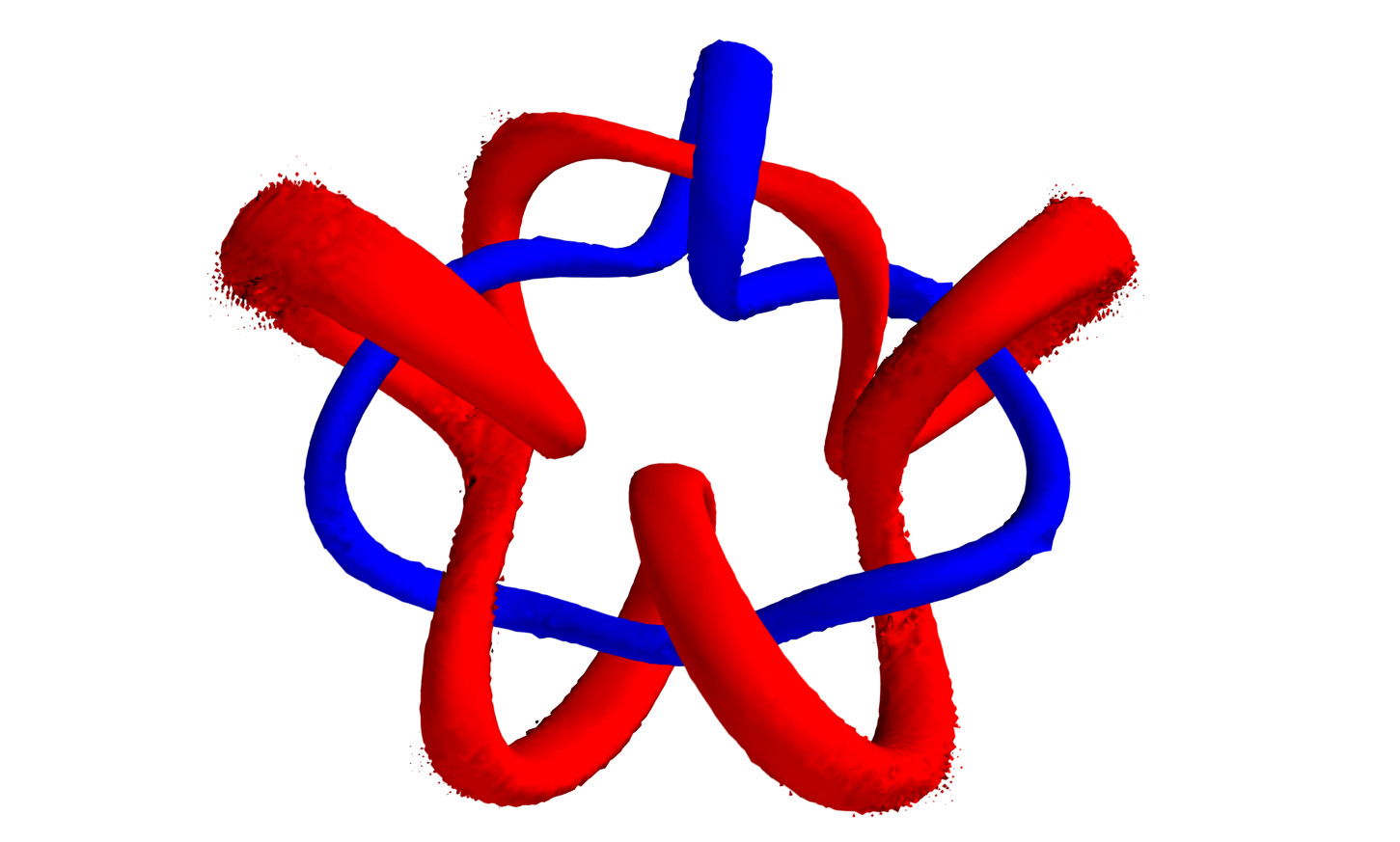} &\includegraphics[height=2cm]{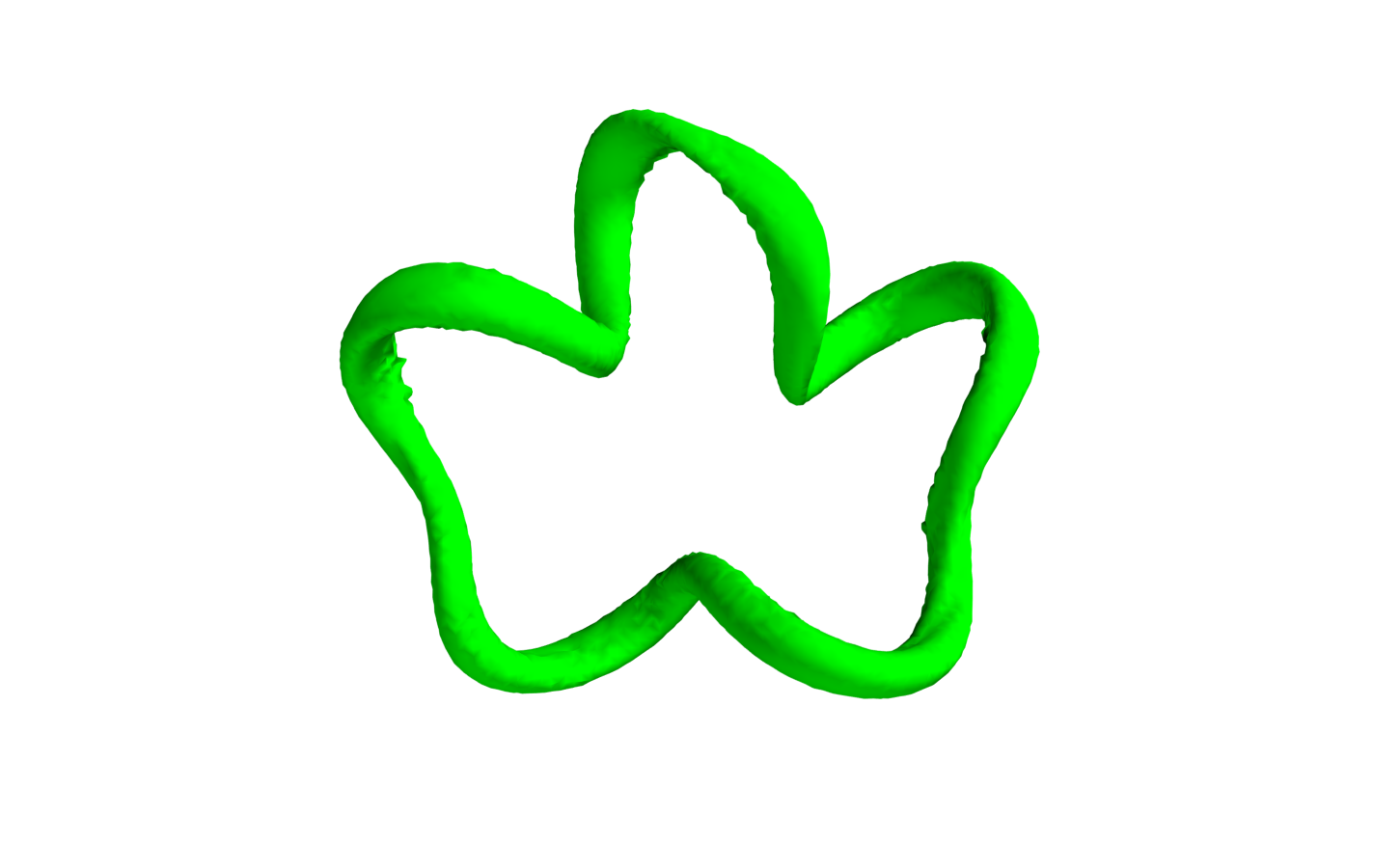} \\
            $7(\CK_{3,2}\between\CK_{3,2})_{\CK_{3,2}}$  &  \includegraphics[height=2cm]{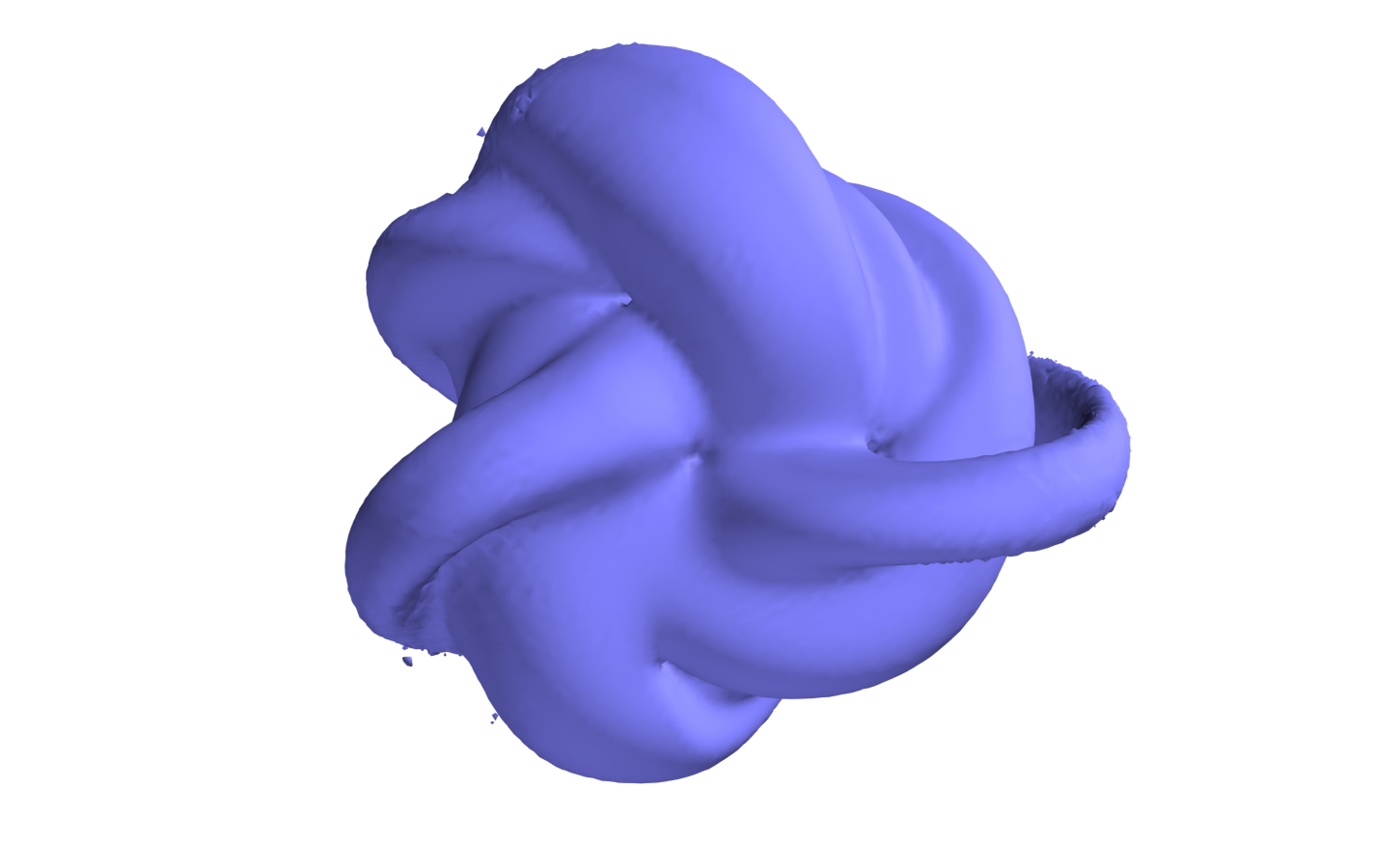}& \includegraphics[height=2cm]{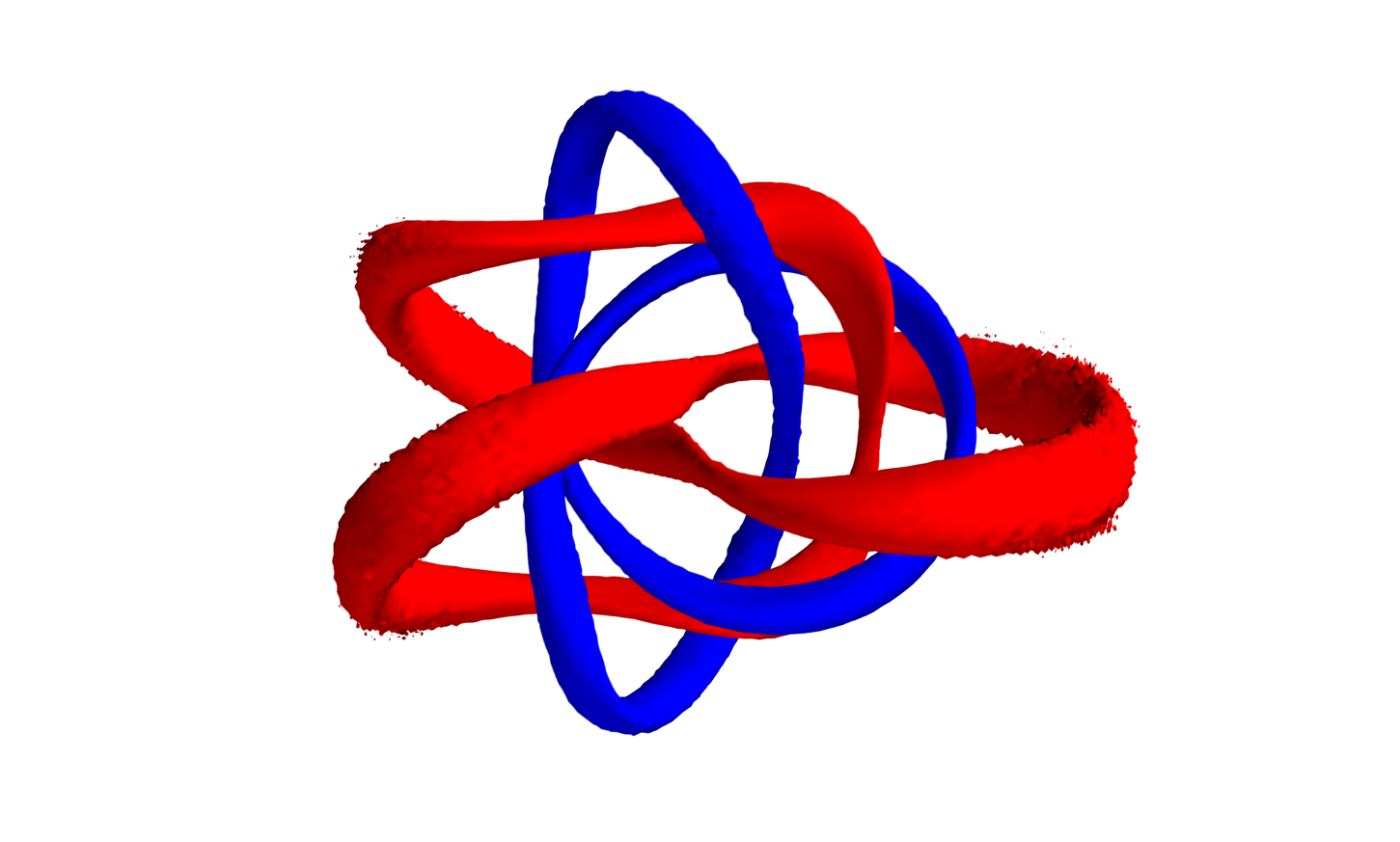}& \includegraphics[height=2cm]{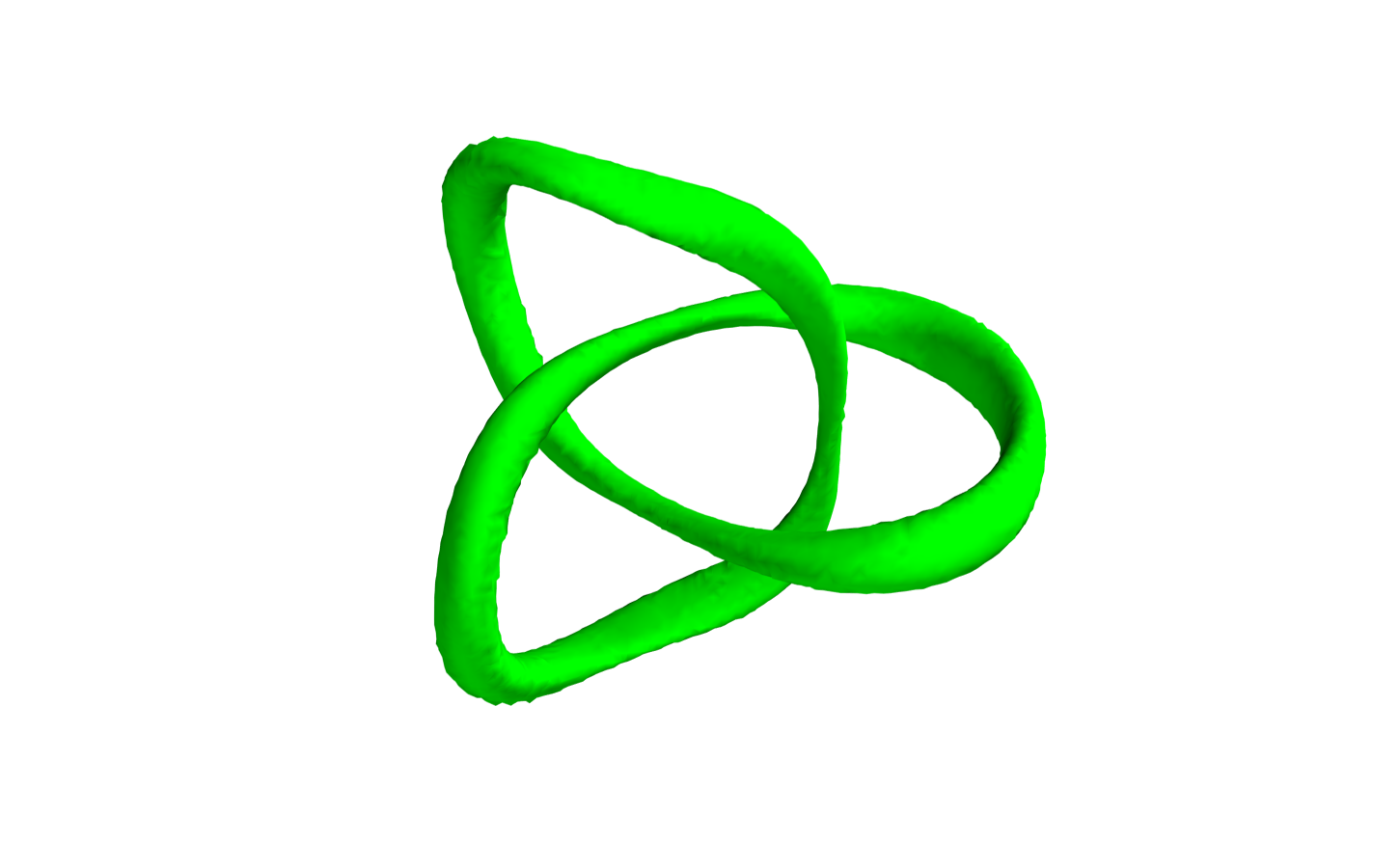} \\
            $7(\CK_{2,3}\between\CK_{2,3})_{\CK_{2,3}}$  &  \includegraphics[height=2cm]{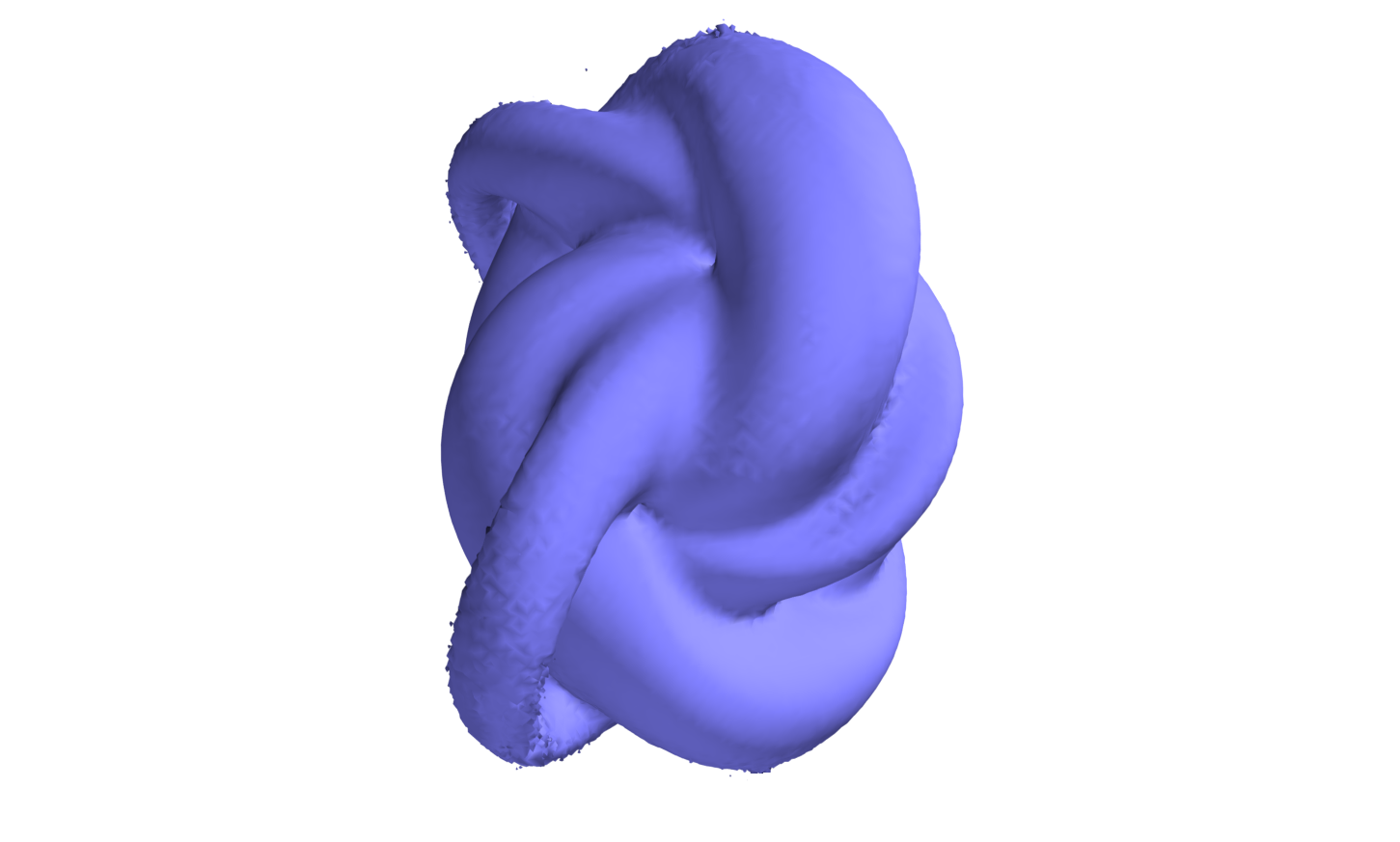}& \includegraphics[height=2cm]{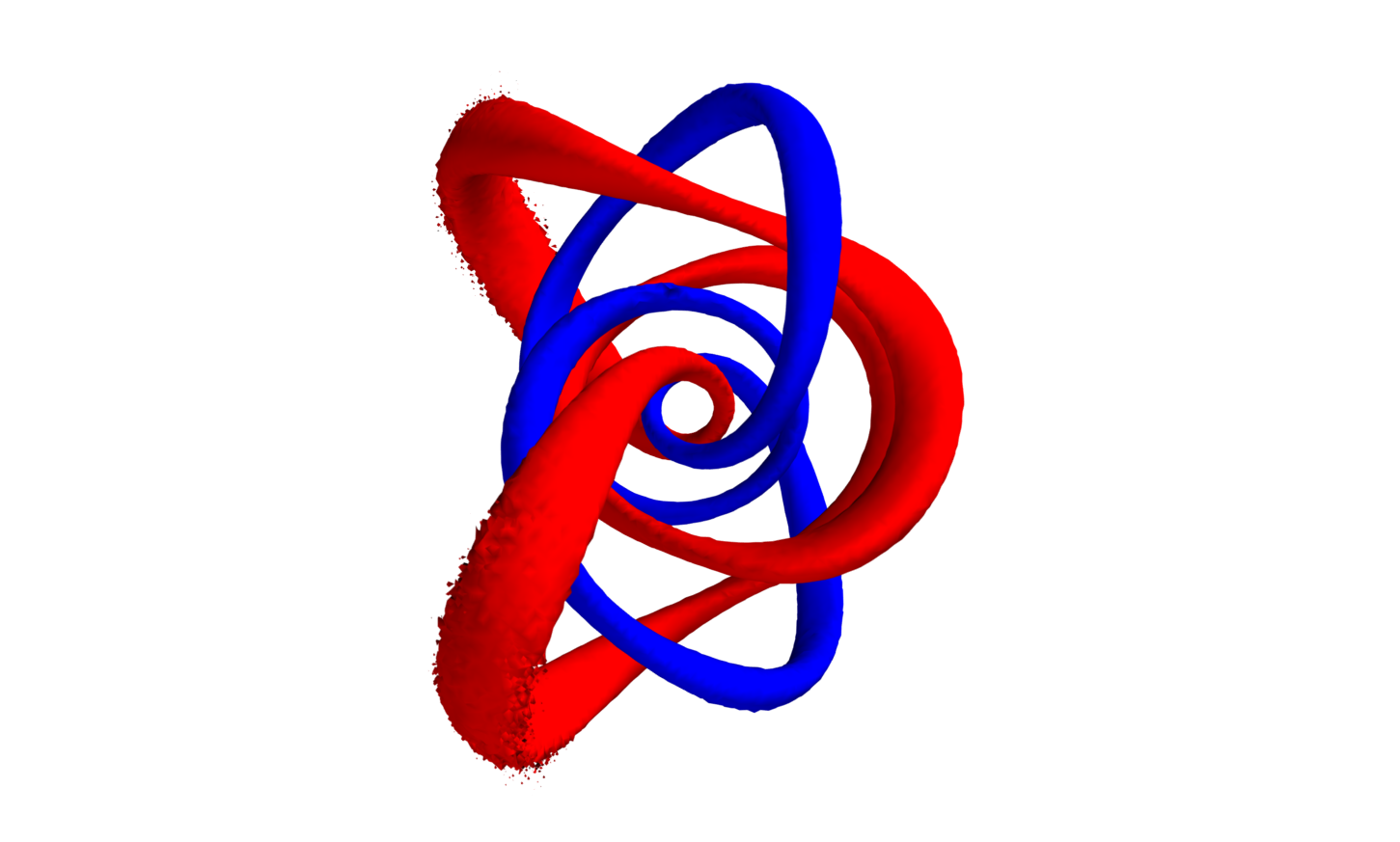}& \includegraphics[height=2cm]{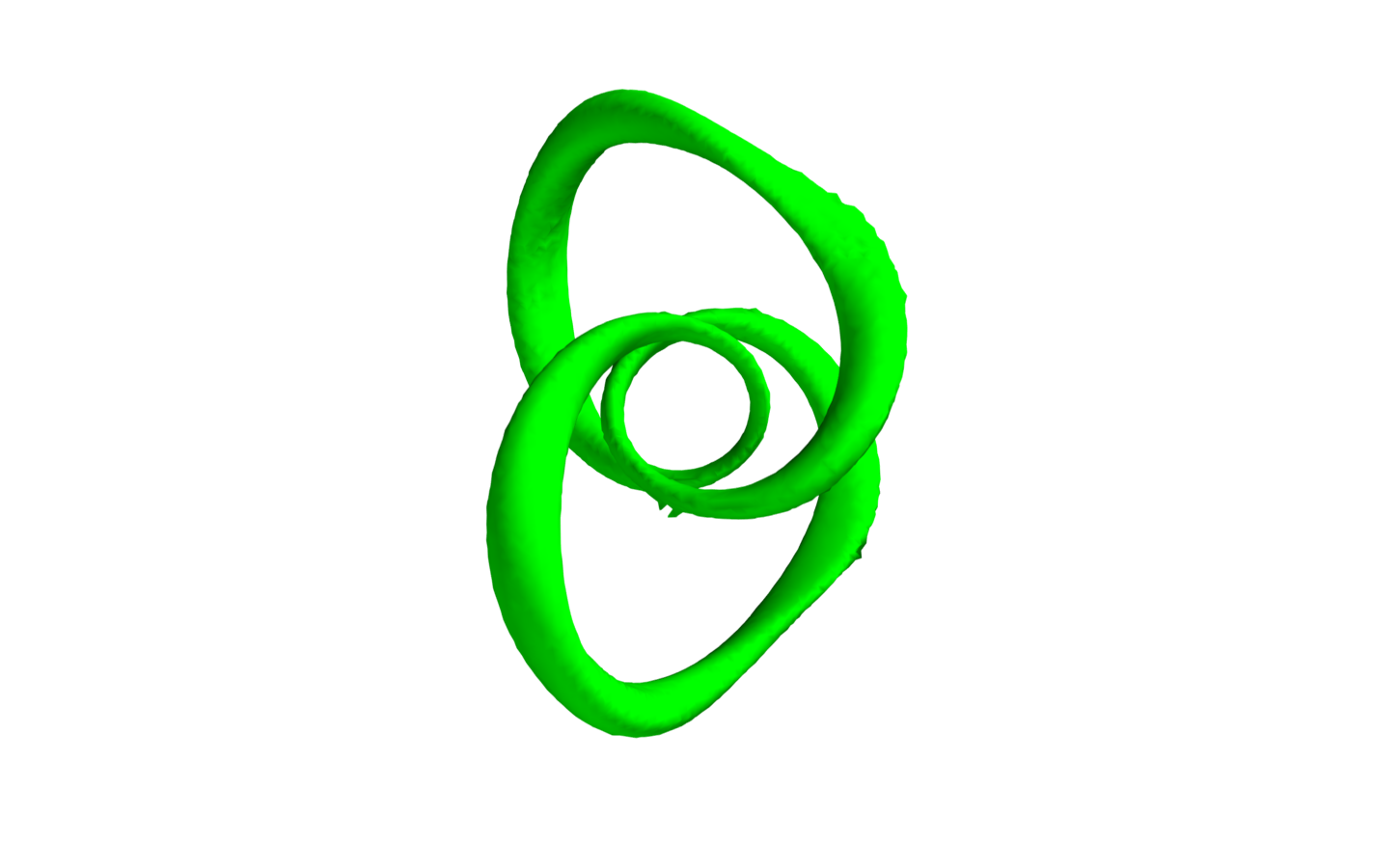} \\
            $7(\widetilde{\CA}_1\between\widetilde{\CA}_1)_{\widetilde\CA_{7,1}}$  &  \includegraphics[height=2cm]{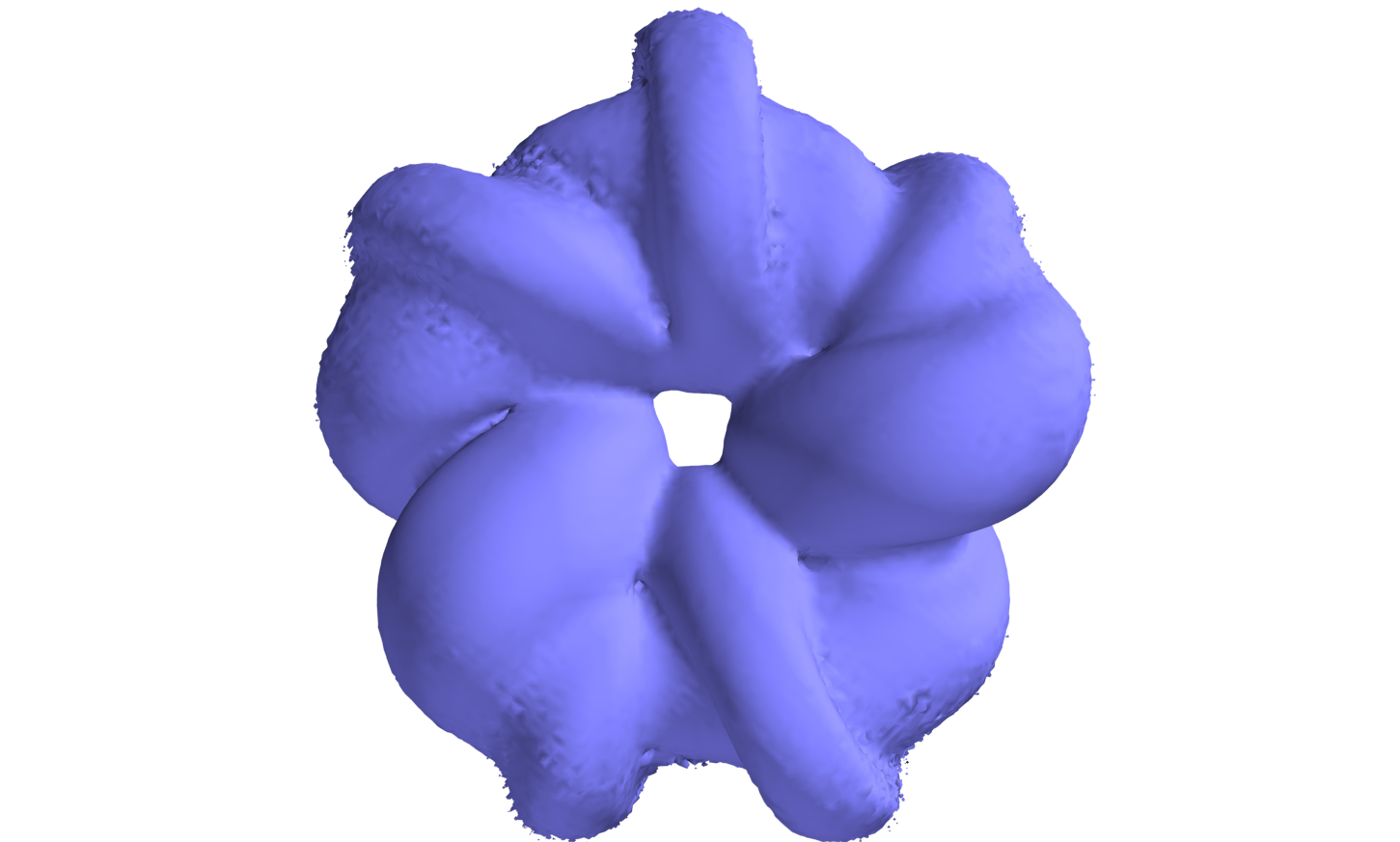}& \includegraphics[height=2cm]{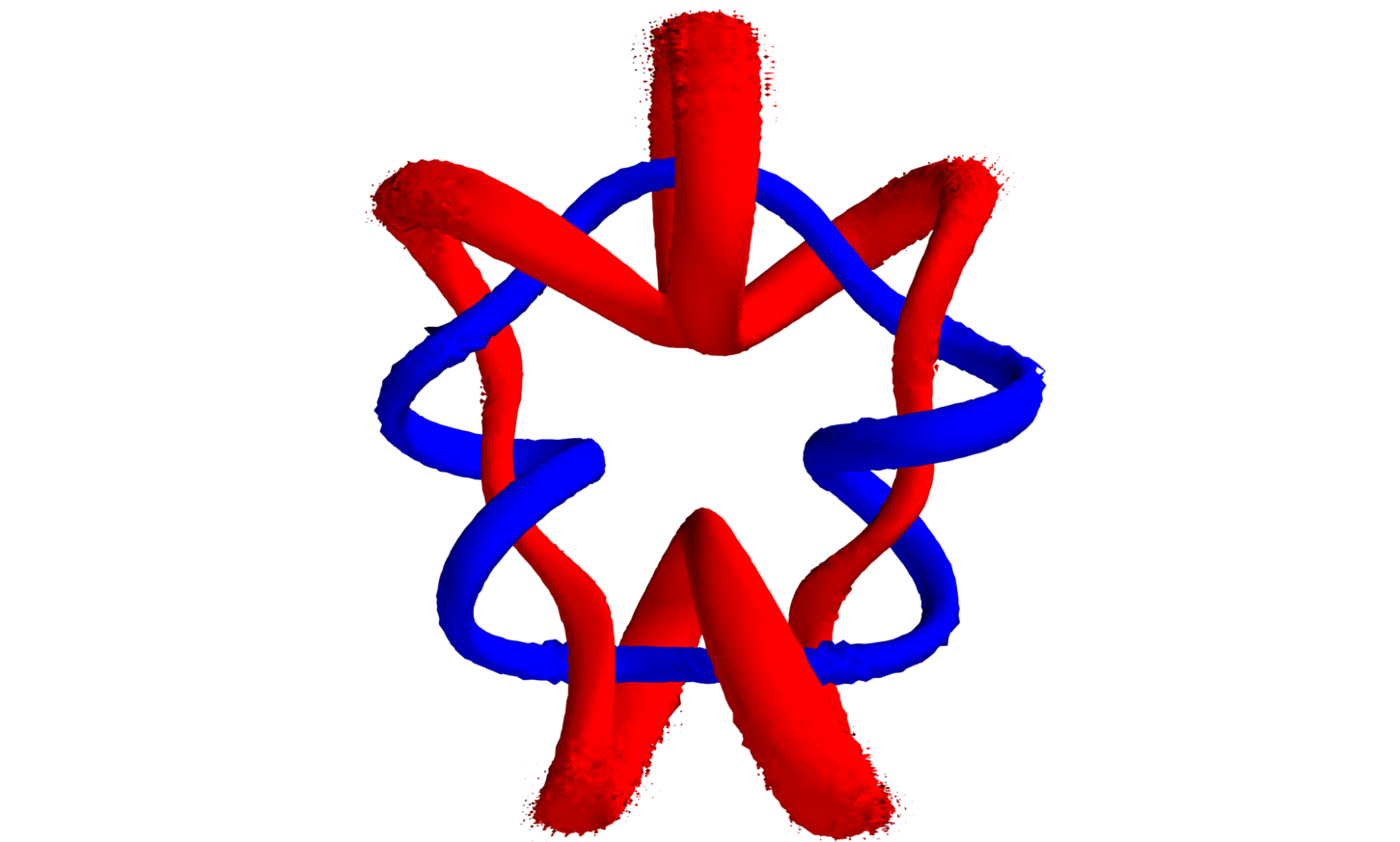}& \includegraphics[height=2cm]{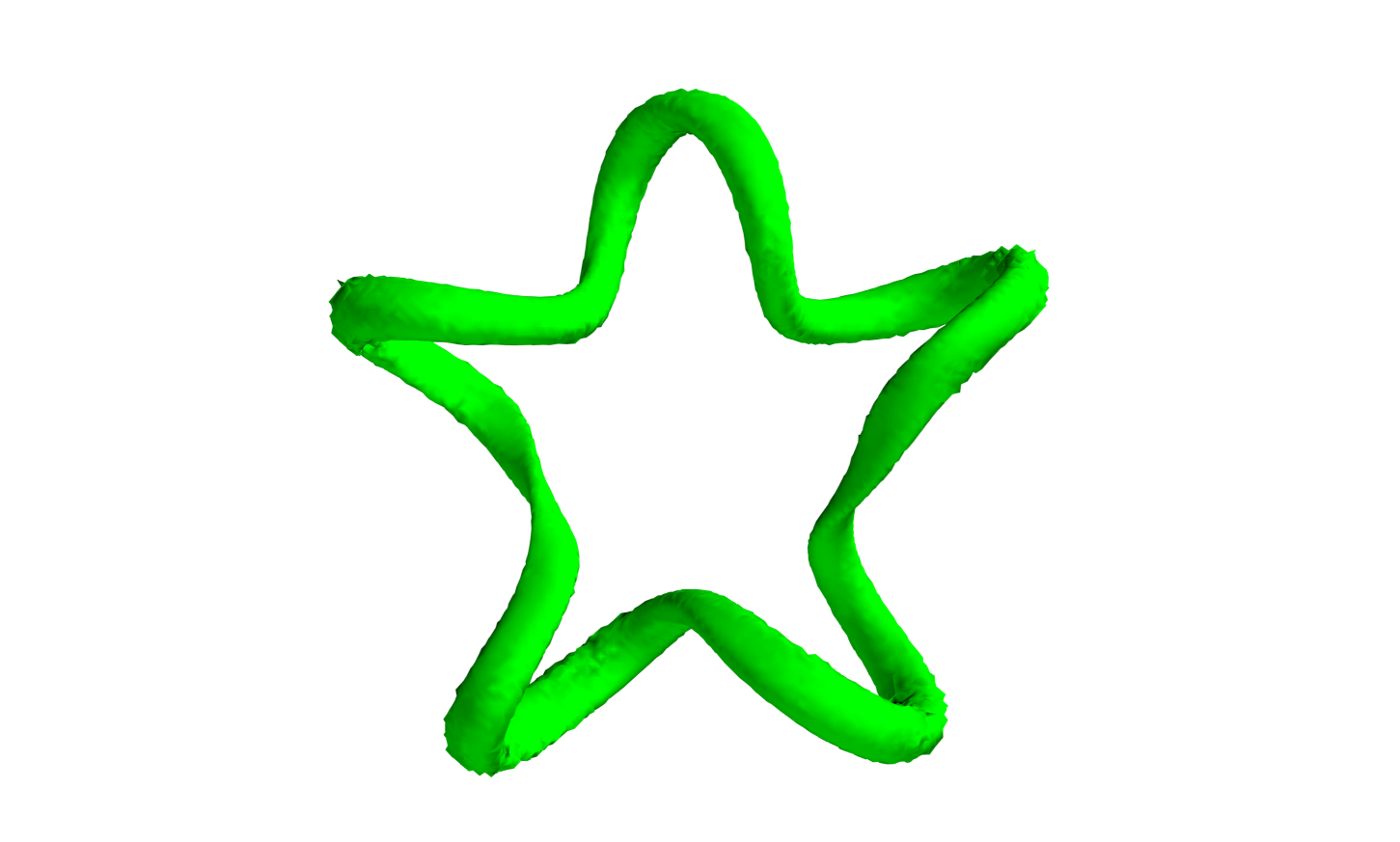} \\
        \end{TAB}
\end{center}
\caption{$\rho_E=15$ isosurface, $\phi_1=\pm0.9$ and $\phi_3=-0.9$ for $Q=1-7$ Hopfions in model with
$c_2=0.5, c_4=1, m=4$ and potential $V=m^2(\phi_1-1/3)^2$.}
\label{hopftable2}
\end{table}

\begin{table}
    \begin{center}
        \begin{tabular}{|c|c|c|c|c|}
            \hline
            & \multicolumn{2}{c|}{$V=m^2\phi_1^2$} & \multicolumn{2}{c|}{$V=m^2(\phi_1-1/3)^2$} \\
            \hline
            $Q$ &Configuration & $E$ & Configuration & $E$  \\
            \hline
            1 & $\CA_{1,1}\rightarrow (\CA_1\between\CA_1)_{\CA_{1,1}}$  & 1.38 & $\CA_{1,1}\rightarrow(\CA_1\between\CA_1)_{\CA_{1,1}}$ & 1.42 \\
            \hline
            2 & $\CA_{2,1}\rightarrow (\CA_1\between\CA_1)_{\CA_{2,1}}$  & 2.23 & $\CA_{2,1}, \CA_{1,2}\rightarrow
            (\CA_1\between\CA_1)_{\CA_{2,1}}$ & 2.26 \\
            & $\CA_{1,2}\rightarrow (\widetilde\CA_1\between\widetilde\CA_1)_{\CA_{1,2}}$  & 2.59 &  &  \\
            \hline
            3 & $\CA_{3,1}\rightarrow (\CA_1\between\CA_1)_{\CA_{3,1}}$  & 3.11 & $\CA_{3,1}\rightarrow
            (\CA_1\between\CA_1)_{\CA_{3,1}}$ & 3.13 \\
            & $\CK_{2,1}, \CA_{1,3}\rightarrow (\widetilde{\CA}_1\between\widetilde{\CA}_1)_{\widetilde\CA_{3,1}}$  & 3.11 &
            $\CK_{2,1}, \CA_{1,3}\rightarrow (\widetilde{\CA}_1\between\widetilde{\CA}_1)_{\widetilde\CA_{3,1}}$ & 3.14 \\
            \hline
            4 & $\CA_{2,2}, \CL_{1,1}, \widetilde\CA_{4,1}\rightarrow (\CL_{1,1}\between\CL_{1,1})_{\CA_{2,2}}$  & 3.92 & $\CK_{2,1}
            \rightarrow (\widetilde\CA_1\between\CL_{1,1})_{\widetilde\CA_{4,1}}$ & 3.98 \\
            & $\CK_{2,1}\rightarrow (\widetilde\CA_1\between\widetilde\CA_1)_{\widetilde\CA_{4,1}}$  & 3.97 & $\CA_{2,2}, \CL_{1,1}, \widetilde\CA_{4,1}\rightarrow (\CL_{1,1}\between\CL_{1,1})_{\CA_{2,2}}$ & 4.02  \\
            & $\CA_{4,1}\rightarrow (\CA_1\between\CA_1)_{\CA_{4,1}}$ & 4.02 & $\CA_{4,1}\rightarrow
            (\CA_1\between\CA_1)_{\CA_{4,1}}$ & 4.02 \\
            \hline
            5 & $\CL_{1,2}\rightarrow (\CL_{1,2}\between\CL_{1,2})_{\CL_{1,2}}$  & 4.69 & $\CL_{1,2}\rightarrow
            (\CL_{1,2}\between\CK_{3,2})_{\CL_{1,2}}$ & 4.74 \\
            & $\CA_{5,1}\rightarrow (\CA_1\between\CA_1)_{\widetilde\CA_{5,1}}$  & 4.94 & $\CA_{5,1}\rightarrow
            (\CA_1\between\CA_1)_{\widetilde\CA_{5,1}}$ & 4.91 \\
            \hline
            6 & $\CA_{3,2}, \CK_{3,2}, \CK_{4,2}, \CL_{2,2}\rightarrow
            (\CK_{3,2}\between\CK_{3,2})_{A_{3,2}}$  & 5.21 &
            $\CA_{3,2}, \CK_{3,2}, \CK_{4,2}, \CL_{2,2}\rightarrow (\CK_{3,2}\between\CK_{3,2})_{\CK_{3,2}}$ & 5.30 \\
            & $\CL_{1,3}\rightarrow (\CL_{1,3}\between\CK_{3,2})_{\CL_{1,3}^{1,1}}$  & 5.43 &
            $\CL_{1,3}\rightarrow (\CL_{1,3}\between\CK_{3,2})_{\CK_{3,2}}$ & 5.48 \\
            & $\CK_{5,1}\rightarrow (\widetilde\CA_1\between\widetilde\CA_1)_{\widetilde\CA_{6,1}}$  & 5.75 &
            $\CK_{5,1}\rightarrow (\widetilde\CA_1\between\widetilde\CA_1)_{\widetilde\CA_{6,1}}$ & 5.80 \\
            \hline
            7 & $\CK_{3,2}\rightarrow (\CL_{1,2}^{2,2}\between\CK_{3,2})_{\CK_{3,2}}$  & 5.91 &
            $\CK_{3,2}\rightarrow (\CK_{3,2}\between\CK_{3,2})_{\CK_{3,2}}$ & 6.04 \\
            & $\CK_{2,3}\rightarrow (\CK_{2,3}\between\CK_{2,3})_{\CK_{2,3}}$  & 6.20 &
            $\CK_{2,3}\rightarrow (\CK_{2,3}\between\CK_{2,3})_{\CK_{2,3}}$ & 6.34 \\
            & $\CK_{5,1}\rightarrow (\widetilde{\CA}_1\between\widetilde{\CA}_1)_{\widetilde\CA_{7,1}}$  & 6.65 &
            $\CK_{5,1}\rightarrow (\widetilde{\CA}_1\between\widetilde{\CA}_1)_{\widetilde\CA_{7,1}}$ & 6.72 \\
            \hline
        \end{tabular}
    \end{center}
    \caption{Initial guesses and final solutions together with the final rescaled energies of the Hopfions in the sectors
    of degrees $Q=1-7$.}
    \label{energytable}
\end{table}

As said above, the Hopfion configuration can be constructed from planar Skyrmions located in the plane transverse to
the direction of the loop in 3d \cite{Kobayashi:2013xoa}. It is known, however that the structure of the multisoliton solution of the
planar Skyrme model is very sensitive to the choice of the potential \cite{Weidig:1998ii,Leese:1989gj,Hen:2007in,Ward:2003un,Jaykka:2010bq}.
In particular, it was proposed to consider the potential \cite{Jaykka:2011ic,Winyard:2013ada}, which explicitly
breaks the $O(3)$ symmetry of the planar Skyrme model to the dihedral group $D_3$. In the baby Skyrme model with the potential
$V=m^2(\phi_1-c)^2$ a single
Skyrmion is splitted into two partons with the same topological charge $Q\simeq1/2$, as
$c=0$ and into constituents with fractional charges $Q=1/3, 2/3$, as $c=1/3$. Hence the appearance of two tubes in the energy density
distribution of the Faddeev-Skyrme model with the potential \re{pot-Nitta}, can be related with
decomposition of the planar solitons. Non-equal distribution of the topological charge between these lumps is reflected in different
thickness of the corresponding location curves $\mathcal{C}_1$ and $\mathcal{C}_{-1}$, see Table \ref{hopftable2}.

In the sector of degree $Q=3$ we also found another solution $3(\widetilde A_1\between \widetilde A_1)_{\widetilde A_1}$
(Table \ref{hopftable}). Since the energies of both solutions are very close, we are not able
to identify which configuration represents a global minimum, our algorithm suggests that the configuration with
hexagonal symmetry of the location curve of the preimage of the point $\vec \phi=(0, 0, -0.9)$ has slightly lower energy.

Considering the structure of the $Q=3$ Hopfion solutions in the \re{lagr} with the
potential \re{pot-Nitta}, we found that
the energy density isosurfaces again represents two buckled tubular loops with interlinking number three, see
Table \ref{hopftable2}.
Thus, as before, the location curves $\mathcal{C}_1 = \vec \phi^{-1}(1,0,0)$ and $\mathcal{C}_{-1} = \vec \phi^{-1}(-1,0,0)$
define the spacial distribution of the energy density.
Similar to the corresponding $Q=1,2$ solutions we discussed above, the energy is not equally distributed
among these tubes. This observation holds for all configurations in the sectors of higher degrees.

At degree four in the model \re{lagr} with potential $V=m^2{\phi_1}^2$ we found solution of three different types, see
Table \ref{hopftable}. The minimal energy configuration is of the type
$4(\CL_{1,1}\between\CL_{1,1})_{\widetilde \CA_{2,2}}$. The energy density isosurfaces of this very symmetric
Hopfion represent four tubular loops interlinked with each other, see the corresponding plot in the Table \ref{hopftable}.

The configuration of the form $4(\widetilde \CA_{1,1}\between \widetilde \CA_{1,1})_{\widetilde \CA_{4,1}}$ has less symmetry, it is a local minimum
of the energy functional. In this sector we also found  the  dihedrally symmetric Hopfion
$4(\CA_{1,1}\between \CA_{1,1})_{\CA_{4,1}}$, its energy is even higher. Similar to the corresponding solutions in the sectors
of lower degrees, for large values of the energy, the energy density isosurface forms a bead necklace with 8 beads and the bounding
planar curve $\mathcal{C} = \vec \phi^{-1}(0,0,-1)$  possessing dihedral symmetry $D_4$.

$Q=4$ Hopfions in the model  \re{lagr} with mixed potential \re{pot-Nitta} are of similar forms, see Table \ref{hopftable2}.
Surprisingly, we found that lowest energy configuration in this sector is
$4(\widetilde\CA_1\between\CL_{1,1})_{\widetilde\CA_{4,1}}$. More symmetric solution $4(\CA_1\between\CA_1)_{\CA_{4,1}}$ then
has almost the same energy as the configuration $4(\CL_{1,1}\between\CL_{1,1})_{\CA_{2,2}}$, so it is difficult to identify which state
is the first local minimum. Remind that the former configuration represents the global minimum energy solution in the massless
model \cite{Sutcliffe:2007ui}.

\subsubsection{Degrees 5,6 and 7}

The $Q=5$ minimal energy solution is $5(\CL_{1,2}\between\CL_{1,2})_{\CL_{1,2}}$, it is a link, which is similar
to the usual minimal energy charge five Hopfion in the massless Faddeev-Skyrme model. However, the structure of the energy density
of this solution in the model with Heisenberg type potential is different, one can clearly identify four linked tubes, see
Table \ref{hopftable}. Confirming the general pattern we also found the dihedrally symmetric configuration
$5(\CA_1\between\CA_1)_{\widetilde\CA_{5,1}}$ with energy about $5 \%$ above the global minimum.

As expected, variations of the parameter $\alpha$ in the potential \re{potential} affects the solutions. In the model with
the potential $V=m^2(\phi_1-1/3)^2$  the ground state configurations is of very unusual type, the location curves
$\mathcal{C}_1 = \vec \phi^{-1}(1,0,0)$ and $\mathcal{C}_{-1} = \vec \phi^{-1}(-1,0,0)$ correspond to the knot $\CK_{3,2}$ and to the
link $\CL_{1,2}$, respectively, see Table \ref{hopftable2}. At the same time, the curve $\mathcal{C} = \vec \phi^{-1}(0,0,-1)$
is a link of two bend tubes $\CL_{1,2}$, thus this new Hopfion with mixed types of location curves can be labeled as
$5(\CL_{1,2}\between\CK_{3,2})_{\CL_{1,2}}$. The local minimum, which we found in this sector, represent the deformation
of the $D_5$ symmetric configuration $5(\CA_1\between\CA_1)_{\widetilde\CA_{5,1}}$.

Considering the Faddeev-Skyrme model with the Heisenberg type potential at $m=4$, we found that in the
sector of degree six the minimal energy has the
double trefoil knot configuration  $6(\CK_{3,2}\between\CK_{3,2})_{A_{3,2}}$.
This is the solution with dihedral
symmetry, see the corresponding plots in Table \ref{hopftable}.
Recall that in the massless Faddeev-Skyrme model the link $\CL_{2,2}^{1,1}$ is the minimal energy charge six solution
\cite{Sutcliffe:2007ui} whereas in the presence of the usual mass term the global minimum is the axially symmetric
Hopfion $A_{3,2}$ \cite{Battye:2013xf,Harland:2013uk}.

Making use of variety of initial configurations in the model with the potential
$V=m^2{\phi_1}^2$, we found two other solutions, which represent local minima. The solution
$6(\CL_{1,3}\between\CK_{3,2})_{\CL_{1,3}^{1,1}}$ yields another example of mixed type configuration, here the
location curves $\mathcal{C}_1$  and $\mathcal{C}_{-1}$ form the trefoil knot and the link, respectively, see
Table \ref{hopftable}. The curve  $\mathcal{C}$ in this case is the usual link $\CL_{1,3}^{1,1}$, which corresponds to
a local minimum in the model with the usual pion-mass type potential. The bent configuration
$6(\widetilde\CA_1\between\widetilde\CA_1)_{\widetilde\CA_{6,1}}$ has even higher energy.

Last solution also exist as a local minimum in the model with potential $V=m^2(\phi_1-1/3)^2$, see
Table \ref{hopftable2}. But global minimum now is represented by trefoil knot Hopfion
$6(\CK_{3,2}\between\CK_{3,2})_{\CK_{3,2}}$, the next by energy configuration is
$6(\CL_{1,3}\between\CK_{3,2})_{\CK_{3,2}}$, which differs from the corresponding $c=0$ solution
by configuration of the position curve. We should mention that for these solutions, curves of the types
$\CK_{3,2}, \CL_{1,3}, \CL_{1,2}, \CA_{3,2}$  are looking very similar to each other, they are
almost indistinguishable in some cases, moreover they possess almost the same energy, so we cannot exclude a possibility that further
more precise numerical minimization scheme may reduce the number of local minima.

Finally, in the sector of degree seven we identify  $7(\CL_{1,2}^{2,2}\between\CK_{3,2})_{\CK_{3,2}}$
as the lowest energy solution in the model with the potential $V=m^2{\phi_1}^2$ at $m=4$.
The curve $\mathcal{C}$, which in the massless limit corresponds to the position curve, has the form of the
trefoil knot, which becomes less symmetric as the mass parameter $m$ increases.
Another double knot
configuration $(\CK_{2,3}\between\CK_{2,3})_{\CK_{2,3}}$ in that case represents a local minimum, the double bend
axial configuration $(\widetilde{\CA}_1\between\widetilde{\CA}_1)_{\widetilde\CA_{7,1}}$ is another metastable
local minimum, see Table \ref{hopftable}. Configurations of the last two types also exist in the model
with potential $V=m^2(\phi_1-1/3)^2$, however the global minimum now has the form of usual trefoil knot, see
Table \ref{hopftable2}.

\section{Compact solutions}
It is known that, according to the scaling arguments of the Derrick's theorem, stable
soliton solitons may exist in the truncated $c_2=0$ submodel of \re{lagr}, which contains only Skyrme term and a potential
\cite{Foster:2010zb,Adam-compact}. This submodel effectively corresponds to the infinite mass limit $m\to \infty$. Interestingly,
these Hopfions usually are solitons with compact support, the configuration
possess no exponentially decaying tail, instead the
field components attain the vacuum values outside some finite domain in space.

In our simulations we considered the $c_2=0$ submodel of \re{lagr} with the symmetry breaking potential \re{potential}.
We fix the scale by setting $m=4$. Since there are usually severe difficulties with the numerical simulations of the compactons,
we make use of the small $c_2$ parameter as a regulator in our numerical scheme.

\begin{figure}[h]
    \begin{center}
        \includegraphics[height=3cm]{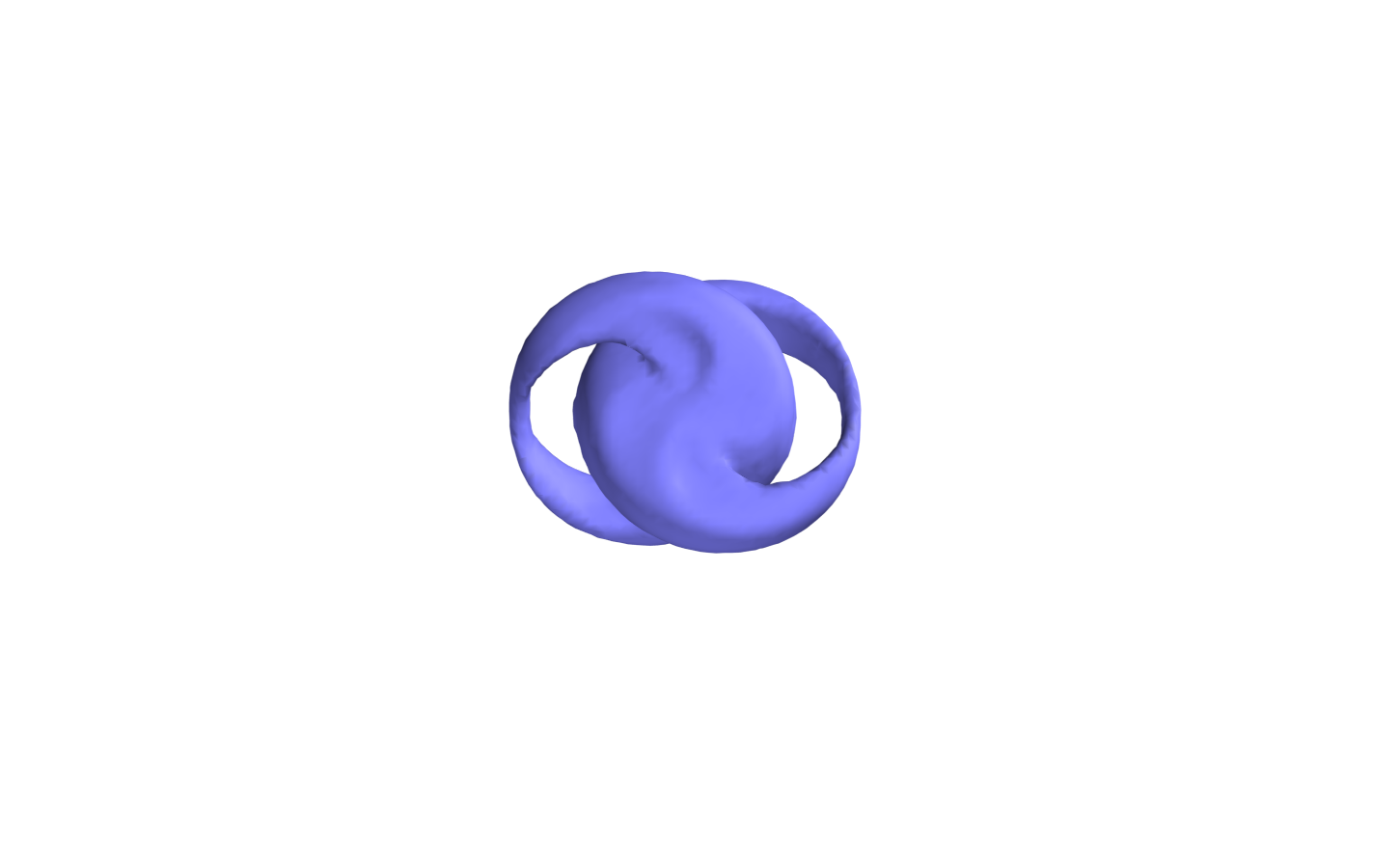}
        \includegraphics[height=3cm]{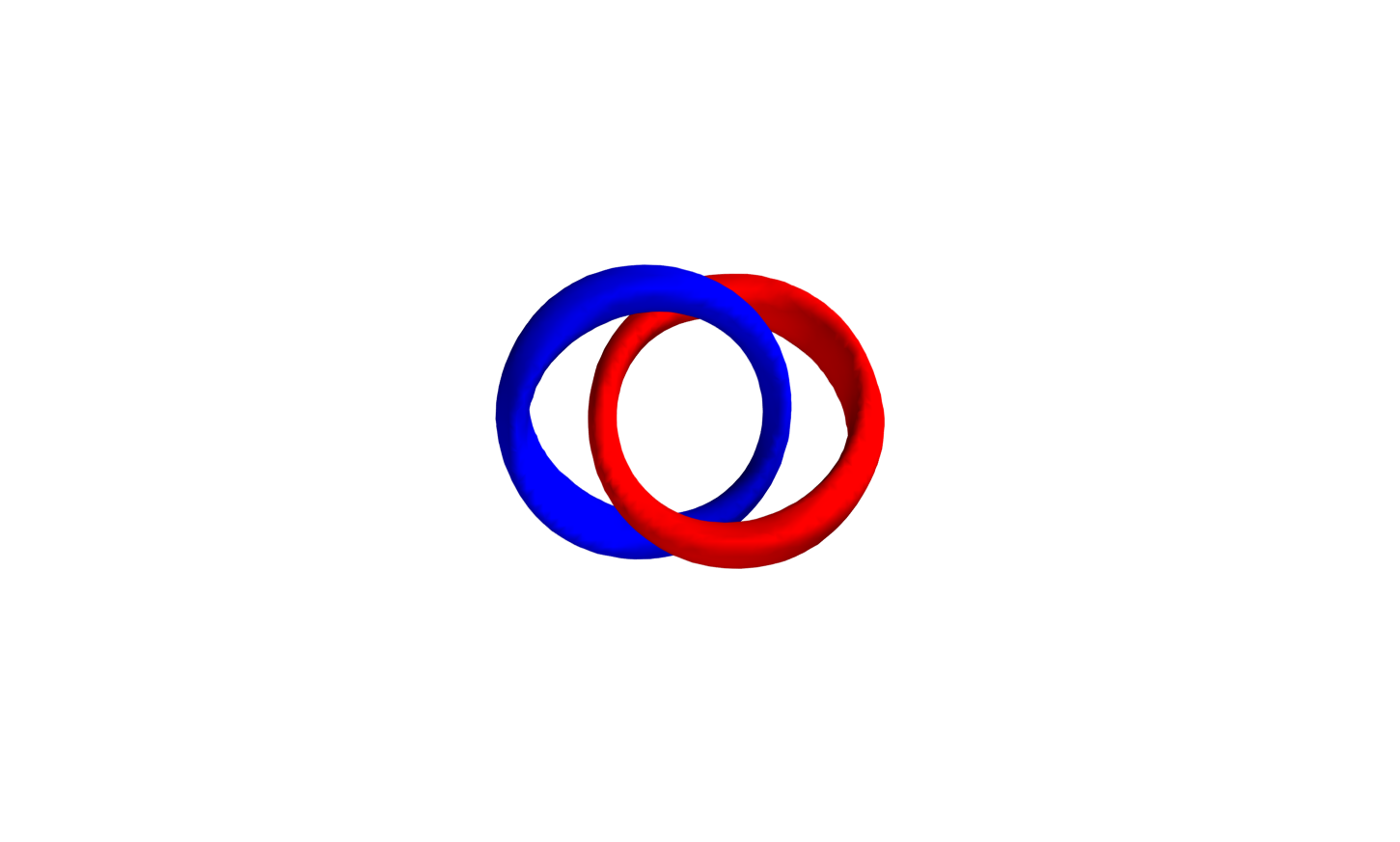}
        \includegraphics[height=3cm]{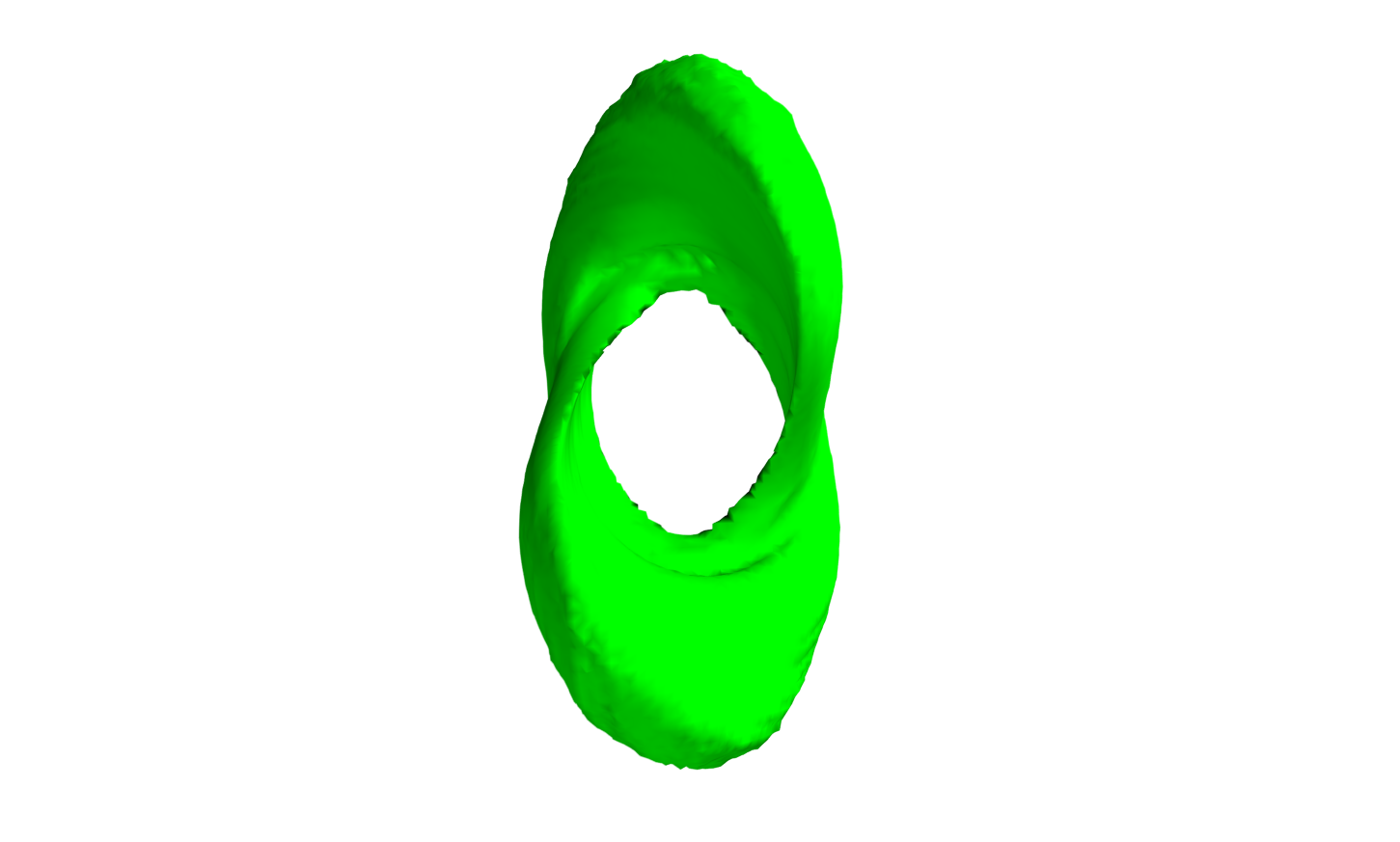}
    \end{center}
    \begin{center}
        \includegraphics[height=3cm]{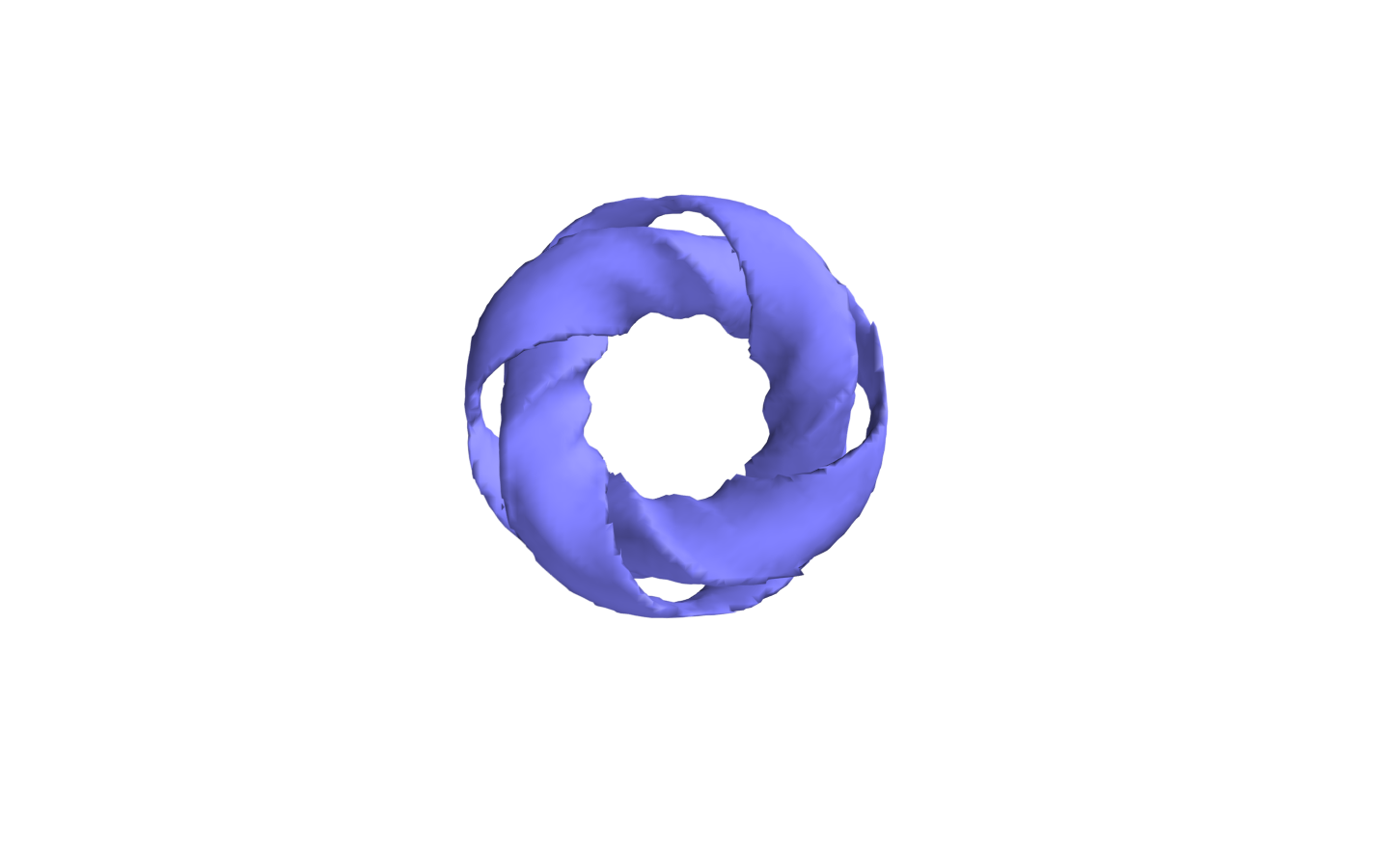}
        \includegraphics[height=3cm]{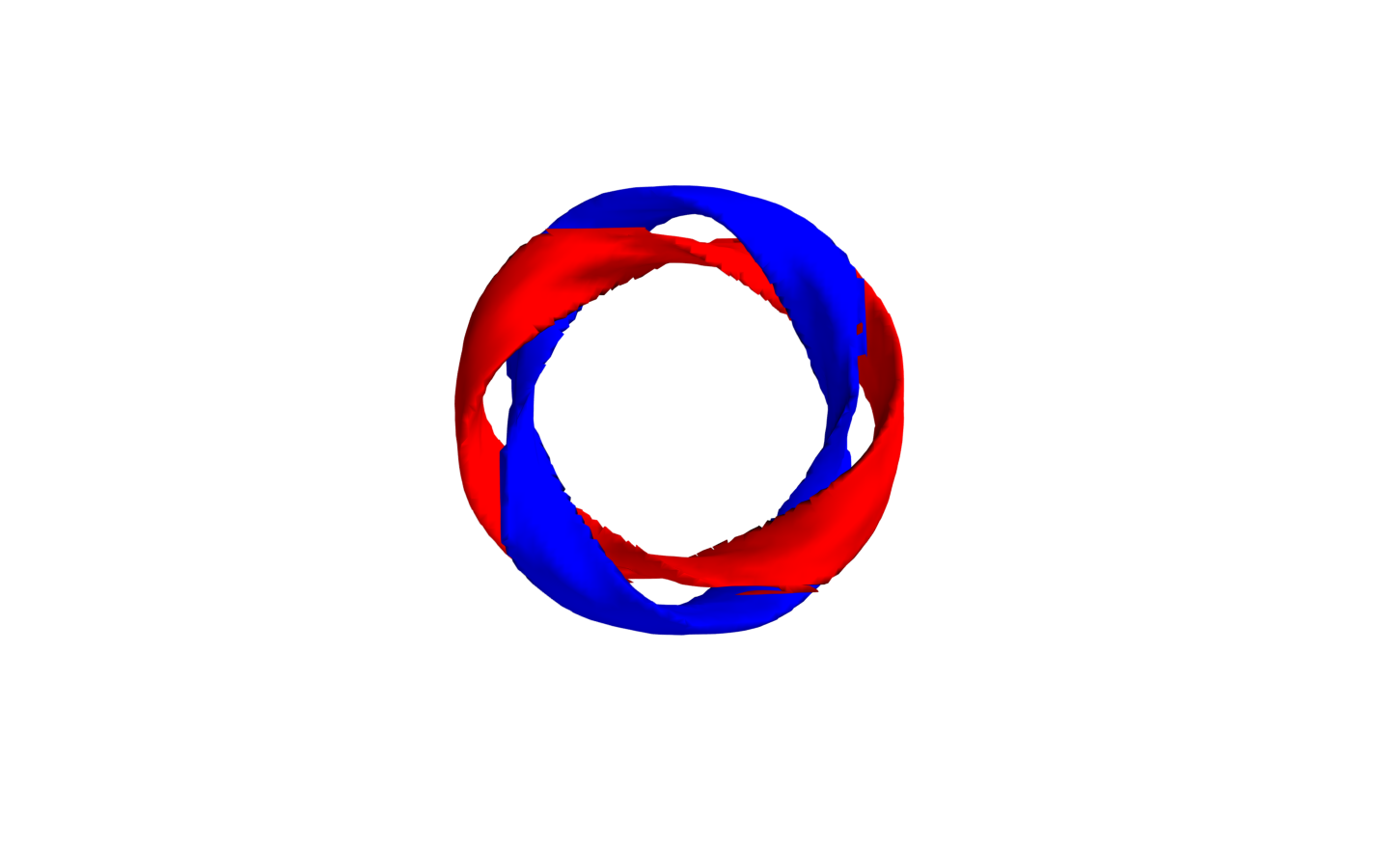}
        \includegraphics[height=3cm]{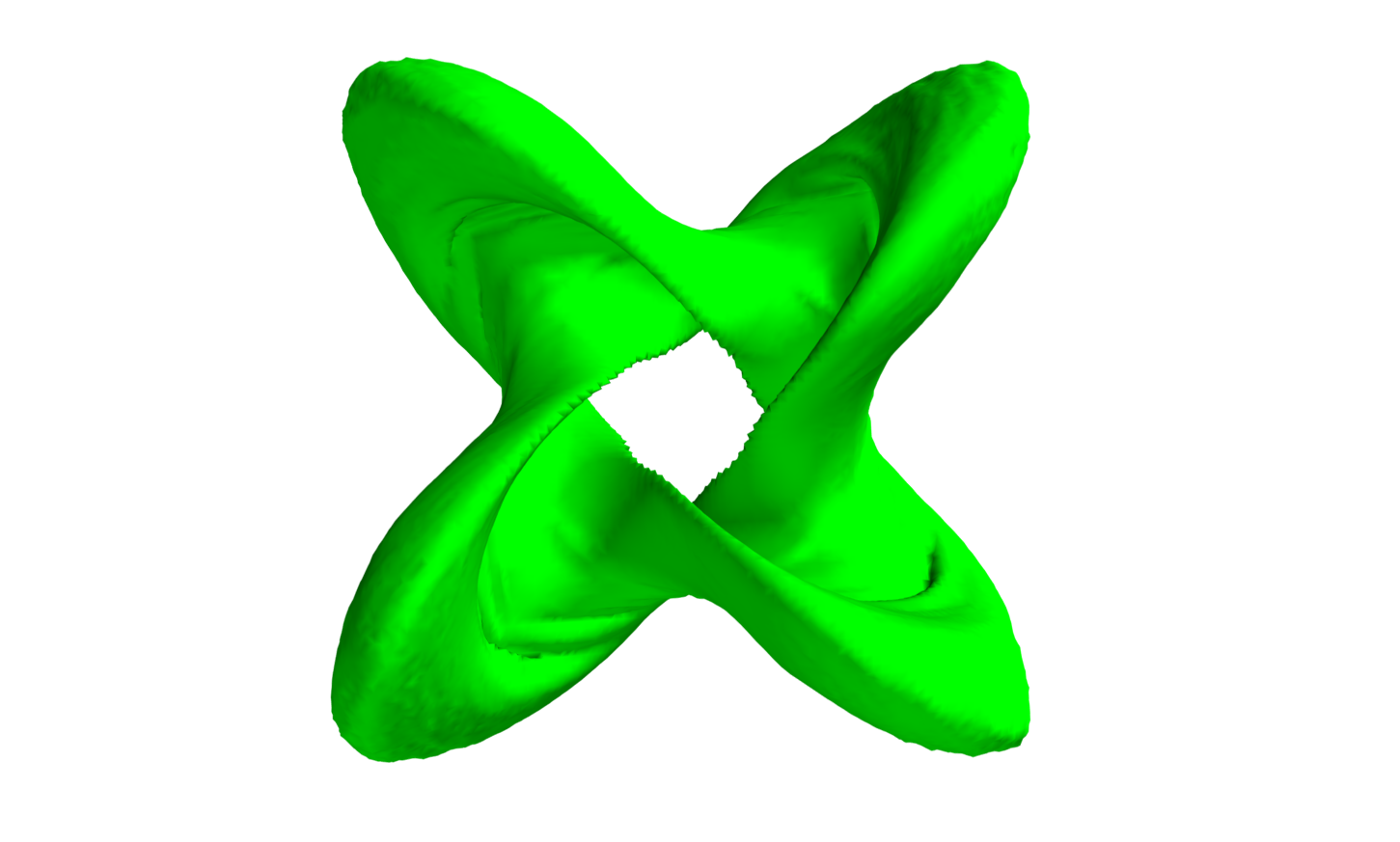}
    \end{center}
    \begin{center}
        \includegraphics[height=3cm]{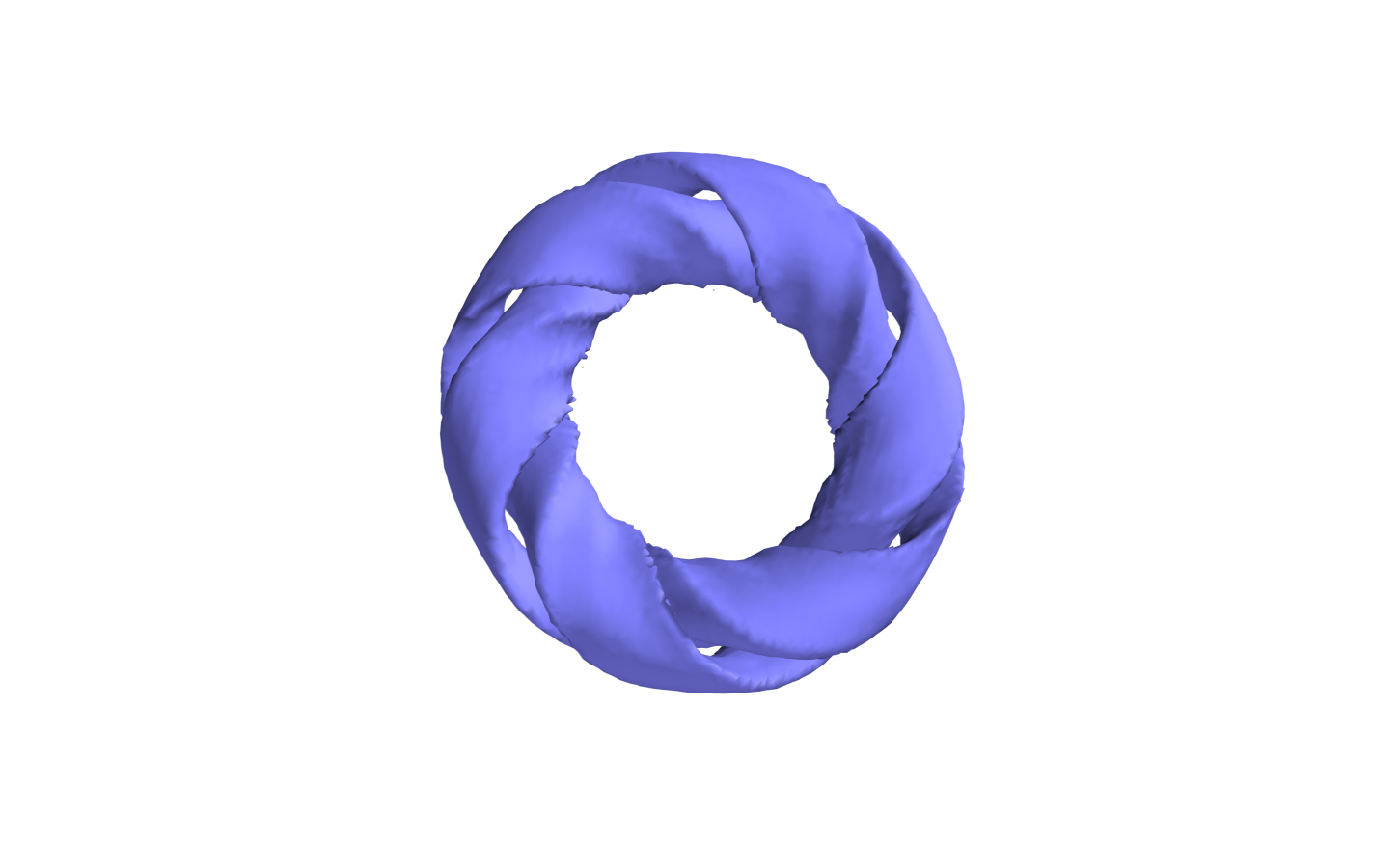}
        \includegraphics[height=3cm]{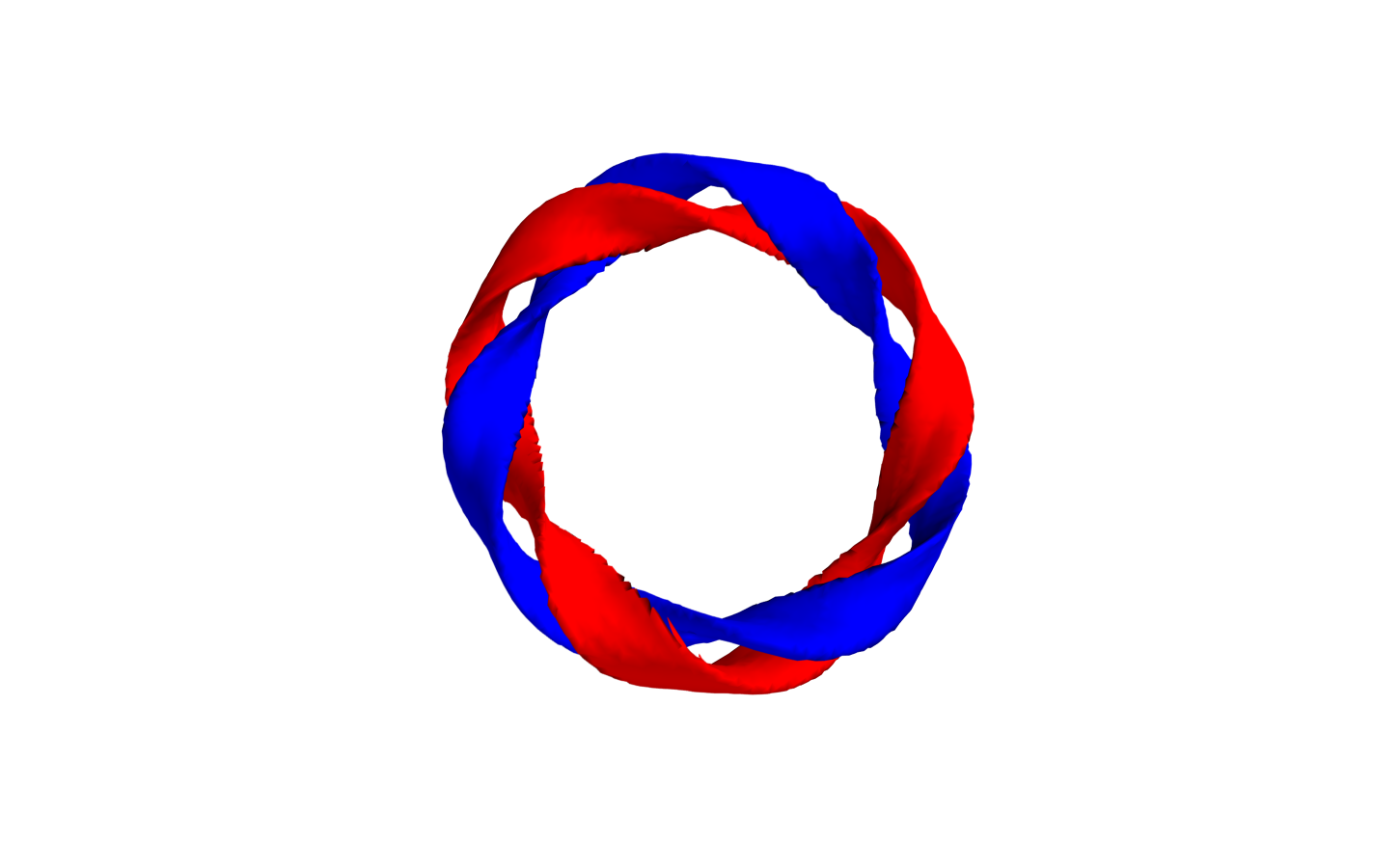}
        \includegraphics[height=3cm]{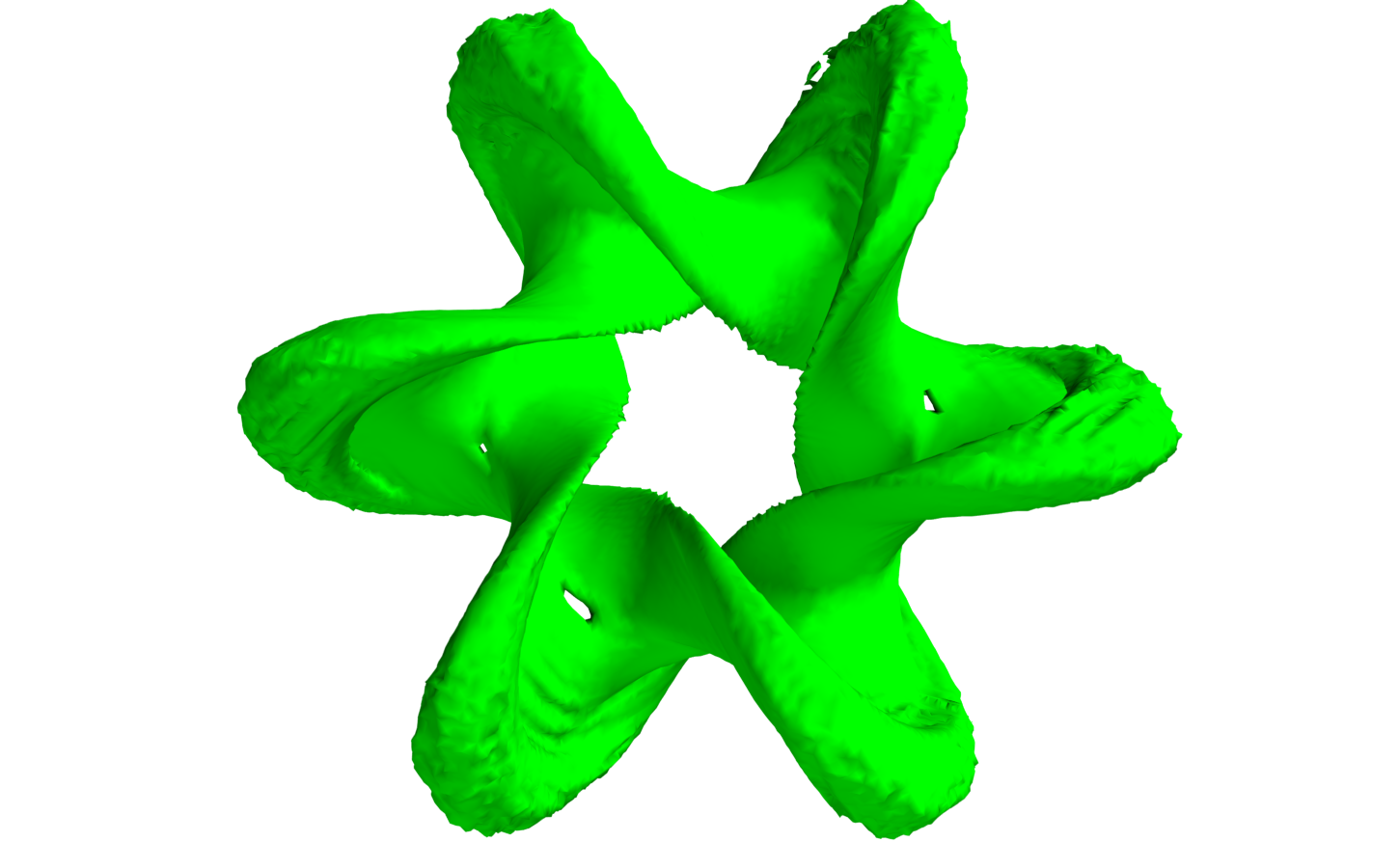}
    \end{center}
    \caption{Isosurfaces showing the energy density distributions $\mathcal{E}=40$ (first column),
    $\phi_1=\pm0.9$ (second column) and $\phi_3=-0.8$ of the $Q=1,2,3$ "compact" Hopfions
    (first, second and third row respectively) in the model \re{lagr} with potential $V=m^2\phi_1^2$ at $c_2=0$, $c_4=1$ and $m=4$.}
    \label{q1q2q3comp}
\end{figure}

In Fig.~\ref{q1q2q3comp} we present the solutions of this submodel with Heisenberg type symmetry breaking potential.
As in the case of the usual pion-mass type potential, these solurions are compactons.
Note that the interior domain, where the fields are located, is not spherically symmetric.

We found that the solutions of the $c_2=0$ submodel are similar to the corresponding minimal energy configurations in the full
Faddeev-Skyrme with potential \re{potential}. On the other hand, the planar Skyrmions in the cross sections of such
compact Hopfions have more prominent dihedral symmetry.
The spacial distribution of the energy
density possess much "sharper" structure, as expected.

As seen in the second column of Fig.~\ref{q1q2q3comp}, the location curves
$\mathcal{C}_1 = \vec \phi^{-1}(1,0,0)$ and $\mathcal{C}_{-1} = \vec \phi^{-1}(-1,0,0)$
define the spacial distribution of the energy density again. The curve $\mathcal{C} = \vec \phi^{-1}(0,0,-1)$
of these Hopfions becomes very geometrical, see the third column of Fig.~\ref{q1q2q3comp}. Note that
$\phi_2, \phi_3$ components of the solutions of the $c_2=0$ submodel have no exponentially decaying tail.

\section{Summary}
The main purpose of this work was to construct  Hopfion solutions of the Faddeev-Skyrme model with symmetry breaking potential
$V=m^2\left(\phi_1\sin\alpha-(1-\phi_3)\cos\alpha\right)^2$, which interpolates between the usual form of the
"pion mass" potential as $\alpha=0$ and the Heisenberg type potential as $\alpha = \pi/2$, hence the vacuum of the model
is a circle $S^1$. Further, we found that, in contrast to the usual massive Faddeev-Skyrme model \cite{Foster:2010zb},
increase of the mass parameter in the model with Heisenberg type potential, makes the energy density distribution more extended,
although it still localizes the fields.

We show that in the general case, the structure of the solutions  is defined by the
location curves $\mathcal{C}_1 = \vec \phi^{-1}(1,0,0)$ and $\mathcal{C}_{-1} = \vec \phi^{-1}(-1,0,0)$, which
correspond to the maximum of the potential. The Hopf index then is the
linking number of these curves.

The usual classification
scheme of the Hopfions, related to the possible types of the curve $\mathcal{C} = \vec \phi^{-1}(0,0,-1)$,
which is the antipodal to the vacuum, becomes less useful in such a case, the spacial distribution of the energy density follows the
location curves $\mathcal{C}_1$ and $\mathcal{C}_{-1}$, which form various structures of two linked knots and links.
Relative thickness of the tubes plotted around these location curves depends on the value of the angular parameter $\alpha$, both
tubes are of the same thickness as $\alpha = \pi/2$, as $\alpha$ decreases, one of the tubes becomes thinner, it disappears as
$\alpha=0$. The remaining tube  $\mathcal{C}_{-1} = \vec \phi^{-1}(-1,0,0)$ written
in the internally rotated field components, which appear in the potential \re{potential}, coincides with the
position curve $\mathcal{C} = \vec \phi^{-1}(0,0,-1)$.
Further, the energy density isosurfaces of the  Hopfion configurations in the model with Heisenberg type potential
may possess dihedral symmetry. In this model, in the sectors of degrees $Q=5,6,7$
we found solutions of new type, for which one or both of these tubes represent
trefoil knots. Further, some of these solutions possess different types of curves  $\mathcal{C}_1$ and $\mathcal{C}_{-1}$.
Considering the $c_2=0$ submodel with Heisenberg type symmetry breaking potential,
we found new compacton solutions with dihedral symmetry.

Our final remark is that, since the Hopfion solution can be thought as an embedding of a two-dimensional
planar Skyrmion configuration as a slice of a circle in
three-dimensional space and consequent twisting of the configuration  along this loop,
the appearance of these tubes in the  model with symmetry breaking potential can be related with
decomposition of the planar solitons into two components with fractional topological charge.
Non-equal distribution of the topological charge between these lumps is reflected in different
thickness of the corresponding location curves.

\section*{Acknowledgements}
Y.S. gratefully
acknowledges support from the Russian Foundation for Basic Research
(Grant No. 16-52-12012), the Ministry of Education and Science
of Russian Federation, project No 3.1386.2017, and DFG (Grant LE 838/12-2).
The parallel computations were performed on the cluster HIBRILIT at LIT, JINR, Dubna.

\begin{small}

\end{small}


\end{document}